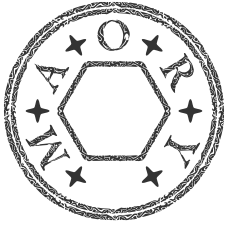

Programme: **E-ELT**

Project: **ELT MCAO Construction – MAORY**

# MAORY science cases white book

**Document Number:** E-MAO-000-INA-PLA-009

**Document Version:** 1

**Document Type:** PLA

**Released On:** 2017-09-01

| | | | |
|---|---|---|---|
| **Owner :** | Giuliana Fiorentino | | |
| **Approved by :** | Emiliano Diolaiti | | |
| **Released by :** | Paolo Ciliegi | | |
| | Name | Signature | Date |

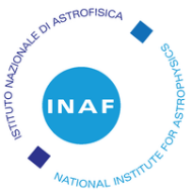
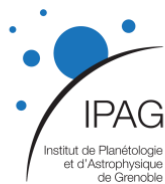
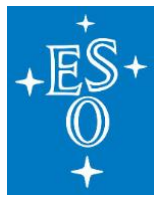

# Authors

| Name | Affiliation |
|---|---|
| Giuliana Fiorentino | INAF - OABO |
| Michele Bellazzini | INAF - OABO |
| Paolo Ciliegi | INAF - OABO |
| Gael Chauvin | IPAG |
| Sylvain Douté | IPAG |
| Valentina D'Orazi | INAF - OAPD |
| Elisabetta Maiorano | INAF - IASFBO |
| Filippo Mannucci | INAF - OAA |
| Michela Mapelli | INAF - OAPD |
| Linda Podio | INAF - OAA |
| Paolo Saracco | INAF - OABR |
| Marilena Spavone | INAF - OACN |

# Change Record from previous version

| Date | Affected Section(s) | Changes / Reason / Remarks |
|---|---|---|
| July 10, 2017 | All | First issue |
| August 17, 2017 | All | Minor changes after review from all the members of the ST |
| June 7, 2019 | All | New science cases added |
| | | |
| | | |

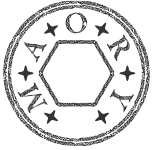

# Contents

# 1. Introduction

## 1.1 Purpose

In the present contribution, we have collected 45 representative scientific cases proposed to exploit the combination of MAORY+MICADO, the first light AO relay + imager/spectrograph at the E-ELT. This effort involves a broad scientific community, mainly from scientists of the two institutes of the MAORY consortium, INAF and IPAG, who answered to a call from the MAORY ST. The main intent of this call was to trigger the interest of the community at large on the MAORY project and on its scientific capabilities (when coupled with MICADO) and to invite scientists to begin thinking real science to be done with these innovative telescope and instruments.

## 1.2 Scope

This document is intended to present the first ideas of the community of the institutes that are designing and building MAORY for the scientific use of MAORY+MICADO.

E-ELT Science Cases have been already thoroughly studied and discussed from a broad perspective (see, e.g., https://www.eso.org/sci/facilities/eelt/science/ ) and many cases are being developed by the MICADO ST. Here we decided to have a complementary approach, that is to focus on real observations, on feasibility, on observing strategies, asking the community to translate their science into real MAORY+MICADO observations, possibly with real targets, serving as examples. The goal of this approach is to find out critical issues that can lead to modifications of the design and/or to elaborate practical solutions to problems that arise only when one has to deal with the real potential and limits of a given experimental apparatus. To this aim we have asked science case proposers to fill a simple (two-pages) standard form including some technical specifications. The form is inspired to the one that is used by the MICADO ST for the same purposes. This will facilitate comparisons, communication and collaboration between the STs of the two consortia, that are working in parallel to the science that can be obtained from the *same* system.

The material collected in this book will provide the basis for discussion and analysis of the science capabilities of MAORY+MICADO by the MAORY ST and, in particular, will drive our program of simulated observations and their analysis, to explore the best way to reach specific science goals. This document is a first exercise and the current version of the book maybe expanded whenever additional interesting science cases are proposed to the MAORY ST. Indeed, it should be seen as a work *constantly in progress* as not only new ideas but also new results from many ``big players'' in the years to come (e.g. Gaia, JWST, Euclid, etc.) can completely change the scene in some field, leading to a blossom of new cases. Finally we stress that the cases presented here does not fully cover the ample range of scientific interests of the MAORY community. Future issues of this document will expand also on this front, as more people from the consortium institutes is involved in the scientific development of the project.

## 1.3 Definitions, Acronyms and Abbreviations

| | |
|---|---|
| AO | Adaptive Optics |
| ADC | Atmospheric Dispersion Corrector |
| arcsec | Arc second |
| arcmin | Arc minute |
| CMD | Color Magnitude Diagram |
| CoI | Co-Investigator |
| ELT | Extremely Large Telescope |
| ESO | European Southern Observatory |
| ETC | Exposure Time Calculator |
| FoV | Field of View |
| GTO | Guaranteed Time Observing |
| INAF - IASFBO | INAF - Istituto di Astrofisica Spaziale e Fisica cosmica di Bologna |
| INAF - IASFMI | INAF - Istituto di Astrofisica Spaziale e Fisica cosmica di Milano |
| INAF - IASFPA | INAF - Istituto di Astrofisica Spaziale e Fisica cosmica di Palermo |
| INAF | Istituto Nazionale di AstroFisica |
| INSU | Institut National des Sciences de l'Univers |
| IPAG | Institut de Planétologie et d'Astrophysique de Grenoble |
| LGS | Laser Guide Stars |
| LP | Legacy Program |
| mas | Milli Arc Second |
| MAORY | Multi Conjugate Adaptive Optics RelaY |
| MCAO | Multi Conjugate Adaptive Optics |
| MICADO | Multi-AO Imaging Camera for Deep Observations |
| µas | Micro Arc Second |
| NGP | North Galactic Pole |
| NGS | Natural Guide Star |
| INAF - OAA | INAF - Osservatorio Astrofisico di Arcetri |
| INAF - OABO | INAF - Osservatorio Astronomico di Bologna |
| INAF - OABR | INAF - Osservatorio Astronomico di Brera |
| INAF - OACN | INAF - Osservatorio Astronomico di Capodimonte |
| INAF - OACT | INAF - Osservatorio Astronomico di Catania |
| INAF - OAPA | INAF - Osservatorio Astronomico di Palermo |
| INAF - OAPD | INAF - Osservatorio Astronomico di Padova |

| INAF - OAR | INAF - Osservatorio Astronomico di Roma |
| INAF - OAT | INAF - Osservatorio Astronomico di Trieste |
| INAF - OATE | INAF - Osservatorio Astronomico di Teramo |
| INAF - OATO | INAF - Osservatorio Astronomico di Torino |
| PCCB | Project Office Configuration Control Board |
| PI | Principal Investigator |
| PSF | Point Spread Function |
| SC | Science Case |
| SCAO | Single-conjugate Adaptive Optics |
| SR | Strehl Ratio |
| ST | Science Team |

# 2. Related Documents

## 2.1 Applicable Documents

The following applicable documents form a part of the present document to the extent specified herein. In the event of conflict between applicable documents and the content of the present document, the content of the present document shall be taken as superseding.

AD1     Title
        Number XXXX Version X

## 2.2 Reference Documents

The following documents, of the exact version shown herein, are listed as background references only. They are not to be construed as a binding complement to the present document.

RD1     MICADO Operational Concept Description
        Number ELT-PLA-MCD-56301-0004 Version 1.0

RD2     MICADO Masks, Stops, and Filters Description
        Number ELT-TRE-MCD-56300-0014 Version 0.1

RD3     PSF for SCAO
        Number E-MAO-100-INA-TNO-006 Version 1D5

RD4     MAORY Sky Coverage in MCAO mode. An exploratory analysis with Galactic Models
        Number E-MAO-000-INA-TN0-002 Version 1D2

RD5     MAORY (E-ELT MCAO) Technical Specification
        Number ESO-254311 Version 1

# 3. MAORY+MICADO science cases overview

## 3.1 A basic description of the MAORY+MICADO system

MAORY is the Adaptive Optic (AO) module that will be installed at the E-ELT at the first light of the telescope. It provides two different types of AO correction, a very high correction over a small FoV (diameter ~10 arcsec, with performances rapidly degrading with distance from the bright natural star used to probe the wavefront – SCAO mode) and a moderate correction over a wide FoV (diameter ~60 arcsec, with pretty homogeneous performances over the whole FoV – MCAO mode).

The applicability of the SCAO mode will be limited by the need of a bright (approximately $V \lesssim 16.0$) star within a few arcsec of the scientific target. In the ideal SCAO case the bright star that is used to sense the wavefront (natural guide star, NGS) coincides with the scientific target (e.g., for exoplanet imaging). On the other hand, the MCAO mode will make use of three NGS (with $H \lesssim 19.0$) to be found within an annular patrol field with inner radius of ~45 arcsec and outer radius of ~100 arcsec. According to the current version of the instrument design, these NGS will allow us to correct low-order modes of the wavefront distortions, while the sensing of six artificial laser guide stars will be used to correct for high-order modes. This will make possible to get AO assisted observations over a large fraction of the sky accessible from Cerro Armazones, meeting the system specification on Sky Coverage (SC $\gtrsim$ 50% over the whole sky; see RD4 for a preliminary analysis of this issue). For further details see RD5.

While MAORY must provide a port for a second instrument, still to be defined, its main goal is to feed the high-resolution NIR imager and spectrograph MICADO, a workhorse instrument for E-ELT. In imaging mode MICADO will provide an option with a wide FoV ($50.4 \times 50.4$ arcsec$^2$) with pixel scale Nyquist sampling the PSF in H and K bands, and a high-resolution option with a $18.9 \times 18.9$ arcsec$^2$ FoV and a pixel scale providing Nyquist sampling of the PSF down to the I band. Long-slit spectroscopy will be available with three different slit widths: (a) *16 mas wide and 4 arcsec long slit* providing a spectral resolution R~8000 for sources filling the slit and 11000 < R < 18000, depending on wavelength, for point sources within the slit; (b) *48 mas wide and 4 arcsec long slit* providing R~2500 for extended sources filling the slit; (c) *20 mas wide and 20 arcsec long slit* with a resolution R~6000 (when filled) specifically suited to observe galaxy nuclei, to include sky simultaneously at larger off-axis distances. This slit is available for the K-band only.

An option on the order sorting filter will allow access to two different spectral ranges 0.8-1.45 μm (IzJ passbands simultaneously) and 1.45-2.4 μm (HK passbands simultaneously). The pixel scale will always be 4 mas. Since the ADC is located after the slit, to avoid wavelength-dependent light losses observations must be performed in parallactic angle. For further details on MICADO see RD1.

## 3.2 Institutes participating and proposed science

The MAORY consortium is formed by two international institutes: the Italian National Institute for Astrophysics (INAF) and the Institute de Planétologie et Astrophysique de Grenoble (IPAG). The activity of INAF is performed in the following six INAF structures: INAF - OABO, INAF - IASFBO, INAF - OAA, INAF - OAPD, INAF - OACN, INAF - OABR. Twelve people affiliated to these institutions form the MAORY ST (http://wwwmaory.oabo.inaf.it/?da_image=286) that coordinated the call for science cases and the assembling of this. The editorial work has been supervised and coordinated by Giuliana Fiorentino (MAORY instrument scientist and deputy Chair of the MAORY ST) and by Michele Bellazzini (MAORY project scientist and Chair of the MAORY ST).

Although this effort is not fully exhaustive of all the kind of science that can be done with MAORY+MICADO, it represents a general overview of the current scientific interests of the community involved in the construction of MAORY. In the following we make a short overview of the community that participated to the MAORY ST call, describing the main research streams proposed and the main modes required to reach the scientific goals. We have decided to trace the requests on the observing modes, on the AO modes, and on the pixel scale/FoV, i.e. all the properties that make MAORY+MICADO unique when compared with current and future facilities. We remember that currently ESO instrumentation is missing an MCAO imager, whereas an AO module of this kind remain almost unique on future giant telescopes and will assure us SR larger than 30% on a substantial portion of sky (larger than 50%, see RD5). The analogous instrument in the northern hemisphere, will be the Narrow Field Infrared Adaptive Optics System (NFIRAOS, Herriot et al. 2014SPIE.9148E..10H), the MCAO facility of the Thirty Meter Telescope, with a smaller FoV (10"-30").

We have collected a total of 45 scientific cases from a broad scientific community. Science cases were proposed by IPAG scientists and by scientists from 14 out of the whole 17 INAF Institutions spread over Italy. The relevant statistic is reported in the left panel of Figure 3.2-1. The distribution of the various science cases according to their broad scientific field is shown in Figure 3.2-1 (right). The main chapters of this book are entitled after the categories used in this Figure.

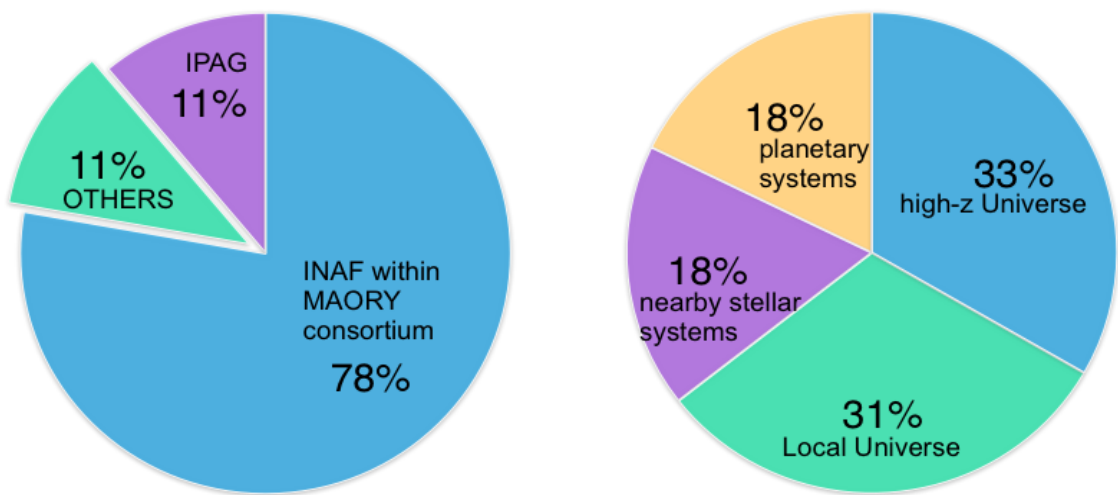

Figure 3.2-1 Affiliation of the PIs (left) and distribution of the different topics proposed, as divided by main streams (right).

*Planetary systems* include cases on our own Solar system, exoplanets and the formation of planetary systems, as well as atmospheric characterization of sub-stellar objects. *Nearby stellar systems* encompasses all the cases on stars and stellar systems within our own Galaxy and its satellites. *Local Universe* refers to all the cases aimed at studying the stellar content and the structure of distant stellar systems that can be at least partially resolved into individual stars. In many cases, they will fall within the range of resolved systems only thanks to the advent of MAORY+MICADO. *High-z Universe* include all the cases addressing the formation of structures and cosmology using the formidable sensitivity and resolution of the E-ELT with MAORY+MICADO to probe the very distant Universe and, consequently, the earliest phases of galaxy formation, as well as high-energy phenomena over the whole range of cosmic distances and times made accessible by E-ELT.

## 3.3 Required observing, AO modes and FoV/Pixel scale

We have grouped the possible observing modes in four categories: IMA- standard imaging; SPEC- standard spectroscopy; IMA/ASTR- imaging with the final goal of high spatial resolution astrometry; IMA/SPEC- combination of imaging and spectroscopy required for some particular science case (e.g., Section 7.6). In Figure 3.3-1 (left panel) we show the distribution of occurrence of the various modes in the considered SCs. In the future, more effort should be dedicated to the development of astrometric cases, as the expected astrometric accuracy of MAORY+MICADO will be unchallenged in the coming decades for a wide range of applications.

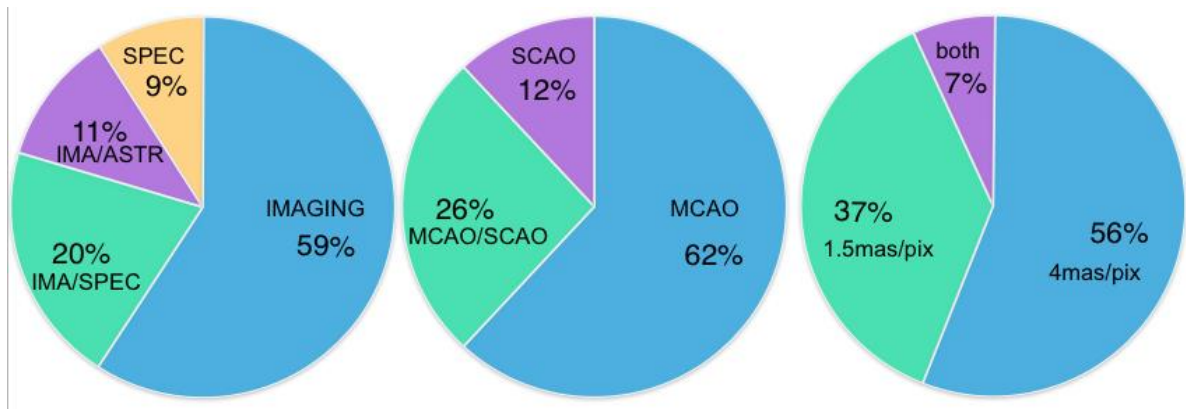

Figure 3.3-1 Charts that show the required observing mode (left), AO mode (center) and pixel scale/FoV (right).

The middle panel of Figure 3.3-1 shows the distribution of AO modes required in this collection of science cases. MCAO takes the largest share, mainly for two reasons: the larger corrected FoV, and the larger portion of sky accessible. The cases for which doubts remain on the most convenient mode are, in fact, cases that would benefit from SCAO performances in terms of Strehl ratio but the accessibility of the relevant targets in this mode is uncertain and MCAO is a viable alternative.

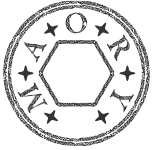

Document Title

Finally, the right panel of Figure 3.3-1 reports the statistic on the pixel scale/FoV requested for the imaging/astrometric mode. There is an obvious trade-off here between an extremely high spatial resolution and the reduced size of the FoV (plus the need for Nyquist sampling the PSF). A small pixel scale, in principle, is less affected by noise and background (very important in near infrared) and provides a better sampling of the PSF. However, the accuracy we need strongly depends on the particular science and it is very often satisfied already by 4 mas/pixel. This latter scale implies a FoV with a sampled area >7 times larger than the 1.5 mas/pixel option.

## 3.4 Legacy Programs and MAORY ST initiatives

Also in view of the future best exploitation of the MAORY GTO time the MAORY ST is working on ideas for *Legacy Programs* (LPs). The key point is to identify sets of observations whose outcome would allow us to address many different scientific problems, such that various scientific communities can benefit from the same dataset. While the process is still at a brainstorming level, some Galactic and extragalactic cases have been discussed in some depth at the ST meeting held in Arcetri in October 2016.

Examples of the LPs discussed in this context are:

- a survey of the 30 Doradus region in the Large Magellanic Cloud complementing with deeper and extremely higher spatial resolution observation the large survey performed with HST (Sabbi et al. 2016, ApJS, 222, 11), with multi-epoch observations of selected fields. This kind of survey would provide breakthrough data on scientific issues ranging from stellar IMF to dynamics of star clusters, from intermediate mass Black Holes to pre-Main Sequence stars, brown dwarfs and sub-stellar objects, taking full advantage from the large set of excellent existing data. See Sect. 5.8.

- On the extragalactic side an obvious option is a deep field, where MAORY+MICADO would provide unprecedented depth and, in particular, unchallenged resolution on compact sources. The impacted research areas include compact and low-mass galaxies, spatially resolved properties of distant galaxies, population gradients at different cosmic epochs, properties of the interstellar medium in 2<z<8 galaxies. etc. Also in this case coordination with existing and future surveys with other facilities should play a key role. See, e.g., Sect. 7.10, 7.11, and 7.12.

More developed and/or additional studies of Legacy Programs will be included in future versions of this document, when available. It is important to stress that, in the present phase, *the aim of the ST is by no means to select GTO programs*, but to elaborate the concept of large programmes that can maximise the scientific return of MAORY+MICADO for the communities of reference.

To keep track of the many science cases that focus on specific area where the MAORY+MICADO is expected to give a breakthrough contribution and the interest from the MAORY community is very high, we tagged them with the label "MAORY ST initiative on… ". This may help future coordination of the efforts in areas of special interest. The active *MAORY ST initiative* labels at this stage are *planetary science*, *compact stellar systems*, and *high-redshift galaxies*.

## 3.5 Contributors

The authors of each case and their affiliations are reported in each case form. We are very grateful for their contribution to the writing of this document. While technical issues prevent to include all the names in the Authors table above, *all of them must be considered as co-authors of this document.*

Here we list the names of all contributors in alphabetic order.

Alcalá J. (INAF - OACN), Amati L. (INAF - IASFBO), Annibali F. (INAF - OABO), Antoniucci S. (INAF - OAR), Bacciotti F. (INAF - OAA), Ballo L. (INAF - OABR), Ballone A. (INAF - OAPD), Balmaverde B. (SNS), Beck, P. (IPAG), Bellazzini M. (INAF - OABO), Berton M. (University of Padova, INAF - OABR), Biazzo K. (INAF - OACT), Bonal L. (IPAG), Bonito R. (INAF - OAPA), Bono G. (University of Tor Vergata), Bortolas E. (INAF - OAPD, University of Padova), Braga V.F. (PUC), Branchesi M. (University of Urbino), Brocato E. (INAF - OAR), Brusa. M. (University of Bologna), Buson L. (INAF - OAPD), Calura F. (INAF - OABO), Cantiello M. (INAF - OATE), Capetti A. (INAF - OATO), Castellani M. (INAF - OAR), Castellano M. (INAF - OAR), Chauvin, G. (IPAG), Chen S. (University of Padova), Cicone C. (INAF - OABR), Ciliegi P. (INAF - OABO), Ciolfi R. (INAF - OAPD), Ciroi S. (University of Padova), Codella C. (INAF - OAA), Comastri A. (INAF - OABO), Congiu E. (University of Padova, INAF - OABR), Corsini E. M. (University of Padova, INAF - OAPD), Costantin L. (University of Padova), Covino E. (INAF - OACN), Covino S. (INAF - OABR), Cracco V. (University of Padova), Cristiani S. (INAF - OAT), Cuomo, V. (University of Padova), Cupani G. (INAF - OAT), Czoske O. (University of Innsbruck), D'Avanzo P. (INAF - OABR), D'Orazi V. (INAF - OAPD), Dall'Ora M. (INAF - OACN), Dalla Bontà E. (INAF - OAPD), Dalessandro E. (INAF - OABO), Davies B. (LJMU), De Pasquale M. (Istanbul University), Della Ceca R. (INAF - OABR), Desidera S. (INAF - OAPD), Di Carlo U. N. (INAF - OAPD, University of Padova), Di Rico G. (INAF - OATE), Douté S. (IPAG, CNRS), Falomo R. (INAF - OAPD), Fantinel D. (INAF - OAPD) Fedele D. (INAF - OAA), Ferraro I. (INAF - OAR), Fiorentino G. (INAF - OABO), Fontana A. (INAF - OAR), Frasca A. (INAF - OACT), Frezzato M. (University of Padova), Gallazzi A. (INAF - OAA), Gargiulo A. (INAF - IASFMI), Garufi A. (Univ. Aut. De Madrid), Giannini T. (INAF - OAR), Giacobbo N. (INAF - OAPD, University of Padova), Giallongo E. (INAF - OAR), Gilli R. (INAF - OABO), Gratton R. G. (INAF - OAPD), Grazian A. (INAF - OAR), Greggio L. (INAF - OAPD), Gruppioni C. (INAF - OABO), Guidi G. (INAF - OAA, University of Firenze), Gullieuszik M. (INAF - OAPD), Herique A. (IPAG), Hunt L. (INAF - OAA), Iannicola G. (INAF - OAR), Iodice E. (INAF - OACN), Kudritzki R. P. (UHawaii), La Barbera F. (INAF - OACN), La Mura G. (University of Padova), Landoni M. (INAF - OABR), Lardo C. (EPFL), Lodato G. (University of Milano), Maiorano E. (INAF - IASFBO), Malesani D. (University of Copenaghen), Manara C. (ESO), Mannucci F. (INAF - OAA), Mapelli M. (INAF - OAPD), Marafatto L. (INAF - OAPD), Marconi M. (INAF - OACN), Martinez-Vazquez C. (INAF - OABO), Masetti N. (INAF - IASFBO), Massari D. (University of Groningen), Meneghetti M. (INAF - OABO), Mercurio A. (INAF - OACN), Merlin E. (INAF - OAR), Mesa D. (INAF - OAPD), Mignoli M. (INAF - OABO), Molinaro R. (INAF - OACN), Monelli M. (IAC), Morelli L. (University of Padova, INAF - OAPD), Moretti A. (INAF - OAPD), Moretti M. I. (INAF - OACN), Musella I. (INAF - OACN), Napolitano N.R. (INAF - OACN), Nicastro L. (INAF - IASFBO), Nisini B. (INAF - OAR), Nonino M. (INAF - OAT), Origlia L. (INAF - OABO), Pagotto I. (University of Padova), Paiano S. (INAF - OAPD), Palazzi E. (INAF - IASFBO), Pancino E. (INAF - OAA), Pentericci L. (INAF - OAR), Pian E. (INAF - IASFBO), Piranomonte S. (INAF - OAR), Pizzella A. (University of Padova, INAF - OAPD), Podio L. (INAF - OAA), Poggianti B. (INAF - OAPD), Portaluri E.

(INAF - OAPD), Pulone L. (INAF - OAR), Quirico E. (IPAG), Rafanelli P. (University of Padova), Raimondo G. (INAF - OATE), Ripamonti E. (INAF - OAPD, University of Padova), Ripepi V. (INAF - OACN), Rizzo F. (MPIA, Garching), Rodighiero G. (University of Padova, INAF - OAPD), Rosati P. (University of Ferrara), Rossi A. (INAF - IASFBO), Saracco P. (INAF - OABR), Savaglio S. (University of Calabria), Schmitt B. (IPAG), Severgnini P. (INAF - OABR), Sollima A. (INAF - OABO), Spavone M. (INAF - OACN), Spera M. (INAF - OAPD), Spiniello C. (INAF - OACN), Stelzer B. (INAF - OAPA), Stratta G. (University of Urbino), Tazzari M. (ESO), Testa V. (INAF - OAR), Testi L. (INAF - OAA, ESO), Tortora C. (University of Groningen), Tosi M. (INAF - OABO), Tozzi P. (INAF - OAA), Trani A. A. (INAF - OAPD, SISSA), Uslenghi M. (INAF - IASFMI), Valenti E. (ESO), Vanzella E. (INAF - OABO), Vegetti S. (MPIA, Garching), Vignali C. (University of Bologna), Zibetti S. (INAF - OAA), Zoccali M. (PUC), Zocchi A. (University of Bologna).

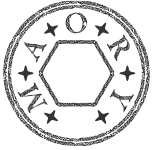

# 4. Planetary systems

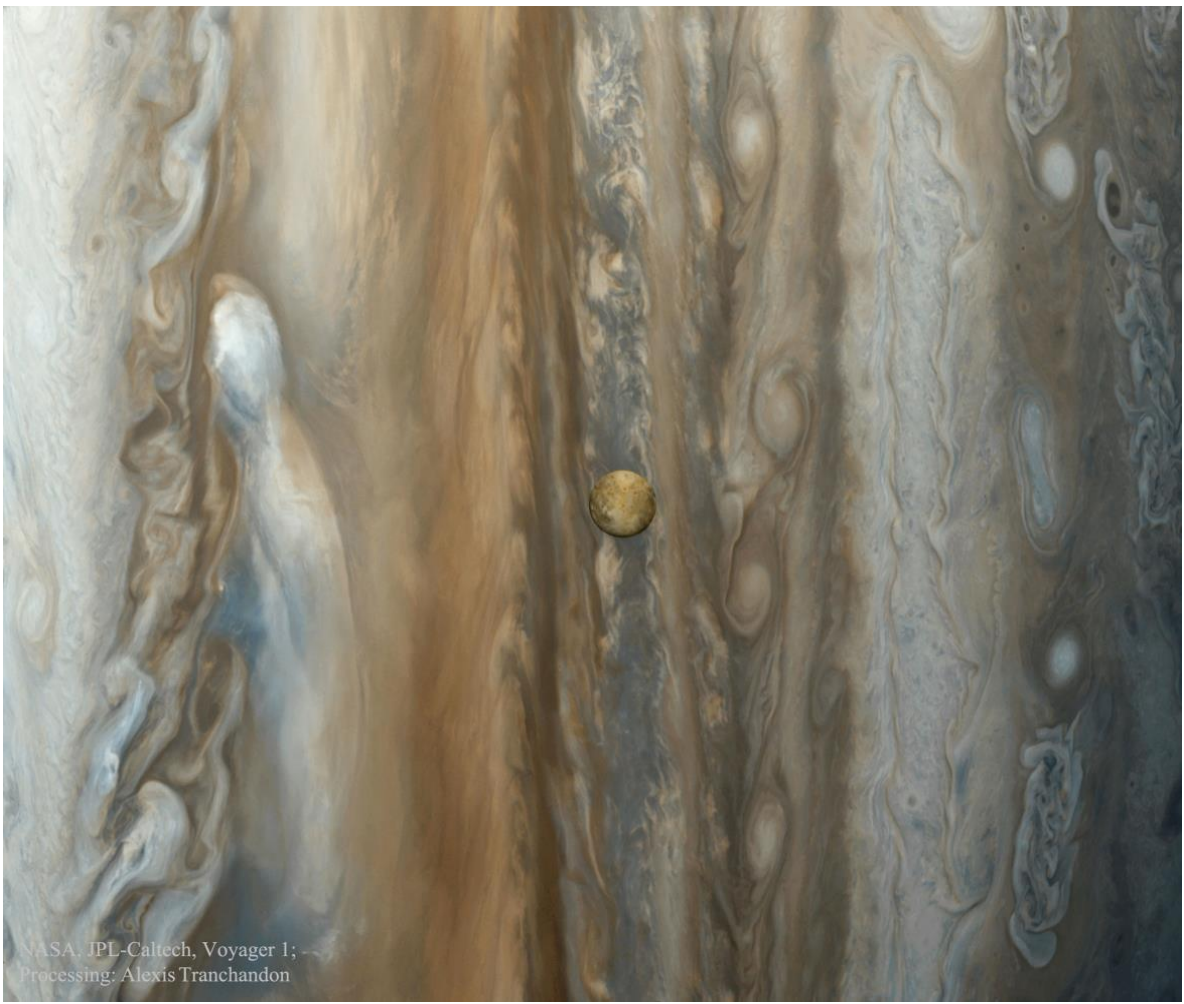

Figure 4-1 Io in front of Jupiter, as seen from Voyager 1. The disc of Io will be sampled by ~7×10⁴ to ~5×10⁵ MICADO pixels in MAORY-MCAO mode, depending on the adopted set-up. With the 50.4"×50.4" FoV the full disc of Jupiter will be imaged in one shot, one pixel sampling ~15 km on the surface of Jupiter.

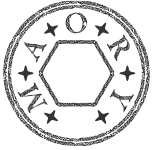

# 4.1 The launching mechanism of jets from young stars and their feedback on protoplanetary disks

**Authors:** *Podio L. (INAF - OAA), Antoniucci S. (INAF - OAR), Bacciotti F. (INAF - OAA), Bonito R. (INAF - OAPA), Codella C. (INAF - OAA), Garufi A. (Univ. Aut. De Madrid), Giannini T., Nisini B. (INAF - OAR) & the JEDI team*

**Brief description of science case:**

The first step to comprehend the formation of our solar system as well as of extra-solar planetary systems is to study the evolution of the planet formation site: the circumstellar disk. To understand how stellar disks lose their mass and form protoplanets is a front line in astrophysics. These processes are believed to be regulated by the simultaneous effects of the mass accretion onto the star and the ejection of matter from the stellar-disk system. However, the interface between jets and disks has been poorly investigated to date because of the very small angular scales involved. ***The goal of this project is to unveil the jet launching mechanism and its feedback on disks by exploiting the unprecedented angular resolution and contrast provided by MICADO+MAORY.***

Observations and theoretical studies suggest that jets are accelerated and collimated by the action of magneto-centrifugal forces on scales of fractions up to tens of AU and may have a deep influence on the properties of the disk region in which planets form (e.g. Ferreira et al. 2006, A&A, 453, 785). In order to explore the jet-launching region, high spatial resolution and high contrast observations are required. ***We propose to study the interplay between accretion, ejection, and disks in young low and intermediate mass stars using MAORY-MICADO with narrow band (NB) filters centred on the typical jet and accretion tracers*** ([Fe II]1.25um, 1.64um, $H_2$ 2.12um, Pa-beta 1.28 um). Observations with NB filters at adjacent wavelengths will be acquired to subtract the stellar continuum emission (J1, H-cont, K-cont). A Classical Lyot Coronograph (CLC) will be used in the case of faint jet emission to increase the contrast. These images will enable us to resolve the jet across its width and close to its driving source, down to 6-10 mas, i.e. ~0.6-1 AU for the closest young stellar objects (d~100 pc). Such unprecedented resolution will allow us to obtain the first direct image of the jet launching region (up to a few AUs), hence to infer crucial jet properties such as the collimation, density, and mass flux rates at its base (e.g., Podio et al. 2006, A&A, 456, 189; Podio et al. 2016, A&A, 593L, 4; Maurri et al. 2014, A&A 565, A110; Antoniucci et al. 2014, A&A, 566, A129; Antoniucci et al. 2016, A&A, 593, L13), which will ***enable us to distinguish for the first time between the different theoretical scenarios.*** We aim at observing bright T Tauri and Herbig stars, that can be directly used as reference stars for the SCAO module in order to reach the optimal resolution. In addition, with the MCAO facility it will be possible to target also fainter/lower mass stars, as well as younger and more embedded Class I objects. This will allow us to explore the variation of the jet structure and properties as a function of the stellar mass and age. The astrometric capabilities of MICADO (~10/50 µas) will also allow us to recover jet proper motions, i.e. to study the jet dynamics on very small scales and investigate the connection of mass ejection with episodic accretion events (see, e.g. Antoniucci et al. 2016, A&A, 593, L13; Ellerbroek et al. 2014, A&A, 563A, 87). Finally, coupling proper motions with additional kinematic information from HARMONI IFU (jet radial velocity and jet rotation around its axis) will provide a complete 3D map of the jet structure and velocities.

Our team has a broad expertise in the study of stellar jets from X-rays to millimeter wavelengths and in the analysis of high angular resolution and high contrast images taken with adaptive optics system such as VLT-NACO, VLT-SPHERE and LBT. We have complementary datasets of several jets in the X-rays (e.g. Chandra), UV/optical/near-infrared (e.g. with HST, VLT-SPHERE/XShooter, Keck-HIRES), and far-infrared/millimeter range (e.g. with Herschel, NOEMA, and ALMA), which will be used to plan the observations (e.g. time estimate) and to reconstruct the jet energetics (from cold to hot gas components). To analyse the data we have dedicated spectral diagnostic tools, models of line emission, and multi-wavelength numerical simulations (e.g. Bacciotti et al. 1999, A&A, 342, 717; Bonito et al. 2007, A&A, 462, 645; Giannini et al. 2013, ApJ, 778, 71; Giannini et al. 2015, ApJ, 814, 52; Nisini et al. 2016, A&A, 595A, 76).

**Keywords:** jets, pre-main sequence, protoplanetary disks, star formation, T Tauri stars

**MICADO Observation mode:** imaging, astrometry.

**MICADO Pixel Scale / Fov:** 1.5mas/px and 20arcsec FoV.

**MICADO Spectral set-up:** none.

**Filters required:** NB filters on the typical jet/accretion tracers: [Fe II]1.25um, 1.64um, $H_2$ 2.12um, Pa-beta 1.28um; NB continuum filters to subtract the stellar continuum.

**Estimate Survey Area/Sample Size/ Number of Images/Epochs:** ~25 sources, 7 images per source (NB imaging in 4 filters for the lines + 3 filters for the adjacent continuum), 2-3 epochs in 1 filter to measure proper motions.

**Average Integration time per image (magnitude of targets; S/N required):** typical target magnitudes: V$\geq$10, H=8-14; typical flux of line emission tracers ~ $1e^{-14}$ — $1e^{-15}$ erg/s/cm$^2$; S/N > 10.

**Observation requirements:** dithering patterns.

**Strehl or EE required:** a good SR is required to reach high angular resolution and adequately subtract the stellar continuum emission (through PSF reconstruction).

**Astrometric Accuracy:** we aim at an astrometric accuracy of the order of 1 mas to study tangential velocity variations of the order of 1 km/s per year in objects at a distance of <250 pc (e.g. Perseus, Taurus).

**SCAO vs. MCAO:** SCAO can be used for an optimal SR on the bright T Tauri and Herbig stars (V$\leq$13-14 to obtain an on-axis SR > 0.6 at 2.2um). MCAO will be instead needed to observe fainter T Tauri stars, as well as younger and more embedded class I sources (V$\geq$16mag), which are currently not accessible with the available adaptive optics instrumentation such as VLT/SPHERE. In this regard, MCAO will allow us to extend our study of the accretion/ejection mechanism and jet-disk connection to lower mass as well as younger objects. This in turn will allow us to investigate the variation of the jet properties with the stellar mass and evolution.

**Comparison with JWST or other facilities:** MAORY+MICADO/HARMONI will allow us to reach an angular resolution of 6-10 mas in the selected jet/accretion tracers, i.e. about an order of magnitude better than what we can reach with VLT and JWST. This will allow us to directly image the jet launching region for the first time thus obtaining direct constraints on the jet feedback on the disk region where planets form.

**Synergies with other facilities (4MOST/MOONS, LSST/ALMA/HARMONI/METIS, HIRES/MOSAIC), but also VLT or other smaller telescope instruments:** our group is already leading present and future observations with VLT/SPHERE, LBT/SHARK, and LBT/LUCI, which will be crucial to plan the observations with MAORY+MICADO.

Moreover, we will complement the E-ELT observations with NOEMA and ALMA data, which will allow us to image all the different jet components at comparable resolution (cold molecular gas at mm wavelengths and warm/hot atomic/ionised gas at NIR wavelengths). The NOEMA and ALMA data will also allow us to have a good characterisation of the disk properties to constrain the jet-disk interplay. Complementary observations with HARMONI will be crucial to recover the jet kinematics (jet radial velocities and rotation velocity around the jet axis) and to obtain the simultaneous coverage of the optical diagnostic lines. By coupling the proper motions measurable with MAORY+MICADO with the radial velocities obtained with HARMONI, we will obtain a 3D map of the jet velocity and structure. Finally, the MAORY+MICADO and HARMONI observations will serve to plan follow-up HIRES IFU observations in the J and H band (R~100 000), which will provide jet maps at unprecedented angular and spectral resolution.

**Simulations made/needed to verify science case or feasibility:** we are working on simulations of jet observations for the AO-feeded instruments on LBT to evaluate the optimal image reconstruction procedures (e.g. La Camera et al. 2014, PASP, 126, 180). These simulations will be exported to the MAORY+MICADO case.

**Origin of the targets:** the targets will be selected among the sample of jet driving sources for which we are acquiring a database of complementary information, and will include Herbig Ae/Be, T Tauri and Class I sources.

**NGS:** the brightest T Tauri and Herbig stars (Vmag < 13-14) can be efficiently used as NGS in the SCAO mode. For fainter targets, we will use the MCAO mode.

**Acquisition:** the jet sources are all bright in the IR so the pointing can be verified with finding charts.

**Calibrations:** 'Standard' calibrations. Accurate PSF is required for image deconvolution (e.g. Antoniucci et al. 2016, A&A, 593, 13) and for rotation measurements to subtract the uneven slit illumination effect (e.g. Bacciotti et al. 2002, ApJ, 576, 222). It is important to know if there are image distortions as these may affect jet structure and rotation measurements.

**Data Processing Requirements**: a detailed PSF knowledge would be desirable.

**Any other comments:** none.

# 4.2 Massive (proto-)planets in protoplanetary disks

**Authors:** *Testi L. (INAF - OAA, ESO), Alcalá J. (INAF - OACN), Antoniucci S. (INAF - OAR), Bacciotti F., Fedele D. (INAF - OAA), Garufi A. (Univ. Aut. De Madrid), Guidi G. (INAF - OAA, University of Firenze), Lodato G. (University of Milano), Nisini B. (INAF - OAR), Podio L. (INAF - OAA), Tazzari M. (ESO) & the JEDI team*

**Brief description of science case:**

Planets are expected to form in protoplanetary disks around pre-main sequence stars. Previous infrared surveys suggest a typical giant planet formation time longer than 2 Myr, based on disk dissipation timescales (e.g. Hernandez et al. 2008, ApJ, 686, 1195). In contrast, new ALMA observations of protoplanetary disks reveal a very different picture, with very young disks (≤ 1 Myr) showing features attributed to disk-planet interaction (e.g. ALMA Partnership 2015, ApJ, 808, L3, Isella et al. 2016, PhysRevLet, 117, 251101, Perez et al. 2016, Science, 353, 1519). These findings could imply that fast and effective planet formation is a common phenomenon. This hypothesis can be checked with infrared high contrast imaging. We propose to carry out a survey to verify (or disprove) the existence of ≥ 1 $M_{Jup}$ planets in the outer regions of disks that have been (are being) observed with ALMA. Our survey will assess quantitatively the frequency of giant planets formed at early times (≤ 1 Myr) and provide constraints on the demographics of planetary systems at birth and on their formation mechanism. Possible no detections will also provide strong constraints on the nature of the planets and their formation history. Smaller planetary bodies, with masses ~ 0.1−0.3 $M_{Jup}$, may still be able to produce some of the observed mm dust depletion rings but would remain embedded in the disk and undetectable by infrared observations and would have not formed via gravitational instabilities (see, e.g. Dipierro et al. 2015, MNRAS, 453, L73). Negative results of our search would thus imply a much more efficient aggregation process than currently thought.

We propose to obtain deep, high-contrast, Ks broad-band imaging with MAORY+MICADO to search for planetary mass companions associated to density sub-structures (mainly gaps and spirals) detected with ALMA in protoplanetary disks. Using angular differential imaging (ADI), MAORY+MICADO should be able to achieve sufficient contrast with the Ks filter to detect a 1 Myr old 1 $M_{Jup}$ hot start planet ($m_{Ks}$=19.1 mag; Spiegel & Burrows 2012, ApJ, 745, 174) as close as ~8 AU from the central star, probing the region of the disk where ALMA detected the structures.

**Keywords:** protoplanetary disks, pre-main sequence stars, T Tauri stars, star formation

**MICADO Observation mode:** standard imaging.

**MICADO Pixel Scale / Fov:** 1.5mas/px and 20arcsec FoV.

**MICADO Spectral set-up:** none.

**Filters required:** Ks broad band continuum filter. At K band, the contrast between planets and their central star should be optimal compared with Y, J, or H bands. Likewise, the extinction at Ks, both towards the star forming region and within the disk itself, is minimal.

**Estimate Survey Area/Sample Size/ Number of Images/Epochs:** ~10-15 targets (protoplanetary disks selected based on ALMA observations), 1 image per target.

**Average Integration time per image (magnitude of targets; S/N required):** For a specific exposure time estimate we need contrast curve of the expected contrast as a function of angular distance from the central star to estimate what is the minimum planet mass detectable at a given distance from source.

**Observation requirements:** Dithering patterns. We will use the ADI technique in the Ks band. In order to optimise and ensure the adequate sky rotation for ADI we ask to observe the targets preferentially at meridian passage.

**Strehl or EE required:** A good SR is required to reach high angular resolution and adequately subtract the stellar continuum emission. This is crucial for high contrast imaging.

**Astrometric Accuracy:** none.

**SCAO vs. MCAO:** Bright T Tauri stars and Herbig stars can be used as NGS in the SCAO mode (V≤13-14 to obtain an on-axis SR > 0.6 at 2.2um), which provides a better AO correction than MCAO. The small FoV (10") is adequate for our purposes as we are interested in the inner disk region (10-50 AU, i.e. 0.1"-0.5" for the close protoplanetary disks located at ~100 pc). On the other hand, MCAO will allow us to search for protoplanets around young stars with V≤13-14mag previously not accessible with the available adaptive optics instrumentation such as VLT/SPHERE. Hence, MCAO will allow us to extend our survey of protoplanets to a much larger sample of young disks.

**Comparison with JWST or other facilities:** MAORY+MICADO will allow us to reach an angular resolution of ~11 mas in the K band (~0.6-1.6 AU for the closest protoplanetary disks, d~50-150 pc), i.e. about an order of magnitude better than what we can reach with VLT and JWST. This should produce an increase of the contrast in the inner disk region (10-50 AU distance from the central star).

**Synergies with other facilities (4MOST/MOONS, LSST/ALMA/HARMONI/METIS, HIRES/MOSAIC), but also VLT or other smaller telescope instruments:** The synergy with ALMA observations is crucial to obtain high angular resolution mm continuum observations of protoplanetary disks and select the best targets for protoplanets detection (i.e. protoplanetary disks showing signatures of planet-disk interaction and planet formation through gravitational instabilities, such as rings, gaps, spirals. Our team already analysed ALMA observations, which allowed us to select a small sample of protoplanetary disks for our survey and more ALMA observations will be available soon. The synergy with METIS is also crucial: in case of protoplanets detection, follow-up observations with METIS will allow us to estimate the planet properties, such as its temperature, gravity and chemical composition.

**Simulations made/needed to verify science case or feasibility:** Detailed simulations are needed to establish the real MAORY+MICADO potential and to better define the filters and the exposure times.

**Origin of the targets:** ALMA observations already taken or to be taken in the next future.

**NGS:** For the brightest targets the source itself can be used as NGS in SCAO mode. For the fainter ones we will use the MCAO mode.

**Acquisition:** The pointing can be verified with finding charts

**Calibrations:** 'Standard' calibrations.

**Data Processing Requirements:** Specific techniques to remove the disk emission and detect the protoplanet are required, such as Angular Differential Imaging (ADI).

**Any other comments:** none.

# 4.3 Exoplanets & Circumstellar Disks

**Authors:** *Gaël Chauvin (IPAG) & Anthony Boccaletti (LESIA/MICADO consortium)*

**Brief description of science case:**

The observation and characterization of exoplanets and planetary architectures is crucial to expand and complete our view of planetary formation and evolution, as well as of the physics of exoplanets. With two decades of exoplanet studies, observations regularly obtained key successes with the discovery of the Hot-Jupiters family, the detection and confirmation of about 2000 exoplanets today (see http://exoplanet.eu), the first glimpse of planetary atmospheres and internal structures, the first images of exoplanets, the discovery of Super-Earths in the Habitable Zone (where water is expected to be liquid) or more recently Earth-mass or Mars-size planets. Within 10 years, the era of large-scale systematics surveys will end thanks to a complete census of exoplanetary systems within 100–200 pc from the Sun. A new Era fully dedicated to the characterization of known systems will start with new generation of extremely large telescope, the Giant Magellan Telescope (GMT; Shectman and Johns 2010), the Thirty Meters Telecope (TMT; Simard and Crampton 2010) and the European Extremely Large Telescope (hereafter EELT; McPherson et al. 2012). With the first Lights foreseen in 2024-2026, the EELT will arrive at a propitious time to exploit discoveries of the upcoming generation of instruments and space missions owing to its capabilities in terms of sensitivity, spatial resolution and instrumental versatility. In this context, MAORY/MICADO offers the possibility to directly image and characterize planets and young planetary systems at the EELT diffraction limit in near-infrared (10mas at K-band). The high-contrast imaging mode of MAORY/MICADO will use the combination of SCAO-correction, coronagraphy and angular differential imaging to bring us a step further with the E-ELT in $1^{st}$ Light to address several outstanding questions still unanswered:

**How do giant planets form?** Understanding how the giant exoplanets form, how they evolve and interact, is critical because they completely shape the planetary system architecture and offer the possibility to form Earth or Super-Earth-like planets, i.e. rocky planets on a stable orbit sustaining Life. Nowadays, there are still fundamental questions that are unanswered. We do not know if there is one formation mechanism or several mechanisms to form giant planets and possibly operating at different timescales, locations and physical conditions. We do not understand the influence of the initial conditions, i.e. the impact of the stellar mass, the stellar metallicity and multiplicity or the effect of the close stellar environment on planetary formation. The synergy between the MAORY/MICADO high-contrast imaging mode and additional techniques like astrometry (*GAIA*) and radial velocity (VLT/ESPRESSO) will offer a unique view of the planetary system architecture at all orbits (for the giant planet population). It includes to characterization of their frequency, multiplicity, distribution of mass and orbital parameters (period, eccentricity) for a broad range of stellar properties (mass, metallicity and age). Observables will be directly confronted to predictions of population synthesis models for various types of formation mechanisms (core accretion, gravitational instability or gravo-turbulent fragmentation).

**What are the initial conditions of planet formation?** The high-contrast mode of MAORY/MICADO combined with the large field of view of the near-infrared camera will offer the unique possibility to explore the properties and morphology of young protoplanetary and debris disks at an unprecedented spatial resolution. Asymmetries (cavity, gap, hole, clump, vortex…) will be actually connected to the various physical processes at play including planetary formation itself. This will allow identifying the key mechanisms of dynamical evolution (planet-disk and planet-planet interactions) of planetary architectures.

**What are the physical properties of young Jupiters and Saturns?** There are still large uncertainties crossing two orders of magnitude to predict basic properties such as the luminosity of young Jupiters and Saturns. We still do not understand how the gaseous component is transferred and accreted from the protoplanetary disk onto the planetary atmosphere. The masses of young imaged exoplanets are currently predicted by non-calibrated evolutionary models. The synergy between the MAORY/MICADO high-contrast imaging mode and additional techniques will offer the unique possibility to simultaneously derive the dynamical mass together with the orbital properties and the luminosity of the young Jupiters and Saturns. This will set stringent constraints on the phase of planetary atmosphere formation and will offer unprecedented tests for current theoretical predictions of giant planet formation and evolution.

The interest of MAORY/MICADO in this field is clearly the gain in angular resolution with respect to the current or near term facilities like GPI/SPHERE and JWST. Given the current estimations of performance and limitation, three particular areas were identified as niches for MAORY/MICADO in the context of planetary systems:

1. **Study of nearby known planetary systems**: Assuming MAORY/MICADO is able to achieve the same level of contrasts as NACO, a beta Pic b-like object becomes easily detectable at ~2 AU instead of 8-10 AU with the capability to extend this minimal distance even closer in (down to 1 AU for a star at 10-20 pc). MAORY/MICADO will have the capability to reach closer physical separations than NACO and even SPHERE on nearby targets (<50 pc). Since planetary systems are found to be compact in RV surveys (discovery of planets at long periods favours the existence of other planets at shorter periods), it is obviously crucial to reach small separations. In addition, this range of physical distances is overlapping the one probed with RV, which now starts to be applicable on young early type stars although active (Lagrange et al. 2012b). The GAIA satellite will also bring new detections or limit of detection in the same range of physical distances as probed with MAORY/MICADO. Therefore, it will become possible to connect these various techniques for the very same systems and derive constraints on the true mass of planets. A more precise calibration of evolutionary models will become feasible hence with the advantage to ultimately perform better spectral characterization (from photometric data or even spectroscopic data).

2. **Planetary systems in young distant associations**: A complete new range of targets will be attainable with MAORY/MICADO given the larger collecting area and the gain in resolution. MAORY/MICADO will be able to search for young planets in distant (>50-100 pc) young star associations, like for instance the Sco-Cen associations which has been fruitful for the search of exoplanets and disks (Bonnefoy et al. 2016; Wagner et al. 2016, Kasper et al. 2015, Wagner et al. 2015, Draper et al. 2016). Typical performance will allow to detect massive planets (>5 MJ) on wide orbits (>10 AU) complementing the current performance of SPHERE in these young associations. This category of observations will provide information to address the question related to how planets form. Observations will be preferentially performed in K band with the additional ability to use narrow bands to derive near IR colors and to put constraints on some atmospheric properties like temperature and surface gravity.

3. **Planet-disk interactions**: The observation of circumstellar disks is strongly connected to that of exoplanets as the latter form in the former. Here again, with the same contrasts than those achieved with NACO, there are several interests to the imaging of protoplanetary and debris disks with MAORY/MICADO owing to the gain in angular resolution and sensitivity. There are hundreds of stars known with IR excesses detected from unresolved photometry in the IR (Beichman et al. 2006 for instance). The modelling of their spectral energy distribution informs us about dust properties but the knowledge of the dust spatial distribution is important to remove some degeneracies. MAORY/MICADO can potentially resolve some of the disks observed recently by Spitzer and still out of reach from SPHERE or even JWST (because they are too small or too faint). Measuring the size of the disks as well as their surface brightness is the key to determine the location of the planetesimal belts, which are the seeds for planets. In addition, near IR colors are important to measure the scattering efficiency and hence can bring constraints on the dust properties (Debes et al. 2009, for instance). For disks that are already known from previous observations, the improvement of angular resolution will be valuable to carry out dynamical studies of structures. Any structures (warps, clumps, offsets, spirals…) deviating from the simple assumption of an axi-symmetrical system is of interest in the context of planetary formation (see Lee & Chiang et al. 2016). Either some of these structures can be indirectly generated by planets or may lead to planet formation. Here the advantage of MAORY/MICADO is to resolve angularly these structures with unprecedented details, but also bring time resolution to determine their dynamics.

**Keywords:** Direct imaging – high-contrast and high angular resolution – exoplanets and circumstellar disks

**MICADO Pixel Scale / Fov:** 1.5mas/px and 20 arcsec FoV, motivated by the maximization of the performances in terms of Strehl ratio and by the small size of the actual targets.

**MICADO Observation mode:** Coronagraphy

**MICADO Spectral set-up:** none

**Filters required:** Narrow band filters: K_coro, H_coro, J_coro, NB-H-cont, CH4-short, CH4-long, H20_K, H20_J, NH3, Br-gamma, Pa-beta

**Estimate Survey Area/Sample Size/ Number of Images/Epochs:** Priority will be given to systems with known imaged planets and young circumstellar disks, young, nearby planetary systems discovered by GAIA and radial velocity surveys, and members of the Sco-Cen regions. Not limit on the sample size but the GTO time available.

**Average Integration time per image (magnitude of targets; S/N required):** depending on the contrast performances achievable with MAORY/MICADO. However, MAORY/MICADO will be used with the coronagraphic mode in angular differential imaging. Therefore, a typical observing sequence of 2hrs per target is required to maximize the field rotation.

**Observation requirements:** small dithering to remove detector defects; targets must be observed at meridian to maximize the field rotation and minimize instrumental aberrations.

**Strehl or EE required:** the best Strehl and spatial resolution performances

**Astrometric Accuracy:** no special requirement

**SCAO vs. MCAO:** SCAO

**Comparison with JWST or other facilities:** Spatial resolution is the key issue here: MAORY/MICADO is much better than JWST or planet imagers on 10m class Telescopes like SPHERE and GPI.

**Synergies with other facilities:** HARMONI that will be devoted to the spectral characterization of exoplanets

**Simulations made/needed to verify science case or feasibility:** Simulations of the contrast performances achievable with the E-ELT Telescope (number of segments, number of missing segments every night…) + MAORY-SCAO + MICADO are mandatory

**Origin of the targets:** Lists available in the literature,

**NGS:** the star itself (all sufficiently bright),

**Acquisition:** Finding charts available.

**Calibrations:** Astrometric field

**Data Processing Requirements:** Angular differential imaging processing as used with SPHERE at VLT,

**Any other comment:** none at this stage

# 4.4 Spectroscopic characterisation of sub-stellar objects

**Authors:** *D'Orazi V., Desidera S., Gratton R. G., Mesa D. (INAF - OAPD)*

**Brief description of science case:**

The physical characterisation of known sub-stellar companions, both brown dwarfs and planets, is becoming increasingly important as discoveries cumulate. Dedicated new instrumentation (such as e.g., SPHERE at VLT or GPI at Gemini-South) exploit high-contrast imaging observations, providing us with direct measurements of the thermal emission of young sub-stellar objects. Approximately 20 young sub-stellar (roughly M ≤ 75 $M_{Jup}$) companions have been found so far. In this respect, determining the fundamental atmospheric properties (i.e., chemical composition, surface gravity, clouds) of these systems is of paramount importance in order to shed light on their origin and evolution. However, photometric observations, though powerful, suffer degeneracy and spectroscopic information is sorely needed to investigate the above mentioned structural properties. Because of their location sufficiently far away from the host stars (see e.g., the Eta Tel system, Lowrance et al. 2000, where the companion is at ~ 4 arcsecond separation), several of these sub-stellar companions can be studied with no need of coronagraphic devices. Conversely, the long-slit spectroscopy (LSS) at medium resolution (R~18000 for point sources in IzJ band), as that foreseen for the MAORY+MICADO facility, will allow us not only to identify broad molecular bands (e.g., $CH_4$, $H_2O$), but also to resolve atomic features, which are critical diagnostics as to atmospheric characterisation. In particular, we can distinguish between young and old brown dwarfs by inspecting at gravity-sensitive spectral features. The radius of field (old) brown dwarfs varies only slightly with mass and age, and therefore the surface gravity is determined by the mass (*log g ~ 5*). Young objects can exhibit significantly lower surface gravities (10–100 times) than the more massive evolved dwarfs of the same spectral type. It has been shown in the literature, that the K I lines at 1.17/1.25 µm and/or the Na I line at 1.14 µm are very efficient gravity sensitive features (see e.g., Allers et al. 2007, ApJ, 657, 511; Allers & Liu 2013, ApJ, 772, 79).

Moreover, LSS observations will allow to shed light on *L-T transition* and on the characteristics of brown dwarfs and planet atmospheres, such as the presence and the structure of their cloud decks. The presence of condensate clouds is one of the most unique features of the ultra-cool atmosphere of directly imaged planets and brown dwarfs. The re-emergence of the 0.99 µm FeH feature in early- to mid-T spectral type has been suggested as evidence for cloud disruption where flux from deep, hot regions below the Fe cloud deck can emerge. The same mechanism could account for colour changes at the L/T transition and photometric variability.

Along with sub-stellar companions, the LSS will be crucial to investigate also the atmospheric properties of the so called "free-floating" objects in young clusters, allowing us to infer information on their possible differences with respect to bound companions.

Finally, we will be able to cover crucial diagnostics of accretion phenomena occurring in very young sub-stellar objects, as for example the case of GQ Lup b (Seifahrt et al. 2007).

**Keywords:** spectroscopy– sub-stellar objects – brown dwarfs – spectral characterisation

**MICADO Observation mode:** Slit Spectroscopy

**MICADO Pixel Scale / Fov:** only 4mas/px is offered for Slit Spectroscopy mode.

**MICADO Spectral set-up:** Slit 16 mas x 4 arcsec. Wavelength coverage: 0.8 – 1.45 µm as a first priority (the J band include critical spectral features).

**Filters required:** none.

**Estimate Survey Area/Sample Size/ Number of Images/Epochs:** As for sub-stellar companions, the target list (approximately 4-5 objects) is based on current information furnished from past and ongoing surveys. In the following years, new discoveries will significantly enlarge the sample. For isolated, field objects the typical sample size is approximately 10 targets per cluster/association.

**Average Integration time per image (magnitude of targets; S/N required):** We have calculated that we will be able to reach SNR=30 in 1 hour of integration for objects with H=21 and SNR=10 for H=23 objects, implying that in very young systems (1-5 Myr) objects less massive than Jupiter will be investigated.

**Observation requirements:** Dithers not crucial for this kind of observations.

**Strehl or EE required:** High SRs are required.

**Astrometric Accuracy:** no special requirement.

**SCAO vs. MCAO:** SCAO is the obvious choice for science cases focussing on spectral characterisation sub-stellar companions; MCAO can be exploited when isolated (not bound) sub-stellar objects in populated clusters are under scrutiny.

**Comparison with JWST or other facilities:** The great advantage with respect to JWST is obviously provided by the higher spectral resolution, which will allow us to resolve several molecular and atomic features.

**Synergies with other facilities:** The complementarity with high-contrast imaging facilities such as e.g., SPHERE and GPI is crucial: these instruments will indeed provide with new targets.

**Simulations made/needed to verify science case or feasibility:** Given the uncertainty related to the expected performances for this observing mode, simulations are needed to investigate the feasibility of our program (inspecting SN, resolution).

**Origin of the targets:** SPHERE and GPI datasets will provide new targets when ELT will be operational.

**NGS:** Magnitudes of the stars hosting sub-stellar companions are usually brighter than V~12, so that no issue for SCAO. For free-floating planets, typical star magnitudes are suitable for MCAO; the crowding of these fields ensures will have at least one star in the FoV.

**Acquisition:** no particular requirement.

**Calibrations:** Standard. No special astrometric requirement.

**Data Processing Requirements:** none.

**Any other comment:** none.

# 4.5 Formation and evolution of the minor bodies of the Solar System

**Authors:** Douté S., Chauvin G., (IPAG, CNRS) et al.

**Description of science cases:**

### 4.5.1 Size, shape, and chemical composition of asteroids and trans-Neptunian objects

Asteroids, Centaurs and trans-Neptunian objects (a.k.a Kuiper Belt Objects: KBO) are the most primitive planetary bodies in the solar system thus reflecting the conditions of its formation and evolution. A preliminary feasibility study indicates that about 950 asteroids of an estimated size more than 40 km could be resolved and characterized from the point of view of size, shape, and chemical composition with MAORY+MICADO. A large observation program could be possible to better understand the collisional history, the large-scale shattering (breccia), and the aging of the surface by space weathering of the objects in the main belt. Composition analyses make the link with meteorites and interplanetary dust. As for trans-Neptunian objects and Centaurs, the focus is on the greatest representatives of this family that could be resolved with MAORY+MICADO. Despite the faintness of these distant objects (magnitude 16 to 24 in the H band) a relatively high signal to noise ratio can be expected for high-resolution spectra thanks to the collective power of the telescope. Thus we can expect to get the composition of ice and some organic as well as to determine the surface temperature by the use of spectral tracers in fairly good conditions. The dwarf planet Pluto, which was the subject of exploration at close range by the New Horizons spacecraft in July 2015, will be a prime target for the ELT. Despite extremely low temperatures, the largest trans-Neptunian object shows a surprisingly high activity due to the high volatility of its surface ices (N2, CH4, CO). At the diffraction limit, Pluto would be resolved in the best case with a pixel of approximately 200 km allowing to monitor the area after New Horizons in very good conditions.

### 4.5.2 High-resolution astrometry as a tool to characterize the physical and orbital properties of asteroids (and KBO).

There is strong evidence that most asteroids in the main belt are rubble piles rather than single coherent objects. Rubble piles are composed of a gravitationally bound collection of smaller bodies and internal void spaces. Collisional fragments may also form binary or multiple systems.

Indeed, many minor planets orbit the sun as gravitationally bound pairs. When the bodies are of comparable size, these are known as binaries, whereas when one body is much smaller than the other, the smaller objet is generally referred to as a satellite (or moon) of the larger objet. Most bound pairs in the asteroid belt lie within the satellite regime whereas many known Kuiper belt pairs are binary (currently of the order of 30 systems have well characterized mutual orbits). Such systems however are usually not long lived. Almost 200 minor planets have been identified as multiple systems. About 2% of all main belt and Trojan asteroids with R>10 km are known binaries or multiplets. Over 10% of the TNO are in multiple systems, most of which have been discovered on HST images.

Bound pairs can be used to determine masses and thus densities of individual minor planets via Kepler's third law. Eventually, more detailed observations of orbits may yield information on the internal mass distribution as well. Once a body's mass is known, its density can be determined if its shape and size are known. A body's density yields invaluable information on its internal structure if its composition is known (e.g. via spectroscopy). The masses of a few of the largest asteroids have been estimated from the perturbations that they cause on the heliocentric orbits of other asteroids and spacecraft tracking also yielded some mass estimates.

Only a tiny fraction of the minor planet population has density measurements.

The distribution of mass ratios and orbital characteristics of multiple systems provide constraints on the origin, collisional history and tidal evolution of minor planets.

*This Science Cases are part of the MAORY ST initiative on planetary science.*

**Keywords:** imaging – spectroscopy – astrometry, asteroids and Kuiper Belt Objects

### 4.5.3 Identification of the best objectives for MAORY+MICADO and other facilities

Based on a review and analysis presented in the Annexe (Section 4.4.5) we summarize in the form of a table what we think are the best objectives for a selection of present and future observing facilities. We distinguish the asteroids (Table 4.5-1) and the KBO (Table 4.4.3-2) as two independent categories of objects.

| ASTEROIDS | SPHERE @VLT | NIRcam/Spec @JWST | MAORY+MICADO @ELT |
|---|---|---|---|
| size/albedo/thermal | | X | |
| shape | X | | X |
| spectral detection of components | | X | |
| spectral mapping of components | | X | X |
| binarity | X | | X |
| density, internal structure by astrometry | | | X |
| dwarf planets (overall investigation) | ? | X | X |

Table 4.5-1 Best use of different observing systems for characterizing the asteroids

| Kuiper Belt Objects | SPHERE @VLT | NIRcam/Spec @JWST | MAORY+MICADO @ELT |
|---|---|---|---|
| size/albedo/thermal | | X | |
| shape | | | X |
| spectral detection of components | | X | |
| spectral mapping of components | | | X |
| binarity | X | | X |
| density, internal structure by astrometry | ? | | X |
| dwarf planets (overall investigation) | | | X |

Table 4.5.3-2 Best use of different observing systems for characterizing the KBO.

Our analysis leads to the conclusion that MAORY/MICADO will be especially competitive to determine precisely the size and the shape of the main asteroids and KBOs (dwarf planets) as well as to map the distribution of certain minerals and ices on their surfaces. Besides, the high astrometric precision of the instrument could be crucial to determine the bulk properties of some objects with companions. As a consequence, the main objectives we propose to pursue with MAORY/MICADO taking advantage of its competitive strengths are:

1. High angular resolution => volume and shape model of a large number of minor objects,

2. High astrometric precision to measure the dynamics of asteroidal and KBO systems => mass, density, and possibly internal structure of the primary and secondary,

3. NIR imaging in multiple narrow band filters => assessing the compositional diversity at the surface of an object,

4. long slit spectroscopy => detailed composition and physical state but mapping challenging because the scanning of the object within the slit must be performed in parallactic angle.

In particular observing the main dwarf planets with MAORY/MICADO will allow an overall investigation of their topography and composition with an unprecedented spatial resolution of the order of 12 km.pixel-1 for the rocky objects (Vesta, Pallas, Hygiea) and 250 km.pixel-1 for the icy objects (Pluto, Haumea, Makemake, Orcus, Quaoar).

## 4.5.4 Representative observations

In the following we propose a series of representative observations that embody the previous objectives (see Table 4.4.4-1).

| Scientific target | Apparent diameter (primary) mas | AO mode | guide object (NGS) | proper motion (mas s⁻¹) | mag. Primary (H band) | apparent separation mas | mag. diff. | obs. mode | spatial res. Km pix⁻¹ |
|---|---|---|---|---|---|---|---|---|---|
| Rocky dwarf planet (Pallas) | 372 | SCAO | =scientific target | 5 | 8 | - | - | SIMG, LSS | 11 |
| Asteroid (1359 Prieska) | 34 | MCAO or SCAO | =star or scientific target | 8 | 15 | - | - | SIMG | 10 |
| Asteroidal pair (762 Pulcova) | 100 | SCAO | =scientific target | 9 | 12 | 446 | 4 | SIMG, AIMG | 13 |
| Asteroidal pair (3673 Levy) | <8 | MCAO | star | 10 | 16 | 13 | 2.8 | AIMG | 8 |
| Icy dwarf planet (Pluto) | 79 | MCAO | star | <1.4 | 16 | | | SIMG, LSS | 230 |
| Icy dwarf planet (make make) with a satellite | 45 | MCAO | star | <1 | 16-17 | 640 | 4.5 | SIMG, LSS, AIMG | 250 |
| Icy dwarf planet (Eris) with a satellite | 50 | MCAO | star | <1 | 18 | 706 | 1 | SIMG, LSS, AIMG | 420 |
| KBO Ixion | 22 | MCAO | star | <1.4 | 20 | - | - | SIMG, LSS | 230 |
| KBO 2001 RZ143 | < 8 | MCAO | star | <1.0 | 23 | 51 | 0 | AIMG, LSS | 300 |

Table 4.5.4-1 Representative scientific cases that can be fruitfully addressed by MAORY+MICADO. Code used for the observing mode: SIMG->Standard Imaging, AIMG->Astrometric Imaging, LSS->long slit spectroscopy.

**MICADO Pixel Scale/Fov:** 1.5mas/px and 20 arcsec FoV, motivated by the maximization of the performances in terms of SR and by the small size of the actual targets.

**MICADO Observation mode:** Standard Imaging, Astrometric Imaging, Long Slit Spectroscopy depending on the science case

**MICADO Spectral set-up:** LSS of planetary minor bodies (most of which being resolved) should preferentially be performed with a slit width of 16 mas to allow some scanning in the dimension perpendicular to the slit. The wavelength coverage is maximum: 0.8-2.4 microns requiring two separate acquisitions: IzJ and HK bands.

**Filters required:** J, H and K seem the best compromise between AO performances and spectral features that are crucial for the scientific goal. Acquisitions in the I and z bands may suffer from a significant decrease in AO performances but are mandatory for the spectroscopic characterization of asteroids.

**Estimate Survey Area/Sample Size/ Number of Images/Epochs:** Based on the analysis presented in section 5.4.5, the asteroidal sample that can be addressed fruitfully with MAORY+MICADO amounts approximately 1000 single objects and 70 multiple systems. When considering the best objectives identified in table 4.4.1 the target list could be reduced to a dozen of asteroids and pairs. As for the KBO, the sample that can be addressed fruitfully with MAORY+MICADO amounts approximately 100 single objects and 80 multiple systems. When considering the best objectives the target list could be reduced to the 6 main KBO (dwarfs icy planets) and a dozen of pairs.

**Average Integration time per image (magnitude of targets; S/N required):** TBC

**Observation requirements:** [refer to table 8 of MICADO OCD]

**Strehl or EE required:** the best spatial resolution performances, the most effective would be the characterization of our scientific targets.

**Astrometric Accuracy:** requirements in terms of astrometric accuracy have yet to be determined by conducting detailed feasibility studies (see below).

**SCAO vs. MCAO:** the required AO mode depends on the scientific target magnitude.

**Synergies with other facilities:** compositional mapping of our planetary science cases by spectroscopy will be achieved more easily with the IFU HARMONI than the MICADO instrument provided that the former could be assisted with an AO system adapted to observing the objects of interest. As noted in Table 4.4.4-1 a MCAO or a LTAO mode is required to resolve faint asteroids and KBO (mag.<15 in H band) because the limiting magnitude for optimum correction in SCAO mode with the scientific target used as an NGS is of the order of 12 to 16 for optimal performances.

**Simulations made/needed to verify science case or feasibility:** The possibility of using extended (D~40-400 mas) scientific targets as SCAO NGS has to be verified. Detailed feasibility studies need to be conducted in particular on the use of high precision astrometry to constrain the internal structure of the minor bodies. This is an innovative and challenging task. Evaluating the capability of the instrument to measure the mutual orbit of a gravitationally bound pair with very high accuracy requires to model the instrumental response as a function of time for a non sidereal system. The feasibility study will rely also on the quality of the inversion scheme of a dynamical physical model.

**Origin of the targets:** Lists available in provided by the Small Bodies Node (SBN) which is part of NASA's Planetary Data System (PDS) and by Wm. Robert Johnston.

**NGS:** minor bodies of the solar system are usually not far from the ecliptic with exceptions among the Near Earth Asteroids and some classes of Kuiper Belt Objects. The scientific target itself can be used as a NGS in the SCAO mode and possibly, in the MCAO mode (to be verified) depending on its apparent magnitude and diameter.

**Acquisition:** Compositional mapping with MICADO requires scanning the object with the slit while working in parallactic angle. A detailed feasibility must be conducted depending on the apparent diameter, magnitude and proper motion of the target. When targeting asteroids one must take care of the non-sidereal proper motion which can be very significant..

**Calibrations:** Standard.

**Data Processing Requirements:** TBC.

**Any other comment:** the typical size of our targets is much smaller than the FoV size (~10-1000 mas vs 20 arcsec).

## 4.5.5 Annexe: expected performance of MAORY+MICADO

There is an interest for evaluating the competitive advantage of MAORY+MICADO and HARMONI against other ground and space observing systems at various time frames. We restrict the comparison to the scientific cases of interest: minor bodies of the Solar System. The main competitors we chose are (i) the SPHERE instrument currently operating on the VLT with its state-of-the-art set of instruments and superior adaptive optics SAXO. (ii) the JWST which will start operations with a high performing instrumental suite for planetary science: NIRcam and NIRspec. In the following section we provide plots and statistics based on data provided by the Small Bodies Node (SBN) which is part of NASA's Planetary Data System (PDS) and by Wm. Robert Johnston. We test the capability of the different instruments (each in two spectral channels) to provide spatially resolved images of the objects and to separate the primary and the secondary in the case of a binary system.

## 4.5.5.1 Expected performances on asteroids.

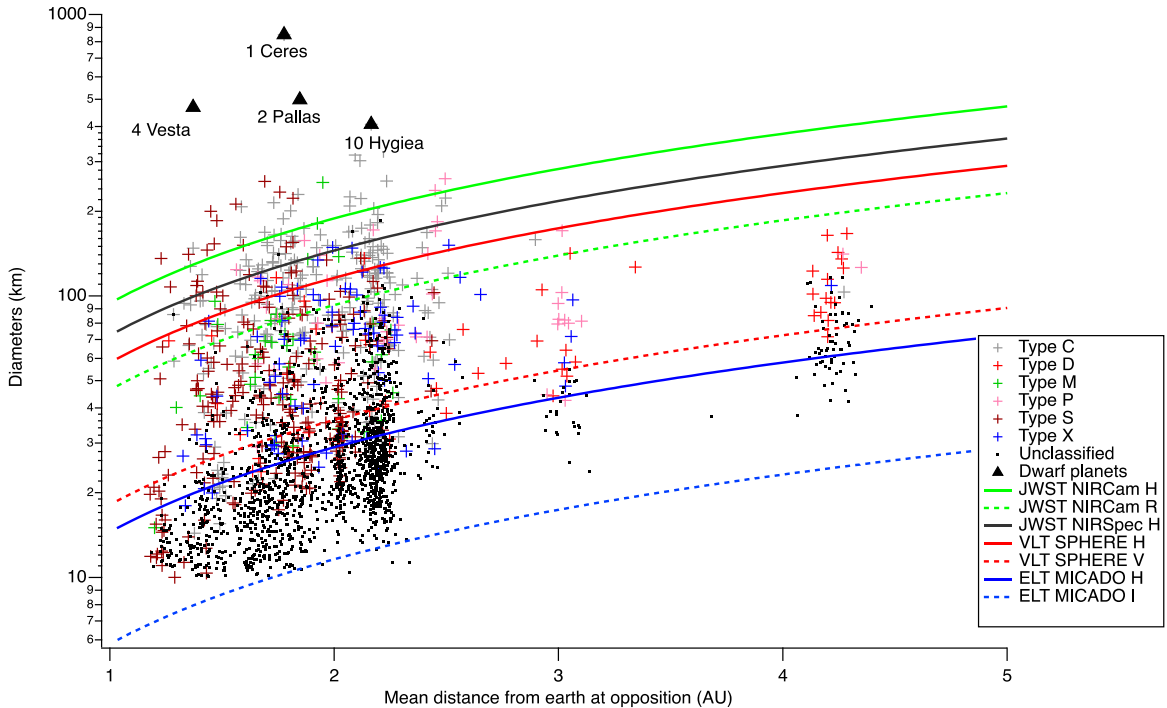

Figure 4.5.5.1-1 Scatter plot of the asteroids as a function of their diameter and of their mean distance from earth at opposition. Different symbols/colors correspond to different types of asteroids: C -> carbonaceous, D -> outer belt and Trojans, M -> stony iron or iron asteroids, P -> Outer and extreme asteroids, S -> stony asteroids, X -> high albedo enstatite asteroids, and U -> unclassified. The biggest asteroids (dwarf planets) are emphasized by a special symbol. Superimposed are monotonic curves corresponding to the resolving power of different combinations of instruments and channels (ELT MAORY+MICADO I 8mas and H 20mas; VLT SPHERE V 25mas and H 80mas; JWST NIRcam R 64mas and H 130mas, JWST NIRSpecH 100mas).

| | JWST NIRSpec H | JWST NIRCam H | JWST NIRcam R | VLT SPHERE H | VLT SPHERE V | ELT MICADO H | ELT MICADO I |
|---|---|---|---|---|---|---|---|
| 1 Ceres | 6 - (130) | 5 - (168) | 10 - (83) | 8 - (104) | 26 - (33) | 33 - (26) | 77 - (11) |
| 4 Vesta | 5- (100) | 4 - (130) | 7 - (64) | 6- (80) | 19 - (25) | 23 - (20) | 59 - (8) |
| 2 Pallas | 4 - (134) | 3 - (174) | 6 - (86) | 5 - (107) | 15 - (34) | 18 - (27) | 45 - (11) |
| 10 Hygiea | 3 - (157) | 2- (204) | 4 - (100) | 3 - (126) | 10 - (40) | 13- (32) | 31 - (13) |
| 511 | 2 - (161) | 1 - (210) | 3 - (103) | 2 - (129) | 8 - (41) | 10 - (33) | 25 - (13) |

| Davida | | | | | | | |
|--------|--|--|--|--|--|--|--|

Table 4.5-4 Resolving power of the different instruments/channels for the biggest asteroids. First number: number of pixels across the diameter (Second number: kilometers per element of resolution).

| | JWST NIRSpec H | JWST NIRCam H | JWST NIRcam R | VLT SPHERE H | VLT SPHERE V | ELT MICADO H | ELT MICADO I |
|--------|-----|-----|-----|-----|-----|-----|------|
| Type C | 57 | 20 | 146 | 104 | 238 | 252 | 270 |
| Type D | 0 | 0 | 7 | 3 | 38 | 47 | 49 |
| Type M | 3 | 3 | 10 | 6 | 34 | 40 | 41 |
| Type P | 7 | 2 | 17 | 13 | 43 | 45 | 46 |
| Type S | 19 | 12 | 50 | 31 | 161 | 203 | 252 |
| Type X | 1 | 0 | 18 | 7 | 67 | 77 | 84 |
| Unclas. | 4 | 0 | 14 | 7 | 333 | 575 | 1614 |

Table 4.5-5 Approximate number of asteroids which apparent diameter is higher or equal to the angular resolution for each combination of asteroid class and instruments/channels.

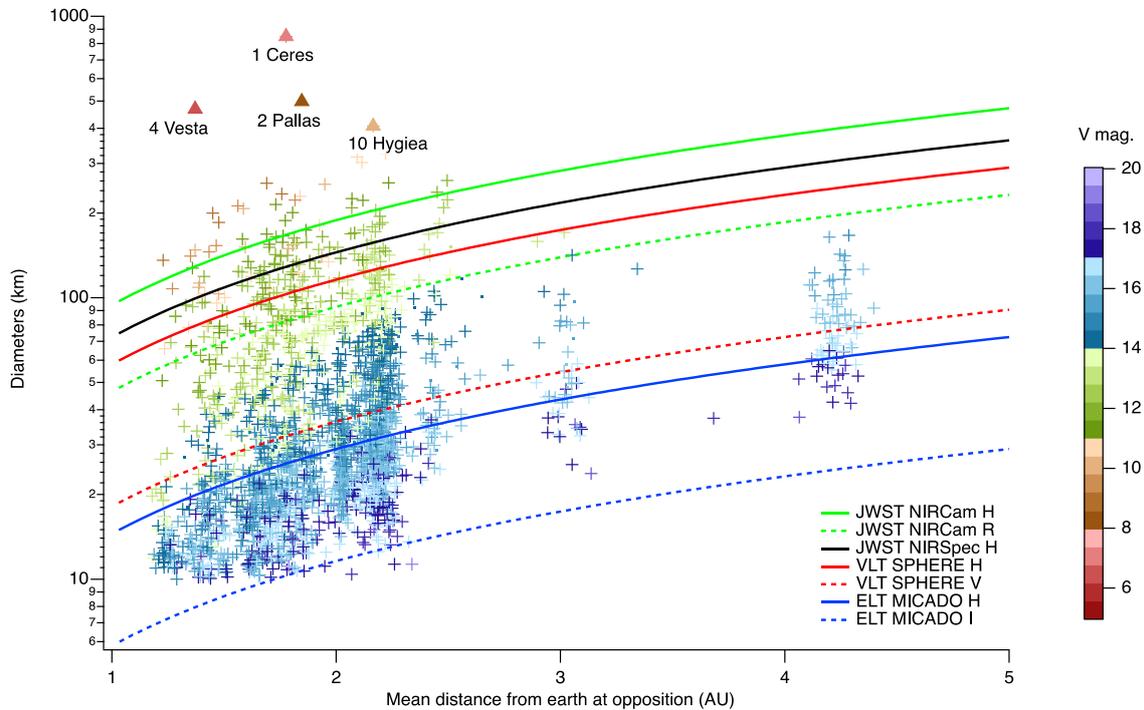

Figure 4.5.5.1-2 Scatter plot of the asteroids as a function of their diameter and of their mean distance from earth at opposition. Different colors correspond to different visual apparent magnitude. The biggest asteroids (dwarf planets) are emphasized by a special symbol. Superimposed are monotonic curves corresponding to the resolving power of different combinations of instruments and channels (ELT MAORY+MICADO I 8mas and H 20mas; VLT SPHERE V 25mas and H 80mas; JWST NIRcam R 64mas and H 130mas, JWST NIRSpecH 100mas).

**In the SCAO mode of SPHERE**, NGS with magnitude H ≤ 15 are preferred to obtain an on-axis SR up to 0.9 at λ = 1.6 μm. Consequently, **only asteroids with diameter approximately > 30 km and with mean distance < 2.5 AU can be used as a guide object** for the SCAO in addition to being the target.

**In the SCAO mode of MAORY**, bright NGS (V ≤ 12) are preferred to obtain an on-axis SR > 0.6 at λ = 2.2 μm. Consequently, **only asteroids with diameter approximately > 70 km and with mean distance < 2.5 AU can be used as a guide object** for the SCAO in addition to being the target. The other asteroids may require the operation of the MCAO.

**In the MCAO mode of MAORY**, approximate faint magnitude limit for a star to be eligible as a NGS is H ~ 18.0-19.0. Consequently, we can imagine using **all asteroids as guide objects** in addition to being the target.

When using asteroids as guide objects one must take care of the non-sidereal proper motion which can be very significant.

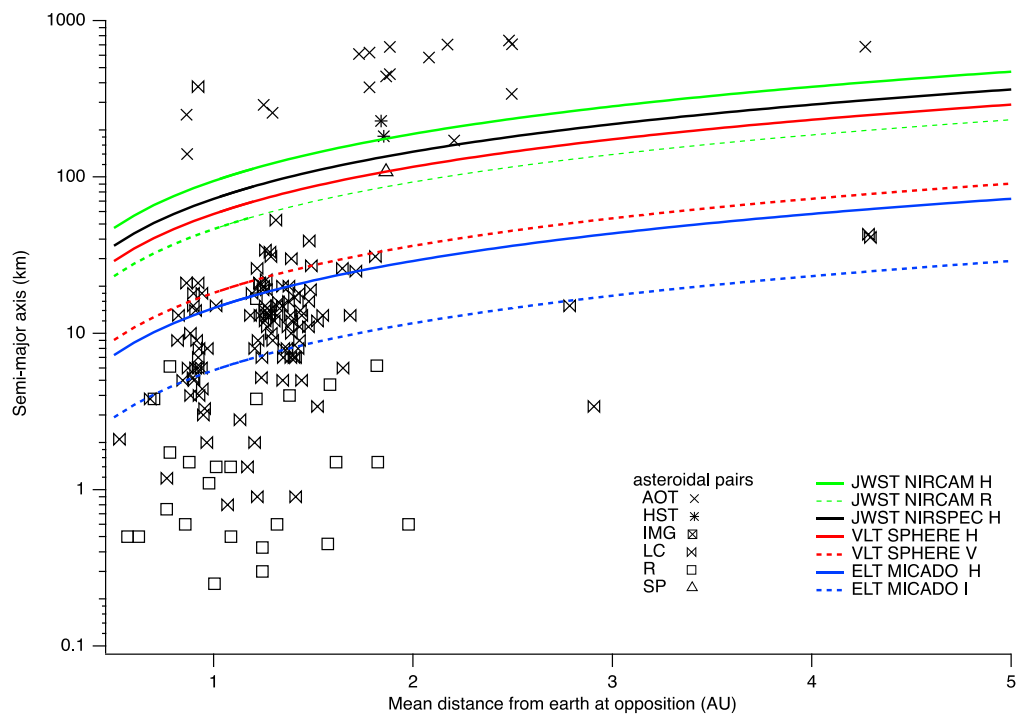

Figure 4.5.5.1-3 Scatter plot of the asteroidal pairs as a function of the primary-secondary separation and of their mean distance from earth at opposition. The symbols indicate the detection method for secondaries (AOT = direct imaging, adaptive optics telescope; HST = direct imaging, Hubble Space Telescope; IMG = direct imaging, other ground based telescope; LC = photometric light curve; R = radar; SP = direct imaging by space probe). Superimposed are monotonic curves corresponding to the resolving power of different combinations of instruments and channels (ELT MAORY+MICADO I 8mas and H 20mas; VLT SPHERE V 25mas and H 80mas; JWST NIRcam R 64mas and H 130mas, JWST NIRSpecH 100mas).

We note that MICADO@ELT will allow the direct imaging of a significant number of asteroidal pairs that are currently suspected from light curve analysis, the limit in terms of mean separation being from 5 to 10 km approximately depending on the geocentric distance. In advance SPHERE@VLT will have explored pairs with mean separation down to 10-50 km depending on the geocentric distance. Of the order of 40 closely bound pairs presently recognized by radar or light curve analysis will remain out of reach whatever the observing system. The JWST instruments will not do better than what is currently achieved in direct imaging with adaptive optics telescopes or the HST.

Figure 4.4.5.1-4 tells us that, for most targets (asteroidal pairs) of interest for MICADO@ELT, the primary and secondary have a difference of apparent magnitude less than 4 except for a few individual targets for which the separation is higher than about 120 km (130 mas).

|  | JWST NIRSpec H | JWST NIRCam H | JWST NIRcam R | VLT SPHERE H | VLT SPHERE V | ELT MICADO H | ELT MICADO I |
|---|---|---|---|---|---|---|---|
| **Resolved Asteroidal pairs** | **30** | **28** | **32** | **32** | **47** | **66** | **121** |

Table 4.5.5.1-3 Approximate number of presently detected asteroidal pairs, which mean separation is higher or equal to the angular resolution for each combination of asteroid class and instruments/channels.

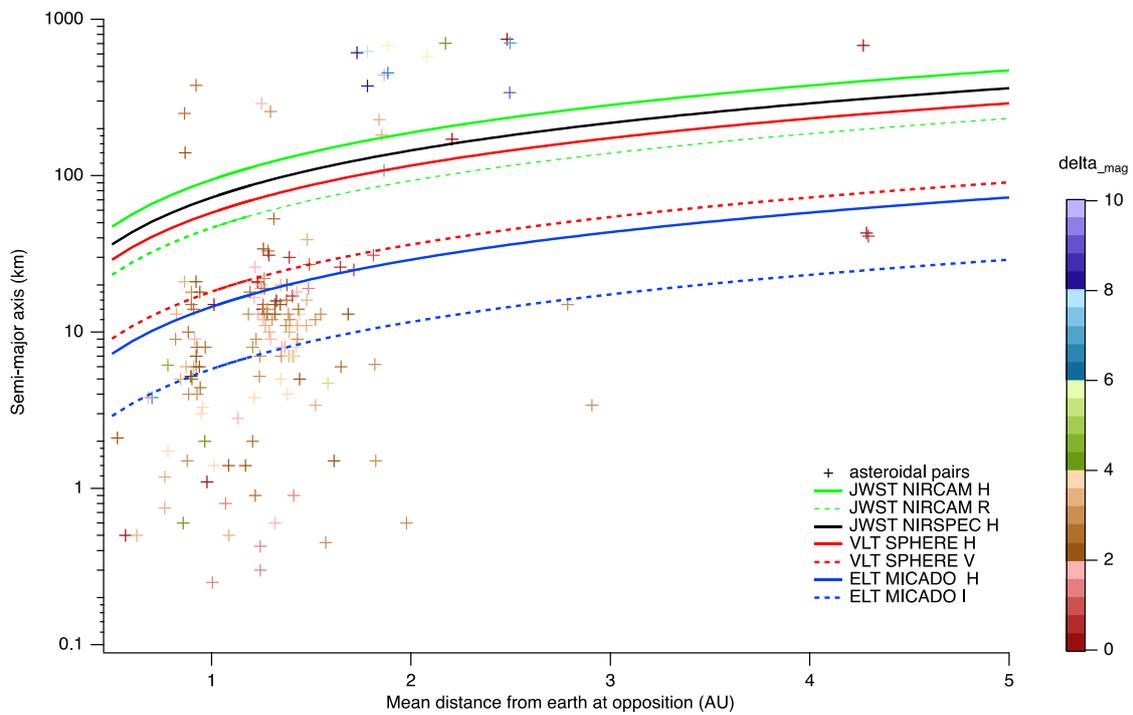

Figure 4.5.5.1-4 Scatter plot of the asteroidal pairs as a function of the primary-secondary separation and of their mean distance from earth at opposition. The color-coding indicates the difference of magnitude between the two objects. Superimposed are monotonic curves corresponding to the resolving power of different combinations of instruments and channels (ELT MAORY+MICADO I 8mas and H 20mas; VLT SPHERE V 25mas and H 80mas; JWST NIRcam R 64mas and H 130mas, JWST NIRSpecH 100mas).

### 4.5.5.2 Expected performances on the KBO

**In the SCAO mode of SPHERE**, NGS with magnitude H ≤ 15 are preferred to obtain an on-axis SR up to 0.9 at λ = 1.6 μm. Consequently, **no KBO, even Pluto, can be used as a guide object for the SCAO in addition to being the target**. Observing the KBO in diffraction-limited mode of VLT requires regular stars in the vicinity: difficult!

**In the SCAO mode of MAORY**, bright NGS (V ≤ 12) are preferred to obtain an on-axis SR > 0.6 at λ = 2.2 μm. Consequently, **no KBO, even Pluto, can be used as a guide object for the SCAO in addition to being the target**. Observing the KBO in diffraction limited mode thus requires the operation of the MCAO.

**In the MCAO mode of MAORY**, approximate faint magnitude limit for a star to be eligible as a NGS is H ~ 18.0-19.0. Consequently, we can imagine **using the main KBO (dwarfs icy planets) as guide objects** in addition to being the target. For the other KBO stars with adequate characteristics are to be found within the patrol field.

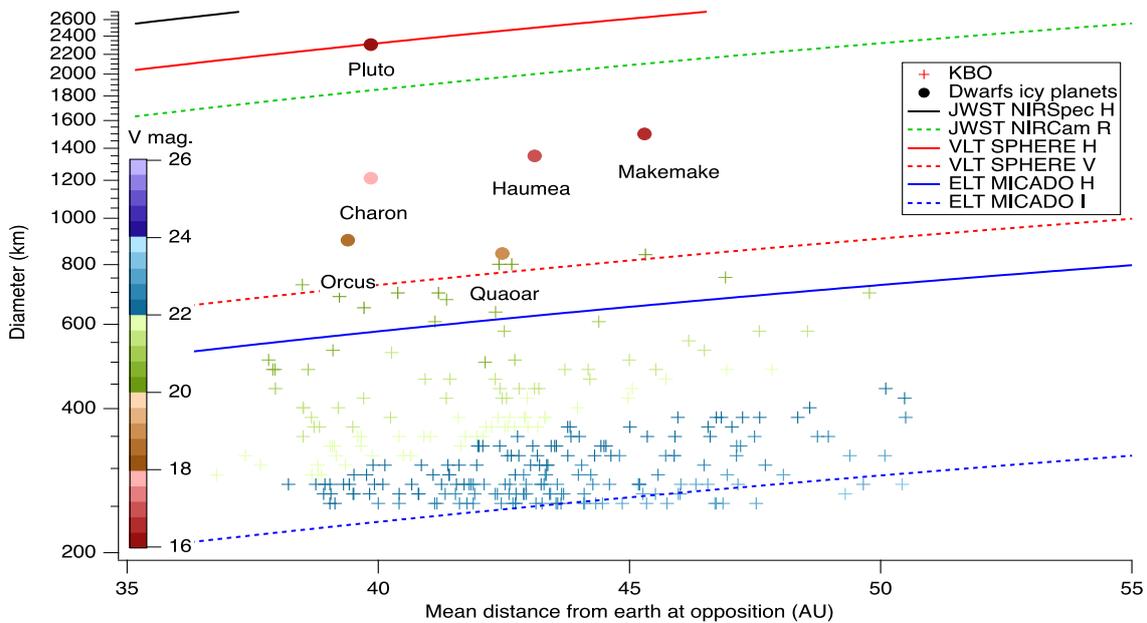

Figure 4.5.5.2-1 Scatter plot of the KBO as a function of their diameter and of their mean distance from earth at opposition. The color-coding indicates the average H apparent magnitude. The biggest KBO (dwarf icy planets) are emphasized by a special symbol. Superimposed are monotonic curves corresponding to the resolving power of different combinations of instruments and channels (ELT MAORY+MICADO I 8mas and H 20mas; VLT SPHERE V 25mas and H 80mas; JWST NIRcam R 64mas and H 130mas, JWST NIRSpecH 100mas).

| | JWST NIRSpec H | JWST NIRCam H | JWST NIRcam R | VLT SPHERE H | VLT SPHERE V | ELT MICADO H | ELT MICADO I |
|---|---|---|---|---|---|---|---|
| **Pluto** | - | - | **1 (1849)** | **1 (2306)** | **3 (722)** | **4 (578)** | **10 (231)** |
| **Makemake** | - | - | - | - | **1 (822)** | **2 (658)** | **6 (263)** |
| **Haumea** | - | - | - | - | **1 (782)** | **2 (626)** | **5 (250)** |
| **Charon** | - | - | - | - | **1 (722)** | **2 (578)** | **5 (231)** |
| **Orcus** | - | - | - | - | **1 (714)** | **1 (571)** | **4 (228)** |
| **Quaoar** | - | - | - | - | **1 (771)** | **1 (617)** | **3 (247)** |

Table 4.5-6 Resolving power of different instruments for the biggest KBO. First number: number of pixels across the diameter (Second number km: per element of resolution). The symbol '-' means object not spatially resolved.

| JWST NIRSpec H | JWST NIRCam H | JWST NIRcam R | VLT SPHERE H | VLT SPHERE V | ELT MICADO H | ELT MICADO I |
|---|---|---|---|---|---|---|
| **0** | **0** | **1** | **1** | **11** | **21** | **205** |

Table 4.5-7 Approximate number of KBO which apparent diameter is higher or equal to the angular resolution for each instrument/channels.

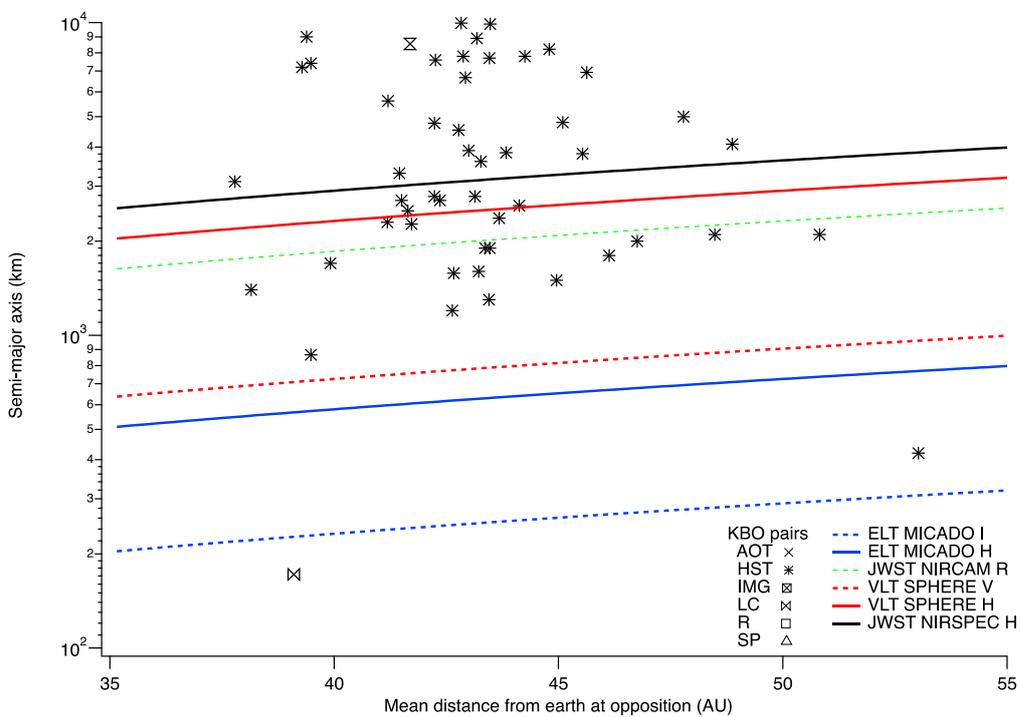

Figure 4.5.5.2-2 scatter plot of the KBO pairs as a function of the primary-secondary separation and of their mean distance from earth at opposition. The symbols indicate the detection method for secondaries (AOT = direct imaging, adaptive optics telescope; HST = direct imaging, Hubble Telescope; IMG = direct imaging, other ground based telescope; LC = photometric light curve; R = radar; SP = direct imaging by space probe). Superimposed are monotonic curves corresponding to the resolving power of different combinations of instruments and channels (ELT MAORY+MICADO I 8mas and H 20mas; VLT SPHERE V 25mas and H 80mas; JWST NIRcam R 64mas and H 130mas, JWST NIRSpecH 100mas).

MAORY+MICADO will be able to separate almost all presently detected KBO pairs but this is also true with SPHERE@VLT and, to a lesser extent, with the JWST suite. However, MAORY+MICADO will certainly open a new window on KBO pairs with separation down to 200-300 km depending on the geocentric distance. Figure 4.4.5.2-3 tells us that, for most targets (KBO pairs) of interest for MAORY+MICADO, the primary and secondary have a difference of apparent magnitude less than 4.

| | JWST NIRSpec H | JWST NIRCam H | JWST NIRcam R | VLT SPHERE H | VLT SPHERE V | ELT MICADO H | ELT MICADO I |
|---|---|---|---|---|---|---|---|
| **Resolved KBO pairs (among the already detected ones)** | **56** | **49** | **66** | **62** | **84** | **84** | **86** |

Table 4.5-8 Approximate number of presently detected KBO pairs, which mean separation, is higher or equal to the angular resolution for each combination of asteroid class and instruments/channels

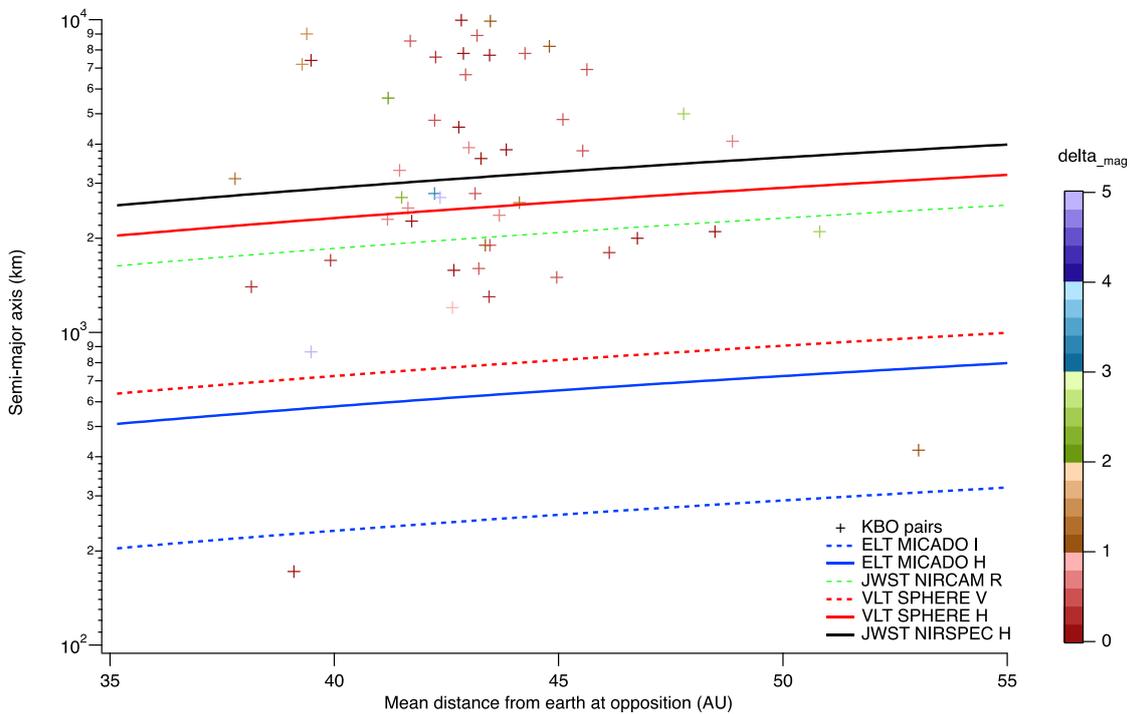

Figure 4.5.5.2-3 Scatter plot of the KBO pairs as a function of the primary-secondary separation and of their mean distance from earth at opposition. The color-coding indicates the difference of magnitude between the two objects. Superimposed are monotonic curves corresponding to the resolving power of different combinations of instruments and channels channels (ELT MAORY+MICADO I 8mas and H 20mas; VLT SPHERE V 25mas and H 80mas; JWST NIRcam R 64mas and H 130mas, JWST NIRSpecH 100mas).

# 5. Nearby stellar systems

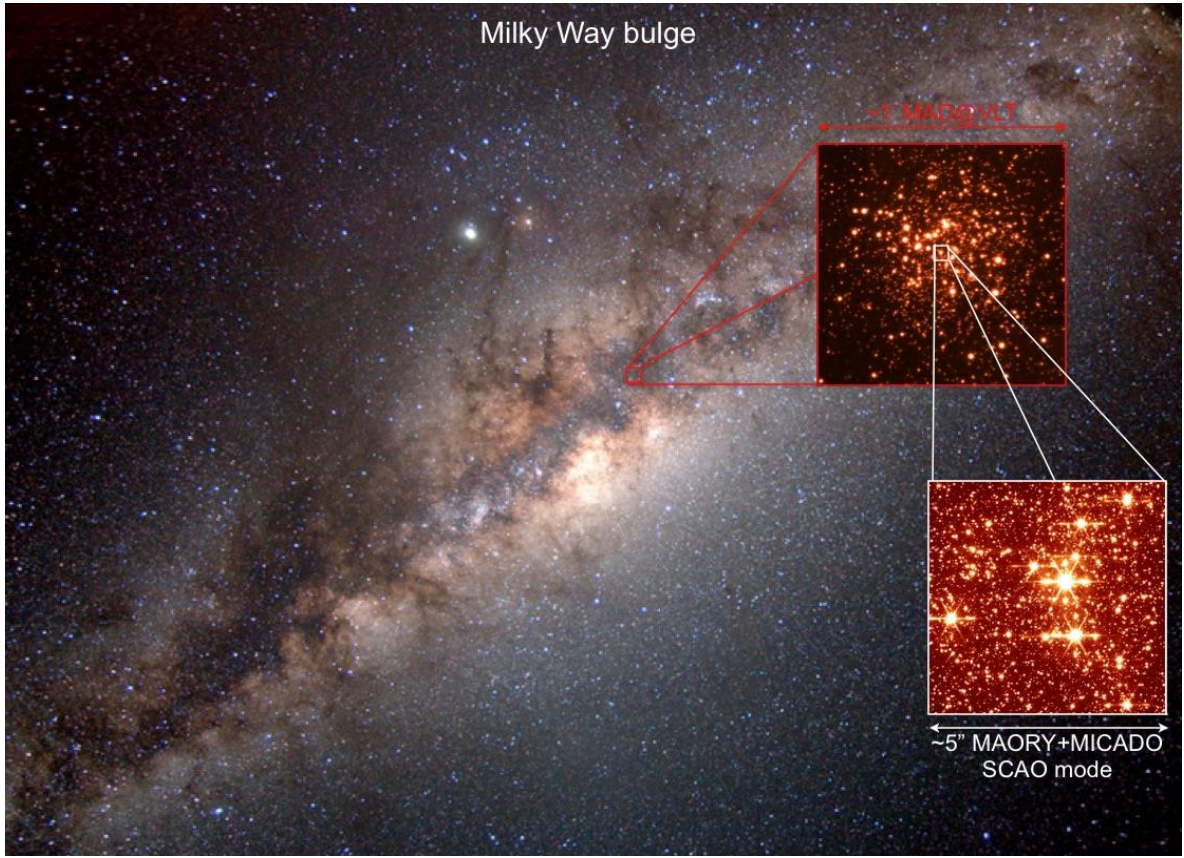

Figure 5-1 Zooming on a candidate building block of the Galactic Bulge with Adaptive Optics. The globular cluster Terzan 5 as imaged by the MCAO instrument MAD@VLT (real image) and by MAORY+MICADO in SCAO mode (Phase-B simulation).

# 5.1 The fraction of white dwarfs in Galactic globular clusters

**Authors:** *Sollima A. (INAF - OABO)*

**Brief description of science case:**

Since the beginning of the stellar astrophysics, globular clusters (GCs; containing millions of stars and formed by a relatively simple stellar population) constituted a benchmark for developing and testing stellar evolution, population synthesis and dynamical models. However, in spite of their proximity, for decades only a fraction of their stellar population has been accessible by observational facilities due to the intrinsic faint magnitude of the least massive Main Sequence (MS) stars and the prohibitive crowding conditions occurring in the core of these stellar systems. The advent of the Hubble Space Telescope (HST) allowed us to study the stellar populations of GCs down to the lower MS reaching in a few cases the hydrogen-burning limit. Dark remnants left by the evolution of stars with masses larger than the mass at the tip of the Red Giant Branch are instead several orders of magnitude fainter than other stars at optical/NIR wavelengths, so that only a fraction of bright White Dwarfs (WDs) are actually observable in a handful of GCs. For this reason, the actual fraction of dark remnants must be assumed as a free parameter, preventing any definitive conclusion on their influence on the cluster dynamics, on the existence of intermediate-mass black hole and on the possible presence of non-baryonic dark matter. This hidden mass has also an impact on the uncertainty of the estimated M/L ratios.

WDs are expected to constitute more than 90% of the baryonic dark mass of GCs, so the estimate of their fraction provides a strong constraint to the actual budget of dark mass. Moreover, the location in the color-magnitude diagram of the WD cooling sequence can be used as a good distance indicator and the peak of the WD luminosity function (LF; where more than 70% of WDs is located) is a sensitive and robust indicator of the age of the cluster, fully independent of the usual clock adopted to age date GCs (the MS turn-off, Bono et a. 2013, A&A, 549, 102), being less sensitive to both metallicity and adopted convective transport theory. Furthermore, because of the significant mass-loss occurring during the planetary nebula-WD transition and two-body relaxation, the radial distribution of WDs traces the cluster relaxation time: recently formed (bright) WDs preserve the concentrated distribution of their progenitors, while older (faint) WDs have a more extended distribution. Therefore, the ratio of the relative fractions of WDs in the center and at the half-mass radius as a function of magnitude allows us to study also the mass segregation timescale in these stellar systems. Finally, the faintest WDs are the remnant of massive progenitor (up to M~8 Ms) and their fraction can be used to constraint the initial mass function in this hidden range. At present, a fairly complete sampling of the WD population has been possible only in two nearby GCs and in one Open Cluster (Richer et al. 2013, ApJ, 778, 104 and references therein) with several hours of exposure time using instruments on board of HST.

WDs are relatively hot stars (4000<T<20000 $^{o}$K), thus they are generally studied from blue to visual passbands. However, the huge collecting power of ELT coupled with the diffraction-limit imaging provided by MAORY+MICADO will make possible to observe large part of the WD sequence in several GCs. In particular, the peak of the WD LF is at $M_J$~14.5, thus brighter than the expected MAORY+MICADO limiting magnitude for GCs at distances within 6.5 kpc. We have simulated the stellar population of a typical GC as based on the phase-A MAORY+MICADO PSF in order to evaluate the feasibility of a complete WD sampling in nearby GCs. This has been done at different distances from the cluster center, thus taking into account the effect of crowding on the detection efficiency. We found that it is possible to sample the WD LF down to its peak with a completeness larger than 50% in the centers of 4 GCs (namely NGC4372, NGC6121, NGC6366 and IC1276) and at the half-mass radius for 20 GCs. The MICADO field of view will sample a large number of WDs in a single pointing: at a distance modulus of $(m-M)_0=13$ we expect to sample ~2000 WDs per pointing with a surface brightness of $\mu_V=19.6$ mag/arcsec$^2$ (see Figure 5.1-1) thus allowing us to perform a robust statistical analysis. The proposed observations (listed below) are tailored on the analysis of the 4 GCs where WDs can be observed both in the centers and at the half-mass radius. As stated above, the same analysis limited to less crowded regions can be extended to twenty Galactic GCs.

**MICADO Pixel Scale / Fov:** 4 mas/px and 50 arcsec FoV

**MICADO observation mode:** standard imaging.

**MICADO Spectral set-up:** None

**Filters required:** J, H. WDs are relatively blue objects (-0.3<J-K<0.5) so filters centered at short wavelengths still providing SRs >0.1 are the most appropriate ones.

**Estimate Survey Area/Sample Size/ Number of Images/Epochs:** Two 50x50 arcsec$^2$ pointings for each of the 4 GCs. No specific epoch is required, the target GCs are between 12h<RA<19h being therefore visible during the same observing period.

**Average Integration time per image (magnitude of targets; S/N required):** The peak of the WD LF of a typical GC is expected to occur at $M_J$~14.5 so, given the distance range of the 4 target GCs, it lies in the range 26.5<J<29.1 and 26.2<H<28.5. A S/N~5 for these magnitudes is required to achieve a completeness of at least 50%. According to the E-ELT ETC, assuming airmass=1.2 and 3 days from the new moon, the exposure time on a single cluster (2 pointing X 2 filters) ranges from 20m (NGC6121) to 15h (IC1276). The entire set of observations would need 32h of observing time.

**Observation requirements:** Being focussed on the analysis of very faint stars, dark nights (<3 nights from the new moon) are required. A good seeing (FWHM<0.8") is also needed to ensure a proper AO correction.

**Strehl or EE required:** Performances have been successfully simulated with SRs >0.3 and >0.1 in J and H bands, respectively. High SRs are advisable to avoid critical overlap between PSF haloes affecting the detection of faint WDs. Enhanced performances in terms of SR will allow significant enlargement of the accessible sample.

**Image Stability Required:** TBD

**Astrometric Accuracy:** not relevant

**SCAO vs. MCAO:** MCAO

**Comparison with JWST or other facilities:** Virtually, a perfect AO correction provides a PSF width that is proportional to the inverse of the telescope diameter with a decrease of the exposure times proportional to the square of the diameter. E-ELT has a diameter ~6 times larger than JWST, thus providing a better angular resolution allowing to study faint WDs even in the extremely crowded centers of the target clusters. The same project could be therefore conducted with JWST only in a fraction of the target GCs with observing times ~2.5 times longer.

**Synergies with other facilities (4MOST/MOONS, LSST/ALMA/HARMONI/METIS, HIRES/MOSAIC), but also VLT or other smaller telescope instruments:** The MICADO

imager will provide data good are enough to achieve the goal of this project. Additional ACS@HST data (already available as part of the "ACS treasury project"; Sarajedini et al. 2007, AJ, 133, 1658) and wide field imaging (Trager et al. 1995, AJ, 109, 218) are valuable to constraint the parameters of dynamical models (present-day mass function, binary fraction, concentration, etc.) needed to interpret observational evidence. No preparatory observations are needed.

**Simulations made/needed to verify science case or feasibility:** We have simulated a synthetic CMD for the central region of a targeted GC, using a typical stellar population ([Fe/H]=-1.3; [alpha/Fe]=+0.4; t=13 Gyr) accordingly to the isochrones from the BASTI database (Pietrinferni et al. 2006, ApJ, 642, 797) and assuming an appropriate distance and surface density profile (see Figure 5.1-1). The surface brightness-completeness relation expected for MAORY+MICADO in the two considered passbands is taken from Schreiber et al. (2014, MNRAS, 437, 2966). New simulations based on the updated MAORY+MICADO PSF will be needed to evaluate the feasibility of this project.

**Origin of the targets:** Targets have been selected among the Harris catalog (Harris 1996. AJ, 112, 1487; 2010 edition).

**NGS:** We will naturally have NGS at disposal from the star clusters. In the analysed regions, stars with magnitudes brighter than H~18 are cluster RGB/SGB stars. Their surface density ranges from $80<\text{arcmin}^{-2}<1350$ in the cluster center to $6<\text{arcmin}^{-2}<100$ at the half-mass radius. Thus, within the NGS patrol field there are $400<N<7500$ and $30<N<500$ potentially available NGS in the center and the half-mass radius, respectively.

**Acquisition:** The two pointings for each cluster will be: 1) on the cluster center; 2) at the half-mass radius. These positions are known with accuracies of a few arcsec. Given the size of the MICADO FoV, standard pointing accuracy is required. Due to the high stellar density, finding charts are useful to set the pointing.

**Calibrations:** Standard calibration can be performed for this project. Relative photometry should be accurate at 0.01 mag at $M_J=13.5$. In principle, the census of WDs can be done also if instrumental magnitudes are available. Absolute calibration is however needed to derive WD masses when compared with theoretical isochrones. WD masses do not significantly vary with magnitude below the WD LF peak, so an accuracy of ~0.1 mag is acceptable. On the other hand, no IR photometry of such depth is available to use secondary calibrators. So, usual observations of standard stars are needed. VHS survey can be useful for calibration purpose. No particular astrometric accuracy is required.

**Data Processing Requirements:** This project requires the determination of magnitudes of very faint stars (J~28) in crowded fields where the star fluxes often overlap. For this reason a detailed PSF knowledge is required. According to the homogeneity of the PSF shape and SR, popular PSF-fitting software (Daophot, Starfinder, PSFex) can be effective in determining magnitudes with the required accuracy.

**Any other comments:** none.

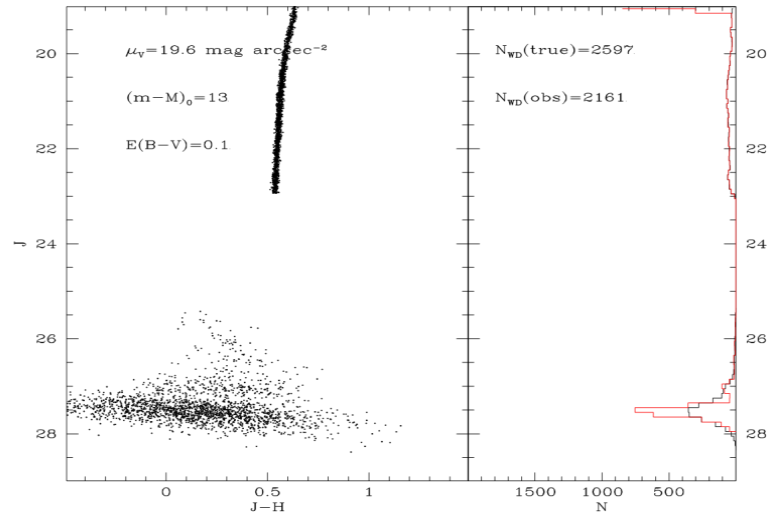

Figure 5.1-1 Left panel: Synthetic CMD of the central region of a typical nearby GC observed with MAORY+MICADO. The adopted input parameters are indicated. Right panel: LF of faint MS and WD sequences. The input and output LFs are represented with red and black histograms, respectively. The number of simulated and detected WDs is also indicated.

# 5.2 Internal kinematics of multiple stellar populations in globular clusters

**Authors:**  *Dalessandro E. (INAF - OABO), Massari D. (University of Groningen), Bellazzini M., Fiorentino G., Origlia L. (INAF - OABO)*

**Brief description of science case:**

Globular clusters (GCs) are true touchstones for Astrophysics and probably some of the most studied cosmic objects. They have always played a pivotal role for Cosmology and in the validation of the stellar evolution t;heory, which is at the very basis of any interpretation (in terms of age, metal content, stellar mass, etc.) of observations of distant galaxies.

During the last two decades, our understanding of the physical mechanisms driving GC formation and early evolution has been seriously challenged by the discovery of multiple stellar populations (MPs; for a review Gratton et al. 2012, A&Arv, 20,50) with spreads in He, many light elements (e.g., Na, O, Al) and even Fe in a few cases. This phenomenon appears to be ubiquitous, as it is observed in basically all GCs in the Galaxy (both in the halo and bulge), in nearby galaxies (like the Magellanic Clouds) and in massive Early Type galaxies (like M87). Various scenarios have been put forward to explain the properties of MPs, however none of them can explain more than a few relevant observations and significant holes in the theories remain. We still lack a self-consistent explanation of the physical processes at the basis of MP formation.

The key to shed new light on the MP phenomenon and on GC formation is a comprehensive description of their properties by means of state-of-the-art photometry and spectroscopy.

Although the long-term dynamical evolution of stars can smooth out the initial structural and kinematical properties of sub-populations to a large extent, some kinematical differences between MPs are expected to be still imprinted in present-day GCs and they may be key to constrain the early phases of GC formation (e.g. D'Ercole et al. 2011, MNRAS, 451, 1304; Bekki et al. 2010, ApJ 724L, 99). Indeed, MPs are found to show difference in their radial distributions and first evidence of different anisotropy radial profiles among MPs have been recently found in a couple of massive and nearby GCs by means of HST proper motions (Richer et al. 2013, ApJ, 771L, 15; Bellini et al. 2015, ApJ, 810L, 13). However, with the exception of a couple of cases so far, the derivation of the kinematic properties of MPs has been proven to be extremely challenging due to the resolution needed to resolve the innermost regions of GCs and the ability to measure accurate proper motions and radial velocities for large and significant samples of stars.

The typical accuracy required to investigate the internal kinematics of globular cluster is of few km/s, since this corresponds to the typical velocity dispersion of these systems. Assuming distances of the order of 10 Kpc, this translates to a required proper motion accuracy of about 0.1 mas/yr. The exceptional performance of MAORY+MICADO will provide positional measurements as accurate as 0.05 mas, that is the requested accuracy with observations separated by only 1 year. Increasing the temporal baseline, or the use in combination with existing HST observations will allow us to extend this kind of investigations to the entire population of Galactic GCs.

As an example of the possible intriguing targets, we suggest the following two GCs hosting well studied multipopulations:

| ID | [Fe/H] | E(B-V) | Dist$_{sun}$ (Kpc) | Multi-populations phase | Possible explanations |
|---|---|---|---|---|---|
| NGC1851 | -1.2 | 0.02 | 12 | 2 Sub giant branches (Milone, et al. 2008,ApJ, 673, 241) | Different CNO (Cassisi et al. 2008, ApJ, 672, 11) |
| NGC2808 | -1.9 | 0.22 | 10 | >3 Main Sequences/Red giant branches (Milone, et al. 2015,ApJ, 808, 51) | He enriched (Bragaglia et al. 2010, ApJ, 720,41) |

Table 5.2-1 An example of the possible intriguing targets where we could distinguish kinematically the well studied multipopulations.

*This Science Case is part of the MAORY ST initiative on compact stellar systems.*

**Keywords:** imaging – astrometry – globular clusters – multi-populations – kinematics

**MICADO Pixel Scale / Fov:** 4 mas/pixel should guarantee a sufficient positional accuracy while ensuring to maximize the field of view, hence the sample of measured stars in the clusters core. However detailed simulations will be used to fine-tune the choice of this parameter.

**MICADO Observation mode:** imaging, astrometry.

**MICADO Spectral set-up:** none.

**Filters required:** Two filters are the minimum request to characterize the cluster CMD and to measure the proper motions for the MP in the cluster. The filter combination has to be verified with detailed simulations since H and K-band will assure us the higher expected SR whereas J-H band the lowest background contamination, the highest sensitivity to the stellar temperature and the smallest diffraction limit.

**Estimate Survey Area/Sample Size/ Number of Images/Epochs:** at least two epochs of observations separated by at least 1 year. Each epoch requires several (at least 5) dithered exposures to exclude photometric artifacts and to determine reasonable positional uncertainties.

**Average Integration time per image (magnitude of targets; S/N required):** we have used the E-ELT ETC (version 6.2.2), with 3X3 pix reference area, airmass<1.5, 5mas/pixel resolution, S/N≥5 on sources with H~27 mag, provides a total integration time per target of ~2500s integration on source, in both H band (60s DIT X 20 NDIT) and J-band (60s DIT X 22 NDIT). Short exposures are also required in order to recover stars saturated in long exposures. The total exposure time for the two clusters, does not include overheads, and is of ~ one hour per epoch. For proper motion studies we need at least two epochs per targets for a total integration time of ~2 hours. We remember here that the chosen clusters are only a demonstrative example.

**Observation requirements:** sub-pixel dithers are required. Positioning is acceptable within 1 arcsec precision, as we will sample the regions around globular clusters center. Scale stability is fundamental as this translates into smaller geometric distortions to correct.

**Strehl or EE required:** We require the best compromise between SR and spatial resolution available.

**Astrometric Accuracy:** 0.05 mas over at least 1 year.

**SCAO vs. MCAO:** MCAO is justified by the larger FoV that will increases stellar statistics.

**Comparison with JWST or other facilities**: MAORY+MICADO provides a 6 times better resolution than JWST at fixed wavelength. This is crucial to resolve the innermost regions of GCs.

**Synergies with other facilities (4MOST/MOONS, LSST/ALMA/HARMONI/METIS, HIRES/MOSAIC), but also VLT or other smaller telescope instruments:**

- synergy with JWST-NIRCAM that could sample the GC central 2 arcmin but with a 6x lower spatial resolution and over a temporal baseline limited to ~5 yr.
- synergy with EELT-HARMONY to measure RVs and some chemical abundances in the same core region sampled by MICADO.

**Simulations made/needed to verify science case or feasibility:** Yes, in order to verify the best pixel scale/filter combinations.

**Origin of the targets:** GCs well known objects with precise coordinates.

**NGS:** globular cluster cores are crowded environment and it should not be an issue to find a suitable asterism for MCAO.

**Acquisition:** pointing can be done using a finding chart.

**Calibrations:** Standard photometric calibration is sufficient for the requested science case. Geometric distortion calibrations will be required and they have a direct impact on the result.

**Data Processing Requirements:** detailed PSF knowledge is not mandatory but will help in the data analysis.

**Any other comments:** none.

# 5.3 The Initial Mass Function, accretion and disk fraction of young stellar objects in low-metallicity environments

**Authors:** *Alcalà J.M. (INAF - OACN), Antoniucci S. (INAF - OAR), Biazzo K. (INAF - OACT), Bonito R. (INAF - OAPA), Covino E. (INAF - OACN), Fedele D. (INAF - OAA), Frasca A. (INAF - OACT, C. Manara (ESO), Nisini B. (INAF - OAR), Podio L. (INAF - OAA), Stelzer B. (INAF - OAPA), Testi L.* (INAF - OAA) *& the JEDI team*

**Brief description of science case:**

An important issue for the understanding of the star and planet formation processes is the determination of the Initial Mass Function (IMF), in particular in the very low-mass and substellar regimes. The easiest time to determine the IMF is early in the life of a stellar cluster, before high-mass stars burn out and dynamical friction segregates masses and lower mass systems are ejected. The investigations in galactic clusters and the field show no strong systematic variations among the derived IMFs. *However, whether the IMF is instead sensitive to the environmental conditions has not yet been well established.* Compared to our detailed knowledge of the galactic stellar and sub-stellar mass function, very little is known about the low-mass (near or below $1M_\odot$) content in low-metallicity environments. Differences between solar and sub-solar metallicity IMFs may arise from the star formation process taking place in low-metallicity regions of the outer Milky Way Galaxy and in the Magellanic Clouds. Another important observational result is that young (about 1-3 Myr) stars in Galactic star forming regions at low metallicity ($Z<0.2\ Z_\odot$) disperse their circumstellar disks more rapidly than their solar-metallicity analogs. This suggests that a low metallicity may induce a higher mass accretion rate and/or a very efficient photo-evaporation of the circumstellar matter, both processes causing a very rapid disk dispersal. As a consequence, most stars forming in a low-metallicity environment should experience disk dispersal at an earlier stage (in less than about 1Myr) than those forming in a solar metallicity environment (where disk dispersal may last up to 5-6Myr). Thus, metallicity may play a pivotal role on circumstellar disk evolution, hence on the time scale available for planet formation and on the (debated) dependence of exoplanet frequency in metal-rich stars. Here we propose to take diffraction-limited, high contrast, deep adaptive optics images in the JHKs bands and in the Pa-beta (1.28micron) and Br-gamma (2.16micron) narrow-band filters of selected young clusters in the Magellanic Clouds in order to: *a)* detect young stellar and substellar objects (YSOs) using broad-band color-magnitude and color-color diagrams, and hence derive the *luminosity function* of the YSO populations in the mass range between 2 and $0.03M_\odot$, *b)* detect Pa-beta and Br-gamma emission in the YSOs hence, *single out the objects with accretion disks*, and *determine their mass accretion rate*, *c)* investigate the *disk fraction* of the clusters through *photometric IR excess, and d)* examine the IMF, disk fraction and accretion properties of the identified YSO populations in *comparison with those of low-mass solar metallicity populations of clusters in the Milky Way Galaxy*.

Our team has a broad experience in the study of star forming regions in general, and in particular on the selection of YSO candidates using photometric criteria in the near-IR, as well as in the determination of stellar and accretion properties.

**Keywords:** magellanic clouds – IMF

**MICADO Pixel Scale / Fov:** 1.5mas/px and 20arcsec FoV. We need the best possible spatial resolution in order to avoid problems related with crowding. A typical low-mass star forming region in the solar neighbourhood like Taurus or Lupus would extend about 2arcsec at the distance of the Magellanic Clouds, but we also expect larger star forming regions. Thus, extreme adaptive optics observations are appropriate.

**MICADO Observation mode:** narrow and broad band imaging.

**Filters required:** J, H, Ks, Pab 1.28um, Brg 2.16um and corresponding adjacent narrow-band filters for stellar continuum subtraction. Observations in the near-infrared are more suitable in order to detect low-mass Pre-Main Sequence (PMS) stars and identify stars with active accretion. Adaptive optics works best in the near-infrared. The Pab and Brg emission lines in YSOs are the best diagnostics of accretion in the wavelength range of sensitivity of the instrument.

**Estimate Survey Area/Sample Size/ Number of Images/Epochs:** We foresee between 5 and 7 single-epoch (20arcsec-FoV ) images in each one of the 7 filters (J,H,K, Pab and Brg + the two corresponding adjacent narrow-band filters for stellar continuum subtraction).

**Average Integration time per image (magnitude of targets; S/N required).**

**Broad-band imaging:** according to the Baraffe et al. (1998) COND3 models, the corresponding absolute magnitudes of a 5Myr old YSO with a mass of $0.03M_\odot$, are 8.2mag, 7.9mag and 7.5mag in the J, H, and K bands respectively. At a distance modulus of 18.5mag, the corresponding observed JHK magnitudes, in absence of extinction, are 26.7mag, 26.4mag and 26.0mag, respectively. Based on the available ETC, we estimate that in 3600sec of exposure (6 dithers of 600sec in each band), a S/N of 29, 20 and 10 is achieved in the three bands, respectively, assuming a M5V-type star, airmass 1.5 and the 5mas/pix configuration (note that the current version of the ETC does not contain the 1.5mas/pix mode).

**Narrow-band imaging:** from our experience the typical fluxes of the Pab and Brg emission lines in a $0.1M_\odot$ accreting YSO in the Lupus star forming region (distance=200pc) are on the order of $1e^{-15}$ - $1e^{-16}erg/s/cm^2$, respectively, meaning $1e^{-20}$ - $1e^{-21}erg/s/cm^2$ at the distance of 50Kpc. A tool to estimate exposure times in the Pab and Brg filters is still missing. Exposures will be needed in the corresponding adjacent narrow-band filters for stellar continuum subtraction.

Observation of several star forming regions is desirable, since their actual member surface densities will vary.

**Observation requirements:** no particular observational requirements are needed.

**Strehl or EE required:** Due to crowding, a good SR (> 0.6 in the Ks band) is required in order to reach the appropriate angular resolution and adequately perform source detection using PSF modelling algorithms.

**Astrometric Accuracy:** standard.

**SCAO vs. MCAO:** we will use MCAO unless we have in the field a source of V≤13-14mag (in principle from 7 to 16mag) to be used as NGS.

**Comparison with JWST or other facilities:** The observations are challenging because of the intra-cluster diffuse background emission and stellar crowdedness due to the small size of a typical star forming region at the distance of the Magellanic Clouds. MAORY+ MICADO will allow us to achieve the necessary angular resolution to minimize the crowding problem with a performance of about an order of magnitude better than VLT or JWST. In addition, thanks to the much better angular resolution, we will be able to perform a more accurate background subtraction, which represents a problem when using JWST+NIRCAM.

**Synergies with other facilities (4MOST/MOONS, LSST/ALMA/HARMONI/METIS, HIRES/MOSAIC), but also VLT or other smaller telescope instruments:** Follow-up observations with METIS and HARMONI are foreseen in order to investigate in much more detail a number of interesting YSOs detected in the MAORY+MICADO observations. Synergies with LSST are also foreseen. For the later, most CoIs of our team are involved in the collaborations on Stars, Milky Way & Local Volume, as well as on the Transients/Variable stars, and on topics regarding YSOs accretion/outflow.

**Simulations made/needed to verify science case or feasibility:** Detailed simulations for this science case have been performed in chapter 10 of the E-ELT Design Reference Document. In short, the simulations take into account a Chabrier IMF, the transformation of the JHK absolute magnitudes in mass using appropriate theoretical isochrones, the distance modulus of the Magellanic Clouds, and the adequate extinction law in a range of extinction values in the K band. The simulations show that nearly complete samples of young stellar and substellar objects, above the deuterium-burning limit (Mstar > 13MJup), can be obtained in star forming regions of the Magellanic Clouds. Nevertheless, simulations using the current design of E-ELT and MAORY+MICADO are desirable.

**Origin of the targets:** The regions to be observed will be selected on the basis of their thermal infrared emission detected in optical and IR surveys by HST, VISTA (VMC), Halpha and JWST-GTO observations of the LMC. The low-mass aggregates will also contain high mass stars, with a mass up to a few solar masses, representing the top end of the mass function, as is the case of well-known star forming regions in the solar neighbourhood. The object density is unknown, as the IMF in such low-metallicity regions is unknown down to the masses of interest. The YSO candidates will be selected on the basis of color-magnitude and color-color diagrams and using theoretical isochrones of the appropriate metallicity to define the YSO locus on these diagrams.

**NGS:** The fields will likely include a number of sufficiently bright stars that can be used as NGS.

**Acquisition:** No particular requirements

**Calibrations:** Standard. The broad-band photometric accuracy for the faintest objects should be below the 0.1mag level (S/N better than 10) in order to confirm their low masses on the basis of their photometric colors. The observations of spectrophotometric standards will be needed to calibrate the absolute flux in the Pab and Brg lines.

**Data Processing Requirements:** knowledge of PSF and/or its variation over the FoV is highly desirable.

**Any other comments:** none.

# 5.4 Deep into the darkness of the Galactic bulge

**Authors:** *Bono, G. (University of Tor Vergata), Fiorentino G. (INAF - OABO), Braga V.F. (PUC), Dall'Ora M., Marconi M. (INAF - OACN), Monelli M. (IAC), Valenti E. (ESO), Zoccali M. (PUC)*

**Brief Description Of Science Case:**

The Galactic bulge includes more than 25% of the total baryonic content of the Milky Way. Recent photometric investigations on the Bulge stellar populations are mainly based on the brightest evolved stars, i.e. few magnitudes below the Horizontal Branch. There is mounting evidence that the Bulge includes two main components. The former one is old and/or metal poor, mainly traced by RR Lyrae stars (RRLs, with age older than 10 Gyr) and by Red Clump stars (RCs, 1-8 Gyr), has a spherical distribution, it is slowly rotating and the radial velocity dispersion has a shallow gradient. The latter one is metal rich, mainly traced by the bulk of RC stars, it is distributed along a bar that flares up into a boxy/peanut structure in its outer region, it is fast rotating and with a steep gradient in radial velocity dispersion (Rojas Arriagada et al. 2014, A&A, 569, 103; Pietrukowicz et al. 2015, ApJ, 811, 113; Kunder et al. 2016, ApJ, 821, L25; Zoccali et al. 2017, A&A, 599, 12). While the bar was very likely formed by dynamical instabilities of the thin disk, the origin of the spheroidal component is not clear yet. These findings have been independently supported by radial distribution of Mira variables. The short period (older) Miras trace the spheroidal component, whereas longer period ones (intermediate ages) follow the main bar (Catchpole et al. 2016, MNRAS, 455, 2216).

Evolved stars (Miras, RCs, RRLs) do not provide detailed age information, therefore, we still lack an extensive description of the two main Bulge components that could constrain their different formation scenarios. To overcome this intrinsic limitation, we plan to use deep and accurate NIR Colour-Magnitude Diagram (CMD) down to the Main Sequence Turn Off (MSTO) of the oldest Bulge stars. In passing we note that MS stars located across TO outnumber by at least two order of magnitude the evolved stellar populations. This means that our analysis is based on statistical significant samples even with modest field of views. However, MS stars are also hampered by some undisputable problems:

a) *Spatial Resolution-* When moving from the outer to the inner bulge the stellar density becomes prohibitive for 8-10m class telescopes;

b) *Depth effect-* The distance of individual stars is unknown and prevent us to build a detailed CMD;

c) *Extinction-* The reddening steadily increases and the selective absorption in the K-band becomes of the order of 3 mag. This means that current-observing facilities do not allow us to unveil MS stars in the Nuclear Bulge. Furthermore, differential reddening makes the interpretation of the CMD even more difficult;

d) *Field contamination-* Bulge stars, along the line of sight, are heavily contaminated by thin and thick disk- stars;

e) *Stellar metallicity-* A clear separation among the different Bulge components should show up also in the chemical composition (alpha enhancement, neutron capture elements). However, it is not likely to have detailed spectroscopic abundances of faint MS stars.

Accurate astrometry and deep photometry of faint Bulge stars do require MAORY+MICADO. We plan to take full advantage of its superb image quality assisted by MCAO and by large field of view to perform a comprehensive photometric analysis of MS stars across the Bulge and the Galactic centre. We plan to trace four independent lines of sight just above and below the Galactic plane to trace the X-shape of the Galactic bulge.

To accomplish this goal, we plan to use a new algorithm we have already developed in optical bands. It is based on a method that was originally suggested for variable stars (Classical Cepheids, Inno et al. 2016, ApJ, 832, 176; RR Lyrae, Braga et al, 2017, in preparation) that allows a simultaneous estimate of both the individual reddening and the distance of individual targets. The novelty is to apply the same method to MS stars and requires accurate magnitudes in at least three different photometric bands. To perform this analysis, we plan to use three broad bands (e.g., z, y, J, K, depending on the expected reddening) together with the narrow band centred on the Wing-Ford (WF) band ($m_{WF} \sim 23\text{-}26$ mag). This diagnostic is strongly correlated with the iron abundance and therefore it will provide us the opportunity to trace the metallicity of the different stellar populations. This approach will allow us to provide a detailed ranking in magnitude and in colour of the different Bulge stellar populations. In particular, we will provide accurate reddening maps, age and metallicity distributions of the surveyed area with unprecedented spatial resolution.

A second epoch will be used to complete our study with an exceptional proper motion study of Bulge stars below the MSTO. MAORY+MICADO spatial resolution is expected to reach the desired positional accuracy (0.05 mas) in about one year to definitely trace the two bulge components ($\Delta\mu \sim 3$mas/year within the low reddening Sagittarius window, see Clarkson et al. 2008, ApJ, 684, 1110; Lagioia et al. 2014, ApJ, 782, 50). A follow up with HARMONI will be used to trace also the radial velocity distribution between bulge clusters and field stellar populations.

*The main aim of this experiment is to drill the Bulge along four different lines of sight to trace radial variations in age and in chemical composition, and in turn, to constrain the impact either of nature or of nurture in Bulge formation and evolution.*

**Keywords:** imaging - photometry – galaxy formation – bulge formation

**MICADO Pixel Scale / Fov:** 4mas/px and 50 arcsec FoV is the ideal scale in order to maximize the stellar statistics.

**MICADO Observation mode:** Standard imaging.

**MICADO Spectral set-up:** none

**Filters required:** *a combination of four broad band filers (e.g., z, y, J, K) will be used. The final choice will depend on the expected reddening of the pointing. We will also use the narrow band centred on the WF band that will give us hints on the individual stellar metallicity.*

**Estimate Survey Area/Sample Size/ Number of Images/Epochs:** 5 pointings per 2 epochs (after more than one year). The five pointings are intended to trace the Nuclear Bulge and its x-shape. The two epochs are needed for a proper astrometric study of stellar proper motions.

**Average Integration time per image** (magnitude of targets; S/N required):

*Broad band photometry:* 3X3 pix reference area, airmass<1.5, 5mas/pixel resolution, S/N≥5 on sources with Ks~27 mag, the E-ELT ETC (version 6.2.2) provides ~3300s (z-band, DIT=30xNDIT=20; J-band, DIT=30xNDIT=30; K-band, DIT=30xNDIT=52;) integration per pointing for a total of ~9 hours not including overheads.

*Narrow band photometry:* we need a more detailed ETC for this estimate.

**Observation requirements** Small dithers will be needed. Single frames will be stacked after proper astrometric calibration.

**Strehl or EE required** high SRs are essential for measuring faint individual stars at large distances.

**Astrometric Accuracy:** no special requirement.

**SCAO vs. MCAO:** MCAO is preferred for the larger field of view and PSF uniformity.

**Comparison with JWST or other facilities:** JWST will be a key facility for low surface brightness studies. To make a detailed analysis we do need the best spatial resolution offered by MAORY+MICADO.

**Synergies with other facilities** (4MOST/MOONS, LSST/ALMA/HARMONI/METIS, HIRES/MOSAIC), but also VLT or other smaller telescope instruments: A follow up with HARMONI will be used to trace also the radial velocity distribution between bulge clusters and field stellar populations.

**Simulations made/needed to verify science case or feasibility:** simulations can help in the final choice of filter combination and pointings.

**Origin of the targets:** VVV survey.

**NGS:** This region is plenty of stars eligible as NGS.

**Acquisition:** no specific constraints.

**Calibrations:** standard for broadband photometry. We plan to use the spectroscopic capabilities of MAORY+MICADO to perform a new solid calibration of the WF index.

**Data Processing Requirements**: standard PSF fitting will be performed across the images.

**Any other comments**: none.

## 5.5 The transition from late to extreme spectral types (MLTY)

**Authors**: *Bono G. (University of Tor Vergata), Dall'Ora M. (INAF - OACN), Fiorentino G. (INAF - OABO), Ferraro I., Iannicola G., Pulone L., Castellani M., Testa V. (INAF - OAR), Marconi M. (INAF - OACN), Monelli, M. (IAC), Valenti, E. (ESO), Zoccali, M. (PUC)*

**Brief Description Of Science Case:**

The current empirical evidence indicates that the Intial Mass Function (IMF) is a power law (Salpeter, 1955, ApJ, 121, 161). Moreover, there is mounting evidence that the slope of the IMF becomes shallower when moving from the low ($<0.5M_\odot$) to the very low ($<0.07M_\odot$) mass regime (Kroupa et al. 2013, in *Planets, Stars and Stellar Systems*, Vol. 5, , p. 115). Although, the physical mechanisms driving star formation has been the cross-road of paramount theoretical and empirical investigations we still lack firm constraints on the IMF at the transition between very low mass stars and brown dwarfs. This open problem is far from being an academic issue, since this transition marks the Minimum Mass Hydrogen Burning (MMHB) limit. Stars less massive than MMHB are entirely dominated by electron degeneracy, and therefore, they will not be able to start the proton-proton chain, thus evolving as Brown Dwarf. The current knowledge on this transition is mainly based on nearby field stars. We still lack detailed information concerning the age dependence and the metallicity dependence. This limitation is a consequence of the fact that these objects are quite faint and cool, so extremely deep near infrared (NIR) observations like those provided by MAORY+MICADO are ideal to observe and characterize them.

The advent of all sky surveys in the NIR (2MASS) and in the MIR (WISE) regime provided some interesting properties:

a) ML transition- when moving from late M-type to L-type stars the stellar spectra are dominated by diatomic metal species (TiO, VO, FeH) incorporated in grains. In particular, the formation of Fe and Si grains causes the formation of optically thick clouds that veil the flux of these stars. Then the NIR colors become fainter ($M_K\sim12mag$) and redder ($J-K\sim2mag$) and display a new bending along the main sequence ($T_{eff}\sim1500$-$2000K$; Saumon et al. 2008, ApJ, 689, 132);

b) LT transition- at lower $T_{eff}$ the clouds start to sink and $CH_4$ supplant CO as the dominant C-bearing molecules. These structures are systematically fainter ($M_K\sim14mag$) and cooler ($T_{eff}\sim1000K$), however, the late T-type stars attain bluer color. This effect is caused by an opacity mechanism: the Collisional Induce Absorption (CIA) of molecular hydrogen ($H_2$) in the atmosphere of these structures.

c) TY transition- for even cooler effective temperatures ($T_{eff}\sim600K$) the $NH_3$ join Water and $CH_4$ absorption. Moreover, the vertical mixing of molecular nitrogen ($N_2$) causes a steady decrease in the NIR flux, but we should be able to detect them either in the Y or in the J band. These are the Y-type stars ($M_K\sim16$-$18mag$).

d) YGP transition- interestingly enough, the current theoretical predictions indicate that at cooler effective temperature we should find giant floating planets (GP, Burrows et al. 2003).

| ID | Cluster type | [Fe/H] | E(B-V) | Age (Gyr) | $\mu_0$ | $MSK_K$ |
|---|---|---|---|---|---|---|
| NGC6253 | old open | +0.36 | 0.20 | 5 | 11.5 | 16.6 |
| Ruprecht46 | old open | -0.04 | 0.07 | 4 | 9.6 | 14.6 |
| NGC104 | globular | -0.72 | 0.04 | 11.7 | 13.2 | 18.3 |
| NGC6121 (M4) | globular | -1.16 | 0.35 | 11.5 | 11.7 | 16.9 |
| NGC6544 | globular | -1.40 | 0.76 | 12.0 | 12.3 | 17.6 |
| NGC6656 (M22) | globular | -1.70 | 0.35 | 12.5 | 12.5 | 17.7 |
| NGC6397 | globular | -2.02 | 0.18 | 13.0 | 11.8 | 16.9 |

Table 5.5-1 Illustrative example of a possible sample of clusters spanning a significant range of metallicity and age.

MAORY+MICADO, with high spatial resolution (4 mas/pix) and MCAO assisted, together with METIS are going to play a crucial role in this context, since they will allow us to trace the quoted transitions in several nearby globular and open clusters (distance modulus range $10 \leq \mu_0 \leq 13$mag, see Table 5.5-1). This means the opportunity to investigate the environment dependence (age and metallicity). In particular, we want to build extremely deep J-K CMD in order to detect and characterize magnitudes and colours of very low mass stars.

Furthermore, we plan to use the narrow band filter centred on the CH4 band (1.66 micron) to properly trace the transition between L-type (dominated by CO bands) and T-type (dominated by CH4 bands) stellar structures. We also plan to provide a new calibration of this diagnostics using intermediate resolution spectra from MAORY+MICADO.

Note that this is a new regime not only for the physics of stellar interiors (fully convective, electron degeneracy) but also for stellar atmospheres, since we need to deal with weather conditions (cloud formation), gravitational settling, grain formations and molecular bands.

**Keywords:** imaging - photometry - low mass stars – spectral transition

**MICADO Pixel Scale / Fov:** 4mas/px and 50 arcsec FoV is the ideal scale in order to maximize the stellar statistics.

**MICADO Observation mode:** standard imaging.

**MICADO Spectral set-up:** none.

**Filters required:** *J and H-band* may be the best choice to obtain the deepest CMDs in order to minimise the sky-background and optimise the S/N ratio (~5). Detailed simulations may help in the final filter choice. We also plan to use the narrow band $CH_4$ to identify characterize T-type stars.

**Estimate Survey Area/Sample Size/ Number of Images/Epochs:** the sample size includes seven clusters spanning a large range of metallicity and age (see Table 5.2-1).

**Average Integration time per image** (magnitude of targets; S/N required):

*Photometry with wide band filters:*

We have used the E-ELT ETC (version 6.2.2), with 3X3 pix reference area, airmass<1.5, 5mas/pixel resolution, S/N≥5 on sources with H~27 mag, provides a total integration time per target of ~2500s integration on source, in both H band (60s DIT X 20 NDIT) and J-band (60s DIT X 22 NDIT).  The total exposure time for the selected sample, does not include overheads, and is of ~ 5 hours.

*Photometry with narrow band filters:* the exposure times are preliminary, they need to be checked with detailed simulations taking account for the expected apparent magnitudes.

**Observation requirements** Small dithers will be needed. Single frames will be stacked after proper astrometric calibration.

**Strehl or EE required** high SRs are essential for measuring faint individual stars at large distances.

**Astrometric Accuracy:** no special requirement

**SCAO vs. MCAO:** MCAO is preferred for the larger field of view and PSF uniformity.

**Comparison with JWST or other facilities:** JWST will be a key facility for low surface brightness studies, so very likely it will complement the proposed study in low density stellar fields and in the outskirts of star clusters.

**Synergies with other facilities** (4MOST/MOONS, LSST/ALMA/HARMONI/METIS, HIRES/MOSAIC), but also VLT or other smaller telescope instruments: We also plan to take advantage of METIS, the first light high resolution imager and spectrograph of ELT. We are interested in this instrument because the spectra of Y-type stellar structures are dominated by molecular nitrogen ($N_2$) and by ammonia ($NH_3$) that can be easily identified in the L and M spectral range.

**Simulations made/needed to verify science case or feasibility:** simulations can help in posing the limit in distance for the targets to be observed.

**Origin of the targets:** already known.

**NGS:** available.

**Acquisition:** no specific constraints; accurate finding charts are available over any the possible target of interest.

**Calibrations:** local standard maybe needed for a proper calibration. A spectroscopic calibration of the photometric narrow band $CH_4$ will be needed. To accomplish this goal we plan to use MAORY+MICADO in spectroscopic mode for the brighter targets.

**Data Processing Requirements**: Good PSF modelling is required; target selection will by necessity take such requirement into account.

**Any other comments**: none.

# 5.6 Unveiling the presence of IMBH in globular clusters with accurate proper motions.

**Authors:** *Fiorentino G. (INAF - OABO), Massari D. (University of Groningen), Dalessandro E., Bellazzini M. (INAF - OABO), Bono G. (University of Torvergata), Dall'Ora M. (INAF - OACN), Monelli M. (IAC), Sollima A. (INAF - OABO), Zocchi A. (University of Bologna)*

**Brief Description Of Science Case:**

In the center of giant galaxies Super Massive Black Holes (SMBHs, with masses larger than $10^6$ Mo) are usually observed and their formation is thought to be connected with the formation of the galaxy itself. One of the channels proposed for the formation of SMBH is growth from Intermediate Mass Black Holes (IMBHs with mass in the range $10^2$-$10^4$ Mo) seeds. However, IMBHs have not been detected yet and they are one of the Holy Grails of contemporary astrophysics. Extrapolation of $M_{BH}$-σ relations for spheroids suggests that globular clusters (GC) may be the right place to search for IMBH (see, e.g., Ferrarese et al. 2006, ApJ, 664, 21; Lanzoni et al. 2007, ApJ, 668, L139, and references therein)

The most direct way to unveil the presence of an IMBH within a GC is to study the internal dynamics of its stars very close to the center (Anderson & Van der Marel, 2010, ApJ, 710, 2032). An inner cusp in the velocity dispersion profile and/or the detection of high velocity stars are the more clear-cut signatures of a central compact mass (see Fig. 2 in Kiziltan, Baumgardt & Loeb, 2017, Nature, 542, 203). In order to unveil these signatures, we need accurate measurements (typical precision ≲2.0 km/s) of both radial (from stellar spectra) and tangential velocities (from PMs) of stars lying within a few parsec from the GGC center. However, radial velocities can only be measured for few bright red giant stars in cluster centers, whereas PMs can be measured for thousands of main sequence stars and they trace two components of stellar motions, thus helping to break the mass-anisotropy degeneracy.

Today, the most severe observational limit on accurate PMs measurements is that even HST cannot penetrate the very center of GCs down to faint magnitude limits (HST diffraction limit, $DL_I$~ 50mas, pixel scale~ 50mas/pix). This is not going to dramatically improve with JWST ($DL_I$~30mas, pixel scale~ 30mas/pix). Thus, although there have been several attempts to detect IMBH in the center of GGCs using PMs with HST (e.g., 47 Tuc, McLaughlin et al. 2006, ApJS, 166, 249; ω Cen, Van der Marel & Anderson, 2010, ApJ, 710, 1063), none of them reached firm conclusions (see also, Kiziltan et al. 2017, for a different approach to the problem). MAORY+MICADO will allow a major step forward, with a significant increase in spatial resolution with respect to HST and JWST. To contrast the MAORY+MICADO performance we consider below the case of the relatively nearby massive GC 47 Tucanae (MGC 104).

47 Tuc is located within 5 kpc from the sun and has a low reddening. According to theoretical model the typical difference in central velocity dispersion in case of presence of a IMBH with a mass 0.5-1% of the GC total mass is of ~3-7 km/sec, this translates in PMs of ~0.1-0.3mas/year. With MAORY+MICADO we plan to reach an accuracy of ~0.05 mas in the tangential velocity with a time baseline of 5 years at the distance of 47Tuc. This will overcome HST and JWST of a factor of 3-6, depending on the adopted wavelength and other factors. The exquisite spatial resolution and sensitivity will allow us to trace the motions of stars down to the H-burning limits in the central pc, thus optimizing the numbers of useful kinematic tracers in the small sphere of influence of the IMBH. This means accuracy in proper motions within the central parsec of 1-3% when ~250 stars are detected.

We stress again that 47 Tuc is only an example of globular clusters we could focus on in the search for IMBHs. As an additional example, another intriguing target at similar distance is Terzan 5, that may be the remnant of one of the building blocks that formed the Galactic bulge (Ferraro et al. 2009, Nature, 462, 483).

In principle, the we will reach an accuracy of 3-5% (3-10%) in proper motions for all the clusters within 10Kpc (50Kpc), since the expected difference in proper motions is ~0.06-0.10mas/year (0.01-0.03 mas/year). This corresponds to ~80 (~120) globulars visible in the Southern sky. *Then, MAORY+MICADO will make possible a definitive claim on the existence of IMBH within these compact stellar systems.*

*This Science Case is part of the MAORY ST initiative on compact stellar systems.*

**Keywords:** imaging – astrometry – galaxy formation – globular clusters – IMBH

**MICADO Observation mode:** Standard imaging.

**MICADO Pixel Scale / Fov:** 1.5mas/px and 20 arcsec FoV is the ideal scale in order to maximize the accuracy in the first parsec from the cluster center.

**MICADO Spectral set-up:** none

**Filters required:** *a combination of J and K filters* will be used. The final choice will depend on the expected instrument performance; it could be also J and H bands.

**Estimate Survey Area/Sample Size/ Number of Images/Epochs:** 1 pointing per 2 epochs (after more than one year).

**Average Integration time per image** (magnitude of targets; S/N required):

*Broad band photometry:* 3X3 pix reference area, airmass<1.5, 5mas/pixel resolution, S/N≥5 on sources with Ks~27 mag, the E-ELT ETC (version 6.2.2) provides ~2500s (J-band, DIT=30xNDIT=30; K-band, DIT=30xNDIT=52) integration per pointing for a total of ~3 hours not including overheads.

**Observation requirements** Small dithers will be needed. Single frames will be stacked after proper astrometric calibration.

**Strehl or EE required** high SRs are essential for measuring faint individual stars, thus reducing the statistical errors, at large distances.

**Astrometric Accuracy:** the maximum precision is required.

**SCAO vs. MCAO:** given the small size of the FoV, we want to perform ad hoc simulations SCAO vs MCAO in order to understand the best compromise.

**Comparison with JWST or other facilities:** JWST will be a key facility for low surface brightness studies, we have already HST observation to survey the external part of the cluster. To make the detailed analysis proposed in the very center of the cluster we do need the best spatial resolution offered by MAORY+MICADO. In the outer field of 47Tucanae, future proper motions provided by Gaia will complete the radial profile in the very external part with comparable precision.

**Synergies with other facilities** (4MOST/MOONS, LSST/ALMA/HARMONI/METIS, HIRES/MOSAIC), but also VLT or other smaller telescope instruments: A follow up with HARMONI or with MICADO in its spectroscopic mode will be used to trace the velocity distribution along the line of sight.

**Simulations made/needed to verify science case or feasibility:** simulations can help in the choice of the best AO flavour and filter combinations to reach the extreme astrometric accuracy.

**Origin of the targets:** literature.

**NGS:** This GC is plenty of stars eligible as NGS.

**Acquisition:** no specific constraints.

**Calibrations:** standard for broadband photometry. High precision astrometric calibration is needed in order to reach the desired accuracy.

**Data Processing Requirements**: standard PSF fitting will be performed across the images.

**Any other comments**: none.

## 5.7 Extreme young star clusters in the Milky Way: the Arches cluster

**Authors:** *Mapelli M., Ballone A., Spera M. (INAF - OAPD), Trani A. A. (INAF - OAPD, SISSA), Giacobbo N., Bortolas E., Di Carlo U. N. (INAF - OAPD, University of Padova)*

**Brief description of science case:**

The Arches cluster is a remarkably dense (central density $\sim 2 \times 10^5$ M$_\odot$/pc^3) and massive (present-day mass $\sim 2 - 4 \times 10^4$ M$_\odot$) young ($\sim$2–3 Myr) star cluster close to the centre of the Milky Way (at $\sim$26 pc projected distance from the Galactic centre). Less than 10 objects similar to the Arches exist in the Milky Way, and the Arches seems to be the most extreme star cluster of its kind (e.g. Portegies Zwart et al. 2010, ARA&A, 48, 431, for a review). Moreover, it is considered the prototype of dense star clusters that might host intermediate-mass black holes born from runaway collisions (e.g. Portegies Zwart et al. 2004, Nature, 428, 724). Together with the nuclear star cluster of the Milky Way, Arches is the ideal laboratory for investigating massive star formation under extreme conditions.

Due to the large amount of extinction ($A_V \sim 30$ mag) along the line of sight, the study of the Arches stellar content is limited to infrared wavelengths. Previous adaptive optics (AO) observations with 8m telescopes (e.g. Stolte et al. 2002, A&A, 394, 459) and space-borne observations with the HST (Figer et al. 1999a, ApJ, 525, 750; Figer et al. 2002, ApJ, 581, 258) suggested that the stellar mass function in the Arches might be top-heavy or truncated at low masses, similar to what observed in the inner parsec of the Galaxy. However, Espinoza et al. (2009, A&A, 501, 563) show that a Salpeter initial mass function cannot be discarded if potential biases of the observations (e.g. the strong differential extinction and the problem of transforming the observations into a standard photometric system in the presence of strong reddening) are taken into account. Other studies (eg Habibi et al 2013, A&A, 556, 26) suggest that the slope of the mass function might be different in the periphery of the cluster (where it is even steeper than the Salpeter mass function) with respect to the central parts (where it is top-heavy), consistent with the effect of mass segregation. Overall, the issue of the Arches initial mass function (whether it is top-heavy or not) is still unresolved, and only 30m class telescopes can shed light on it.

As to the kinematics, recent measurements of the proper motions in the Arches cluster (based on a three years baseline and performed with the laser-guide star AO system at Keck, Clarkson et al. 2012, ApJ, 751, 132) show that the internal velocity dispersion is $0.15 \pm 0.01$ mas/yr $= 5.4 \pm 0.4$ km/s (assuming a distance of 8.4 kpc), consistent with a mass of $1 - 1.2 \times 10^4$ M$_\odot$ inside 0.4 pc. The velocity dispersion measurement is based only on stars with magnitude K'<18. A more accurate measurement of proper motions is crucial to assess the dynamical mass of the cluster and to check for the presence of a newly born intermediate-mass black hole at the centre of the Arches cluster.

*Photometric observations of the Arches in two bands (H and K, see the technical specifications) with MICADO+MAORY will enable us to study the mass function significantly below the magnitude Ks~16, corresponding to a stellar mass of ~2–3 M$_\odot$ (which is the completeness limit of Espinoza et al. 2009). Astrometric observations of the Arches in K-band will allow us to reconstruct the velocity dispersion profile of the Arches cluster, providing a dynamical mass measurement of the cluster (independent from photometry) and hunting for the presence of an intermediate-mass black hole.*

*This Science Case is part of the MAORY ST initiative on compact stellar systems.*

**Keywords:** imaging – photometry – astrometry – proper motions – intermediate-mass black holes – dense young star clusters – Galactic centre

**MICADO Pixel Scale/Fov:** 1.5mas/px and a 20" FoV is required to resolve stars in the crowded field at the centre of the Arches cluster.

**MICADO Observation mode:** (1) Standard Imaging for photometry; (2) astrometric imaging for proper motions

**MICADO Spectral set-up:** none

**Filters required:** (1) for photometry H and K (or Ks) are the best choice given the large amount of extinction ($A_V \sim 30$ mag) along the line of sight, (2) for astrometry, the K (or Ks) band should be preferable.

**Estimate Survey Area/Sample Size/ Number of Images/Epochs:** At the distance of the Galactic centre 1" corresponds to ~0.04 pc. The core of the Arches is thus well within the 20" FoV. About 10 hr are requested to observe the central 0.4 pc of the Arches plus other 4 fields around it, assuming 1 hr exposure time per each observation per each of the two 2 bands (H and K).

**Average Integration time per image (magnitude of targets; S/N required):** 1h is required to reach K=25 (Vega) with S/N~30 and K=26 with S/N~10. 1h is required to reach H=26 (Vega) with S/N~25 and H=27 with S/N~10. However, detailed simulations are needed to take into account the correct exposure time, given the extreme crowding of the field and considering the correct size of the pixel for these observations (1.5mas). These estimates have been obtained on May 21, 2017, with the ESO-ELT ETC, assuming a px scale of 5 mas, a M0V star as a template and a 1 px X 1 px reference area for computing the S/N ratio.

**Observation requirements:** Small dithers + Large dithers [refer to table 8 of MICADO OCD]

**Strehl or EE required:** the spatial resolution is a key point in these crowded field. Thus, the higher the SR, the better the result of this observation.

**Astrometric Accuracy:** (1) standard for photometry; (2) long-term astrometric accuracy of ~50 μas for proper motions (to be checked with simulations).

**SCAO vs. MCAO:** The option of SCAO might be viable but must be checked with dedicated simulations. MCAO is certainly possible in this FoV.

**Comparison with JWST or other facilities:** The spatial resolution of MAORY+MICADO is the key point, exceeding that of JWST by a factor of ~6.

**Synergies with other facilities:** AO observations with Keck and VLT are available and can be used as first epochs for proper motions (eg Espinoza et al. 2009; Clarkson et al. 2012). Also, HST observations are available (Figer et al. 1999a, ApJ, 525, 750; Figer et al. 2002, ApJ, 581, 258).

**Simulations made/needed to verify science case or feasibility:** Simulations are needed to evaluate the possibility of using SCAO, and to obtain accurate estimates of the limiting magnitudes for H and K in both SCAO and MCAO, accounting for the crowding of the region. The filter choice has to be double-checked with simulations.

**Origin of the targets:** available in the literature, e.g. Espinoza et al. (2009).

**NGS:** Several stars in the periphery of the Arches cluster can be used as NGS.

**Acquisition:** no special requirements. Finding charts available.

**Calibrations:** (1) standard for photometry; (2) astrometric for proper motions.

**Data Processing Requirements:** The PSF modelling should be feasible with stars in the FoV.

**Any other comment:** The Quintuplet cluster (slightly older and less dense than the Arches) is a similar interesting target for the same scientific goals as discussed in this science case. Observing the Quintuplet together with the Arches might give invaluable insights on the formation of these two peculiar objects: did they originate in the central molecular zone (e.g. Stolte et al. 2014, ApJ, 789, 115)?

# 5.8 Stellar population and dynamics of 30 Doradus in the Large Magellanic Cloud

**Authors: :** *Mapelli M., Ballone A., Spera M. (INAF - OAPD), Trani A. A. (INAF - OAPD, SISSA), Giacobbo N., Bortolas E., Di Carlo U. N. (INAF - OAPD, University of Padova), Fiorentino G., Bellazzini M. (INAF - OABO)*

**Brief description of science case:**

Dense young star clusters and star forming regions outside the Milky Way are a unique science case for E-ELT. Young star clusters are the places where most stars (and especially massive stars) form in the local Universe (Lada & Lada 2003, ARA&A, 41, 57). They host the most massive stars we know (>100 M$\odot$, Crowther et al. 2010, MNRAS, 408, 731; Crowther et al. 2016, MNRAS, 458, 624), and give us a unique opportunity to investigate the initial mass function (Kroupa 2001, MNRAS, 322, 231) as well as the dynamics of stars: dense star clusters are the only astrophysical environment where stellar collisions are likely (Portegies Zwart et al. 1999, A&A, 348, 117). Moreover, young star clusters are thought to be a cradle of massive (>20 M$\odot$) stellar black holes and intermediate-mass black holes (IMBHs, Mapelli 2016, MNRAS, 459, 3432), and might provide an invaluable key to understand detections by ground-based and space-borne gravitational wave detectors (e.g. Abbott et al. 2016, Phys. Rev. Letters 116, 1102).

The region of 30 Doradus in the Large Magellanic Cloud (LMC) is the most spectacular star forming region in the Local Group and has significantly lower metallicity (Z~0.008) than the Solar neighbourhood. It is only ~50 kpc away from our Sun, which will allow us to resolve stars down to ~0.1 M$\odot$ with a less than 5 hour pointing of MAORY+MICADO, even if most regions in 30 Doradus are extremely crowded. Proper motions can be measured in 30 Doradus with a ~2–5 km/s accuracy, enabling us to investigate the presence of an IMBH at the centre of the most massive young star clusters. Moreover, R136 in 30 Doradus is the youngest star cluster with a claim of global rotation (Hénault-Brunet et al. 2012, A&A, 545, L1). This claim is based on a small sample of 36 O-type stars. The signature of rotation (~3 km/s) is likely too weak for proper motion measurements, but can be tested with the HARMONI IFU.

**The main goals of an E-ELT campaign on 30 Doradus are (1) to reconstruct the colour magnitude diagram (CMD) of several clusters in the region (a 0.1 M$\odot$ star has a magnitude J ~ 26 at the distance of the LMC; current CMDs obtained with HST are complete down to magnitude J ~ 22, Sabbi et al. 2016, ApJS, 222, 11), and (2) to obtain proper motions with a precision of ~2–5 km/s (to test the presence of IMBHs in R136 or other dense star clusters). These goals can be achieved with the expected (1) photometric and (2) astrometric requirements of MAORY+MICADO, respectively.**

**Follow-up observations with HARMONI are important to strengthen the constraints on IMBHs and to probe the claim of rotation in the R136 cluster.**

The extension of 30 Doradus (40' x 25') makes it prohibitive to survey the entire region with E-ELT. Thus, areas of particular interest should be selected. These include the main clusters in the region (NGC2070 including R136, Hodge 301, NGC2060) and at least one area without massive clusters.

*This Science Case is part of the MAORY ST initiative on compact stellar systems.*

**Keywords:** imaging – photometry – astrometry – proper motions – extragalactic young star clusters

**MICADO Pixel Scale / Fov:** 1 arcsec corresponds to ~0.24 pc at the distance of 30 Doradus. Thus, 20 arcsec correspond to ~ 5 pc, which is sufficient to enclose the half-mass radius of R136. Thus, 1.5mas/px and 20arcsec FoV are the best option for the central regions of the most crowded clusters, motivated by the maximization of the performances in terms of spatial resolution ratio into the crowding of the central regions of star clusters in 30 Doradus. 4 mas/px and 50 arcsec FoV might be a better choice for the less crowded regions outside the clusters or in their outskirts, but this must be checked with dedicated simulations.

**MICADO Observation mode:** (1) Standard Imaging for photometry; (2) Astrometric imaging for astrometry

**MICADO Spectral set-up:** none

**Filters required:** (1) J and H are optimal for the photometry. It is important to produce colour-magnitude diagrams of the observed population, to get constraints on their age and metallicity. (2) H is likely the best choice for the astrometry (to measure proper motions), because it offers a high SR than J band, even if with the drawback of a higher background. Dedicated simulations will enable us to decide which band is the best for astrometry.

**Estimate Survey Area/Sample Size/ Number of Images/Epochs:**

At least 10 pointings, 3 of them with 20" x 20" area (centred on NGC2070, NGC2060, Hodge301), the remaining with 50"x50". Assuming 1 hr per pointing, we estimate 20hr of total observing time for 10 pointings in each of the two requested bands.

For the regions where we plan to study proper motions (the three 20" x 20" pointings on NGC2070, NGC2060, and Hodge301) we also need a second epoch on a ~5 year baseline (required to achieve the expected precision). This implies that we will need 3 additional hours in a second epoch for these three pointings.

In addition, the Tarantula survey (Sabbi et al. 2016, ApJS, 222, 11) observations of the entire 30 Doradus area are available, providing possible first-epoch data for proper motion estimates, but only in the less crowded regions (dedicated simulations will be used to assess this point).

**Average Integration time per image (magnitude of targets; S/N required):** with 1h per filter S/N~10 is reached at H=27, S/N~30 at H=26. These values allow us to reconstruct the colour-magnitude diagram with sufficient precision (precision of ~0.05 mag in J-H at H=26) down to stars with J~26 These estimates have been obtained on Feb 8, 2017, with the ESO-ELT ETC, assuming a px scale of 5 mas (the minimum available on the current version of the ESO-ELT ETC), a K5V star as a template and a 1 px X 1 px reference area for computing the S/N ratio.

**Observation requirements:** Small dithers [table 8 of MICADO OCD]

**Strehl or EE required:** a high SR is essential, because the best spatial resolution is crucial, especially for the core of the main star clusters.

**Astrometric Accuracy:** (1) standard accuracy for the photometry; (2) ~50 µas for proper motions will likely be needed.

**SCAO vs. MCAO:** While SCAO might be viable for some targets (eg stars with V magnitude <13-14 mag are present in the region of R136), it is not likely for most of the pointings. There will be no issues for MCAO: plenty of 30 Doradus stars can be used as NGS in the MCAO configuration.

**Comparison with JWST or other facilities:** the spatial resolution will allow us to resolve single low-mass stars even in the most crowded fields (eg. center of R136) where HST does not have access. The expected spatial resolution of MAORY+MICADO is a factor of ~ 6 better than that of JWST.

**Synergies with other facilities (4MOST/MOONS, LSST/ALMA/HARMONI/METIS, HIRES/MOSAIC), but also VLT or other smaller telescope instruments:** 30 Doradus was the target of several recent surveys, in particular the Tarantula treasury project with HST. Synergy with HARMONI is essential to assess the issue of rotation and to put constraints on the existence of IMBHs in the most massive star clusters (eg R136).

**Simulations made/needed to verify science case or feasibility:** accurate simulations of the main clusters can be done with available tools. Simulations in the SCAO mode are needed to check the possibility of using SCAO for the main clusters.

**Origin of the targets:** Tarantula survey (Sabbi et al. 2016, ApJS, 222, 11, and references therein).

**NGS:** plenty of stars in the 30 Doradus region can be used as NGS.

**Acquisition:** no special problem or requirement. Finding charts are available.

**Calibrations:** (1) Standard for the photometry, (2) astrometric for the proper motions.

Several ground-based (VMC survey, Cioni et al. 2011, A&A, 527, 116) and space-borne (HST data, the Tarantula survey, Sabbi et al. 2016, ApJS, 222, 11) observations of the entire 30 Doradus area are available, providing a wealth of photometric calibrators, and possible first-epoch data for proper motion estimates.

**Data Processing Requirements:** The PSF modelling should be feasible with stars in the FoV.

**Any other comment:** none.

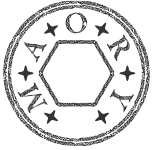

# 6. Local Universe

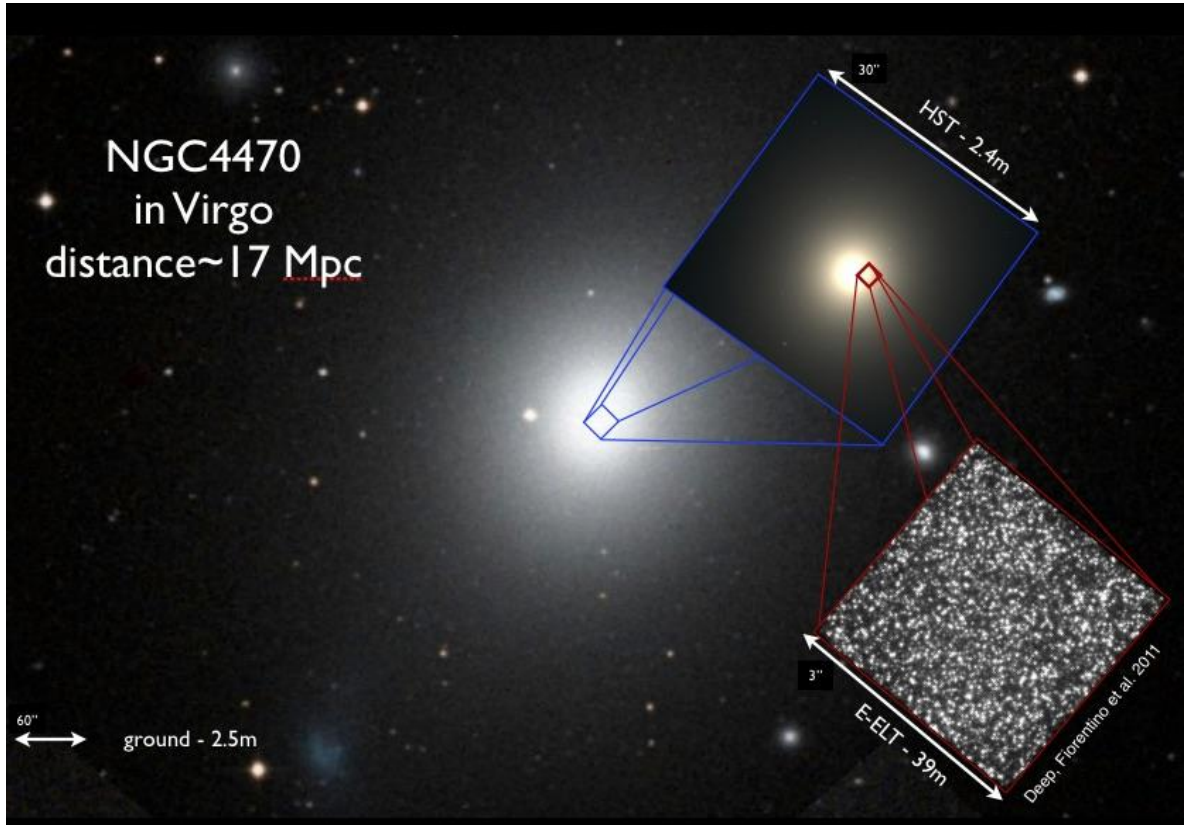

Figure 6-1 Zooming on a giant elliptical galaxy in Virgo, with HST and with MAORY+MICADO in MCAO mode (phase A simulations).

# 6.1 Revealing the missing population of intermediate-mass BHs in dwarf galaxies.

**Authors:** *Balmaverde B. (SNS) and Capetti A. (INAF - OATO)*

**Brief description of science case**:

The $M_{BH}$-$\sigma$ relation (Ferrarese & Merritt, D., 2000, ApJL, 539, L9; Marconi, A. & Hunt, L.K, 2000, ApJL, 589, L21) is one of the strongest arguments in favor of a coevolution of the supermassive black holes and their host galaxies. The feedback, i.e. the kinetic and luminosity energy released by the BH, is invoked to explain the properties of galaxies in the local Universe (Di Matteo, T., Croft, R.A.C., Springel, V., & Hernquist, L., 2003, ApJ, 593, 56). However, we need to establish whether this law, that predicts a ratio between the BH and the galactic bulge mass of $\sim 10^{-3}$, is universal or it breaks at some mass. In fact there is a missing link between stellar mass and supermassive black holes that we currently observe in the Universe. Intermediate-mass black holes (IMBHs), with masses from $\sim 10^2$ to $10^6$ $M_\odot$, are predicted by the $M_{BH}$-sigma relation at the center of low luminosity galaxies, but their existence and properties remain uncertain. Furthermore, with the current observational uncertainties, we cannot constrain different models for the formation of primordial BHs (light or heavy seeds).

The detection of IMBHs is challenging because the signatures of these objects are elusive: most of the candidates are weakly active and so they might not have detectable electromagnetic counterparts. In fact, up to date, there are less than ten IMBHs candidates, one of the most promising is the $\sim 10^3$ $M_\odot$ black hole in the globular cluster in NGC104 (Kiziltan, Baumgardt & Loeb, 2017, Nature, 542, 203). The existence of IMBHs can only be proven modelling the kinematic of star and gas in the central region of dense stellar system with high resolution spectroscopic observations. This technique has been intensively applied in the local universe to directly measure the masses for about $\sim 30$ BHs. However, the direct measurements are limited to high BH masses ($M_{BH} > 10^8 M_\odot$) that do not represent the demography of more typical black holes and that completely miss the IMBH population. In other cases, the masses are estimated using an indirect and more uncertain methods (e.g. via the virial theorem that relate the mass of the black hole to the width of broad lines, e.g. Xiao, T., Barth, AJ, Greene, J.E., et al, 2011, ApJ, 739, 28). Almost all studies target nearby elliptical galaxies, that are relatively dust free and have BHs massive enough that their gravitational spheres of influence is resolved with HST on Mpc scale (R=0.11 $(M_{BH}/10^8)$ $(200 \text{ km s}^{-1}/\sigma)^2$ (20 Mpc/D)", see Barth, AJ, 2004 Cambridge University Press, 21).

We propose here to map the Paschen $\beta$ emitting gas in the near infrared band with MAORY+MICADO to model the kinematic of the gas around the BH in a sample of dwarf galaxies. This line has the great advantage of being unabsorbed by dust and flux from the AGN does not dominate the continuum. With MAORY+MICADO we could increase the spatial resolution by a factor of ~30, we will finally be able to reveal the bulk of the missing IMBH population.

**Keywords:** narrow band imaging – galaxy formation – dwarf galaxies – IMBH

**MICADO Pixel Scale / Fov:** 4 mas/px, the only choice allowed in spectroscopic mode.

**MICADO Observation mode:** Long slit spectroscopy.

**MICADO Spectral set-up:** 50 mas x 4 arcsec. Wavelength coverage 0.8-1.45 µm

**Filters required:** none.

**Estimate Sample Size:** ~10 targets

**Average Integration time per image:** for a Pa β (1282 Å) of 1 mJy to have a S/N of 10 in 4X4 spaxels, we need ~5700s, according to the E-ELT ETC.

**Observation requirements:** none

**Strehl or EE required:** the science goal would benefit from the highest resolution achievable. In particular, we would resolve the sphere of influence of a $10^5\,M_\odot$ (assuming $\sigma=50$ km/s from the $M_{BH}$-$\sigma$ relation) at 5Mpc we need a spatial resolution of at least 7mas. To resolve a BH of $10^4\,M_\odot$ with the same resolution (assuming $\sigma=25$ km/s) we need to select targets at distances lower than 2Mpc (Barth 2004).

**Astrometric Accuracy:** standard

**SCAO vs. MCAO:** SCAO mode

**Comparison with JWST or other facilities:** the spatial resolution of MAORY+ MICADO is about 6 times larger than JWST, pushing the limits of our BH mass detection limit a factor 6 lower (at any given distances).

**Synergies with other facilities (4MOST/MOONS, LSST/ALMA/HARMONI/METIS, HIRES/MOSAIC), but also VLT or other smaller telescope instruments:** none

**Simulations made/needed to verify science case or feasibility:** none

**Origin of the targets:** From the All-Sky Catalogue of nearby galaxies (Karachentsev, I.D., et al., 2013, VizieR Online Data Catalogue, 110) we can consider the 108 dwarf galaxies within 11 Mpc (all with H and K band magnitude from the 2MASS survey catalogue).

**NGS:** available, from our large catalogue we could select targets with a H<16 NGS within few arcsec.

**Acquisition:** The pointing can be verified with a finding chart

**Calibrations:** standard

**Data Processing Requirements:** standard

**Any other comments:** none

## 6.2 RSGs as Cosmic abundance probes

**Authors**: *Lardo C. (EPFL), Davies B. (LJMU), Kudritzki R. P. (UHawaii), Fiorentino G., Bellazzini M. (INAF - OABO)*

**Brief Description of Science Case**:

The observed trend between galaxy mass and its metallicity is a unique diagnostic of the mechanisms driving galaxy evolution (Tremonti et al. 2004, ApJ, 613, 898). In general, determinations of central metallicity and the radial abundance gradient of star-forming galaxies are obtained using statistical or 'strong line' methods, which rely on empirical relationships between different optical emission line ratios and metallicities determined via direct method. However, this simplified emission line analysis is affected by intrinsic and poorly understood uncertainties (Kewley & Ellison 2008, ApJ, 681, 1183).

A study by Davies, Kudritzki & Figer (2010, MNRAS, 408L, 31) introduced the J-band diagnostics, spanning the narrow spectral range between 1.15-1.22 μm, as a powerful tool to *directly* measure stellar metallicities in extra-galactic red supergiants (RSGs). Low resolution (at R of a few thousand) observations in the J-band, where AO correction are more effective compared to shorter wavelengths, were demonstrated to provide reliable estimate of metallicity of stars. The J-band method was initially introduced by Davies et al. (2010) for individual RSGs in the Milky Way and has been extended to several nearby galaxies spanning a broad range in mass and metallicity (e.g., Gazak et al. 2015, ApJ, 805, 182; Patrick et al., 2016, MNRAS, 458, 3968). Robust metallicity determinations accurate to ~ 0.1 dex can be recovered, providing excellent agreement with other high-precision abundance tracers such as blue supergiants (Bresolin et al. 2001. ApJ, 548L, 159; Kudritzki et al. 2008, ApJ, 681. 269) and direct measures in HII-regions (Andrews & Martini 2013, ApJ, 765, 140) in a wide mass and metallicity regime.

The potential of this region for direct abundance estimates from individual of RSGs in galaxies beyond the Local Volume with ELT was further investigated by Evans et al. (2011, BSRL, 80, 456) who demonstrated that for AO supported spectrographs magnitudes out to ~30 Mpc could be reached with sufficient signal-to-noise (SNR ~100) in two observing nights. This will secure direct metallicity determinations well beyond the Virgo and Fornax Clusters. The dynamics of a given region of the host galaxy can also be mapped, as observations at R ~10 000 will provide velocity resolution of 30 km s$^{-1}$ (Evans et al. 2011).

The J-band technique has been also applied to unresolved super star clusters (SSCs) rather than individual stars. Unresolved SSCs are dominated by hundreds of RSGs as soon as they are older than ~6 Myr and their integrated spectra have the spectral appearance in the J-band of a single RSG RSG (Gazak, J. Z. et al. 2014, ApJ, 788, 58 and ApJ, 787, 142). Hence, the J-band technique can be used to measure metallicities of unresolved SSCs and extend at a limiting distance ten times greater than for individual stars. indeed, accurate metallicities have been obtained in the Antennae at a distance of 20 Mpc using KMOS/VLT (Lardo et al. 2015, ApJ, 812, 160). A significant larger volume of the local universe will be accessible for quantitative investigations of the chemical evolution of galaxies with J-band observations using MAORY+MICADO.

Estimates of stellar metallicity well beyond our Local Group, extending from Virgo to Coma clusters, would also have a relevant impact on the cosmic distance scale that relies on SNIa calibrated on Classical Cepheids (Riess et al. 2016, ApJ, 826, 56). There is indeed no general agreement concerning the metallicity dependence of Cepheids pulsation properties, thus it would be ideal to know their chemical composition at very large distances (> 20Mpc, see also M. Marconi et al. MAORY science case) in order to minimise systematic errors on SNIa calibration.

**Keywords:** spectroscopy – metallicity – Red Super Giants

**MICADO Observation mode:** Standard spectroscopy.

**MICADO Pixel Scale / Fov:** 4mas/px and 50 arcsec FoV is the only offered mode.

**MICADO Spectral set-up:** Narrow slit spectroscopy for point source, the expected resolution is 18'000 in IzJ broad band.

**Filters required:** IzJ bands simulateusly setup.

**Estimate Survey Area/Sample Size/ Number of Images/Epochs:** Galaxies with $\delta \leq 20°$, i.e. those observable from Cerro Armazones at reasonable altitudes ($\geq 45°$) within the 30 Mpc volume (e.g. the limiting distance for direct abundance determinations from single RSGs) are >1000 (Evans et al. 2013). A larger volume can be sampled by observing SSCs from integrated light once the technique has been extensively tailored and tested on relatively close systems for which direct metallicity determinations are available (i.e. the two groups of galaxies in Sculptor at a distance of ~1.9 Mpc and 3.6-3.9 Mpc; e.g. Gazak et al. 2015). We will include the galaxies where Cepheids will be searched for using MAORY+MICADO, if any.

**Average Integration time per image** (magnitude of targets; S/N required): RSGs have absolute magnitudes in the range between $-8 < M_J < -11$. Evans et al. 2011 show that magnitudes down to J ~21 mag could be reached with sufficient signal-to-noise (SNR ~100) in two nights of observing ($t_{exp}$=10h). J=21 (at R~10 000) corresponds to an impressive distance of ~25 Mpc for the most luminous RSGs (e.g., well beyond the Sculptor and Cen A group).

**Observation requirements:** no special requirement.

**Strehl or EE required** high SRs are essential for measuring RSGs at very large distances.

**Astrometric Accuracy:** no special requirement.

**SCAO vs. MCAO:** MCAO is preferred for sky coverage issue.

**Comparison with JWST or other facilities:** The J-band technique can possibly be extended to JWST NearIR Spectrometer (NIRSpec) data. NIRSpec is a near-infrared (wavelengths of 0.6 to 5.3 μm) multi-object spectrograph (over 100 sources) capable of several modes of observation including a novel multi-object mode in a field of view of ~ 3.5 × 3.5 arcmin. NIRSpec will provide three spectral resolving powers (R = 100, 1000 and 2700). Point-source targets down to a magnitude of $J_{VEGA}$ = 23.1 can be observed with the grating providing the highest resolution (R=2700 at 1.15 μm) with a S/N = 10 (two-pixel resolution element) in an integration of approximately 28 h. This suggests a limiting magnitude of J = 18.7 (to reach a S/N ~ 100) which, combined with the large multiplex capabilities of the spectrograph, could obtain large samples of RSGs in galaxies out to the Virgo Cluster if the maximum selectable spectral resolving power of R=2700 is proven to be sufficient for absorption line spectroscopic studies. A possible synergy between MAORY+MICADO and JWST is to use the space spectrograph for closer galaxies (limiting magnitude J~18.7, for multiple objects at the same time with enough S/N) and the ground one for farther galaxies (J~21) extending the observable volume to distances of ~200 Mpc.

**Synergies with other facilities** (4MOST/MOONS, LSST/ALMA/HARMONI/METIS, HIRES/MOSAIC), but also VLT or other smaller telescope instruments: Accurate metallicity measurements from RSG observations through the J-band methods have been performed in the Magellanic Clouds and beyond (e.g., NGC 6822, WLM, M 83, M 31, NGC 55, NGC 300, Antennae merging galaxies) with VLT/XShooter, VLT/KMOS, and Keck/MOSFIRE (see Davies et al. 2015, Msngr, 161, 3, for a summary). The J-band method will also have applications with the e.g., VLT/MOONS, JWST/NIRSpec, TMT/IRIS-IRMS (Evans et al. 2011). The design of E-ELT/MOS has the potential of direct metallicity estimates RSGs well beyond the Virgo and Fornax Clusters via J-band technique (Evans et al. 2011, 2012, SPIE, 8446E, 7KE).

**Simulations made/needed to verify science case or feasibility:** simulations can help in order to estimate detailed exposure times.

**Origin of the targets:** already known.

**NGS:** Given the nature of the targets we will select, late-type galaxies, MCAO mode will assure us the presence of eligible NGS.

**Acquisition:** no specific constraints.

**Calibrations:** standard for broadband spectroscopy.

**Data Processing Requirements**: standard.

**Any other comments**: none

# 6.3 Chemical composition of globular clusters from the Local Group to Virgo/Fornax galaxy clusters.

**Authors:** Pancino E. (INAF - OAA) and Fiorentino G. (INAF - OABO)

**Brief description of science case:**

The science case is focused on *integrated light spectroscopy*, to derive detailed chemical information and radial velocities, of *distant globular clusters (GCs)* going from the outskirts of the Local Group to Virgo and Fornax galaxy clusters, i.e., at distances of 1-20 Mpc.

The fundamental scientific questions are two, and interrelated. One is related to the chemical evolution and assembly history of galaxies, using chemical abundances of the GC populations as tracers in their external regions. The other one is linked to the origin of the puzzling Multiple Populations (MPs) found in GCs at all distances, and about the possibility that only GCs formed in particular enviroments – i.e, in different host galaxies and at different distances from their host galaxy – possess different degrees of internal abundance variations.

Concerning the Galactic history, the abundance of alpha-elements enhancements such as [Ca/Fe] has been used to infer the star formation rate of different Galactic components (see Colucci et al., 2013, ApJ, 733L, 36). Also, the low [Mg/alpha] values found in some extragalactic GCs have been used to reverse-engineer the supernova yields and the formation history of external galaxies (see references in Pancino et al., 2017, A&A, 601, 112).

Concerning the GCs internal chemical variations, so far extragalactic clusters beyond 1 Mpc have only been studied with low resolution spectroscopy (R<10000). At those resolutions, [N/Fe] is the ideal tracer element because it is known to vary by more than one order of magnitude and its molecules can be studied even at low resolution (see Schiavon et al., 2013, ApJ, 776L, 7). The spectral resolution achievable with MAORY+MICADO (R~10000) opens up the entire space of varying elements (C, N, O, Na, Mg, Al, He, s-process elements) enabling the correlation between the internal variations of globulars and their position around the host galaxy, and later, with the properties of their host galaxy itself.

This project is particularly suited for MAORY+MICADO because the globular clusters appear sufficiently point-like at those distances, so that integrated light spectroscopy techniques (Larsen et al., 2012, A&A, 546, 53, Colucci et al., 2017, ApJ, 834, 105) can be applied to obtain representative mean abundances for the clusters. They are faint (approximately V~20 mag) and spread around the main galaxies over large areas (a few degrees), therefore single slit spectroscopy is more indicated. The closest globulars in the Southern hemisphere, such as those in the Local Group (WLM, Fornax dwarf, LMC, SMC) can also be observed either with integrated light techniques or resolved in stars with HARMONI, thus bridging very different distance scales for an unprecedented study.

**Keywords:** spectroscopy – metallicity – globular clusters

**MICADO Pixel Scale / Fov:** 4mas is the only pixel scale possible in spectrographic mode.

**MICADO Observation mode:** Standard spectroscopy.

**MICADO Spectral set-up:** the combination of the IJ range (0.770 – 1.450) and the HK range (1.400 – 2.500) provides a rich spectroscopic domain containing, in the form of atomic lines and molecular bands, all of the major nucleosynthetic sources (iron-peak elements, alpha-elements, r- and s-process elements, p-capture elements).

**Filters required:** none

**Estimate Survey Area/Sample Size/ Number of Images/Epochs:** a couple of hundreds of globulars to cover different host galaxy distances in the Virgo cluster (and possibly even in the Fornax one).

**Average Integration time per image (magnitude of targets; S/N required):** the targets in the Virgo cluster have typically V~18-20 mag, and the minimum S/N>30 should be achieved, requiring very roughly 5 min of exposure, with a total time for the programme of ~20 hours for 200 clusters. In the Fornax cluster the typical cluster magnitudes are ~20-23 mag, so only the brighter clusters can be observed in another ~20 hours. To go significanlty deeper, one should increase ten times the total exposure time (~200 hours).

**Observation requirements:** for the spectroscopic observations, the main requirement is to have a stable enough slit centering and guiding, to avoid spoiling the resolution by more than ~10% approx. A spectral resolution lower than expected would diminish the amount of atomic lines that can be measured with sufficient accuracy. An increase of the S/N could partially compensate this loss for the more isolated atomic lines.

**Strehl or EE required:** while SCAO could increase the S/N ratio of the obtained spectra, it would impose a limiting magnitude that is too bright (V~15 mag). Therefore, we opt for MCAO. A high S/N is nevertheless important for us.

**Image Stability Required:** the PSF should remain stable to 10% for 5 minutes to allow the maximum attainable resolution.

**Astrometric Accuracy:** sufficient to keep the objects centered in the slit (to 10%)

**SCAO vs. MCAO:** MCAO (see above)

**Comparison with JWST or other facilities:** JWST does not have the collecting area (6.5m vs. 39 m) to reach those distant clusters with the required resolution (R of a few 1000 vs 10000).

**Synergies with other facilities (4MOST/MOONS, LSST/ALMA/HARMONI/METIS, HIRES/MOSAIC), but also VLT or other smaller telescope instruments:** it would be highly desirable to connect GCs across the whole distance scale using 4MOST, MOONS, and existing archival spectra for the Galactic ones, HARMONI for the ones up to a few Kpc (integrated light to have higher S/N ratio), and MICADO for the very distant ones. Some overlap would be desirable in terms of the same clusters observed with different instruments, and with different techniques (single star spectroscopy vs. integrated light spectroscopy).

**Simulations made/needed to verify science case or feasibility:**

**Origin of the targets:** There is a plethora of literature catalogues for selecting targets.

**NGS:** Galactic coordinates of Fornax (l;b)=(237;-54) and Virgo (l;b)=(283;74) clusters should assure us a skycoverage of ~50% accordingly to current estimation.

**Acquisition:** no particular requirement.

**Calibrations:** standard calibrations - meaning at least one flux standard to obtain at least the correct spectral shape - are sufficient. Telluric standard stars (hot, nearly featureless, possibly fast rotating stars) would be useful.

**Data Processing Requirements:** a 1D, extracted and wavelength calibrated spectrum, possibly calibrated in flux as well, is the starting point of the analysis. No PSF knowledge is required and possibly the only non-standard calibration requirement would be the observation of telluric standard stars.

**Any other comments:** none.

# 6.4 Metallicity gradients in the giant elliptical galaxy NGC 3379 through planetary nebulae

**Authors:** Annibali F., Fiorentino G., Tosi M. (INAF - OABO)

**Brief description of science case:**

How and when did giant early-type galaxies (ETGs) assembled the bulk of their stars is still a matter of strong debate. Formation paradigms for ETGs are centered around two main antipodal hypotheses: the monolithic dissipative collapse and the hierarchical merging scenario. The first scenario supposes that massive ETGs formed the bulk of their stars in a single violent burst of star formation (SF) at high redshift (z≥2), and evolved quiescently thereafter. On the other hand, the hierarchical merging hypothesis predicts that giant ETGs formed through a continuous assembly of smaller units, reaching their final stellar masses in more recent epochs (z≅1).

Studies of metallicity gradients in present-day ETGs are a powerful diagnostic on their formation: a monolithic dissipative collapse is expected to produce steep gradients (e.g. Carlberg 1984, ApJ, 286, 403; Tantalo et al. 1998, A&A, 335, 823; Pipino et al. 2008, A&A, 484, 679; see Figure 6.4-1 of this science case), while mergers should dilute them (White 1980, MNRAS, 191, 1; Kobayashi 2004, MNRAS, 347, 740). Previous studies have shown that metallicity gradients are indeed present in ETGs (e.g. Annibali et al. 2007, A&A, 463, 455), but they appear somewhat flatter than predicted by pure dissipative collapse, suggesting that both in-situ formation and mergers possibly play a role. Unfortunately, due to the lack of giant ETGs within the Local Group (the closest cases being the "peculiar" E/S0 galaxy Cen A, at a distance of D≅3.8 Mpc, and the "normal" giant elliptical NGC 3379, at D≅10 Mpc), the majority of these results rest on integrated light studies, which are notoriously affected by a strong age-metallicity degeneracy. Abundance determinations of individual planetary nebulae (PNe), characterized by bright emission lines, have been attempted in Cen A and in more distant ETGs using optical spectroscopy on 8-10 m telescopes (e.g. Mendez et al. 2005, ApJ, 627, 767; Walsh et al. 2012, A&A, 544, 70), but have not been able to detect faint *auroral* lines (like [OIII] λ4363) that are needed to infer the electronic temperature and to robustly determine the element abundances. Furthermore, none of those studies were able to target PNe in the most crowded central galaxy regions (r<$R_e$), where an early in-situ collapse should produce strong metallicity gradients (see Figure 6.4-1).

We propose to exploit the high performance of MAORY+MICADO in imaging and spectroscopic modes to identify PNe and derive their abundances in the crowded center of the giant elliptical NGC 3379 at D≅10 Mpc. This project builds on a tight synergy with JWST/NIRSpec, which, with its lower angular resolution but larger field of view, will be better suited to study PN abundances in more external regions of NGC 3379 (see Figure 6.4-2). However, the MICADO FoV is large enough to sample a significant metallicity variation in the central region of NGC 3379 (Figure 6.4-1). A follow-up of the selected PN candidates will be performed with MICADO in spectroscopic mode to infer chemical abundances.

There are many emission lines present in the 0.8-2.5 μm spectral region, most notably those due to H and He recombination lines, plus several forbidden metal lines such those of [SIII] and [FeII] (see Figure 6.4-3). These lines can act as diagnostic tools to probe the physical conditions inside the nebula, (temperature, density, and excitation). Using the ELT ETC, we estimate that we need ~0.5 h exposure time for each PN spectrum to reach our goals (see details on the technical page). Following the approach of Rudy et al. (2001, AJ, 121, 362), we will derive the Sulfur (and possibly Fe) PN abundances. To select PNe within the center of NGC 3379, we will perform MAORY+MICADO pre-imaging in selected filters, as illustrated in Figure 6.4-3. The PN metallicity gradients derived from the combined ELT+JWST analysis will be compared with theoretical predictions for different formation scenarios of giant ETGs. Our study will furthermore provide a valuable complement to studies of photometric metallicity gradients in giant ETGs planned with JWST and ELT (see e.g. Greggio et al. 2012, PASP, 124, 653; Schreiber et al. 2014, MNRAS, 437, 2966).

**Keywords:** Early-type galaxies - metallicity gradients – planetary nebulae – imaging - spectroscopy

**MICADO Observation mode:** Pre-imaging + Spectroscopy

**MICADO Pixel Scale / Fov:** 4mas/px and 50 arcsec FoV

**MICADO Spectral set-up:** R=10,000, IzJ and HK bands.

**Filters required:** Pre-imaging with MICADO to select candidate PNe will be performed in two filters: **xY2** ($\lambda_{cent}$=1.08 $\mu$m, $\Delta\lambda$=0.09 $\mu$m), centered on the bright emission line HeI$\lambda$10830, and **xJ1** ($\lambda_{cent}$=1.2$\mu$m, $\Delta\lambda$=0.11$\mu$m), sampling an adjacent region with no prominent emission lines (see Figure 6.4-2). Spectroscopy at the position of the candidate PNe will then be performed with R~10,000 in the 0.8-2.4 $\mu$m range.

**Estimate Survey Area/Sample Size/ Number of Images/Epochs: TBD**

**Average Integration time per image** (magnitude of targets; S/N required):

Imaging Exposure time: we need a more detailed ETC, including narrow band filters, to give a reliable estimate of the exposure time [at D~10 Mpc, flux in HeI $\lambda$10830 is ~1.5 10$^{-17}$ erg/s/cm$^2$];

Spectroscopy exposure time: Shifting the typical spectrum of a bright PN to a distance of D~10 Mpc, and using the ELT ETC, we find that bright [S III] $\lambda\lambda$9069,9532 lines can be measured with a S/N~10 in ~0.5 h. Therefore, we need ~0.5 h net exposure time for each PN.

**Observation requirements:** none.

**Strehl or EE required:** the spatial resolution is a key point in crowded field. Thus, the higher the SR is, the better the result of this observation.

**Astrometric Accuracy:** relative astrometric accuracy is important, this is why pre-imaging is required.

**SCAO vs. MCAO:** MCAO

**Comparison with JWST or other facilities:** MICADO observations are ideal (and sufficient) to address our science goal, as well illustrated in Figure 6.4-1: in fact, the MICADO FoV allows for sampling of the central R≤Re galaxy region (NGC 3379's Re is 42"), within which "monolithic" collapse models predict a large metallicity variation (as high as ~0.6 dex for the simulation displayed in Figure 6.4-1). JWST/NIRSpec will complement these results allowing to study PNe in the periphery. However, only the MICADO spatial resolution will be sufficiently high to resolve the PN population toward the most central, crowded galaxy regions.

**Synergies with other facilities:** The exquisite spatial resolution of MAORY+MICADO system is ideal to obtain PN spectra in the most crowded central galaxy regions; on the other hand, JWST/NIRSpec will be better suited (with its larger field of view, but lower angular resolution) to target PNe in more external, less crowded regions of NGC 3379. The ELT- JWST synergy will allow to trace the PN abundance gradients from the center up to large galacto-centric distances.

**Simulations made/needed to verify science case or feasibility:** simulations may be needed to understand the spectroscopic feasibility.

**Origin of the targets:** The PN candidates will be selected through MICADO pre-imaging.

**NGS:** MCAO should assure the needed sky coverage to select a reliable asterism of NGS for the Galactic coordinates of NGC 3379, i.e. (l;b)~(233, 57).

**Acquisition:** no special requirements.

**Calibrations:** Flux standard calibration.

**Data Processing Requirements**: standard techniques will be applied.

**Any other comments**: additional requirements/issues.

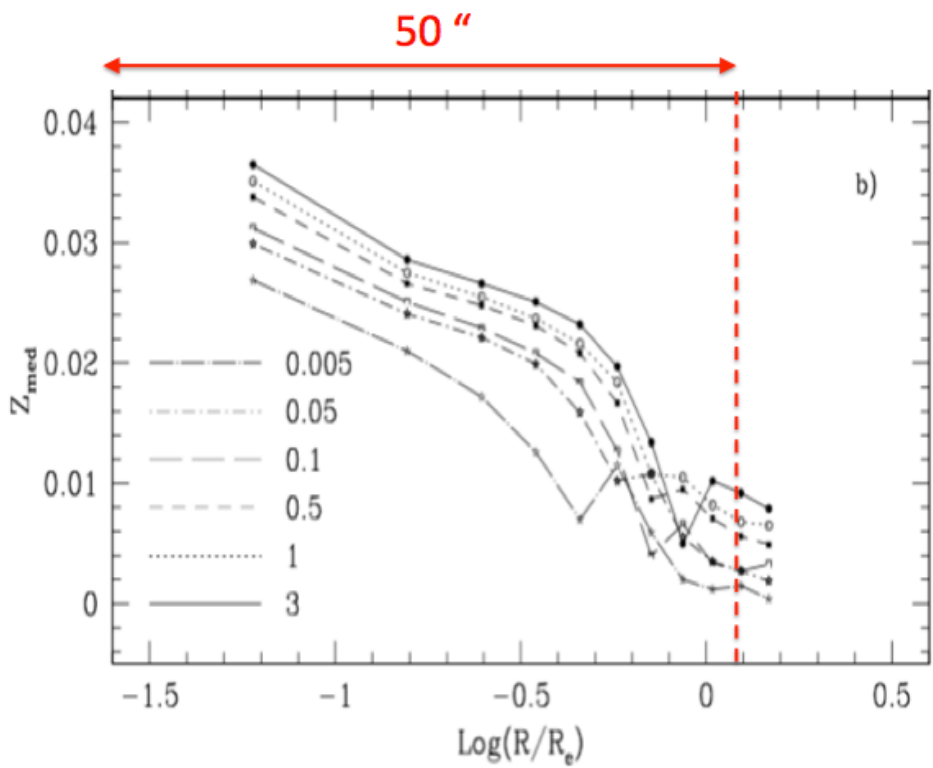

Figure 6.4-1 From Tantalo et al. 1998, A&A 335, 823, simulation of an elliptical galaxy formed through collapse into the dark matter potential well. The figure shows, for the simulated galaxy, the behaviour of metallicity as a function of R/Re, where Re is the half-light radius. The different curves are for different adopted M/L ratios as indicated. The dashed line corresponds to a radius of 50" (equalling the MICADO FOV) adopting Re=42" for NGC3379. This illustrative case shows that strong metallicity gradients are expected in a "monolithic" formation scenario. One MICADO pointing is sufficient to sample a metallicity variation as large as ~0.5 dex. Simulations of giant ellipticals formed via mergers predict significantly flatter metallicity gradients (see e.g. Kobayashi 2004, MNRAS 347, 740; di Matteo et al 2009, A&A 499, 427), the exact trend depending on the details of the merger event.

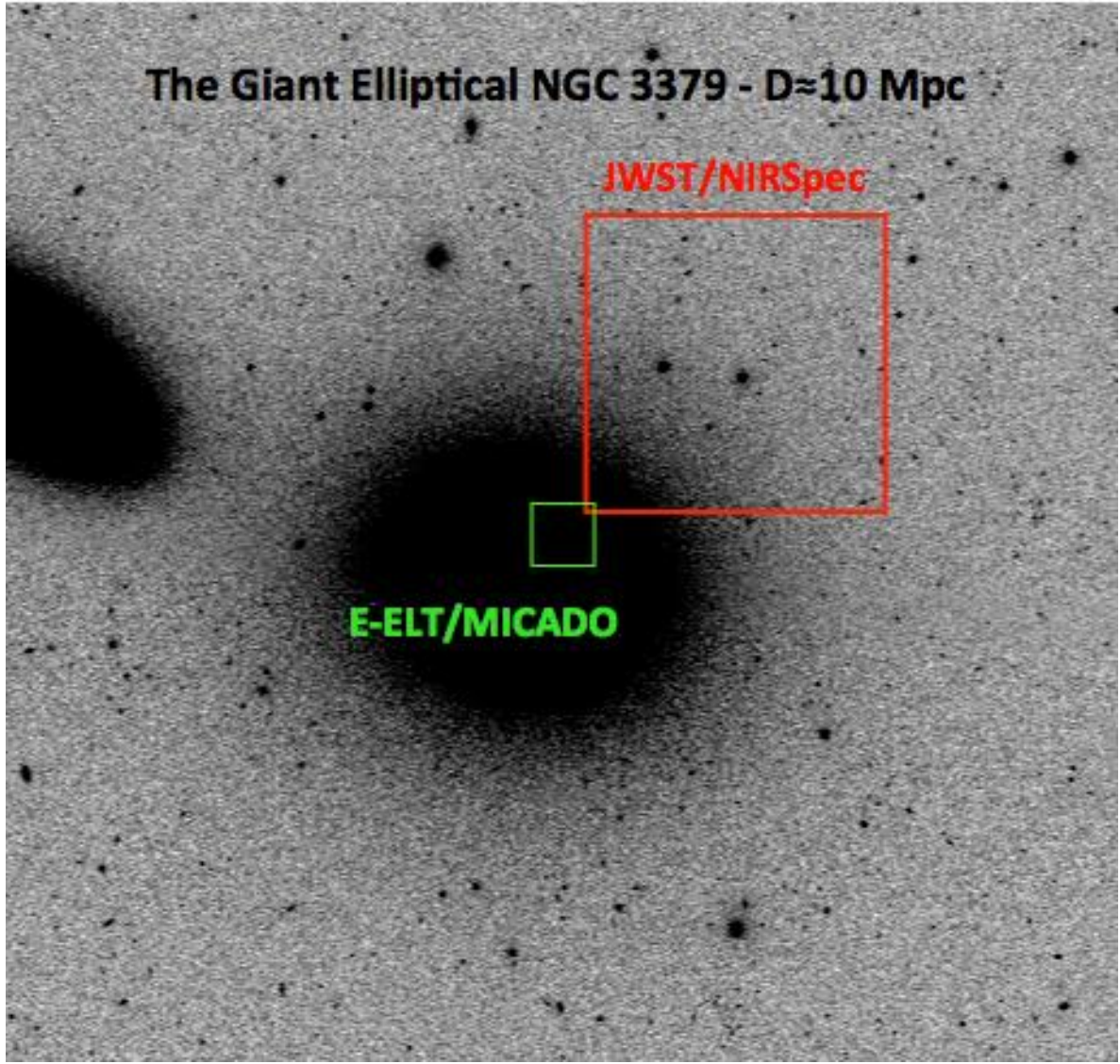

Figure 6.4-2 SDSS i-band image of the giant elliptical NGC 3379 at ~10 Mpc distance with superimposed the MICADO field of view. For comparison, the FoV of JWST/NIRSpec (~4'x4'), that will be used to study PNe in the outer galaxy regions, is also shown.

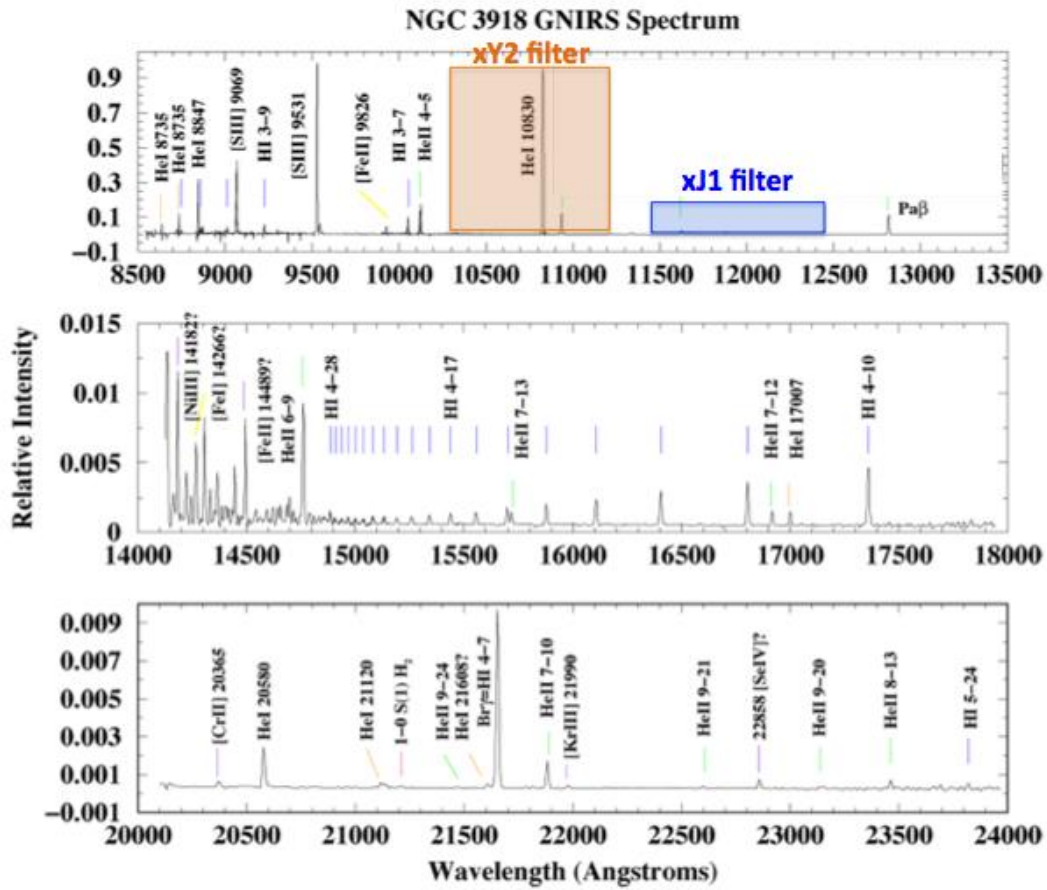

Figure 6.4-3 As an example, near-infrared spectrum of the planetary nebula NGC3918 acquired with Gemini/GNIRS. The PN spectrum shows, besides Pa and Br hydrogen lines, also prominent emission lines of He I, He II, and [S III]. On the spectrum, we have superimposed two MICADO filters (xY2 and xJ1) sampling respectively the bright HeIλ 10830 emission line and an adjacent region with no prominent emission lines. MICADO pre-imaging in the two filters will be used to select candidate PNe in the center of NGC 3379.

# 6.5 Hunting the oldest stars deep into the crowd of giant galaxies beyond the Local Group

**Authors:** *Fiorentino G. (INAF - OABO), Bono G. (University of Tor Vergata), Dall'Ora M., Marconi M. (INAF - OACN), Martinez-Vazquez C. (INAF - OABO), Monelli M. (IAC), Moretti M. I., Musella I., Ripepi V. (INAF - OACN), Valenti E. (ESO), Zoccali M. (PUC)*

**Brief Description Of Science Case:**

A full understanding of how giant galaxies form and evolve is still missing. Our Local Group (LG, distances within 1 Mpc) hosts only two giant galaxies, the Milky Way (MW) and M31. These galaxies have been studied extremely in detail using spectroscopy and photometry of their individual stars. These studies seem to support the hierarchical scenario of galaxy formation where giants are formed by smaller bricks. We want to observe all the giants in the nearby groups of galaxies visible from the southern hemisphere (see Table 6.5-1) in order to make a comparative study between their ancient resolved stellar populations and those in our LG. This will further support (or disprove) current galaxy formation theory.

The ancient stars can be traced by low mass stars that have indeed ages comparable with the Hubble time. The most precise and detailed studies of this old stellar population can be carried out using deep photometry reaching at least two magnitudes below the old Main Sequence Turn Off level (MSTO(V band)~+5mag) and are limited to our LG. This constraint is not going to dramatically change with the advent of JWST neither with giant ground based telescopes, since these telescopes will operate in near and mid Infrared (MSTO$_{K\ band}$~3mag). The range of giant galaxies of the Local Universe explored down to the MSTO is expected to remain quite unchanged, thus still restricted to the MW and M31.

We propose to focus on evolved low, thus old (>10Gyr) mass stars burning helium in their centre instead MSTO stars. Old stars, in their central helium burning phase are located on the horizontal branch (HB), and are still quite common. They are much brighter than their central hydrogen burning counterpart. Given the HB magnitude level (HB$_V$~+0.5; HB$_K$~-2.5), the universe surveyed by HST (within ~2Mpc, Da Costa et al., 2010, ApJ, 708, 121) will be extended up to a factor of 3 by reaching distances out to ~6Mpc. Among HB stars, the well-known RR Lyrae stars can play a crucial role since their detection and characterisation (mean magnitudes, period and amplitude distributions) can provide crucial information on the early epochs of galaxy assembling (Fiorentino et al. 2015, 2017). Furthermore, through the use of consolidate theoretical diagnostics such as the metallicity dependent infrared PL relations we will be able to constrain the galaxy chemical enrichment (Marconi et al. 2015, Martinez-Vazquez et al. 2016). Our recent studies of the MW (Fiorentino et al. 2015) and of M31 (Monelli et al. 2017) support recent hierarchical cosmological models pointing out that small dwarf galaxies (M<$10^8$ Mo) cannot have majorly contributed to giant galaxy assembly. RR Lyrae stars belonging to small dwarfs have different pulsation (amplitudes and periods) properties when compared with those of the bulges/halos of the MW/M31.

**We propose to use the unique high resolution allowed by MAORY+MICADO to explore individual stars down to the HB level in the most crowded regions (bulges/cores) of four giant galaxies** (see Table 6.5-1). This sample includes 3 spirals and one peculiar elliptical galaxy, NGC5128. This latter galaxy is a rare, close, example of an elliptical merging galaxy hosting a confirmed young population (Classical Cepheids) and it is extremely interesting by itself.

| Group | Galaxy | MB (mag) | D (Mpc) | morph. Type |
|---|---|---|---|---|
| Cen A | NGC5128 | -20.77 | 3.7 | S0 |
| Cen A | NGC4945 | -20.51 | 3.6 | SB |
| M83 | NGC5236 | -20.43 | 4.5 | SAB |
| Scl-filament | NGC253 | -21.37 | 3.9 | SAB |

Table 6.5-1 An illustrative example of the galaxie, and their morphological type, that we will be penetrate deep into their cores with MAORY+MICADO.

This systematic study, by exploring all the giants in the nearby galaxy groups, will also allow us to determine the environmental impact on galaxy formation. The proposed data will have an astrophysical legacy value since MAORY+MICADO will be the only instrument, for the next decades, able to penetrate these regions with a promised accuracy of few hundreds of magnitudes. We will provide the deepest Colour Magnitude Diagrams (CMDs) ever for their, otherwise inaccessible, dense galactic components. The synergy of a detailed CMD study, including the variable star population, will provide the most comprehensive picture of how the central regions of giant galaxy formed and evolved posing strong constraints on current cosmological models.

**Keywords:** imaging - photometry – galaxy formation

**MICADO Observation mode:** Standard imaging.

**MICADO Pixel Scale / Fov:** 4mas/px and 50 arcsec FoV is the ideal scale in order to maximise the stellar statistics.

**MICADO Spectral set-up:** none.

**Filters required:** J and H-band may be the best choice to obtain the deepest CMDs in order to minimise the sky-background and optimise the S/N ratio. The search for variable stars will be conducted in J-band, where the amplitude of the variation is larger. Detailed simulation can help in the final filter choice.

**Estimate Survey Area/Sample Size/ Number of Images/Epochs:** one pointing per galaxy is the minimum required. The sample size is made of four galaxies (see Table 6.5-1). The need of time series requires at least 16 epochs in J and 5 epochs in H band. Detection will be conducted in J band, and the use of light curve templates in H band will make sufficient the collection of only five epochs in this filter.

**Average Integration time per image** (magnitude of targets; S/N required): the E-ELT ETC (version 6.2.2) provides, assuming 3X3 pix as reference area, airmass<1.5, 5mas/pixel resolution, and for a S/N~5 on each source: 1) with J~30 mag, ~19'080s integration on source (for 16 epochs, each epoch made of 30s DIT x 40s NDIT); 2) with H~28.5 mag ~9'000s integration on source (for 5 epochs, each epoch made of 30s DIT x 60s NDIT).

**Observation requirements** small dithers will be needed. Single frames will be stacked after proper astrometric calibration.

**Strehl or EE required** high SRs are essential for measuring faint individual stars at large distances.

**Astrometric Accuracy:** no special requirement

**SCAO vs. MCAO:** MCAO is preferred for the larger field of view and PSF uniformity.

**Comparison with JWST or other facilities:** JWST will be a key facility for low surface brightness studies. The proposed project will greatly benefit of a complementary study to be carried out with JWST to survey the halos of the selected giants. This will allow us to give constraints on the link between the ancient stars of the different galactic components.

**Synergies with other facilities:** integrated IR spectroscopic measurements possible with HARMONI or MICADO in spectrographic mode will be investigated in detail, since they could provide fundamental, complementary, information to the stellar population characterisation.

**Simulations made/needed to verify science case or feasibility:** simulations can help in posing the limit in distance for the HB to be observed.

**Origin of the targets:** coordinates available in literature (Karachentsev et al., 2004, AJ 127, 2031).

**NGS:** not available yet. LSST catalogues or VHS will provide them. Since the sky coverage analysis should not be a concern for the specific case considered here.

**Acquisition:** no specific constraints; accurate finding charts are available over any the possible target of interest.

**Calibrations:** standard. No special astrometric requirement. The presence of photometric calibrators in the field has to be verified (VHS may help). The relative photometry precision derived with the ETC (0.01-0.05 mag in J-H color) is sufficient to reach the main science goal.

**Data Processing Requirements**: Good PSF modelling is required; target selection will by necessity take such requirement into account.

**Any other comments**: sky subtraction to be investigated.

# 6.6 Mapping the mass and metallicity distribution in the cores of Ellipticals in the Virgo Cluster

**Authors:** *Greggio L., Falomo R., Gullieuszik M. (INAF - OAPD)*

**Brief description of science case:**

In spite of the tremendous progress achieved in the last decades on both theoretical and observational studies, we still lack a clear and undisputed picture of how galaxies form, grow and assemble. In particular, for Elliptical galaxies, we do not know whether their assembly occurs preferentially via dry merging, involving mostly stars, or wet merging, involving stars and gas, accompanied by star formation (e.g. Ciotti et al. 2007, ApJ 658, 65). Other formation scenarios consider the occurrence of wet merging at high redshift, followed by prevailing dry merging at lower redshift (Oser et al. 2010, ApJ 725, 2312), or in-situ star formation producing the inner regions at early epochs, and the growth of the external parts at later times.

Different formation paths imprint upon different stellar mass and metallicity gradients over the galactic radii. Currently, with a few exceptions, these quantities are estimated from the integrated light, which can only yield a global, luminosity weighted, information on the mass and the metallicity. Conversely, tight constraints on the formation models can be obtained from the detailed metallicity distribution, its peak and extension, and its trend with radius. From the analysis of the Color Magnitude Diagram (CMD) of the bright Red Giant Branch (RGB) stars ($M_I \leq -2$ mag) one derives the stellar mass, and the metallicity distribution via star counts (e.g. Rejkuba et al., 2014, ApJ 791, L2).

In order to draw a clear picture of how elliptical galaxies form, we need to study a sample of early type galaxies spanning a wide luminosity range and to analyze their spatially resolved CMDs. This will allow us to compare the properties of the various objects, testing model predictions of dry/wet merging for different galaxy sizes.

Down to a limiting magnitude of J=28 mag, at the distance of the Virgo Cluster we sample the CMD of stars brighter than $M_J$=-3.3 mag, which is ~2 mag fainter than the RGB Tip. This gives ample room to derive the metallicity distribution from the color distribution of the bright RGB stars (see Figure 6.6-1, left). This task is best accomplished with a wide color baseline, so that we aim to derive observations down to J=28 mag and I=29.5 mag.

Based on Schreiber et al. (2014, MNRAS 437, 2966) we estimate that 2 hr exposures in each band are more than adequate to reach this goal, and that accurate metallicity distributions can be derived in regions with surface brightness fainter than $\mu_V$~20.5 mag/arcsec$^2$. Scaling according to the pixel size, NIRCam@ JWST (32 mas/px) will deliver a comparable photometric performance in regions with $\mu_V$>25 mag/arcsec$^2$. Figure 6.6-1(right) shows the surface brightness profiles of a few Virgo Ellipticals with -23.0 <$M_V$<-17.0; the shaded area highlights the region where the excellent resolution of MAORY+MICADO will yield unique access to the resolved RGB stars. To cover this region ~2 pointings will suffice for the lower luminosity objects (e.g. N4515), while for the most massive, central dominating galaxies (e.g. M87) we will need ~25 pointings to explore the large high surface brightness area. Finally, we point out that these CMDs will also allow us to test the star formation history occurred in the last few Gyrs, as traced by the possible presence of AGB and Red Supergiant stars. The excellent resolution of MAORY+MICADO will give us access to the resolved stellar populations inhabiting the inner regions of Virgo Ellipticals, thereby enabling a thorough investigation of the 2D map of the stellar mass and metallicity where most of the galaxy mass resides.

**Keywords:** imaging – photometry – Elliptical Galaxies

**MICADO Pixel Scale / Fov:** 4 mas/px and 50 arcsec FoV.

**MICADO Observation mode:** Standard imaging.

**MICADO Spectral set-up:** none.

**Filters required:** I and J are the best combination to derive the metallicity from the color of RGB stars.

**Estimate Survey Area/Sample Size/ Number of Images/Epochs:** The galaxies in Figure 6.6-1 (right) range from dwarf to giant ellipticals, and different effective radii. To derive a thorough 2D mapping of this sample requires 1 field in IC3565, 2 fields in N4515 and N4387, 10 fields in N4473, 20 fields in N4636 and 25 fields in the giant Es M49 and M87. This sums up to a total of 85 pointings. Assuming an integration time (on target) of 2 h per filter, the total exposure time needed amounts to (85x2x2) =340 hrs.

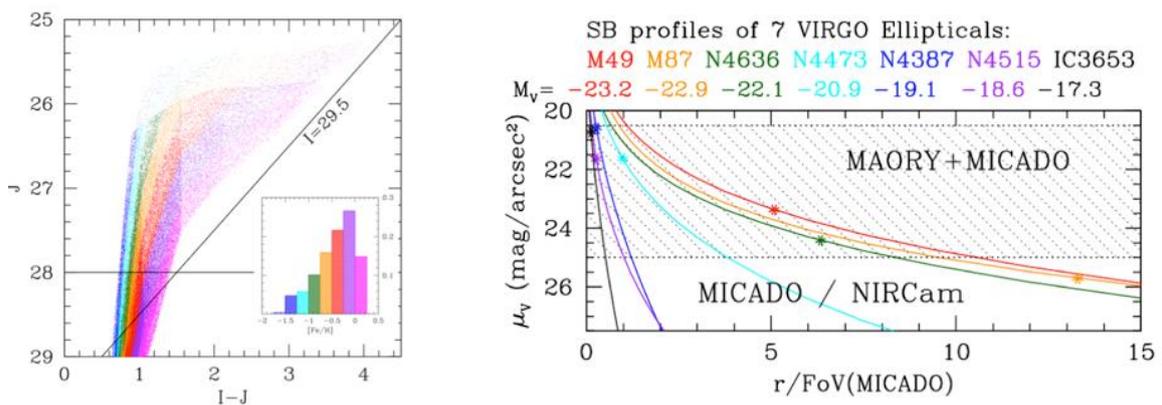

Figure 6.6-1 Left: CMD of a stellar population with a flat age distribution (inset). From Schreiber et al. (2014, MNRAS, 437, 2966). Right: Surface brightness profiles as from Sersic parameters in Kormendy et al. (2009, ApJS, 182, 216) of early type galaxies in Virgo spanning a wide range of absolute magnitudes, labelled on top. The asterisks mark the position of the effective radius. The radial coordinate is plotted in unites of the MICADO FoV (adopted of 50 arcsec).

**Average Integration time per image (magnitude of targets; S/N required):** we estimate that 2h integration in J and in I bands allow us to derive the CMDs with the adequate depth. This is based on detailed simulations performed with the MAORY+MICADO set up that was adopted in the MICADO phase A study (Schreiber et al. 2014). A more accurate estimate will be possible once the expected performance of the instruments will be updated and the MCAO PSF available.

**Observation requirements:** Since the fields are very crowded, large dithers will be needed to be able to measure the sky. The total exposure time could then be up to a factor of two larger than mentioned above.

**Strehl or EE required:** crowded fields photometry requires the best performance achievable, both as SRand as EE.

**Astrometric Accuracy:** no special requirement

**SCAO vs. MCAO:** MCAO; SCAO could be preferable in the very central regions, but there will not be a natural point source in the field.

**Comparison with JWST or other facilities:** The very high surface brightness regions are too crowded to derive accurate and deep photometry with the large pixel of the JWST detector. NIRCam will instead be more efficient than MICADO in the more external regions, due to the 10 times wider FoV.

**Synergies with other facilities:** Coverage of the more external regions will be possible and efficient with JWST. The combination of MICADO and NIRCam will yield a 2D mapping of the stellar content over the total extension of Ellipticals in VIRGO. Notice that RGB stars contain the fossil record of the star formation history over the whole Hubble time. Thus, these combined observations will give access to the spatially resolved star formation ever occurred in the galaxy.

**Simulations made/needed to verify science case or feasibility:** The feasibility of this science case has been tested with detailed simulations which adopted the specifications envisaged for the MAORY+MICADO system in the phase A study, and the results published in Schreiber et al. (2014). New simulations will be necessary once the performance of the instrument is updated.

**Origin of the targets:** Astronomical catalogues. The Virgo Cluster is very well studied.

**NGS:** Galactic coordinates of the Virgo galaxy cluster, (l;b)=(283;74), should assure us a skycoverage of ~50% accordingly to current estimation.

**Acquisition:** The targets are very well known nearby galaxies. We anticipate no problems to pointing the fields of interest.

**Calibrations:** The required photometric accuracy for this science case is estimated of 0.05 – 0.1 mag. This implies that secondary photometric standard stars in the field can be used to set the photometric calibration.

**Data Processing Requirements:** Standard PSF photometry package. The cores of the PSF will be extracted from the brightest members of the target galaxies. To model the PSF wings, we should have a sufficient number of relatively bright (foreground) stars. If not, we will resort to short exposure observations of an adjacent field.

**Any other comment:** none.

## 6.7 The stellar content of massive Globular Clusters: giant clusters or nuclear remnants?

**Authors**: Bellazzini M., Fiorentino G. (INAF - OABO), Mapelli M. (INAF - OAPD)

**Brief description of science case:**

Massive Globular Clusters (GCs) are suspected to be the nuclear remnants of ancient dwarf galaxies whose low-density stellar and dark-matter envelope have been stripped by the tidal field of the giant galaxy they are associated to. A key observational test for this hypothesis is to search for the signature of Supernova-driven chemical enrichment in the cluster stars, as this will proof that there was an epoch in which the system was sufficiently massive ($10^7$ $M_\odot$) to retain the ejecta of these very energetic polluters. This signature is a significant star-to-star spread in the abundance of α-elements and iron-peak elements that is not seen in *normal* GCs and is observed in some of the most massive among them (e.g. ω Cen, G1), while it is a defining property of galaxies (Willman & Strader, 2012, AJ, 144, 76). The presence of this abundance spread is most easily revealed in the Color Magnitude Diagram (CMD) of a cluster as a color spread at any given magnitude along the Red Giant Branch (RGB). The sensitivity of color to metallicity increases from the basis to the Tip of the RGB. According to Valenti et al. (2004, MNRAS, 351, 1204) a metallicity range from [Fe/H] = -2.1 to [Fe/H] = -0.6 (as observed in ω Cen) corresponds to a J-H color spread of ~0.2 mag four magnitudes below the Tip and ~0.5 mag at the RGB Tip; at this level the metallicity range from [Fe/H] =-2.1 to [Fe/H]=-1.4 corresponds to 0.2 mag. Hence the accurate J, H photometry of the cluster RGB allows to firmly detect and quantify the metallicity spread that expected in these systems (Carretta et al. 2010, A&A, 520, 95, see their Fig. 7, in particular). However Massive GCs are relatively rare objects in the Local Group: in the Milky Way there is only one old cluster with $M_V \le -10.0$; in M31 there are five. Therefore, the goal of characterising them as a class, for instance by determining the fraction of them showing a spread in metallicity, has been hampered by the lack of a proper sample. ELT+MAORY+MICADO will greatly enlarge the volume where globular clusters can be resolved into individual stars, providing access to the stellar content of a significant number of massive GCs in various galaxies within ~5 Mpc from us. As an illustrative example, *we focus here on the 32 confirmed globular clusters with $M_V \le -10.0$ hosted in the nearby (D=3.8 Mpc) giant elliptical Centaurus A* (NGC 5128, see Woodley et al. 2007, AJ, 134, 494). These clusters will be resolved by MAORY+MICADO ~5 times better than HST is currently doing with M31 GCs, thus giving us access to the stellar content in ample coronae around a small unresolved center. With 1h+1h MCAO mode observations in J and H we will reach the Red HB level and, in the brightest 4 mag of the RGB, we will have the photometric accuracy that is required to successfully detect the metallicity spreads we are looking for. An accurate modelling of synthetic RGBs using artificial stars experiments will provide a sensitivity to metallicity spreads of order 0.1-0.2 dex rms. The RGB bump will provide an additional constraint, with precious leverage also on He abundance.

*This Science Case is part of the MORY ST initiative on compact stellar systems.*

**Keywords:** imaging – photometry – extragalactic globular clusters

**MICADO Pixel Scale / Fov:** 1.5mas/px and 20 arcsec FoV, motivated by the maximization of the performances in terms of SR and by the small size of the actual targets.

**MICADO Observation mode:** Standard Imaging

**MICADO Spectral set-up:** none

**Filters required:** J and H seems the best compromise between AO performances and sensitivity of the color index to temperature, that is crucial for the scientific goal.

**Estimate Survey Area/Sample Size/ Number of Images/Epochs:** assuming 2h per cluster, a sample of fifteen $M_V \leq -10.0$ clusters in Cen A would require ~30h of observing time, excluding overheads and not accounting for imaging sky far from the targets.

**Average Integration time per image (magnitude of targets; S/N required):** with 1h per filter S/N~10 is reached at H=27, S/N~30 at H=26, 4 mag below the RGB Tip, and S/N~800 at the Tip (H=22). This gives a precision in the J-H color of ~0.05 mag at H=26 and <0.01 at the RGB Tip. The S/N~5 level is at H~28.0, reaching below the Red HB level. These estimates have been obtained on Feb 8, 2017, with the ESO-ELT ETC, assuming a px scale of 5 mas (the smallest scale currently available in the ETC), a K5V star as a template and a 3 px X 3 px reference area for computing the S/N ratio.

**Observation requirements:** Small dithers + Large dithers [refer to table 8 of MICADO OCD]

**Strehl or EE required:** the best spatial resolution performances the most effective would be the test: a larger portion of the cluster light is resolved into stars and, consequently, the stellar content analysis is based on larger and more precisely measured samples.

**Astrometric Accuracy:** no special requirement

**SCAO vs. MCAO:** MCAO is the only viable mode as the circumstance of a target within ~10" of a bright star (allowing SCAO) is rare of non-existing

**Comparison with JWST or other facilities:** Spatial resolution is the key issue here: ELT+MAORY+MICADO is much better than JWST in this respect (by a factor of >3, at high SR, comparing MICADO observations in H band to JWST observations in I band).

**Synergies with other facilities:** Targets positions and integrated magnitudes/ colors are already known. MICADO spectroscopy of the brightest clusters stars (identified with MICADO imaging) is possible according to the ELT ETC, hence spectroscopic abundance analysis may be a natural follow up.

**Simulations made/needed to verify science case or feasibility:** Simulations of the most and less compact among the possible targets will be very useful to study in detail the results that can be actually achieved in terms of number of RGB stars resolved and of sensitivity to abundance spreads. The half-light radius of the targets ranges from ~3 pc to ~12 pc, the H band resolution element (FWHM of the diffraction core) samples 0.16 pc at the distance of Cen A.

**Origin of the targets:** Lists available in the literature, e.g. Woodley et al. (2007, AJ, 134, 494), Georgiev et al. (2009, MNRAS, 392, 879).

**NGS:** Galactic coordinates of NGC 5128 are (l,b)~(309,19) hence the sky coverage should not be a concern for the specific case considered here.

**Acquisition:** the characteristic size of GCs in Cen A is ~1 arcsec hence there should be no problem in getting a useful pointing within the 20"X20" FoV. Finding charts available.

**Calibrations:** Standard. No special astrometric requirement. The presence of photometric calibrators in the field has to be verified (Euclid survey will not reach such low Galactic latitudes). The relative photometry precision derived with the ETC (0.01-0.05 mag in J-H color from the RGB Tip down to 4 mag below) is sufficient to reach the main science goal. A 0.02 rms accuracy in the photometric zero points would allow us to perform useful comparisons with RGB templates of local GCs and/or with theoretical models.

**Data Processing Requirements:** The PSF modelling should be feasible with stars in the FoV. Pre-reduced stacked images are the natural starting point for the scientific analysis. This is a classical PSF-photometry in crowded stellar fields case.

**Any other comment:** Since the typical size of our targets is much smaller than the FoV size (~1-2 arcsec vs 20 arcsec) most of the images will sample the stellar population within ≤ 200 pc X 200 pc (projected) portions of the parent galaxy at various distances from the center. This will provide stellar population tomography of the nearest giant elliptical galaxy, a very valuable product in itself. Note that the actual size of the final FoV available for scientific analysis will be determined by the size of the large dithering (i.e. it would be the intersection of the "A" and "B" pointings set up to get sky subtracted A-B and B-A images).

# 6.8 Unveiling nearby galaxies in the Zone of Avoidance

**Authors:** *Bellazzini M. (INAF - OABO)*

**Brief description of science case:**

A full census of the galaxies in the Local Group and its surroundings is mainly limited by the sensitivity of our surveys to very faint and low surface brightness objects. A significant limitation of a completely different nature is provided by the strong obscuration of a portion of sky of a few degrees around the Galactic Plane by interstellar dust, concentrated on the plane of the Disc. This region is generally known as the Zone of Avoidance (ZOA). Galaxies in the ZOA are usually found via HI observations that are not affected by interstellar extinction (see e.g. Staveley-Smith et al. 2016, AJ, 151, 52, and references therein; SS16 hereafter). The extreme collecting power and spatial resolution (as well as the NIR spectral range) of MAORY+MICADO can provide a mean to go through the thick clouds of the ZOA and find and/or characterize the stellar counterparts of galaxies found by HI surveys. This will allow us to obtain accurate and redshift-independent distances to these systems and to study their stellar content, that are currently out of reach. To illustrate the case, we consider here the nearest systems found by the recent *Parkes HI Zone of Avoidance Survey* (SS16). In this catalog there are thirteen galaxies with heliocentric radial velocity $V_\odot \lesssim +800$ km/s, corresponding to velocities in the Local Group reference frame $V_{LG} < 550$ km/s (see Fig. 5 of McConnachie 2012, AJ, 144, 4, to put these numbers in the proper context). The (roughly) estimated distance of these galaxies range between 1.5 Mpc to 7.3 Mpc, the foreground reddening range is $0.2 < E(B-V) < 5.3$. A few of them have a known stellar counterpart (the most notable case being the Circinus galaxy), but none of them has a distance estimate based on a stellar standard candle and eight of them have no (known) stellar counterpart at all. For all of them MAORY-MICADO observations will resolve stars down to at least ~4 magnitudes below the Red Giant Branch (RGB) tip in ≤ 1 h exposures in J and K, for example. Accurate distances will be derived from the RGB tip in NIR magnitudes (Bellazzini et al. 2004, A&A, 424, 199), the Star Formation History will be obtained by comparison with synthetic Color Magnitude Diagrams. Note that the lack of a stellar counterpart would imply that the system is a dark galaxy, an extremely interesting object per se (see, e.g., Taylor et al. 2017, arXiv:1701.05361).

**Keywords:** imaging – photometry – Local Volume – Galaxies: distance – Galaxies: stellar content

**MICADO Pixel Scale / Fov:** 4.5mas/px and 50 arcsec FoV, motivated by the typical size of the targets (radius ~0.5 arcmin to a few arcmin).

**MICADO Observation mode:** Standard Imaging

**MICADO Spectral set-up:** none

**Filters required:** J and H or J and K will provide the sensitivity to temperature required to obtain meaningful CMDs, in addition to low sensitivity to extinction and high AO performances, e.g., I band.

**Estimate Survey Area/Sample Size/ Number of Images/Epochs:** assuming 2400s per target, a sample of ten ZOA galaxies would require ~7h of observing time, excluding overheads and not accounting for imaging sky far from the targets.

**Average Integration time per image (magnitude of targets; S/N required):** The apparent magnitude of the RGB Tip of the thirteen nearby SS16 galaxies llies in the following ranges: $19.6 \leq K^{tip} \leq 23.2$, $19.9 \leq H^{tip} \leq 23.6$, and $21.2 \leq J^{tip} \leq 25.2$, assuming the absolute magnitude of the tip at [M/H]=-0.5 from the relations by Bellazzini et al. (2004; note that at higher metallicities the tip will be intrinsically brighter). According to ESO-ELT ETC, assuming a px scale of 5 mas (the smallest scale currently available in the ETC), a K5V star as a template and a 3 px X 3 px reference area for computing the S/N ratio, we find that $S/N \geq 5$ in J and H is reached for stars at H=27.0 with a total exposure time of 1200s (splitted into 20 NDIT of 60 s each; simulations performed on Feb 14, 2017). Hence with a 2 X 1200s time-on-target run, <0.2 precision photometry can be reached down to 2-6 mag below the RGB tip, $\leq 0.05$ at the tip, sufficient for the main science goals. One pointing per galaxy will be sufficient, in most cases, to get an adequate sample of member stars.

**Observation requirements:** Small dithers + Large dithers

**Strehl or EE required:** good SR is required.

**Astrometric Accuracy:** no special requirement

**SCAO vs. MCAO:** MCAO is required because of sky coverage and field width issues.

**Comparison with JWST or other facilities:** Similar performances can be obtained with JWST. However new candidates will be found by ongoing and future (more sensitive) HI survey and since the end of the JWST mission (2023) ELT will become the only option to reach this goal. It is possible that the nearest and less obscured targets can be fruitfully observed with AO-assisted 8m class telescopes, but this should be considered as a lucky circumstance.

**Synergies with other facilities:** Targets positions are / will be provided by ZOA HI surveys. MICADO spectroscopy of the brightest stars (identified with MICADO imaging) is possible according to the ELT ETC, hence spectroscopic abundance analysis may be a natural follow up. HARMONI observations can be also very interesting.

**Simulations made/needed to verify science case or feasibility:** Not critical. Simulations for stellar populations in the same range of distance may be applicable. The effects of foreground contamination of the CMDs should be analysed for each individual target. The effect of a large number of (foreground) saturated stars within the FoV is to be verified.

**Origin of the targets:** Lists available in the literature, e.g. SS16.

**NGS:** All the targets are at very low Galactic latitudes, by definition. Hence the sky coverage should not be a concern for the specific case considered here. On the other hand, saturation of foreground stars in the field may be a problem.

**Acquisition:** For targets without a known stellar counterpart, the centers of the HI clouds must be blindly targeted. In the other cases finding charts will be available.

**Calibrations:** Standard. No special astrometric requirement. The presence of photometric calibrators in the field has to be verified (Euclid survey will not reach such low Galactic latitudes). The relative photometry precision derived with the ETC and a 0.05 rms accuracy in the photometric zero points are sufficient to reach the main science goal.

**Data Processing Requirements:** The PSF modelling should be feasible with stars in the FoV. Pre-reduced stacked images are the natural starting point for the scientific analysis. This is a classical PSF-photometry in crowded stellar fields case.

**Any other comment:** none

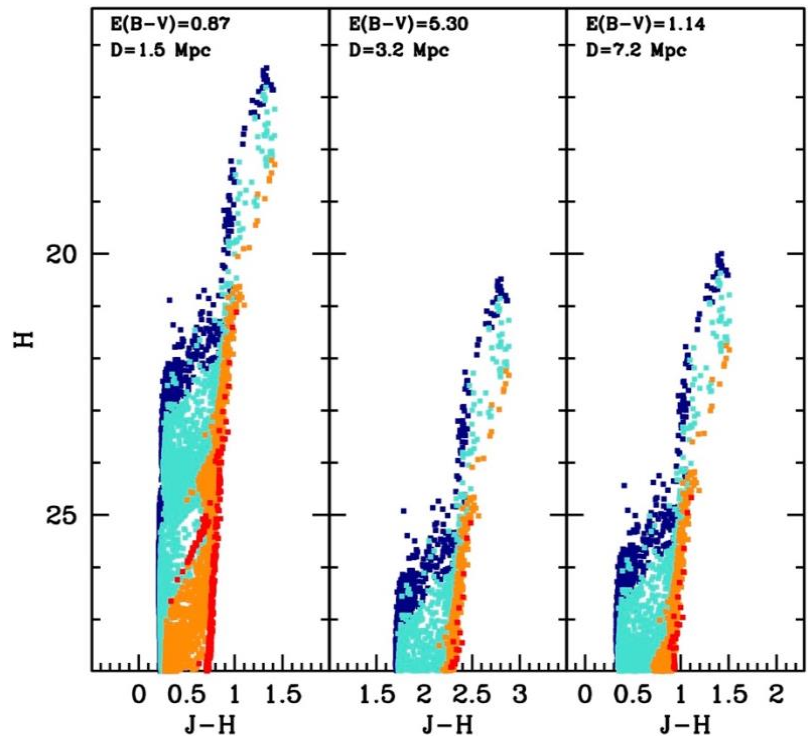

Figure 6.8-1 Synthetic Color Magnitude Diagrams spanning the range of distance and interstellar extinction of the ZOA galaxies sample considered here. The CMD has been obtained with the BASTI synthetic CMD tool, adopting the star formation history of the nearby metal-poor dwarf irregular Sextans A. Colors code the age of the stars. Blue: 10-100 Myr; Cyan: 100 Myr-1 Gyr; Orange: 1 Gyr-10 Gyr; Red >10 Gyr.

# 6.9 Ω and the Hubble diagram around z~0.3 with SBF

**Authors:** *Cantiello M., Raimondo G. (INAF - OATE), Brocato E. (INAF - OAR), Di Rico G. (INAF - OATE)*

**Brief description of science case:**

Observations of distant supernovae led to the discovery of the accelerating expansion of the universe (Riess et al., 1998, AJ, 116, 1009; Perlmutter et al., 1999, ApJ, 517, 565). Surface brightness fluctuations (SBFs) are the only other high-precision luminosity distance indicator capable of reaching distances needed to provide interesting cosmological constraints.

The SBF signal arises from the Poisson fluctuations of the brightest stars in a stellar population, primarily RGB stars, even in unresolved stellar populations (Tonry J., et al., 1990, AJ, 90, 1416). Distant galaxies appear smooth compared to nearby ones because the number of stars per resolution element is much larger, making the stellar counts fluctuations a smaller fraction of the total luminosity. To accurately measure fluctuations it is important to remove or correct for other contaminating sources of variance, including globular clusters (GCs) and background galaxies.

To date, the typical overall uncertainty on distances with SBF in **optical** bands is ≤0.10 mag (≤5% on linear distances; Blakeslee J., et al, 2009, ApJ, 694, 556; Cantiello M., et al., 2013, A&A, 552, 106). A large fraction of the total error comes from the zeropoint calibration (~0.08 mag), based on Cepheid distances. In a near future, it is expected that the results by the astrometric satellite Gaia will reduce such component to the error budget to a negligible factor (<0.02 mag).

Measuring SBF in the near-IR has several advantages: the increased contrast with GCs and background galaxies reduces the level of contamination; the seeing is better, which enables more accurate SBF measurements in galaxies at previously unreached distances in modest integration times; the extinction is much lower than in optical bands.

The use of HST instruments in both optical and near-IR has pushed forward the accuracy and the distance limit, reaching distances beyond 100 Mpc (Jensen J., et al. 2015, ApJ, 808, 91; Biscardi I., et al., 2008, ApJ, 678, 168). It is expected that the JWST will also be a key facility for SBFs: the combination of low background, large aperture, and high spatial resolution at NIR wavelengths, will enable accurate SBF measurements in galaxies at previously unreached distances in relatively small integration times.

The similar sensitivity between NIRCam/JWST and MAORY+MICADO at ELT, coupled with the ~6 times better best spatial resolution of MICADO, highlights the key role of ELT in exploring a range of distances beyond the limits reachable by JWST.

With ~1h integration time in Ks-band for galaxies at z~0.3, thanks to the much better detection of SBF, the more accurate detection and photometry of the GCs population, and the negligible effects from dust, the high-resolution MCAO images by MAORY+MICADO will allow measuring SBF with ≤0.15mag error. Such accurate distance measurements, coupled with precise k-corrections from stellar population models, will be able to provide the first independent check to the results on cosmic acceleration and other cosmological parameters, based on a luminosity distance indicator completely independent from SN Ia.

**Keywords:** imaging - photometry - cosmological distance scale - dark energy

**MICADO Observation mode:** Standard imaging

**MICADO Pixel Scale / FoV:** 4mas/px and 50×50 arcsec$^2$ FoV is the ideal scale, motivated by the need of a uniform PSF over the area.

**MICADO Spectral set-up:** None

**Filters required:** Ks-band will be ideal because of the lowest contamination from GCs and dust in the galaxy, and highest SBF amplitude.

**Estimate Survey Area/Sample Size/ Number of Images/Epochs:** Assuming statistical uncertainties on distances of 15% per galaxy, ~15 galaxies per 0.05 z bin size between $0.25 \leq z \leq 0.35$, will allow to ensure a statistical uncertainty of ~4% per redshift bin. A sample size of ~45 galaxies is sufficient to reach the proposed goal. With ~30/20 galaxies (i.e. 10/7 galaxies per 0.05 z bin), the expected statistical uncertainty will be 5%/6%, respectively

**Average Integration time per image:** GCs are the main bias to the SBF signal, and are harder to detect than SBF in Near-IR bands. Reaching a S/N≥3 at ~2 mag brighter than the GCs luminosity function (GCLF) turn-over magnitude, will ensure GCs bias below ~3%. Assuming GCLF peak at $Mv=-7.4$ mag, or $Mk\sim-8.5$ ABmag, a distance modulus of m-M~40.5 mag (i.e. at z~0.3 assuming $H_0=73$ km/s/Mpc), a template G2V spectrum, 3pix X 3 pix reference area, airmass<1.5, 5mas/pixel resolution, S/N≥3 on sources with K~30 ABmag (i.e. ~2 mag brighter than the expected GCLF turn-over), the ELT ETC (version 6.2.2) provides ~3800s integration on source (60s DIT X 64 NDIT).

**Observation requirements:** Small and large dithers will be needed. Single frames will be stacked after proper astrometric calibration. Reference astrometric frames will be needed from high-resolution imaging.

**Strehl or EE required:** high SRs are essential for measuring SBF to galaxies at the largest distances.

**Astrometric Accuracy:** No special requirement.

**SCAO vs. MCAO:** MCAO is preferred for the larger field of view and PSF uniformity. Specific targets, yet to be found, could benefit from the superior SCAO resolution.

**Comparison with JWST or other facilities:** Because of its PSF stability, and negligible overheads, JWST will be the key facility for SBF studies in the next future. Nevertheless, MAORY/MICADO, with slightly better sensitivity of JWST/NIRCam, will have up to ~6 times better spatial resolution in Near-IR at high SRs. Hence, larger distances, which are key for the proposed science case, will be reached with ELT. *Similar programs are under study by TMT science teams.*

**Synergies with other facilities:** High-resolution imaging, needed for astrometric calibration, will benefit from Gaia and, to fainter magnitudes, from LSST astrometry. Preparatory/pilot studies, for the optimization of observing strategies and data analysis procedures, will also take advantage of the forthcoming ERIS/VLT system.

**Simulations made/needed to verify science case or feasibility:** no simulation has been done, yet. *Stellar population models will be required for calibration of SBF models,* *and to model the k-correction terms.* Our team includes researchers with well-recognized experiences in this field.

**Origin of the targets:** Taking, as an example, as reference "A Catalog of 132,684 Clusters of Galaxies Identified from SDSS III" (Wen Z.L., et al., 2012, ApJS, 199, 34), ~30.000 galaxy clusters are catalogued in the redshift range of interest, 0.25≤z≤0.35.

**NGS:** Of the ~30.000 galaxy clusters in the redshift regime of interest, ~1.400 have galactic latitude |b|<20°. Hence, NGS coverage should not be a concern.

**Acquisition:** No specific constraints; accurate finding charts are available over any possible target of interest.

**Calibrations:** Good reference astrometric stars are needed, if unavailable from any other archive, non-AO images of the same observed fields for SBF analysis will be taken with MICADO/MAORY. PSF spatial uniformity is mandatory for the project.

**Data Processing Requirements**: Good PSF modeling is required; target selection will by necessity take such requirement into account.

**Any other comments**: The pass-bands proposed for the observations, require the imaging of the sky far from the target. A classical OBJECT-SKY-OBJECT-OBJECT-SKY large-dithering pattern will be essential to account for sky variability for sky subtraction. The SKY blocks will be in non-AO mode. Overheads for the observational strategy have to be computed.

# 6.10 A Beacon in the Dark

**Authors:** *Dall'Ora M. (INAF - OACN), Fiorentino G. (INAF - OABO), Bono G. (University of tor Vergata), Marconi M., Moretti M.I., Musella I., Ripepi V. (INAF - OACN)*

**Brief description of science case:**

MAORY+MICADO will give us the unique opportunity to expand the space volume for which stellar populations studies, on a star-by-star basis, will be feasible. As a matter of fact, in the Local Group we are forced to limit our studies to a couple of giants spirals, a dwarf spiral, a dwarf elliptical, and several dwarf spheroidals and irregulars. Moreover, the fact that the Local Group is a small *ensemble* of galaxies does not tell us too much on the effects of the environment on the stellar populations, especially in presence of major merging events. This means that, with MAORY+MICADO, we can probe the stellar populations over a variety of galactic morphological types, especially giant ellipticals, and in a variety of environments. For example, the Virgo cluster would be an excellent laboratory, since it hosts a large number of galaxies ($\approx$ 1500), and it is close enough ($\approx$ 18 Mpc, Fouque et al. 2001, A&A, 375, 770) that individual stars can be resolved and measured by MAORY+MICADO (see Deep et al. 2011, A&A, 531, 151). We stress that this is potentially the very first time that we can investigate stellar populations in early type galaxies and possibly in the bulge of external galaxies.

**The secondary distance indicators.** MAORY+MICADO will also expand the range for which primary distance indicators (especially Classical Cepheids and TRGB) will be available, and this will allow us to extend the calibration of important secondary distance indicators, such as Tully-Fisher (TF), and both type Ia and II supernovae (SNe). Importantly enough, Cepheids and TRGB constitute complementary fundamental classes of standard candles (e.g. Tammann et al. 2008, ApJ, 679, 52), since they are produced by different parent stellar populations. This means that they can provide independent calibrations of the luminosity of the secondary distance indicators. In particular, Cepheids are the product of a young stellar component, and they can be found only in spiral and irregular galaxies, while the TRGB can be observed in all the morphological types, being produced by the omnipresent old component ($\geq$ 10 Gyr). On the observational side, Cepheids require time-series data, which can be observationally expensive for those with the longest periods, while the TRGB can be observed with a single snapshot. Moreover, it is bright enough ($M_k \sim -5.7$ mag, with a well-characterized dependence on the metallicity), to be observed with MAORY+MICADO well beyond the Virgo and Fornax clusters (at $\mu \sim 31.3$ mag and $\mu \sim 31.8$ mag, respectively), and approaching the Coma cluster ($\mu \sim 35$ mag).

The scientific goal of this project is therefore to observe selected galaxies within the Virgo and Fornax clusters, in order to study their stellar populations at the luminosity level permitted by the distance. This will allow us also to get homogeneous and precise distances by a number of primary distance indicators, mainly TRGB. The inferred distances will be of great interest in the calibration of secondary distance indicators, such as the Tully-Fisher relation.

They will be also of invaluable interest for those systems where a type Ia or a type II supernova have already exploded, allowing also a direct calibration of these standardisable candles. Indeed, while the importance of type Ia SNe for cosmological studies it is well established, relatively recent studies indicate that type II SNe are affordable distance indicators (e.g. Hamuy & Pinto 2002, ApJL, 556, 63; Rodriguez et al. 2014, AJ, 148, 107), up to cosmological distances (e.g. Nugent et al. 2006, ApJ, 645, 841; Poznanski et al. 2009, ApJ, 694, 1067; de Jaeger et al. 2017, ApJ, 835, 166). Moreover, at higher redshifts, it is expected the type II SNe would outnumber type Ia SNe (e.g. Hopkins & Beacom 2006, ApJ, 651, 142). However, too few type Ia and II SNe exploded in systems for which Cepheids and/or TRGB distances are available. Thus, we still miss a sample statistically robust against poorly known systematic effects, and it is highly desirable to increase the number of type Ia and II SNe that can be calibrated.

**Is it observationally feasible?** AO is mandatory in order to pinpoint the underlying stellar populations and to characterize their Cepheids and their TRGB. However, the tremendous resolution of E-ELT can be fully exploited in two modes: 1) SCAO mode- when there is at least a bright star, in the luminosity range $7 \leq V \leq 16$ mag, and in a circle of 10" in the field of view (and preferably in the centre); MCAO mode- when there are three natural guide stars of $H \leq 18.5$ mag (preferably symmetrical disposed) in an annulus of an external diameter of 180". However, the expected sky coverage of the MCAO at the typical Galactic coordinates of the, say, Virgo and of the Fornax clusters is of the order of 50%, and therefore only ~ 50% of the galaxies have close foreground stars that meet these requirements.

**Supernovae as beacons.** The simple idea of this case study is to use possible supernovae (regardless of the type), as bright reference stars to make available the SCAO mode for galaxies where no MCAO is feasible, because of the lack of bright foreground stars. Indeed, the luminosity of type Ia SNe range typically from $M_V$ ~ −17 mag to $M_V$ ~ − 20 (Smartt et al. 2015, A&A, 579, 40), while type II SNe have luminosities usually between $M_V$ ~ −15 mag and $M_V$ ~ −18 mag. Type Ib and Ic events can be even brighter. This means that with this technique we can close the loop of the AO module up to a distance modulus of $\mu$ ~ 34 mag, well beyond the Virgo and Fornax clusters.

**How many beacons do we expect?** Current estimates of the SNe rate per volume unit (see Li et al. 2011, MNRAS, 412, 1473) measure ≈ 0.2 type Ia events per year, and ≈ 0.4 core-collapse (type II, Ib and Ic) events per year. SFH-inferred rates expect ≈ 0.2 type Ia events per year and ≈ 0.6 core-collapse events per year (Hopkins & Beacom 2006). More tuned rate estimates per mass unit expect up to 2 events per year (regardless of the type, M. Della Valle, *priv. comm.*). Overall, we expect ≈ 1-2 events per year, in random galaxies.

**Is this a scientific case?** Even if, this case is aimed at using the SCAO-MAORY performances under peculiar circumstances, it describes the only possibility we have, for galaxies for which there are no field stars bright enough to use standard MCAO mode, to achieve the following scientific goals:

a) To measure the distances of the host systems with the TRGB. This will extent the sample of galaxies for which TRGB distances are available, and that can be used to calibrate possible type Ia and II supernovae. This is of particular interest, as we could use an independent estimate of $H_0$, based on type II SNe, to double check the tension between the current tension between the type Ia- and Planck-based estimates of $H_0$ (e.g. Bernal et al. 2016, JCAP, 10, 019, and references therein). However, this discrepancy could be alleviated by taking into account populations effects on the calibration of the Ia SNe, as recently discussed in the literature (e.g. Rigault et al. 2015, ApJ, 802, 20). Therefore, an independent distance scale would be highly valuable;

b) To study the underlying stellar populations, at the deepest luminosity level permitted by the distance and/or crowding.

Moreover, since SCAO mode will be available, the SN will be in the field of the scientific detector, allowing us to collect data of exquisite quality.

**MICADO Pixel Scale / Fov:** 1.5mas/px and 20arcsec FoV.

**MICADO Observation mode:** Standard Imaging

**MICADO Spectral set-up:** none

**Filters required:** JHK filters, in order to fully characterize the stellar populations, and to correct the position of the TRGB for the population effects (see Jang & Lee 2017, ApJ, 835, 28).

**Estimate Survey Area/Sample Size/ Number of Images/Epochs:** due to the intrinsic nature of this project only one field, possibly slightly dithered, will be observable.

**Average Integration time per image (magnitude of targets; S/N required):** according to the E-ELT web pages, a 5hr integration will reach (in the Vega system) J = 29.9 mag, H= 29.4 mag, and K = 28.0 mag at S/N=5. This means that, at the Virgo distance, stars with absolute magnitudes $M_J \sim -1$ mag, $M_H \sim -1.5$ mag and $M_k \sim -3$ mag will be observable, well below the TRGB (from $M_j \sim -5$ mag to $M_k \sim -5.7$ mag). However, due to its nature of "ToO-like" program, these are only indicative exposure times, that will be calibrated as a function of the distance of the host galaxy and of the stellar populations available at a given distance.

**Observation requirements:** small dithers

**Strehl or EE required:** the spatial resolution, and the photometric accuracy, will be calibrated on the basis of the distance (i.e. the crowding) of the target.

**Astrometric Accuracy:** the positions of the supernovae are usually precisely measured, and can be used with confidence as input for E-ELT.

**SCAO vs. MCAO:** SCAO is the only possible mode for this project

**Comparison with JWST or other facilities:** Spatial resolution is the key issue here: ELT +MAORY+MICADO is much better than JWST in this respect (by a factor of >3, at high SR, comparing MICADO observations in H band to JWST observations in I band).

Synergies with other facilities: current transient surveys (i.e. Pan-STARRS, ASAS-SN, CRTS) are very effective to reveal new supernovae. Moreover, the forthcoming LSST spot new transients up to r ~ 24 mag, with a two weeks' cadence. It goes without saying, that this project will also benefit a lot of first light integral field spectrograph available at E-ELT, and that will also be assisted by SCAO.

**Simulations made/needed to verify science case or feasibility:** while existent simulations (Deep et al. 2011) demonstrate the possibility to resolve individual stars at the distance of Virgo, it would be of great interest to understand how well we can pinpoint individual bright stars, at the Cepheids and/or TRGB level, up to 100 Mpc (i.e. at the Coma cluster distance).

**Origin of the targets:** on the basis of the position of the SNe

**Acquisition:** the precision of the pointing is a key issue, since SCAO gives its best performances with the star in the centre of the field

**Calibrations:** Standard. No special astrometric requirement.

**Data Processing Requirements:** This is a classical PSF-photometry in crowded stellar fields case.

**Any other comments:** none.

# 6.11 Resolved stellar population of nearby Nuclear Star Clusters

**Authors**: Gullieuszik M., Mapelli M., Greggio L., Falomo R., Ballone A., Portaluri E., Spera M. (INAF - OAPD), Trani A. A. (INAF - OAPD, SISSA), Giacobbo N., Bortolas E., Di Carlo U. N. (INAF - OAPD, University of Padova)

**Brief description of science case:**

Most galaxies host massive compact nuclear objects, supermassive black holes (SMBHs) and/or nuclear star clusters (NSCs). While SMBHs are found at the center of high-mass galaxies, NSCs are found in most intermediate- and low-luminosity galaxies of all Hubble types (e.g. Côté et al. 2006, ApJS, 165,57). In the intermediate mass range (M~$10^{9-11}$ M$_\odot$), nuclear star clusters and super-massive black holes tend to coexist in the same galaxy (Graham & Spitler 2009, MNRAS, 397, 2148), the most famous example being our own Galaxy. Moreover, NSCs seem to follow the same scale relations as super-massive black holes (Ferrarese et al. 2006, ApJ, 644, L21; but see Scott & Graham 2013, ApJ, 763, 76 for a different result). Thus, collecting information about nuclear star clusters is crucial in order to shed light on their formation, on their connection with supermassive black holes, and on the co-evolution between a galaxy and its central compact object (e.g. Ferrarese et al. 2006, ApJ, 644, L21; Nayakshin & Power, 2010, MNRAS, 402, 789; Arca-Sedda et al. 2016, MNRAS, 456, 2457).

NSCs are 3−4 mag more luminous than Milky Way globular clusters, but they share similar sizes (a few parsecs, see e.g. Böker et al. 2004, AJ, 127, 105). Their estimated total mass ($10^6$ -$10^7$ M$_\odot$; Böker 2010, ASSP, 15, 99) is at the high end of the globular clusters' mass function.

Resolving single stars in nuclear star clusters is a powerful tool toward understanding their formation, the connection with the host galaxy, and the black hole. Currently, the only nuclear star cluster where we are able to resolve single stars is the one in the Milky Way. Is the nuclear star cluster of the Milky Way a "typical" nuclear star cluster? To answer this question, it is necessary to resolve stellar population in other nearby nuclear star clusters. To date, ages, metallicities, and in general, star formation histories are inferred from integrated colours, indices, and/or spectral energy distribution fitting techniques (see e.g., Walcher et al. 2006 ApJ, 649, 692; Seth et al. 2006, AJ, 132, 2539; Rossa et al. 2006, AJ, 132, 1074).

The extraordinary sensitivity and spatial resolution of MAORY+MICADO will be the perfect combination to resolve individual stars in nearby NSCs. In a feasibility study based on simulations carried out with AETC (aetc.oapd.inaf.it), we showed that it will be possible to obtain precise photometry of main-sequence turn off stars up to 2 Mpc for old stellar populations and up to 4-5 Mpc for intermediate/young stellar populations (Gullieuszik et al. 2014, A&A, 568, 89). These observations will provide direct and precise measurements of the star formation history and reliable estimates of the metallicity of the stellar populations in nearby NSCs. These data are fundamental to answer the basic questions: how old are these objects? Did they form stars in a continuous way, or undergo separated episodes of SF? In the latter case, with which duration? A clearer picture of the formation and evolution of NSCs will allow us to tackle the question about the link between the formation of the central compact objects and the parent galaxy.

In addition to photometry, long-slit spectroscopy with MAORY+MICADO will provide unique information on the kinematics of NSCs: whether (a fraction of) their stars belong to a rotating flattened structure, such as in the Milky Way and Andromeda. Moreover, long-slit spectroscopy of stellar light and ionized gas inside the influence radius of a Milky Way-like black hole (≤1 pc) might provide a dynamical mass measurement of the hidden black hole (if any).

We based the selection of the target for this science case on the catalogs of NSC of Georgiev & Böker (2014 MNRAS 441 3570) and Carson et al (2015, AJ, 149, 170); we selected as possible targets 4 nearby galaxies that are observable from Cerro Armazones: NGC300 (d=2.0Mpc), NGC7793 (d=3.4Mpc), NGC247 (d=3.6 Mpc) and NGC 5254 (d=4.4 Mpc). The target selection could however benefit from the outcome of future surveys of the inner regions of nearby galaxies.

*This Science Case is part of the MAORY ST initiative on compact stellar systems.*

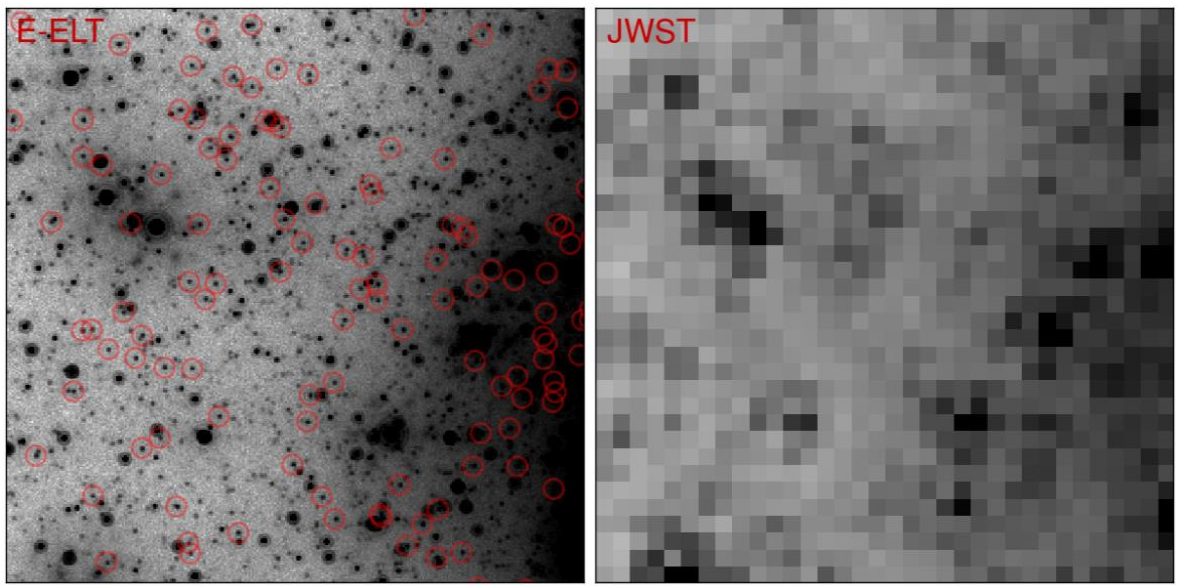

Figure 6.11-1 *1" x 1" E-ELT (left panel) and JWST (right panel) images at 6 Re from the centre of a 1 Gyr old NSC at 2 Mpc. In the left panel the red circles show J = 28 mag stars, corresponding to the MSTO magnitude. From Gullieuszik et al. (2014).*

**Keywords:** imaging – photometry – long-slit spectroscopy – resolved stellar populations – galactic nuclei

**MICADO Pixel Scale / Fov:** 4mas/px and ~50arcsec FoV

**MICADO Observation mode:** Standard Imaging + slit spectroscopy for the kinematics

**MICADO Spectral set-up:** Slit (width x length) = 16 mas x 4 arsec is required to obtain sufficient spectral resolution (R~10000), but dedicated simulations accounting for the surface brightness profile of the nuclear star cluster are required to study the feasibility.

**Filters required:** J and H

**Estimate Survey Area/Sample Size/ Number of Images/Epochs:** The tidal radius of NSCs is of a few arcseconds, therefore a single pointing is large enough to cover the whole extension of each NSC and to obtain photometry for the host galaxy stellar population. A single epoch observation in 2 filters (J and H) will fit our science requirements.

**Average Integration time per image (magnitude of targets; S/N required):**

*Photometry:* In Gullieuszik et al. (2014) we found that 3h exposure time in J and H band would provide the required photometric accuracy. This result should be confirmed using new simulations with updated MAORY+MICADO prescriptions.

Spectroscopy: To estimate the requirement for spectroscopy we assumed that the star cluster is not resolved in its central few parsecs (which is quite likely) and that ~1 – 10 % of the total light of the nuclear star cluster is in this unresolved central part (which is probably an underestimate, considering the nuclear star cluster of the Milky Way). Assuming a target magnitude of ~14 in I, we expect that the unresolved part has magnitude I<~16 (Vega). Requiring an S/N~1000 over the spectrum (ensuring that the main absorption lines are resolved with S/N>~3), we derive an exposure time of ~ 7 hr per each single target in J band. This estimate has been obtained on May 30, 2017, with the ESO-ELT ETC, assuming a blackbody flux distribution (with T~11000 K), resolution R~10000, MCAO mode, and a radius of circular S/N reference area ~15 mas (consistent with the selected slit width). We note that the current version of the ESO-ELT ETC is not up-to-date. According to the MICADO Operational Concept Description (document ELT-PLA-MCD-56301-0004), MICADO's spectrograph will operate in two bands: IzJ and HK, which are not included in the ETC. We expect the IzJ band will be the most suitable for our observations, because of the higher spectral resolution. Our calculations will be updated with the IzJ band, as soon as the ESO-ELT ETC is updated.

**Observation requirements**: Due to the high crowding it will be likely necessary of take offset sky exposure to obtain a reliable background subtraction.

**Strehl or EE required:** The feasibility of this science case was assessed on simulated observations based on MICADO PhaseA study specifications. A new set of simulations with updated specifications is required to confirm our preliminary results.

**Astrometric Accuracy:** No special requirements.

**SCAO vs. MCAO:** MCAO to take full advantage of the large corrected FoV and obtain stable correction across the whole FoV.

**Comparison with JWST or other facilities:** The spatial resolution of JWST would allow us to resolve just the outskirt of nearby NSCs. The extraordinary spatial resolution of MAORY+MICADO is required to resolve stellar populations in the inner crowded regions (see Figure 6.11-1).

**Synergies with other facilities (4MOST/MOONS, LSST/ALMA/HARMONI/METIS, HIRES/MOSAIC), but also VLT or other smaller telescope instruments:** surveys of nearby galaxy central regions will provide important information on the stellar populations of NSC host galaxies. No other ground based or space telescope will have the spatial resolution required to resolve single stars in extragalactic NSCs.

**Simulations made/needed to verify science case or feasibility:** It will be important to perform at least a new sub-set of the simulations carried out in Gullieuszik et al. (2014) using the updated MAORY+MICADO PSFs and specifications.

**Origin of the targets:** list of nearby NSCs from the literature (e.g. Georgiev & Böker 2014 MNRAS 441 3570, Carson et al. 2015, 149, 5) and form future surveys of nearby galaxies.

**NGS:** NGC 247, 300, 7793, and 5254 have galactic coordinates (l,b)~ (114,-84) DEG, (299,-79) DEG, (5,-77) DEG, and (321, 49) DEG, respectively. Thus, the probability of finding at least three NGSs with magnitude H≤19 is ≤0.6, based on a statistical argument, even with the widest proposed patrol field. This might be an issue and must be checked with the analysis of existing observations or new observations.

**Acquisition:** acquisition should not be a problem since the NSC will be clearly visible. Finding charts available.

**Calibrations:** Standard. No special astrometric calibration is required. Absolute photometric calibration is required to compare the observed CMDs with theoretical isochrones.

**Data Processing Requirements:** PSF modeling can be obtained from stars in the FoV. The stacked images are the starting point for the scientific analysis.

**Any other comments:** Additional possible interesting targets include Henize 2 – 10 and NGC 4212. The former is a starburst dwarf galaxy with a peculiarly massive black hole, which seems not to be associated with a nuclear star cluster (Reines et al. 2011, Nature, 470, 66). Unfortunately, it is also quite distant from the Milky Way (~9 Mpc). The latter is at ~3.5 Mpc, is a bulgeless spiral, has a nuclear star cluster, and might host an intermediate-mass black hole (≤$10^5$ M$_\odot$, eg Graham & Scott 2013, ApJ, 764, 151). Its location in sky (DEC ~ +15 degrees) is less suitable for observations with the E-ELT than for the other three galaxies in our sample.

# 6.12 Star formation and super-star clusters in galaxy mergers: the Antennae galaxies

**Authors:** *Mapelli M., Ballone A., Spera M. (INAF - OAPD), Trani A. A. (INAF - OAPD, SISSA), Giacobbo N., Bortolas E., Di Carlo U. N. (INAF - OAPD, University of Padova)*

**Brief description of science case:**

Galaxy mergers have long been known to trigger bursts of star formation in gas-rich galaxies (Arp 1969, A&A, 3, 418; Toomre & Toomre 1972, ApJ, 178, 623). However, several questions remain open: what are the properties of star formation in merger-triggered starbursts? Does most of the merger-induced star formation occur in super-star clusters (Bastian et al. 2009, ApJ, 701, 607)? Are these super-star clusters structurally different from the young star clusters we observe in the Milky Way and other "unperturbed" galaxies? Are these young super-star clusters the low-redshift analogous of globular cluster progenitors (Renaud et al. 2017, MNRAS, 465, 3622)? The impossibility of resolving the internal structure of star clusters in galaxy mergers prevented us from answering these critical questions.

The Antennae galaxies (NGC 4038/39, Duncan 1923, ApJ, 57, 137) are the ideal laboratory for star formation in merging galaxies, because they are the closest example to the Milky Way (distance ~20Mpc). The highest resolution achieved by optical observations (~0.1" that corresponds to ~9 pc), obtained with the HST, allows us to identify young star clusters as bright knots (with completeness limit at V~24, Whitmore et al. 1999, AJ, 118, 1551). HST data allowed us to study the integrated colours of star clusters (providing a rough estimate of the ages), to approximately quantify their size, and to discriminate between compact super-star clusters and super-star clusters that are composed of several knots (sub-clusters?). Only for the three brightest objects it was possible to draw a surface brightness profile and to give some upper limit on their core radii (Fig. 20 of Whitmore et al. 1999).

**The unique spatial resolution of MAORY+MICADO at E-ELT (about a factor of 10 better than that of HST) will allow us to obtain measurements (rather than upper limits) of core and tidal radii of super-star clusters. This will enable us to understand the structure of super-star clusters in the Antennae system.** Moreover, the photometric accuracy of MAORY+MICADO will allow us to update the integrated colour-magnitude diagrams, reducing the uncertainties considerably, and giving far better constraints on the cluster age. We expect to obtain colour-magnitude diagrams for individual stars in the field and in some of the less crowded clusters (the latter hypothesis needs to be checked with dedicated simulations). **This will enable us to obtain powerful constraints on star formation in merging galaxies and on the nature of super-star clusters.**

The results of near-infrared photometry obtained with E-ELT will be combined with information from other wavelengths, such as sub-millimetric (e.g. with ALMA, Whitmore et al. 2014, 795, 156), mid- and far-infrared (e.g. with Spitzer, Brandl et al. 2009), optical imaging (with HST, Whitmore et al. 1999), and spectroscopy (with Gemini, Bastian et al. 2009, ApJ, 701, 607). This will provide us with a complete multi-wavelength identikit of a merging galaxy.

*This Science Case is part of the MAORY ST initiative on compact stellar systems.*

**Keywords:** imaging – photometry – super-star clusters – galaxy mergers

**MICADO Pixel Scale / FoV:** At the distance of the Antennae 1" corresponds to ~99 pc, thus 4mas/px corresponds to ~0.4 pc/px. A single pixel will be of similar size as the core of the densest star clusters. Thus, 4 mas/px and a 50 arcsec FoV are sufficient for studying the structure of star clusters in Antennae.

**MICADO Observation mode:** Standard Imaging

**MICADO Spectral set-up:** none

**Filters required:** J and H seems the best compromise between AO performances and sensitivity of the color index to temperature, that is crucial for the scientific goal.

**Estimate Survey Area/Sample Size/ Number of Images/Epochs:** The central parts of the Antennae (excluding the extreme tails) measure ~5x3 arcmin (~30x20 kpc). Thus, ~20 pointings (with a 50" FoV) are required to cover the entire region in each required band. Assuming 1h per pointing, ~40 hours of observing time are required, excluding overheads and not accounting for imaging sky far from the targets.

**Average Integration time per image (magnitude of targets; S/N required):** with 1h per filter S/N~10 is reached at H=27 (J=27.5), S/N~30 at H=26 (J=26.5). This gives a precision in the J-H color of ~0.05 mag at H=26. These estimates have been obtained on May 1, 2017, with the ESO-ELT ETC, assuming a px scale of 5 mas, a K5V star as a template and a 1 px X 1 px reference area for computing the S/N ratio.

**Observation requirements:** Small dithers + Large dithers [refer to table 8 of MICADO OCD]

**Strehl or EE required:** the spatial resolution is a key point in these crowded field. Thus, the higher the SR is, the better the result of this observation.

**Astrometric Accuracy:** no special requirement

**SCAO vs. MCAO:** MCAO is the only viable mode as there are no bright enough stars (allowing SCAO) in the Antennae FoV.

**Comparison with JWST or other facilities:** Spatial resolution (which is the key point to resolve star cluster structures) is better by a factor of ~3 than that of JWST.

**Synergies with other facilities:** Several observations (both imaging and spectroscopy) with other facilities in nearly all possible frequency range are available: ALMA (Whitmore et al. 2014, 795, 156), Spitzer (Brandl et al. 2009), HST (Whitmore et al. 1999), Gemini (Bastian et al. 2009, ApJ, 701, 607), Chandra (Fabbiano et al. 2001, ApJ, 554, 1035).

**Simulations made/needed to verify science case or feasibility:** Simulations are needed to evaluate the possibility of resolving single stars within the star clusters of the Antennae.

**Origin of the targets:** available in the literature, e.g. Whitmore et al. (1999).

**NGS:** Galactic coordinates of NGC4038 are (l,b)~(287,42) DEG; hence we expect a probability >0.7 of finding at least three NGSs with magnitude H≤19, based on a statistical argument. We envisage a search for NGS should be performed in the near future.

**Acquisition:** no special requirements. Finding charts are available.

**Calibrations:** Standard. No special astrometric requirement.

**Data Processing Requirements:** The PSF modelling should be feasible with stars in the FoV.

**Any other comment:** none.

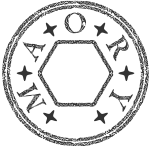

# 6.13 High-redshift galaxies "in the backyard": SBS0335-052E and W

**Authors:** Mapelli M., Ballone A. , Spera M. (INAF - OAPD), Hunt L. (INAF - OAA), Trani A. A. (INAF - OAPD, SISSA), Giacobbo N., Di Carlo U. N., Bortolas E., Ripamonti E. (INAF - OAPD, University of Padova)

**Brief description of science case:**

Extremely metal deficient galaxies (XMDs) are nearby galaxies classified as galaxies with metallicity ≤1/20 solar (Kunth & Oestlin 2000, A&ARv, 10, 1). Some of them are forming stars at extreme rates (~0.01-1 M$\odot$/yr), when considering their very low stellar mass (<$10^9$ M$\odot$). Thus, these are unique objects, allowing us to study the properties of the Early Universe "in the backyard" (Izotov et al. 1990, Nature, 343, 238).

Several tens of galaxies have been catalogued as XMDs so far (Ekta & Chengalur 2010, MNRAS, 406, 1238), but only 3 of them are at distance <100 Mpc: I Zw18 in the northern hemisphere, and the galaxy pair SBS 0335-052E and W in the southern hemisphere. I Zw18 is the closest (18 Mpc) and best known XMD. Similar to their more famous cousin I Zw18, SBS 0335-052E and W are among the most metal-poor galaxies known, with 12+log(O/H)~7.2-7.3 (Izotov et al. 2006, A&A, 459, 71) and 12+log(O/H)~7.1 (Papaderos et al. 2006, A&A, 454, 119), respectively. Their distance from us is ~54 Mpc, about three times larger than that of I Zw18, while the projected distance between SBS 0335-052E and W is ~20 kpc. SBS 0335-052E and W are invaluable targets not only for their low metallicity, but also because they both belong to the class of blue compact dwarf galaxies (with extremely high star formation rate density). The star formation rate (SFR) of SBS 0335-052E is ~ 0.4 M$\odot$ /yr, in the upper range of SFRs observed in blue compact dwarf galaxies (Thuan et al. 1997, ApJ, 477, 661). The high SFR of SBS 0335-052E was proposed to be the result of a recent interaction with the nearby Scd galaxy NGC 1376. Moreover, HST images show that SBS0335-052E hosts at least six super-star clusters, which can provide information on star formation in extreme environments (Thompson et al. 2006, ApJ, 638, 176). Finally, SBS 0335-052E and W are anomalously X-ray bright: each of them hosting one point-like source with X-ray luminosity (assumed isotropic) >$10^{39}$ erg/s, i.e. two so-called ultraluminous X-ray sources (Thuan et al. 2004, ApJ, 606, 213). The X-ray luminosity of I Zw18, SBS0335-052E and W is significantly in excess with respect to the SFR – X-ray luminosity relation (e.g., Ranalli et al. 2003, A&A, 399, 39; Grimm et al. 2003, MNRAS, 339, 793). This has been interpreted as hint of a physical connection between ultra luminous X-ray sources and metal poor binary systems (e.g., Mapelli et al. 2010, MNRAS, 408, 234). Hence, SBS 0335-052E and W are key galaxies to shed light on several open astrophysical questions.

The only XMD for which we were able to resolve single stars and to study star clusters in detail is IZw18. HST observations of I Zw18 (Aloisi et al. 2007, ApJ, 667, L151) allowed to reconstruct a color-magnitude diagram (CMD) in the V and I bands for the brightest stars (I~22-27) with a photometric accuracy of ~0.2. Thanks to these observations, an accurate Cepheid distance (Fiorentino et al. 2010, ApJ, 711, 808) to IZw18 was possible and constrained the existence of an old population (2-13 Gyr) besides the young one.

*Thanks to its higher spatial resolution, MAORY+MICADO will be able to resolve the upper part of the CMD of SBS0335-052E and W (I~24-29mag). The reconstruction of the brightest stars CMD in SBS0335-052E and W (resolving massive main sequence stars*

*and bright giant stars) will enable us to derive their actual star formation history, and will provide unique information on these extreme metal-poor galaxies. Finding an old stellar population in SBS0335-052E and W (as the one that was found in IZw18) will be a strong support for the idea that even XMDs underwent early episodes of star formation before their current burst.*

**Keywords:** imaging – photometry – low metallicity – blue compact dwarf galaxies – extragalactic young star clusters

**MICADO Pixel Scale / Fov:** The size of both targets (~0.2' x 0.2') makes it advisable to choose 1.5mas/px and 20arcsec FoV.

**MICADO Observation mode:** Standard Imaging

**MICADO Spectral set-up:** none

**Filters required:** J and H. It is important to produce colour-magnitude diagrams of the observed population, to get constraints on their age and metallicity.

**Estimate Survey Area/Sample Size/ Number of Images/Epochs:** 2 pointings with 20" x 20" area (one on SBS0335-052E the other on SBS0335-052W) per each of the two required filters. This would require ~28h of observing time (see below), excluding overheads and not accounting for imaging sky far from the targets.

**Average Integration time per image (magnitude of targets; S/N required):** ~7 hours are required to obtain S/N~10 at H=28 (Vega) and S/N~3 at H=29 (Vega). These estimates have been obtained on Feb 17, 2017, with the ESO-ELT ETC, assuming a px scale of 5 mas (the minimum available on the ETC), a K5V star as a template and a 1 px X 1 px reference area for computing the S/N ratio.

**Observation requirements:** Small dithers + Large dithers [refer to table 8 of MICADO OCD].

**Strehl or EE required:** a high SR is essential, because the best spatial resolution is crucial.

**Astrometric Accuracy:** no special requirement.

**SCAO vs. MCAO:** MCAO is the only viable mode, as bright stars (allowing SCAO) do non-exist in the field.

**Comparison with JWST or other facilities:** the spatial resolution of MAORY+ MICADO is needed to resolve stars in these compact galaxies.

**Synergies with other facilities (4MOST/MOONS, LSST/ALMA/HARMONI/METIS, HIRES/MOSAIC), but also VLT or other smaller telescope instruments:** VLA very high spatial resolution, high frequency radio continuum observations (Johnson, Hunt & Reines 2009, AJ, 137, 3788), ALMA data (Hunt et al. 2014, A&A, 561, 49), deep optical spectra (with the 3.6 m at ESO, Papaderos et al. 2006, A&A, 454, 119), HST NICMOS images in I band (GO-9360, PI: R. Kennicutt, see Thompson et al. 2006, ApJ, 638, 176 ), and HST imaging in the F220W, F330W, F435W, F550M, F791W, F160W (J), F187N (Paschen-alpha), and F205W (K) filters (Reines, Johnson & Hunt 2008, AJ, 136, 1415) are available for SBS0335-052.

**Simulations made/needed to verify science case or feasibility:** simulations with different assumptions for the underlying stellar populations (ages and star formation history) are required to test the feasibility.

**Origin of the targets:** e.g., Izotov et al. 1990, Nature, 343, 238

**NGS:** Galactic coordinates of SBS 0335-052 are (l,b)~(191,-45) DEG; hence we expect a probability >0.7 of finding at least three NGSs with magnitude H≤19, based on a statistical argument. We envisage a search for NGS should be performed in the near future.

**Acquisition:** no special problem or requirement.

**Calibrations:** Standard. No special astrometric requirement. The presence of photometric calibrators in the field has to be verified.

**Data Processing Requirements:** The existence of stars in the FoV to do PSF modelling needs to be checked. Synthetic PSF might be needed.

**Any other comment:** none.

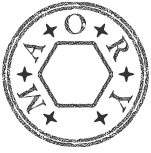

## 6.14 The Forming Nuclear Stellar Disk in NGC 4486A

**Authors:** Morelli L. (University of Padova, INAF - OAPD), Portaluri E. (INAF - OAPD), Pizzella A. (University of Padova, INAF - OAPD), Buson L. (INAF - OAPD), Corsini E. M. (University of Padova, INAF - OAPD), Costantin L., Cuomo, V.(University of Padova), Dalla Bontà E. (INAF - OAPD), Pagotto I. (University of Padova), Rodighiero G. (University of Padova, INAF - OAPD)

**Brief description of science case:**

Nuclear stellar disks (NSDs) are faint spinning structures with scalenght of few tens of parsec (10-50 pc) residing in the center of galaxies. Pizzella et al. (2002, ApJ, 573, 131) and Ledo et al. (2010, MNRAS, 407, 969) observed NSDs in 20% of galaxies interdependently from their morphological type. NSDs are characterized by a smaller scale length and a higher central surface brightness with respect to the large kiloparsec scale disks typical of lenticular and spiral galaxies (Morelli et al 2004, MNRAS, 354, 753). NSDs are with Nuclear Star Clusters and Super Massive Black Holes the most common components residing in the center of galaxies. To resolve and investigate such small objects, subarcsec resolution is needed.

A viable way to form NSDs is from gas funnelled into the nucleus, either via bar-driven secular infall (Scorza & van den Bosch 1998, MNRAS, 300, 469) or by external accretion (Corsini et al. 2012, MNRAS, 423L, 79). In both scenarios, the gas is efficiently driven towards the galactic center, where first it settles as it dissipates into a disk of gas and then it turns into stars as density rises. However, few traces of ionized gas and dust are actually observed only in a few nuclear disks making difficult to understand the process that originates the NSD and to probe the predicted scenario.

NGC 4486A is an S0 galaxy in the Virgo cluster at a distance of 17 Mpc. This galaxy is a unique case for investigating the NSD formation and evolution. In fact, NGC 4486A hosts in its center a stellar and a gaseous/dust disk in the same spatial area (Kormendy et al. 2005, AJ, 129, 2636). It probably represents an intermediate stage of the NSD formation, with the disk of gas already in place that has just started to form stars. Therefore, at this stage we can study directly the processes transforming the gaseous disk into a stellar disk.

The spatial resolution of MAORY+MICADO combined with the huge collecting area of the telescope mirror will allow us to derive the photometric properties of the NSD in J, H, K-bands avoiding possible dust and gas obscuration and contamination. In particular, we will derive in different bands the central surface brightness, scale length, inclination, and position angle of the nuclear stellar disks adopting the photometric decomposition method introduced by Scorza & Bender (1995, A&A, 293, 20). In this way, we can directly compare the structure of the stellar disk with the structure of the gaseous disk that is crucial to test the formation processes. If the NSD is forming from the gas disk we expect they have similar structural properties. Furthermore, from the decomposition in different bands we can estimate the stellar population gradients in the NSD and thus put constraints on the timescale and properties of the star-formation in the disk. An absence of colour gradients in the stellar population of the nuclear disks would suggest a star formation that homogeneously occurred along their radial profile. An inside-out formation scenario is, instead, expected to produce colour gradients.

In addition, we plan to observe an I-band image that will allow us to derive the amount of gas and dust, and to compare the optical and infrared properties of the NSD. The quality of these observations will give a new perspective of the mechanism of formation of the NSD and of co-evolution between the NSD and the hosting galaxy. For the first time we will be able to explore these processes while they are acting instead of to retrace them from the end resulting NSD. From the technical point of view this galaxy is particularly suitable for SCAO observation having a bright (11.5 Mag in V band) star at about 2-3 arcsec distance from the target.

**Keywords:** Galaxies: bulge – galaxies: elliptical and lenticular, cD – galaxies: photometry – galaxies: structure – galaxies: nuclei

**MICADO Pixel Scale / Fov:** at the distance of NGC 4486A (17 Mpc), 1" corresponds to ~82 pc. As Kormendy et al. (2009, ApJ, 69,142) measured an effective radius of 600 pc, 1.5 mas/px and 20" FoV will be more than adequate to study the nucleus of the galaxy.

**MICADO Observation mode:** standard imaging.

**MICADO Spectral set-up:** none

**Filters required:** I, J, H, K filters will be useful to exploit fully the AO performances and to overcome the dust obscuration

**Estimate Survey Area/Sample Size/ Number of Images/Epochs:** a total ~40min on target of exposure time would be enough.

**Average Integration time per image (magnitude of targets; S/N required):** $S/N \geq 50$; surface brightness (V)=20.0 Exposure time on target: K=1500s; H=320s; J=35s; I=40s.

**Observation requirements:** Small dithers

**Strehl or EE required:** the spatial resolution is a key ingredient, therefore a good SR will help this kind of study

**Astrometric Accuracy:** no special requirements

**SCAO vs. MCAO:** SCAO seems a viable option as the star TYC 877-423-1 (V = 11.59 mag) is at a 2 arcsec distance from the center of the galaxy.

**Comparison with JWST or other facilities:** The extraordinary spatial resolution of MICADO+MAORY is the key point to prefer E-ELT instead of JWST in the study of galactic nuclei.

**Synergies with other facilities: MUSE@VLT** operating in Narrow Field Mode will give the 2D spectral map with a sub-arcsec resolution that can complement the photometric information in the optical range.

**Simulations made/needed to verify science case or feasibility:** hydrodynamical simulations of the formation of the NSD and to test their stability against the major merging will be extended following Sarzi et al. 2015 (MNRAS, 453, 107) and Portaluri et al. 2013 (MNRAS, 433,434).

**Origin of the targets:** NGC4486A is a low-luminosity ($M_B=-17.77$ mag) E2 galaxy belonging to the Virgo cluster.

**NGS:** TYC 877-423-1 with V=11.59 mag is available at RA=12 30 57.726 and DEC=+12 16 13.35, i.e. 2 arcsec close to the target.

**Acquisition:** The finding charts are available. Therefore no special requirements.

**Calibrations:** Standard. No special requirements.

**Data Processing Requirements:** The PSF modelling should be feasible with the star present in the FoV.

**Any other comment:** none.

## 6.15 Direct calibration of cosmological distances from Cepheids

**Authors:** *Marconi M., Musella I., Ripepi V. (INAF - OACN), Bono G. (University of Tor Vergata), Dall'Ora M. (INAF - OACN), Fiorentino G. (INAF - OABO), Molinaro R., Moretti M.I. (INAF - OACN)*

**Brief description of science case:**

Classical Cepheids (CCs) are central helium burning pulsating stars (M=3÷13Mo, $M_V$=-2÷-7, P=1÷100d). Their characteristic Period-Luminosity (PL) and Period-Luminosity-Color (PLC) relations make them the most important primary distance indicators to calibrate the extragalactic distance scale (see e.g., Riess et al. 2016, ApJ, 826, 56). Indeed, the most used secondary distance indicators are calibrated using CCs and they provide an estimate of the Hubble constant $H_0$~73 km/sec/Mpc with 2.4% of claimed uncertainty. This is in contrast with the recent value of ~67±1 km/sec/Mpc obtained from the analysis of the cosmic microwave background (Planck coll. 2014, A&A, 571, 16). To reconcile the inconsistency between these values, we need more accurate calibrations of the different steps of the cosmic distance ladder, and/or distance indicators directly able to reach cosmologically significant distances (e.g. the recently discovered Ultra Long Period cepheids, ULPs).

During the last few decades, many efforts have been performed both from the theoretical (Marconi et al. 2005, ApJ, 632, 590; Bono et al. 2010, ApJ, 715, 277) and the observational (Sakai et al. 2004, ApJ, 608, 42; Macri et al. 2006, ApJ, 652, 1133; Mager et al. 2013, ApJ, 777, 79; Fausnaugh et al. 2015, MNRAS, 450, 3597) point of view to investigate and clarify the dependence of Cepheid PL and PLC on systematic effects such as metallicity variations, the non-linearity on the whole predicted period range and possible noncanonical phenomena affecting the Mass-Luminosity relation.

Even if no general consensus has been reached so far on the debated issue of the metallicity dependence of CC properties, both observations and theoretical results point out the power of Near-Infrared (NIR) and Mid-Infrared (MIR) filters for this class of pulsating stars. From the observational point of view it has been widely demonstrated in the literature (Madore & Freedman 1991, PASP, 103, 933; Kervella et al. 2004, A&A, 416, 941; Persson et al. 2004, AJ, 128, 2239; Inno et al. 2013, ApJ, 764, 84; Ripepi et al. 2012, MNRAS, 424, 1807; 2016, ApJS, 224, 21) that the intrinsic width of the observed PL relation is significantly reduced when moving from the optical to the NIR bands. Our nonlinear pulsation models support these results, predicting that the intrinsic width of the instability strip, the non-linearity and the chemical composition effects are all significantly reduced when moving from the optical to NIR and MIR (see e.g. Marconi et al. 2005; Bono et al. 2010).

In this context, the observational capabilities of MAORY+MICADO, that will include the NIR filters J and K, will be fundamental to extend the extragalactic distance scale of CCs, with crucial impacts on the calibration of secondary distance indicators and on our knowledge of $H_0$. To this purpose, we plan to observe CCs in a sample of galaxies with distances ranging from Virgo to Coma, possibly containing SNIa or other secondary distance indicators. Moreover, for the longest period (about 100 d), brightest CCs and ULPs, the expected limit magnitude in K (~26 mag) will allow us to reach the distance of the Coma cluster (~100 Mpc) and in turn to directly calibrate the Hubble flow, skipping the use of any secondary distance indicator. This procedure would allow us not only to constrain, for the first time in one step, the Hubble constant, but also to provide an independent test of the accuracy of various types of secondary distance indicators. In this context, we also plan to explore the accuracy of Type II SNe as distance indicators.

**Keywords:** distance scale – Cepheids

**MICADO Pixel Scale / Fov:** 4mas/px and 50 arcsec FoV motivated by the size of the target.

**MICADO Spectral set-up:** The feasibility has to be verified.

**MICADO Observation mode:** Standard imaging

**Filters required:** J and K.

**Estimate Survey Area/Sample Size/ Number of Images/Epochs:** 6 fields/6 galaxies/102 images/for each galaxy, we need 12 phase points in J and 5 in K. We need to cover the light curves in J because for the more distant galaxies, we do not have a previous identification of the Cepheids in the optical bands, and we need to identify and characterize (to find periods, mean magnitudes and amplitudes) the variables. This is easier in J band since the amplitude of the variation decreases with the wavelength. For the K filter, thanks to the small light curve amplitude in these bands, 5 phase points are enough to determine the mean magnitudes.

**Average Integration time per image (magnitude of targets; S/N required):** We need to reach K=26 mag with S/N=20. In K using the E-ELT Imaging mode version 6.2.2 for DIT=30s, we obtain NDIT=133 for a total exposure time of 66.5 m. In J, for DIT=60 s, we obtain NDIT=20 for a total exposure time of 20 m.

**Observation requirements:** To be tuned using "ad hoc" simulations.

**Strehl or EE required:** We observe to distances where stars are resolved and hence we need the best spatial resolution compatible with MCAO.

**Image Stability Required:** To be verified with dedicated simulations how much the observation should be stable with respect to the seeing conditions and the MAORY performance.

**Astrometric Accuracy:** no special requirement.

**SCAO vs. MCAO:** MCAO is the only possible mode for this science case.

**Comparison with JWST or other facilities:** We remind that even if many of the selected targets might be also observed by JWST, the MAORY+MICADO superior spatial resolution will allow us to reach fainter limit magnitudes (by more than 0.5 mag in J) due to reduced crowding limitations. We note that crowding at these far away distances is a crucial issue (see Riess et al. 2011, ApJ, 730, 119).

**Synergies with other facilities (4MOST/MOONS, LSST/ALMA/HARMONI/METIS, HIRES/MOSAIC), but also VLT or other smaller telescope instruments:**

Our team is also involved both in the data processing and scientific exploitation of the CCs and RR Lyrae stars observed by Gaia and in proposed LSST scientific cases devoted to CCs as distance indicators. The unprecedented accuracy of Gaia parallaxes in the Milky Way and of LSST astrometric results in the Local Volume will represent a crucial step in the calibration of Cepheid PL and PLC relations. On the other hand we are also involved in the ESO public survey "The VISTA near-infrared YJKs survey of the Magellanic System" (VMC, P.I.: M. R. Cioni) that is providing a clear picture of the pulsation properties of the Magellanic Cepheids in the near-infrared regime. The observational and theoretical tools we have developed in the Magellanic Clouds will be extended to investigate the properties of Cepheids in the selected external galaxies, located in the Virgo to Coma distance range. We do not need any specific preparatory observations.

**Simulations made/needed to verify science case or feasibility:** YES.

**Origin of the targets:** LSST will provide very likely catalogues to find NGS and refine the pointings.

**NGS:** targets will be selected taking into account the required NGS. Given the required MCAO mode, the skycoverage should not be an issue.

**Acquisition:** We will have no problems with the pointing and it is possible to verify it with a finding chart.

**Calibrations:** Standard. No special astrometric requirement. The presence of photometric calibrators in the field has to be verified. We would need both relative and absolute astrometric precision.

**Data Processing Requirements:** The PSF modeling should be feasible with stars in the FoV. Pre-reduced stacked images are the natural starting point for the scientific analysis. This is a classical PSF-photometry in crowded stellar fields case.

**Any other comments:** none.

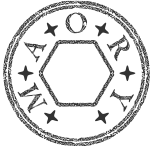

# 7.  High redshift Universe

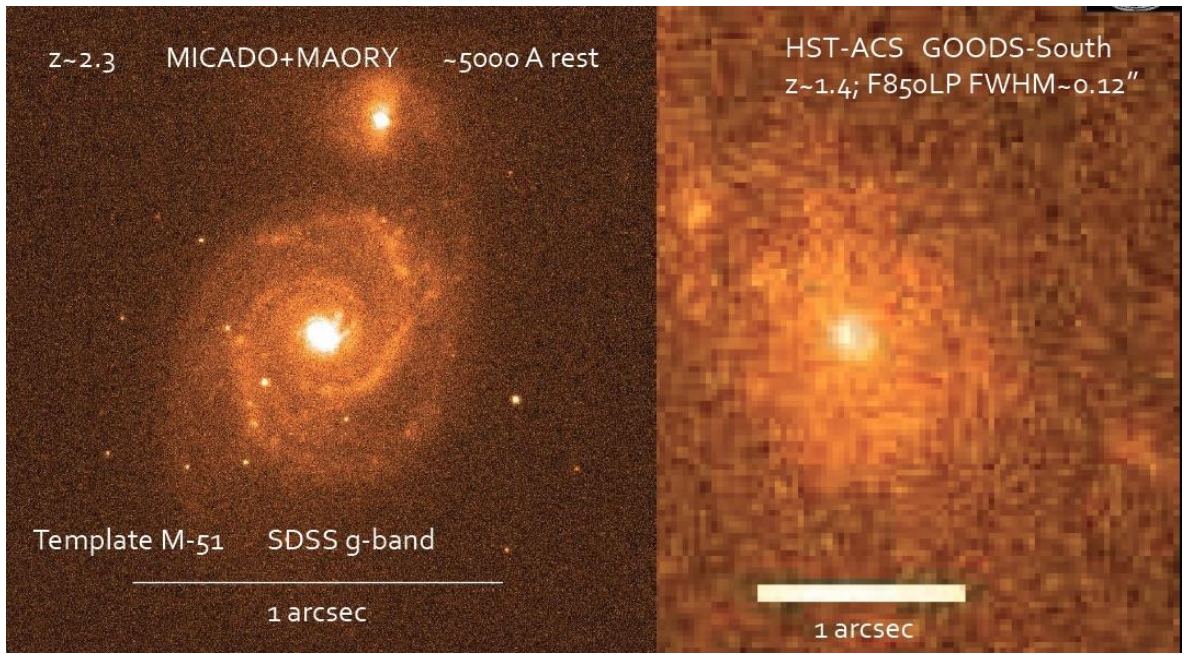

Figure 7-1 The simulated image of a giant spiral at z~2.3 as seen by MAORY+MICADO in SCAO mode is compared to a real image of a similar galaxy at z~1.4. (Figure by P. Saracco).

# 7.1 Strong lensing with MAORY: Resolving galaxies in the distant Universe

**Authors:** *Spiniello C., Napolitano N.R., La Barbera F. (INAF - OACN), Tortora C. (University of Groningen), Meneghetti M. (INAF - OABO), Vegetti S., Rizzo F. (MPIA, Garching), Czoske O. (University of Innsbruck)*

**Brief description of science case:**

The period between redshift 2 and 4 represents a key epoch in the formation history of galaxies. This is the time when most of the galaxy dark and stellar mass is assembled and most of the metals are produced (e.g. Hopkins & Beacom 2006, ApJ, 651, 142; Rudnick et al., 2006, ApJ, 650, 624).

However, despite the progresses made in the understanding the properties of z ~ 2−4 galaxies, there are yet important questions to be clarified (see Shapley A.E., 2011, ARAA, 49, 525, and reference therein for a review on the subject) and some others still waiting for an answer: *What is the role of feedback processes in the regulation of star formation? Is the relation between the gas surface density and the star formation rate (SFR) surface density on sub-galactic scale the same than those obtained from global averages? What is the mechanism that regulates the mass assembly of high-redshift galaxies? How do high-redshift galaxies evolve over time and what is their relation with the galaxies we observe today?*

The combination of strong gravitational lensing and spatially resolved kinematics analysis has proven to be a powerful tool to gain strong insight on high-redshift galaxy structure, with the same level of details that is currently achieved at lower redshift. In fact, the magnification provided by gravitational lensing has allowed scientists to push the spectroscopical studies to higher redshifts, i.e. up to z~2-3, and put more stringent constraints on the galaxy formation processes in these early phases of their evolution (e.g. Shirazi et al., 2014, MNRAS, 440, 2201; Wuyts et al., 2014, ApJ, 781, 61; Jones et al. 2010, MNRAS, 404, 1247).

Image quality is a primary factor to build a precise model of the lensing systems but spectroscopy is equally important. In fact, it is crucial to have the best mapping of the lensed galaxy light distribution (e.g. high spatial resolution imaging) and, at the same time, obtain the spatially resolved stellar kinematics for the lens galaxy which, in combination with the lensing constraints, may allow us to derive the total mass distribution of these galaxies with unprecedented details. If one could add to these constraints the stellar populations of lens and source (emission lines tracers and emission lines ratios to study star formation histories, optical and NIR absorption lines and line strengths to study the passive stellar population of massive early-type) it would be possible to build plausible star formation histories and then reconstruct the full baryons and dark matter assembly of the lens+source pairs. Unfortunately, all the current (ground) instruments suffer from one of both of these two main limitations:

1. The spatial resolution of imagers is not good enough to resolve the arc in multiple images created by the source, thus limiting the accuracy of the lensing models. The only way to obtain a detailed reconstruction of the source is to obtain HST multi-band images.
2. The covered field-of-view of IFU spectrographs is too small or the slits spatial sampling is not good enough to perform proper spatially resolved spectroscopy.

*AO Imaging.* From the imaging and modeling point of view, for the first time, with MAORY it will not be necessary to acquire HST multi-images data to robustly model the lens and source galaxies, thanks to the AO. The great novelty is that the reconstruction of the source will be performed directly on ground-based E-ELT images. MAORY will allow us to constrain the light distribution and morphological properties of high-redshift galaxies with a high spatial resolution that is today only possible at lower redshift or with prohibitive integration times.

Until today the ground based models did not have high-enough resolution to reconstruct the source without assuming ad-hoc parametric models or rescaling the properties observed in the lensed plane by a constant magnification parameter. In fact, the magnification is not a constant number throughout the source plane, but can significantly change over very small areas if the source lies close or across to the critical curves (e.g. Vegetti & Koopmans, 2009, MNRAS, 400, 1583). Thanks to the extremely high spatial resolution of MAORY (AO), no prior assumption will be made about the morphology of the source galaxy and its physical properties will be directly recovered on the source plane, taking into account for a spatially dependent magnification (see Rybak et al., 2015a, MNRAS, 453, L26; Rybak et al., 2015b, MNRAS, 451, L40, for more details about problems in properly reconstructing the source, especially when the source lies close or across to the critical curves).

***MICADO spectroscopy***. Detecting optical emission lines of bright gravitationally lensed star forming high redshift galaxies (through optical emission lines, (e.g. [OIII], Balmer lines,etc) will be possible with the combination of MAORY AO and MICADO slit spectrograph. This will allow us to map the metallicity distribution of high-z lensed galaxies and the amount of mass outflows and inflows constraining in this way the role of feedback processes in the formation and evolution of galaxies.

Finally, accessing star formation rates and metallicity distributions at the scale of few 100 pc will shed light on the relationship between the internal small-scale properties of these galaxies and the galaxies that we observe today.

***Science goals***. Our scientific goals require the knowledge of the source and lens redshift as well as high resolution observations to spatially resolve the source and to build accurate lens models which will then allow to trace intrinsic surface brightness profiles and morphology of high redshift lensed galaxies. Up to now this has always required the combination of different capabilities from different instruments. MAORY+MICADO will allow us to carry on spatial analysis for detailed lensing models and spectroscopic studies to constrain kinematics, dynamic and stellar population without any further follow up.

O***ur technique will be fully self-consistent and will involve only one single instrument.***

In conclusion, thanks to MAORY+MICADO we will be able to study in details galaxies at redshift 2-4 to understand if the star formation at these redshift is dominated by a continuous star formation activity or rather by intense merger-driven bursts. We will also investigate possible evolution in the metallicity mass relation.

Finally, at the same time, it will be possible to perform spatially resolved kinematics and stellar population of lens galaxies up to z~1 to derive their ages, metallicities and to test redshift evolution of the very recently observed stellar Initial Mass Function variation (e.g. Tortora, Romanowsky & Napolitano, 2013, ApJ, 765, 8; Spiniello C. et al. 2012, ApJ, 753, L32; Spiniello C. et al., 2014, MNRAS, 438, 1483; La Barbera et al., 2010, MNRAS, 402, 2335 )

Our team, mainly based in Naples, is in a great position to achieve the goal here proposed: we have extensive expertise in gravitational lens modeling, stellar population analysis and spectroscopical observations and data reduction.

**Keywords:** imaging – spectroscopy – strong lensing

**MICADO Pixel Scale / Fov:** 1.5mas/px and 20arcsec FoV, motivated by the small size of high redshift galaxies.   4mas/px is the only option in spectroscopic mode.

**MICADO Spectral set-up:** 50 mas × 4 arcsec; R~2500.

**Filters required:** H, J and K are crucial to constrain stellar population properties of sources in the considered range of  redshift (e.g. the Hβ, [OIII], OI, Hα, [NII], [SII]) emission lines.

**Estimate Survey Area/Sample Size/ Number of Images/Epochs:** A reasonable statistical sample to start with is of about 100 galaxies, to cover a wide parameter space in redshift, star formation efficiency and mode (starburst vs non-starburst), masses and luminosity.

**Average Integration time per image (magnitude of targets; S/N required):** Currently we are unable to provide a detailed observing strategy. However, based on the CASSOWARY Survey's number, where magnitudes are of the order of 19< r <23 for star forming galaxies up to z=3-4, we estimate that we can reach comparable S/N observing on average 1.5 h for each target. From the ESO-ELT ETC, assuming a px scale of 5mas (the smallest currently available), we calculate that in a 3×3 px region, we will reach a S/N~10 (enough for kinematical analysis) at H=27 and S/N~30 (better for stellar population analysis) at H=26, typical magnitudes of star forming galaxies at z~3 (Bauer et al. 2011)

**Observation requirements:** Large dithering. Precise positioning would not be crucial, since the galaxies and the strong gravitational arcs cover usually less than 10 arcsec.

**Strehl or EE required:**  The highest possible spatial resolution.

**Image Stability Required:**  No special requirement.

**Astrometric Accuracy:** No special requirement.

**SCAO vs. MCAO:** Both can be considered. When a bright star is within ~10" from the lens system, SCAO may be preferred.

**Comparison with JWST or other facilities:**  Spatial resolution is much better for EELT+MAORY+MICADO.

**Synergies with other facilities (4MOST/MOONS, LSST/ALMA/HARMONI/METIS, HIRES/MOSAIC), but also VLT or other smaller telescope instruments:** A possible synergy with IFU instruments, such as MUSE to better study the spatial properties of the lenses on larger scales. Important synergy with ALMA and other radio facilities: our data would represent a NIR and OPTICAL counterpart for the increasing radio observations of star forming galaxies.

**Simulations made/needed to verify science case or feasibility:** Simulations of the imaging and spectroscopic observations will be performed using SimCADO, the MICADO instrument data simulator (O. Czoske is co-responsible for the development of SimCADO within the MICADO consortium).

**Origin of the targets:** Lenses will be selected from existing lensing survey, such as the CASSOWARY, which provide also information about the redshifts and the luminosities. We are also member of LSST for a project that includes strong lenses at z~1.

**NGS:** We expect to collect hundreds of targets in KiDS and other will be gathered in LSST era, so we will have the opportunity to select the targets having suitable NGS asterisms.

**Acquisition:** The characteristic size of the typical system is a few arcsec hence there should be no problem in getting an useful pointing within the 20"× 20" FoV. Finding charts will be provided.

**Calibrations:** Standard.

**Data Processing Requirements:** Slit-spectra are desired final products. Possibly telluric-corrected and flux calibrated.

# 7.2 Probing Super Massive Black Holes in AGN with high resolution observations

**Authors:** *Congiu E., Berton M. (University of Padova, INAF - OABR), Ciroi S., Cracco V., La Mura G. (University of Padova), Marafatto L. (INAF - OAPD), Frezzato M., Chen S., Rafanelli P. (University of Padova).*

**Brief description of science case:**

One of the main properties of active galactic nuclei (AGN) is the mass of the central super massive black hole (SMBH). A technique used to measure the BH mass exploits the stellar velocity dispersion ($\sigma_s$) under the assumption of a virialized system (e.g. Kormendy et al., 1995, ARA&A, 33, 581). Stellar kinematics is indeed very effective, because it is only determined by the gravitational potential, while the gas is also subject to viscous forces (Peterson 2014, SSR, 183, 253-275). However, to obtain accurate BH masses, $\sigma_s$ should be measured at least inside the BH radius of influence $R_{BH}$ (Peterson 2014, SSR, 183, 253-275), defined as the radius where the gravitational potential of the BH dominates over that of the host galaxy. To resolve $R_{BH}$ a very high spatial resolution is required. The radius of influence of a $10^8$ $M_\odot$ BH surrounded by stars with $\sigma_s = 200$ km s$^{-1}$ is ~10 pc, therefore even with the Hubble Space Telescope it is possible to resolve this radius only in objects within ~20 Mpc from us.

The BH mass in AGN is easily measured through the kinematics of the broad line region (BLR) ionized gas, again assuming a virialized system and mostly a shape factor that accounts for the geometry of the BLR. Even if there is some evidence that the BLR is a thick disk with a radius of several lights days (Brewer et al, 2011, ApJ, 733, L33; Pancoast et al., 2012, ApJ, 754, 49), its shape is in general unknown and requires very long reverberation mapping observations and detailed modeling to be understood. Therefore, having an independent measurement of the BH mass through stellar kinematics is fundamental to verify the values obtained with the method applied until now, and to put constraints on the shape factor. Unfortunately, the BLR can be studied only in type 1 AGN. The BH masses in type 2 AGN have been indirectly derived through several assumptions and hypotheses, that make their values rather uncertain. In some cases the BH-$\sigma_s$ relation was applied, but never with $\sigma_s$ measurements obtained at $R_{BH}$. Seyfert 2 galaxies appear to host SMBHs with a range of mass similar to those of Seyfert 1, but a clear answer is possible only having high spatial resolution stellar kinematics. This is even more important in case of LINERs, that belong to the type 2 AGN class and are expected to host less massive SMBHs, close to the stellar cluster values.

The spatial resolution of E-ELT, combined with the high performances provided by MAORY, will allow, at the shortest wavelength, to resolve $R_{BH}$ within distances up to 100 times larger than present-day observatories. In the worst case, the predicted FWHM = 11.2 mas at 2.20 micron corresponds to a resolution of ~ 5 pc at a distance of 100 Mpc, which is enough to resolve $R_{BH}$ in several AGN.

The long-slit spectroscopy of MICADO, will be perfect for this kind of analysis. The 50 mas slit can provide the spectral resolution needed to obtain accurate measurements requiring relatively short observational time.

As an example, we focus here on the observations of three objects, a type 1 AGN (Mrk 1239), a type 2 AGN (NGC 4903) and a LINER (NGC 5124). They are all nearby sources, but their distances (85, 68, and 55 Mpc respectively) make impossible to resolve their $R_{BH}$ with present-day instruments. Their luminosity is however high enough to get high S/N spectra with short integration time. All these objects will be used as a test for future extended surveys.

**Keywords:** Spectroscopy – active galactic nuclei – supermassive black hole

**MICADO Pixel Scale / Fov:** 4 mas/px and 50 arcsec FoV, this is the only configuration available in spectroscopic mode.

**MICADO Observation mode:** Long-slit spectroscopy

**MICADO Spectral set-up:** 50mas x 4arcsec slit in the HK range

**Filters required:** none

**Estimate Survey Area/Sample Size/ Number of Images/Epochs:** This sample is composed by 3 objects. One spectrum for each object with a total integration time of 1800s is required, for a total integration time of 1.5 hours, excluding overheads.

**Average Integration time per image (magnitude of targets; S/N required):** S/N 20 is reached in the H band with 1800s of integration for a target with an H surface brightness of 15 mag arcsec$^{-2}$. This is a typical surface brightness for bright normal galaxies (e.g. Gavazzi et al., 1996, AAS, 120, 521-578) and we used it as an estimate of the surface brightness of our objects once neglected the AGN contribution. This is a good assumption because the adaptive optics will concentrate the light from the AGN in a very small PSF. These estimates were obtained on February 20$^{th}$, 2017 with the ESO-ELT spectroscopic ETC, using a power law with a spectral index $\alpha = 0$ and a spectral resolution $R = 2500$.

**Observation requirements:** Target will cover the entire dimension of the slit, therefore it will not be possible to dither. Sky subtraction will be performed observing empty regions of sky using the sky offset mode.

**Strehl or EE required:** In order to look in the inner part of an AGN the best possible spatial resolution is required. Only in this way it is possible to both concentrate the light of the AGN and to measure the stellar velocity dispersion as close as possible to the black hole.

**Astrometric Accuracy:** no special requirements

**SCAO vs. MCAO:** Mrk 1239 is characterized by an extremely luminous and compact source at the centre of its nucleus. In a nearby object this source can be bright enough to be used as NGS for the SCAO and it will be at the centre of the observed field, in the best possible position. The remaining part of the sample (NGC 4903 and NGC 5124) should be observed using the MCAO.

**Comparison with JWST or other facilities:** The spatial resolution provided by JWST is not sufficient to perform this kind of analysis

**Synergies with other facilities (4MOST/MOONS, LSST/ALMA/HARMONI/METIS, HIRES/MOSAIC), but also VLT or other smaller telescope instruments:** The only telescope able to perform such a study is the E-ELT, however synergies with future facilities such as the Square Kilometer Array (SKA) will be possible to study the inner core of AGN in radio.

**Simulations made/needed to verify science case or feasibility:** an update of the ETC with galaxies and AGN templates will be useful to have a more precise idea of the performance of the telescope. Simulated observations will be used to develop an observation strategy which takes into account the complex shape of the PSF.

**Origin of the targets:** We selected three different type of AGN as test for future surveys. The targets are well known AGN, whose distance is too large to allow precise nuclear stellar velocity dispersion estimates with present-day instruments. They are, however, among the closest AGN that are observable from the southern hemisphere. All the targets have been selected to respect the adaptive optics requirements.

**NGS:** The bright and compact nucleus of Seyfert 1 galaxies can be used as a NGS for the SCAO. The other 2 objects have at least 3 nearby stars with an H magnitude less than 15 mag that can be used as NSG.

**Acquisition:** The targets are nearby and well known objects and there should not be any problem with pointing. Finding chart will be available.

**Calibrations:** *Standard.*

**Data Processing Requirements:** files for a standard IR reduction are needed. Standard stars for flux calibration and telluric correction are needed.

**Any other comments**: Due to the characteristics of the instrument, observations need to be performed in parallactic angle. However, the angle should stay fixed during the whole observing time for each object.

## 7.3 The parsec scale of warm dust around active galactic nuclei

**Authors:** *Gruppioni C. and Ciliegi P. (INAF - OABO)*

**Brief Description of science case:** Active galactic nuclei (AGN) are thought to play a major role in the formation and the evolution of galaxies. The AGN phase provides mechanisms for feedback from the supermassive black hole (SMBH) to its hosting galaxy and the intergalactic medium. Therefore, a thorough knowledge of the accretion process in AGN is required to understand their influence on the formation and evolution of galaxies. Little is known about the accretion process on parsec scales, although models predict that a toroidal distribution of warm molecular gas and dust surrounding the central engine, the so-called dusty torus, is a key component of AGN. In fact, the torus besides playing an important role in fuelling the AGN activity (forming the passive reservoir of material for the accretion onto the SMBH and/or driving the accretion towards the BH; e.g., Hopkins et al. 2012) is supposed to be responsible for the orientation-dependent obscuration of the central engine (e.g., Antonucci & Miller 1985; Antonucci 1993; Urry & Padovani 1995). With a face-on torus, a direct view of the central engine is possible (type 1 AGN), while an edge-on torus prevents the direct view towards the centre (type 2 AGN). Although this scenario is supported by several observational evidence, indicating that type 2 sources host the same central engine as type 1 AGN, but the appearance is different due to the torus orientation, there are also observations that challenge this simple picture.

The dust in the torus, responsible for absorbing the optical/UV light, is heated by the photons emitted by the inner accretion disk: the innermost dust is close to its sublimation temperature and mainly emits in the near-infrared, while the dust at larger distances is at lower temperatures and emits in the mid- and far-infrared (Barvainis 1987). Therefore, direct and unbiased observations of the torus are best carried out at infrared wavelengths.

However, the dust distributions are very compact: they are essentially unresolved by single-dish observations even with the largest currently available telescopes (e.g. Horst et al. 2009; Ramos Almeida et al. 2009, Martínez-Paredes et al. 2015; Garcìa-Bernete et al. 2016). So far, only by employing interferometric methods in the infrared it has been possible to resolve the nuclear dust distributions in AGN. To date, few interferometric studies of the nuclear dust distributions of individual galaxies have been carried out, finding an indication of the presence of a hot, parsec-sized disk surrounded by warm dust in the two brightest local Seyferts: NGC 1068 (e.g., Jaffe et al. 2004, Poncelet et al. 2007) and the Circinus galaxy (e.g. Tristram et al. 2007). However, the first study with a statistically significant sample of AGN shows a rather diverse picture of the dust distribution with quite large differences between the dust distributions in individual galaxies (e.g., a warm component detectable in near-/mid-IR, and a cold component detectable in far-IR/sub-mm/mm; Burtscher et al. 2013). The cold toroidal dust distribution in the nucleus of NGC 1068 has been recently resolved - for the first time - with ALMA (sub-mm wavelength), both in continuum and CO line (i.e., molecular gas) on a 7-10 pc scale (Garcia-Burillo et al. 2016).

Geometrical torus models consider either smooth dust distributions (e.g., Krolik & Begelman 1988; Granato & Danese 1994; Schartmann et al. 2005; Fritz et al. 2006; Feltre et al. 2012) or clumpy (Krolik & Begelman 1988; Nenkova et al. 2002; Schartmann et al. 2008; Hönig & Kishimoto 2010), both providing good fit to the current data: therefore, the physical picture of the torus remains unclear. The degeneracies in the models can, at least partially, be broken by resolving the dust distributions.

Due to the small angular size of this dusty torus, direct studies of its properties require very high angular resolution. The main disadvantages of the current near-/mid-IR interferometers with respect to direct imaging are: 1) they consist of two telescopes only, with a single baseline providing a very poor sampling of the ($u$,$v$)-plane recovering an angular scale only (more observations with different baselines are needed to have an acceptable, though not complete, coverage); 2) the baseline must be short enough to be able to image the requested maximum recoverable scale (i.e., structure larger than $\lambda/D_{min}$ cannot be recovered, the so-called ``resolve-out''), hence, e.g., low surface brightness flux at large scales is completely lost; 3) Michelson interferometers with large apertures offer the highest angular resolution, but minimal field-of-view (FOV, <100mas).

The MAORY Adaptive Optics applied to MICADO will overcome these problems, providing imaging at very high resolution (4 mas or 1.5 mas/pixel) combined with large FOV (50 or 20 arcsec), and will therefore be ideal to study the inner parts of AGN. In order to investigate the physics of the AGN by understanding how the properties of the dusty tori are related to feeding and obscuration, we need to resolve the matter distribution on (sub-) parsec scales. We therefore aim at characterising a statistical sample of AGN tori by resolving for the first time their inner (hotter) dust emission through direct imaging in the near-infrared, on pc/sub-pc scales.

This will allow us to reconstruct real images of the torus from the inner edge to the main body with resolutions of a few milli-arcseconds that are needed to decipher the physics at work.

Through high spatial resolution imaging in the near-IR, we will investigate the inner structure of the dusty tori surrounding the central BHs in local AGN, observing the hot dust in the very central regions, on (sub-)pc scales (e.g., clumps) corresponding to a scale of 0.2 pc/mas in the sky considering the "local" nature of these objects (z_mean = 0.01). We aim at separating the toroidal structure from the host galaxy and the surrounding stellar emission. At z=0.01 the observed size of dusty tori is about 0.05 - 0.1 arcsec in diameter (assuming a dusty torus with a diameter of about 10-20 pc).

A test case could be the local Seyfert NGC 5135, an infrared-luminous galaxy (LIRG) at z = 0.013693, for which we have recently analised the ALMA data in Bands 6 (1.3 mm) and 9 (550 μm), either continuum and CO lines, detecting the cold (external) component of the dusty torus (unresolved on a 50 pc scale). A peculiar kinematic of the molecular gas in the host galaxy, with shocks along the inner bar (fuelling the central BH) overimposed to a disk rotation and radial motions (i.e., inflows towards the nucleus), have been observed. Higher resolution observations are planned with ALMA in order to resolve the cold and molecular component of the dusty torus. With MAORY+MICADO we will resolve the inner and hotter part of the dusty torus, and from the combination of the ALMA and E-ELT observations we will be able to study in detail the torus structure and physics.

**Keywords:** imaging – photometry – inner regions of local AGN

**MICADO Pixel Scale / Fov:** 1.5mas/px and 20 arcsec FoV, motivated by the maximization of the performances in terms of SR and by the small size of the actual targets.

**MICADO Observation mode:** Standard Imaging MICADO

**Spectral set-up:** none

**Filters required:** J, H and K: we need to investigate the colour, the dust temperature and different scales of different dust components (if any).

**Estimate Survey Area/Sample Size/ Number of Images/Epochs:** A sample of 30 target is required in order to have statistically significant results.

**Average Integration time per image (magnitude of targets; S/N required):** With the ESO-ELT ETC, assuming a px scale of 5 mas (the smallest scale currently available in the ETC) and a 3px × 3px reference area for computing the S/N ratio we obtain integration time lower than 10 minutes for each target to reach S/N > 10 over the nuclear region, where the expected magnitude is ~11.11 in K band (Peng et al. 2006). In fact, we estimate that in 10 min observation we will be able to reach 25 mag with S/N ~20). Considering to observe in three different bands (J, H and K) in order to study the different dust component and temperature of the tori, a total of 30 minutes for target is required. Considering dithering and overheads, a total of 45 minutes for target must be considered. A sample of 30 targets can be observed in about 23 hrs of observing time.

**Observation requirements:** Assuming to observe with the smallest pixel scale (1.5 mas/pixel) to maximize the resolution, we have a FoV of 20x20 arcsec. The target proposed are very large in size (1-2 arcmin) and in the observed region of 20x20 arcsec only the central, very crowded region is observed. Large dithers are therefore required in order to estimate the sky background. Moreover, a very high dinamical range is required in these observations in order to disentangle tori from nuclear emissions.

**Astrometric Accuracy:** no special requirements.

**SCAO vs. MCAO:** Given the very small size of the region of interest (sub-arcsec for the observed size of the dusty tori) we propose to use the SCAO mode in order to maximise the SR value and to obtain the best spatial resolution performances. The SCAO mode observation will be guided with the nuclei of these AGN (K~11, V~14). The feasibility of this kind of observations (AO Observation guided with the nuclei of AGN) has been already demonstrated with the FLAO system at LBT.

**Comparison with JWST or other facilities:** Spatial resolution is the key issue here: ELT+MAORY+MICADO is much better than JWST in this respect (by a factor of >3, at high SNR, comparing MICADO observations in H band to JWST observations in I band).

**Synergies with other facilities:** ALMA sub-arcsec resolution image in the sub-mm/mm will resolve the cold dusty/molecular torus component, complementing the observation with ELT+MAORY+MICADO, resolving the warmed dust contribution. JWST/MIRI will observe (but not resolve) the intermediate/warm dust component.

**Simulations made/needed to verify science case or feasibility:** The requested observation require a very high dynamical range and a very high spatial resolution, in order to directly observe the torus emission and to distinguish it from the surrounding stellar emission. As an example, while the total K-band magnitude of NGC5135 is ~8.8, for the AGN emission Peng et al. (2006) estimated ~11.11. However, by disentangling the torus contribution from the stellar one, we estimate an AGN torus luminosity of about $4x10^5$ times fainter than the total stellar emission in the K band, thus 10 magnitudes fainter: we expect the K magnitude of the torus only to be about 18.8 mag. Detailed and dedicated simulation combining AGN model and MAORY-MICADO SCAO PSF are crucial in order to check the feasibility of these observations.

**Origin of the targets:** Lists available in the literature, e.g. Gruppioni et al. (2016, MNRAS), Rosenberg et al. (2012); original sample from Rush, Malkan & Spinoglio (1993, ApJ,).

**NGS:** Observation in SCAO Mode. The nuclei of the Seyferts are very bright (e.g., for NGC5135, K~11, V~14, B~15) and can be used as guide star for the SCAO mode

**Acquisition:** The targets are very bright and finding-charts are available in different optical/near-infrared bands, so we do not expect any problems during the acquisition process. Moreover, the size of the interested region (AGN dusty torus, i.e., < 1 arcsec) are well within the SCAO FoV.

Calibrations: Standard. No special astrometric requirement.

**Data Processing Requirements:** Detailed PSF modelling for SCAO mode is crucial for these observations.

**Any Other Comments:** none

# 7.4 Exploring the Early Universe with Gamma-Ray Bursts

**Authors:** *Maiorano E., Amati L., Masetti N., Nicastro L., Palazzi E., Pian E., Rossi A., (INAF - IASFBO),* Hunt L. (INAF - OAA), *Stratta G. (University of Urbino), Savaglio S. (University of Calabria),* also on behalf of the CIBO collaboration.

## Brief description of science case

Gamma-ray bursts (GRBs) are bright flashes of high-energy radiation. They are so luminous to be detected up to very high redshifts. The goal of this project is to use the brightness of GRBs and their association with star forming regions within galaxies to shed light on the high redshift Universe. From now on we will consider only long GRBs (LGRBs), i.e. GRBs with $T_{90} > 2s$ (Kouveliotou et al. 1993, ApJ, 413, L101) associated with the death of massive stars. These major stellar explosions are unique powerful tracers of the star formation rate (SFR) up to z~10–12, may be signatures of pop-III stars and act as beacons of light bringing to the fore all kinds of galaxies from nearby objects to the faintest and farthest sources in the Universe. Detectable up to z ~10–12 thanks to their huge X/gamma-ray luminosities, absorption *spectroscopy* of the fading NIR afterglow emission and *deep imaging* (and possible emission spectroscopy) of their host galaxies can be used to explore the re-ionization era, to measure the cosmic star-formation rate, the number density and properties of low-mass galaxies, the neutral hydrogen fraction, the escape fraction of UV photons, the cosmic chemical evolution.

To reach this goal, we intend to observe both afterglows (AG) and host galaxies (HG) of a number of high redshift GRBs (z>6), by performing deep imaging and spectroscopy, using E-ELT+MAORY+MICADO. These observations can fully characterize the properties of star-forming galaxies over the whole cosmic history.

High-redshift candidates are those GRBs with the Lyman-α dropout redwards of the I-band, corresponding to a redshift z > 6. Very high-z GRBs can be easily distinguished by the very sharp break produced at Lyman-α dropout. GRBs, having an intrinsic power-law emission, can provide a much more reliable photometric redshift than Lyman break galaxies (LBGs).

In addition, the sinergy with future foreseen high-energy satellites devoted to the GRBs science (like the French-Chinese SVOM to be launched in 2021, the THESEUS mission proposed for ESA call M5, and the already approved X-ray mission ATHENA with superb spectroscopic capabilities) will provide redshifts and luminosities that are essential to optimise the time-critical follow-up (Yuan et al. 2016, SSR, 202, 235). Thanks to this strategy, the selection of the highest priority targets will be possible for the most appropriate observational strategy. The present detection rate of GRBs at z > 6 is about 0.7–0.8 /year, thanks to, and limited by, the Swift satellite capabilities and follow-up possibilities with ground and space facilities. This rate will likely increase to ~2–4 / year at the beginning of the next decade, thanks to next generation space facilities like SVOM and EP (Einstein Probe), and may grow up to several tens of events per year at the end of the '20s if THESEUS, or an analogue mission concept, will be selected and realized by ESA of other space agencies.

The proposed MAORY+MICADO observations will be able to address the following key questions. In parenthesis is given the target of proposed observations (AG or HG of the GRB).

## 1 - *The Lyman continuum escape fraction (AG spectroscopy)*

High-S/N afterglow spectroscopy reveals the neutral hydrogen column along line-of-sight to the GRB. Since the opacity of the medium to far ultraviolet (FUV) photons depends on this column, a statistical sample of afterglows can be used to infer the average escape fraction ($f_{esc}$) over many lines of sight, specifically to the locations of massive stars dominating global ionizing radiation production (see Figure 7.4-1 - Right). Useful constraints have so far only been possible at $z = 2–4$, indicating an upper limit of $f_{esc} < 7.5\%$ (e.g., Fynbo et al. 2009, ApJS, 185, 526), observations with MAORY+MICADO of the AG of GRBs at $z > 6$, will provide much more precise constraints on the epoch of reionization.

## 2 - *The build-up of metals, molecules and dust (AG spectroscopy)*

Bright GRB afterglows with their intrinsic power-law spectra provide ideal backlights for measuring not only the hydrogen column, but also obtaining abundances and gas kinematics probing to the hearts of their host galaxies (e.g., Hartoog et al. 2015, A&A, 580, A139). In addition, the imprint of the local dust law, and the possible detection of $H_2$ molecular absorption, provides further detailed evidence of the state of the host interstellar medium (ISM) (e.g. Friis et al. 2015, MNRAS, 451, 167). Thus they can be used to monitor cosmic metal enrichment and chemical evolution to early times, and search for evidence of the nucleosynthetic products of even early generations of stars (Pop-III). Thanks to MAORY+MICADO spectra, abundance determinations will be possible through simultaneous measurement of metal absorption lines and modelling the red-wing of Lyman-α to determine host HI column density, potentially even many days post-burst (see Figure 8.4-1 - Left).

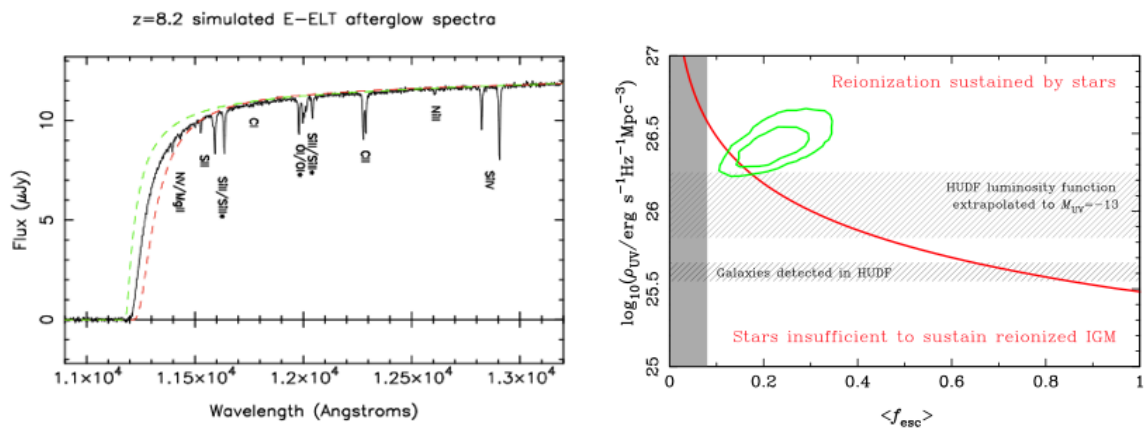

Figure 7.4-1 *Left:* Simulated ELT 30 min spectrum of a faint GRB afterglow observed after ~1 day. The S/N provides abundance determinations from metal absorption lines, while fitting the Lyman-α damping wing simultaneously fixes the IGM neutral fraction and the host HI column density, as illustrated by the two extreme models, a pure 100% neutral IGM (green) and best-fit host absorption with a fully ionized IGM (red). *Right:* The UV luminosity density from stars at z~8 and average escape fraction <$f_{esc}$> are insufficient to sustain reionization unless the galaxy luminosity function steepens to magnitudes fainter than $M_{UV}$=-13 (grey hatched region), and/or <$f_{esc}$> is much higher than that typically found at z~3 (grey shaded region). Even in the late 2020s, <$f_{esc}$> at these redshifts will be largely unconstrained by direct observations. The green contours show the 1-σ and 2-σ expectations for a sample of 25 GRBs at z~7-9 for which deep spectroscopy provides the host neutral column and deep imaging constrains the fraction of star formation occurring in hosts below the JWST limit (Robertson et al. 2013, ApJ, 768, 71). The input parameters were $log_{10}(\rho_{UV})$=26.44 and <$f_{esc}$>≥0.23, close to the (red) borderline for maintaining reionization by stars.

### 3 - Topology of reionization (AG spectroscopy and HG imaging)

One of the last uncharted astrophysical epochs is the time between recombination and the end of the first phase of star formation. Massive stars, possible progenitors of LGRBs, are likely to be a significant contributor to the ionizing photons at high redshift, thus GRBs are expected to occur also during the reionization epoch. Unlike luminous QSOs which ionize the intergalactic medium around themselves, GRBs do not suffer from the "proximity effect" (Fan, Carilli & Keating 2006, ARA&A, 44, 415). This makes GRBs cleaner probes of the Gunn-Peterson absorption used to estimate the degree of ambient ionization. With high-S/N afterglow spectroscopy the Lyman-α red damping wing can be decomposed into contributions due to the host galaxy and the intergalactic medium (IGM). The latter provides the hydrogen neutral fraction and so measures the progress of reionization local to the burst. With samples of few tens of GRBs at high redshift, we can begin to statistically investigate the average and variance of the reionization process as a function of redshift (e.g., McQuinn et al. 2008, MNRAS, 388, 1101).

### 4 - Population III stars (HG imaging & spectroscopy)

The multiwavelength properties of GRBs with a Pop-III progenitor are only predicted on the expected large masses, zero metallicity of these stars. Even the detection of a single GRB from a Pop-III progenitor would put fundamental constraints on the unknown properties of the first stars.

### 5 - High redshift star formation induced by galaxy interactions (HG imaging & spectroscopy)

Nearby absorbers have already been identified in the spectrum of particularly bright GRB afterglows. They indicate possible galaxy interaction which triggers star formation (SF), and are more frequent at high redshift. The fraction of absorbers in the spectra of GRB afterglows is ~5 times larger than in QSO spectra. Thus GRBs are a better tool to understand the bound between galaxy interaction and star formation in the high redshift Universe. We propose to use the high angular resolution and sensitivity achievable with MAORY+MICADO for both imaging and spectroscopy to study the morphological properties and the UV emission of the host of high-z GRBs which show two strong nearby absorbers in the AG spectrum. These observations will allow us to test the hypothesis that galaxy interactions at high redshift induce the formation of very massive stars and GRB progenitors.

### 6 - HG search (HG imaging)

Even to the depth achieved in the Hubble Ultra-deep Field (HUDF), we only know that the faint-end of the luminosity function (LF) at z > 6 approaches a power-law of slope α = 2, with an unconstrained cut-off at low luminosities, which affects the total luminosity integral. Although currently limited by small-number statistics, early application of this technique has confirmed that the majority of star formation at z~6 occurred in galaxies below the effective detection limit of HST (Tanvir et al. 2012, ApJ, 754, 46; McGuire et al. 2016, ApJ, 825, 135). Since the exact position and redshift of the galaxy is known via the GRB afterglow, GRB hosts search and observations are more efficient than equivalent deep field searches for Lyman-break galaxies. We thus propose to conduct deep searches with MAORY+MICADO for the hosts of GRBs at high-z to derive the LF at z >6.

## Observational Strategy 1: Afterglow follow up - request for ToO

**Keywords:** Target of Opportunities - spectroscopy - long gamma ray bursts - Afterglow

**MICADO Pixel Scale / Fov:** 1.5mas/px and 20 arcsec FoV, motivated by the small size of the actual targets.

**MICADO Observation mode:** Standard spectroscopy

**MICADO Spectral set-up:** slit 1, 16 mas × 4 arcsec, spectral coverage: IzJ+HK.

**Filters required:** none

**Estimate Survey Area/Sample Size/ Number of Images/Epochs:** Since 2020 the expected rate of GRBs at z>6 is >5 per year. Considering visibility constraints, we expect to have ~2-3 targets/yr.

**Average Integration time per image (magnitude of targets; S/N required):** Assuming for the spectrum the worst case of an AG at z=10, with H~26 we need ~1.5hr exposure time to reach S/N~3. These estimates have been obtained on March 28, 2017, with the ESO-ELT ETC, assuming power law index -1, radius of circular S/N reference area of 9.9 mas, spectral resolution of 11000 and air mass of 1.5 arcsec

**Observation requirements:** Small dithers + Large dithers [refer to table 8 of MICADO OCD]
**Strehl or EE required:** for our purpose the most important parameter is the sensitivity.
**Astrometric Accuracy:** no special requirement.
**SCAO vs. MCAO:** MCAO is the only viable mode. as the circumstance of a target within ~10" of a bright star (allowing SCAO) is rare or non-existing.
**Comparison with JWST or other facilities:** The sensitivity is the key issue here: ELT+MAORY+MICADO is comparable with JWST in this respect. (to be checked)
**Synergies with other facilities:** Target positions will be provided by high-energy satellites available at that time and/or optical telescopes.
**Simulations made/needed to verify science case or feasibility:** No simulations are needed in addition to the ones obtained from the ETC tool.

**Origin of the targets:** Lists will be available in the GCN Circulars or ATels.

**NGS:** GRBs Latitude distribution (to be checked).

**Acquisition:** our targets are point sources so 20" FoV will be selected. Finding charts available.

**Calibrations:** Standard. No special astrometric requirement. The presence of photometric calibrators in the field has to be verified. The relative photometry precision derived with the ETC is sufficient to reach the main science

**Data Processing Requirements:** Instrumental pipeline and/or pre-reduced data

**Any other comment:** Quick response to the activations

## Observational Strategy 2: Host galaxies observations

**Keywords:** photometry - spectroscopy - long gamma ray bursts - host galaxies

**MICADO Pixel Scale / Fov:** 1.5mas/px and 20 arcsec FoV

**MICADO Observation mode:** Standard Imaging and spectroscopy

**MICADO Spectral set-up:** slit 2, 50mas x 4arcsec. IzJ+HK. The possibility to have a slit with adjustable orientation would enhance the scientific return (e.g. AG+HG).

**Filters required:** J, K plus an additional filter according to the redshift

**Estimate Survey Area/Sample Size/ Number of Images/Epochs:** At present the sample of GRB's HGs at z>6 is limited to 9 events. In the late 20's the rate of GRBs at z>6 is >5 per year. Considering visibility constraints, we expect to have ~2-3 targets/yr.

**Average Integration time per image (magnitude of targets; S/N required):** Assuming the worst case for the imaging of an HG at K=30, with ~4 hr exposure time a S/N~3 is reached. These estimates have been obtained on March 28, 2017, with the ESO-ELT ETC, assuming a px scale of 5 mas (the smallest scale currently available in the ETC), for a point source and a 10px X 10px reference area for computing the S/N ratio and airmass of 1.5 arcsec. If the HG is brighter than K~26, we request a 2.5hr spectrum to reach S/N>5. These estimates have been obtained on March 28, 2017, with the ESO-ELT ETC, assuming powerlaw index -1, radius of circular S/N reference area of 9.9 mas, spectral resolution of 2500 and airmass of 1.5 arcsec.

**Observation requirements:** Small dithers + Large dithers

**Strehl or EE required:** the best spatial resolution performances the most effective would be the test

**Astrometric Accuracy:** no special requirement

**SCAO vs. MCAO:** MCAO is the only viable mode as the circumstance of a target within ~10" of a bright star (allowing SCAO) is rare or non-existing

**Comparison with JWST or other facilities:** Spatial resolution is the key issue here: ELT+MAORY+MICADO is much better than JWST in this respect (by a factor of >3, at high SR, comparing MICADO observations in H band to JWST observations in I band).

**Synergies with other facilities:** Target positions will be provided by high-energy satellites available at that time and/or optical telescopes.

**Simulations made/needed to verify science case or feasibility:** No simulations are needed in addition to the ones obtained from the ETC tool.

**Origin of the targets:** Lists will be available in the GCN Circulars or ATels.

**NGS:** GRBs Latitude distribution (to be checked)

**Acquisition:** the characteristic size of GRB host gal. is ~1 arcsec hence there should be no problem in getting a useful pointing within the 20" FoV. Finding charts available.

**Calibrations:** Standard. No special astrometric requirement. The presence of photometric calibrators in the field has to be verified.The relative photometry precision derived with the ETC is sufficient to reach the main science goal.

**Data Processing Requirements:** The PSF modelling should be feasible with stars in the FoV.

**Any other comment:** none.

## 7.5 Characterization of electromagnetic counterpart candidates of multi-messenger compact binary coalescence systems

**Authors:** Maiorano E., Amati L., Masetti N., Nicastro L., Palazzi E., Pian E., Rossi A., (INAF - IASFBO), Brocato E., Piranomonte S. (INAF - OAR), Mapelli M., Ciolfi R. (INAF - OAPD), Covino S., D'Avanzo P. (INAF - OABR), De Pasquale M. (Istanbul University), Branchesi M., Stratta G. (University of Urbino), Savaglio S. (University of Calabria), Malesani D. (University of Copenaghen), also on behalf of the CIBO and GRAWITA collaborations.

### Brief description of science case

The coalescence of compact binary systems composed of two neutron stars (NSs), a stellar mass black hole (BH) and a NS, or two BHs is the primary target for ground-based gravitational wave (GW) detectors. In addition, NS-NS and NS-BH mergers are also expected to be associated with electromagnetic (EM) counterpart signals across the entire EM spectrum. In particular, they are thought to be the progenitors of short gamma-ray bursts (SGRBs) and kilonova/macronova transients. In the new era of gravitational-wave astronomy, the present science case aims at exploiting the synergy between GW and EM observations, targeting the most promising multi-messenger sources, i.e. NS-NS and NS-BH mergers.

SGRBs are identified as those GRBs with duration less than about two seconds and with harder spectra with respect to the more frequent long GRBs (LGRBs) (Kouveliotou et al. 1993, ApJ, 413, L101). The study of their afterglows (AGs) and host galaxies (HGs) provides critical information about their explosion properties and progenitors. Current theoretical models as well as observations support the association of SGRBs with NS-NS or NS-BH merger systems (Eichler et al. 1989, Nature, 340, 126; Narayan et al. 1992, ApJ, 395, L83; Nakar 2007, Phys. Rev. 442, 166; Berger 2014, ARA&A, 52, 43). Observations carried out in the last decade show that these events, occurring in both early and late-type galaxies, are associated to relatively old ($\geq$ 50 Myr) stellar populations, and do not show evidence of associated supernovae. In addition, SGRBs have systematically larger offsets from their host galaxies centers than LGRBs, in line with the idea of different progenitors.

Indeed, one of the key predictions of the compact object merger model is that systemic natal kicks may lead to substantial offsets between the birth and explosion sites of these systems (e.g., Fryer et al. 1998, ApJ, 496, 333; Fryer, 2004, ApJ, 601, L175). Observationally, some short GRBs at such large offsets can appear to be host-less because their projected locations will extend much beyond the visible extent of typical galaxies. The distribution of natal kick velocities of compact objects and compact object binaries is still an open issue, ranging from typical velocities of few hundred km/s (Hobbs et al. 2005, MNRAS, 360, 974) to less than ~30 km/s in the vast majority of the systems (Beniamini & Piran, 2016, MNRAS, 456, 4089). Besides, dynamical kicks, due to close gravitational encounters, are also expected to contribute to the ejection of compact object binaries from their birthplace (e.g. Mapelli et al. 2011, MNRAS, 416, 1756). For these reasons, it is crucial to assess the impact of natal kicks and dynamics on SGRBs. In particular, it is important to establish that the claimed host-less GRBs do not in fact lie on top of very faint system, hitherto undetected (e.g. the very faint host galaxy of GRB

070707: Piranomonte et al. 2008, A&A, 491,183).

Another predicted electromagnetic signature of NS-NS/NS-BH binary mergers is the so-called "kilonova" (KN). Hydrodynamical simulations have shown that, around $10^{-4}$–$10^{-2}$ M$\odot$ of material may become tidally unbound and expelled (e.g. Rosswog, 2005, ApJ, 634, 1202). This material will assemble into heavy elements via r-process. Kilonova emission is thought to originate from the radioactive decay of these newly formed r-process elements (e.g. Li & Paczynski 1998, ApJ, 507, L59). Computations of the opacities connected to r-process material indicate that the bulk of kilonova emission is expected to peak in the NIR on a timescale of a few days (Kasen, Badnell & Barnes 2013, ApJ, 774, 25; Grossman et al. 2013, MNRAS, 439, 757; Tanaka & Hotokezaka 2013, ApJ, 775, 113). The colour and brightness of kilonova light curves are therefore sensitive markers of the ejecta composition, and can be used to gain insight into the nature and the physics of compact object mergers (Kawaguchi et al. 2016, ApJ, 825, 52; Kasen, Fernandez & Metzger 2015, MNRAS, 450,1777). Kilonova emission is predicted to be isotropic and for this reason it is a promising electromagnetic counterpart of the merging of NS-NS/NS-BH systems, alternatively to the beamed SGRBs for which the chances of a joint GW-SGRB detection are rather low (Ghirlanda et al. 2016, A&A, 594, A84). A kilonova counterpart is thus expected to be found among the transient sources discovered in the GW-localized sky regions (after the detection of a NS-NS/NS-BH GW event) or identified as an emerging NIR component from the rapidly decaying afterglow emission of short GRBs (Yang et al. 2015, NatCo, 6, 7323; Jin et al. 2015, ApJL, 811, 22; Jin et al. 2016, NatCo, 7, 12898; Tanvir et al. 2013, Nat, 500, 547).

In this context we intend to exploit the unprecedented spatial resolution and sensitivity of MAORY+MICADO to perform observations to shed light on the EM counterparts of NS-NS and NS-BH binary systems. In particular we propose to ask for deep observations to search and characterize the faint host galaxies of SGRBs and "late" ToOs (1-10 days after the burst) to detect the kilonova emission.

## 1 - *Deep observations to characterize the host galaxies of SGRBs (imaging)*

NS-NS/BH mergers are expected to be found in the most massive galaxies, but the way the progenitor was born can also play a role. In particular, one should also consider the conditions that lead to the episode of star formation that generated the stellar population which the progenitor belongs to (Leibler & Berger, 2010, ApJ, 725, 1202; Fong et al., 2013, ApJ, 769, 56). Indeed, different channels for the formation of the progenitors lead to different star formation histories, and thus morphology and contribution of the star clusters. Moreover, one of the main uncertainties on the SGRB progenitors is the time between the progenitor formation and the GRB explosion (*delay-time*). Thus, by constraining the age of the host stellar population, it is possible to get a reliable indication on the age of the progenitor. A first explorative study on a sample of SGRB host galaxies has been carried out with HST, providing constraints on their offsets and on how their locations track the host galaxy light distribution (Fong, Berger & Fox, 2010, ApJ, 708,9). However, up to now studies of the mass and age refer to the whole host galaxy, since the angular resolution (considering the redshift z~0.5-1) of the available facilities is not sufficient to measure age and mass of the multiple stellar components in the hosts. The high angular resolution and high sensitivity of MAORY+MICADO will allow us to resolve the morphology of the handful of SGRB hosts located at redshift z<0.2. Thus, we can study in detail any relation between the SGRB site explosion and the surrounding environments and estimate the mass and absolute age of star clusters that we will identify (e.g. Bono et al. 2010, ApJ, 708, L74).

## 2 - Deep observations in the region of host-less SGRBs to search for faint host galaxies (imaging)

The short GRB offsets normalized by host-galaxy size are larger than those of long GRBs, core-collapse SNe, and Type Ia SNe, with only 20% located at $<\sim 1$ $r_e$ ($r_e$ being the galaxy effective radius, which accounts for the range of HG sizes and any systematic trends in these sizes between the various GRB and SN populations) and about 20% located at $>\sim 5$ $r_e$. These results are indicative of natal kicks or an origin in globular clusters, both of which point to compact object binary mergers (Berger, 2014, ARAA, 52, 43). The inferred kick velocities are $\sim 20$-140 km s$^{-1}$, which is in reasonable agreement with Galactic NS-NS binaries and population synthesis models. About 20% of well localized SGRBs ($\sim 7\%$ of the total number of SGRBs) have been classified as host-less (Fong et al. 2013, ApJ, 769, 56). Several galaxies are present in the deep optical-NIR HST images of the region of these "host-less" SGRBs, but they sometimes have a high probability of chance coincidence because of non-optimal X-ray localization precision (2-3 arcsec). Thus, these galaxies cannot be identified as hosts of these events. In this context we propose to first consider a small sample of few ($<10$) accurately localized (via radio and/or optical afterglow detection) host-less SGRBs and investigate if these might be inside very faint host galaxies or may have been ejected by natal kicks or dynamics. Distinguishing between these two scenarios gives constraints on the formation of NS binaries. Note that, currently, this is only an indicative target sample and it will be increased as the detections of more SGRBs will be carried out.

## 3 - Search for kilonova emission (late ToO: imaging & spectroscopy)

Kilonova emission evidence relies so far on one case. Indeed the near-IR excess detected with HST observations following the SGRB 130603B (Tanvir et al. 2013, Nat, 500, 547) was interpreted as kilonova emission and the first direct evidence that SGRBs originate from NS-NS/ NS-BH mergers. Late-time optical and near-IR observations place limits on the luminosity of optical and near-IR kilonova emission following this SGRB of $\sim$ few x 10$^{40}$ erg s$^{-1}$ (Fong et al. 2016, ApJ, 833, 151). Other two suggestive pieces of evidence for SGRB/KN association has been proposed for SGRB 050709 and SGRB 060614 (Yang et al. 2015, NatCo, 6, 7323; Jin et al. 2015,ApJL, 811, 22; Jin et al. 2016, NatCo, 7, 12898). Given current kilonova models, deep optical and near-IR AG observations at 1-10 days after a SGRB to depths of $\approx 23$-24 ABmag are necessary to probe a meaningful range of parameter space. This highlights the key role of instruments like ELT+MAORY+MICADO in performing meaningful searches for EM counterparts to gravitational wave sources, through the dual possibility of looking for KN evidence in the SGRB NIR-afterglows or, alternatively, among the EM candidates found in the 1-10 deg$^2$ sky regions of GW-localized NS-NS/NS-BH mergers, that will be followed-up as soon as their position will be available via GCN or ATel.

## Observational Strategy 1: SGRBs with confirmed host galaxies

**Keywords:** photometry - spectroscopy - short gamma ray bursts - host galaxies

**MICADO Pixel Scale / Fov:** 1.5mas/px and 20 arcsec FoV and 4mas/px and 50"x50" arcsec FoV, depending on the size of the galaxy.

**MICADO Observation mode:** Standard Imaging

**MICADO Spectral set-up:** none

**Filters required:** J, K to characterize the mass and age of the star clusters within the host galaxy

**Estimate Survey Area/Sample Size/ Number of Images/Epochs:** In the 2004-2016 period, we had 5 hosts with redshift less than 0.2 (about 20% of the total SGRBs with measured redshift) (4 HGs from Fong et al. 2015, 1 from http://www.mpe.mpg.de/~jcg/grb160821B.html); 2 images/1 epoch per object.

**Average Integration time per image (magnitude of targets; S/N required):** In order to resolve faint star cluster inside the close SGRBs host galaxies, we reach K~30.5 mag with 2.5 hr exposure time at S/N ~3. The half-light radii of the targets are less than 10 pc at z=0.2, so we can consider them as point sources. These estimates have been obtained on March 24, 2017, with the ESO-ELT ETC, assuming a px scale of 5 mas (the smallest scale currently available in the ETC), for a point source and a 3px X 3px reference area for computing the S/N ratio and air mass of 1.5.

**Observation requirements:** Small dithers + Large dithers [refer to table 8 of MICADO OCD]

**Strehl or EE required:** the best spatial resolution performances the most effective would be the test.

**Astrometric Accuracy:** no special requirement.

**SCAO vs. MCAO:** MCAO is the only viable mode as the circumstance of a target within ~10" of a bright star (allowing SCAO) is unlikely.

**Comparison with JWST or other facilities:** Imaging with MAORY+MICADO and its superb spatial resolution provides a unique opportunity to characterize the environment of the host galaxies and possibly the SGRB location (e.g. presence of star forming knots close to the SGRB location, localization of star clusters in the hosts). ELT+MAORY+MICADO is much better than JWST in this respect (by a factor of >3, at high SR, comparing H-band MICADO observations to JWST I-band imaging).

**Synergies with other facilities:** Targets positions and integrated magnitudes/colors will be known from SGRBs detection with high-energy satellites available at that time (maybe Swift, SVOM, THESEUS) combined with accurate afterglow localizations. Forthcoming facilities (such as JWST) will provide a list of possible hosts to be followed-up with ELT with its higher spatial resolution.

**Simulations made/needed to verify science case or feasibility:** No simulations are needed in addition to the ones obtained from the ETC tool.

**Origin of the targets:** Lists available in the literature e.g. Fong et al. 2013, ApJ, 769, 56 + future studies.

**NGS:** GRBs Latitude distribution (to be checked)

**Acquisition:** the characteristic size of GRB host galaxy is larger than 1 arcsec, hence there should be no problem in getting a useful pointing within the 20" FoV and 50"X50" FoV , depending on the size of the galaxy. Finding charts available.

**Calibrations:** Standard. No special astrometric requirement. The presence of photometric calibrators in the field has to be verified. The relative photometry precision derived with the ETC is sufficient to reach the main science goal.

**Data Processing Requirements:** The PSF modelling should be feasible with stars in the FoV.

**Any other comment:** targets (the most peculiar ones, in particular) can be possibly observed in spectroscopic mode with HARMONI/JWST in order to better characterize the region of GRB.

## Observational Strategy 2: Confirmed hosts-less SGRBs - search for their faint galaxies

**MICADO Pixel Scale / Fov:** 1.5mas/px and 20 arcsec FoV

**MICADO Observation mode:** Standard Imaging

**MICADO Spectral set-up:** none

**Filters required:** K

**Estimate Survey Area/Sample Size/ Number of Images/Epochs:** between 2004-2016 we have 6 host-less SGRBs and 6 "inconclusive" HGs (Fong et al. 2013). By 2024, that is 7 years from now, we expect a sample of ~12 targets. We request 1 images/1 epoch per object.

**Average Integration time per image (magnitude of targets; S/N required):** Considering limiting magnitude for the host-less of K~25, it is possible to reach S/N~5 with 1.8 hr exposure time. These estimates have been obtained on March 24, 2017, with the ESO-ELT ETC, assuming a px scale of 5 mas (the smallest scale currently available in the ETC), for an extended object of 1 arcsec and airmass of 1.5.

**Observation requirements:** Small dithers + Large dithers [refer to table 8 of MICADO OCD].

**Strehl or EE required:** for our purpose the most important parameter is the sensitivity.

**Astrometric Accuracy:** no special requirement.

**SCAO vs. MCAO:** the circumstances of a target within ~10" of a bright star (allowing SCAO) should be considered in order to select SCAO or MCAO mode.

**Comparison with JWST or other facilities:** The sensitivity is comparable with JWST in this respect, but MAORY+MICADO could better characterize the hosts.

**Synergies with other facilities:** Targets positions will be known from SGRBs detection with high-energy satellites available at that time (maybe Swift, SVOM, THESEUS) combined with optical AG precise localizations. Some targets can be observed in spectroscopic mode with HARMONI/JWST depending on the brightness.

**Simulations made/needed to verify science case or feasibility:** No simulations are needed in addition to the ones obtained from the ETC tool.

**Origin of the targets:** Lists will be available in the literature.

**NGS:** GRBs Latitude distribution (to be checked)

**Acquisition:** the characteristic size of putative host galaxy (likely above z~1) is expected to be well within 20" FoV. Finding charts available.

**Calibrations:** Standard. No special astrometric requirement. The presence of photometric calibrators in the field has to be verified. The relative photometry precision derived with the ETC is sufficient to reach the main science goal.

**Data Processing Requirements:** The PSF modelling should be feasible with stars in the FoV.

**Any other comment:** If any host galaxy will be found, spectrum will be possibly requested.

### Observational Strategy 3: Request for late ToO to search kilonova emission

**MICADO Pixel Scale / Fov:** 1.5mas/px and 20 arcsec FoV, motivated by the small size of the actual targets.

**MICADO Observation mode:** Standard Imaging and spectroscopy

**MICADO Spectral set-up:** slit 1, 16 mas × 4 arcsec, IzJ+HK

**Filters required:** I, H. The kilonova emission is best characterized comparing lightcurves obtained with two bands separated in wavelength to test the color evolution; thus, we ask for I and H band: we choose H due to the better sensitivity of the instrument in this filter with respect to the K one.

**Estimate Survey Area/Sample Size/ Number of Images/Epochs:** the rate should be the same of NS-NS, NS-BH merging systems that is still very uncertain of the order of few tens per year. We will consider well-localized SGRBs at z<1 and/or kilonova candidates from GW trigger, so we expect about 3 similar targets/yr. For each event, we propose to have photometric monitoring (I, H filters) of 10 epochs with 2 days cadence starting from 2 days after the trigger (see integration time) and spectral acquisition for brightest events (I<24). Anyway the strategy will be tailored based on the source brightness.

**Average Integration time per image (magnitude of targets; S/N required):** For S/N~10, I=28, and H=25 are reached with integration time of 10 min and 15 s respectively. These estimates have been obtained on March 23, 2017, with the ESO-ELT ETC, assuming a px scale of 5 mas (the smallest scale currently available in the ETC), for a point source and a 3px X 3px reference area for computing the S/N ratio and airmass of 1.5. For the spectrum of the sources with I<24 we need less than 2 hr exposure time to reach S/N~10. These estimates have been obtained on March 23, 2017, with the ESO-ELT ETC, assuming $T_{Blackbody}$=9000K, radius of circular S/N reference area of 10 mas, spectral resolution of 4500 and airmass of 1.5 arcsec.

**Observation requirements:** Small dithers + Large dithers [refer to table 8 of MICADO OCD]

**Strehl or EE required:**

**Astrometric Accuracy:** no special requirement

**SCAO vs. MCAO:** the circumstances of a target within ~10" of a bright star (allowing SCAO) should be considered in order to select SCAO or MCAO mode

**Comparison with JWST or other facilities:** The sensitivity is the key issue here: ELT+MAORY+MICADO is comparable with JWST in this respect.

**Synergies with other facilities:** high-energy satellites available at that time, optical telescopes, ground-based GW detectors are the facilities that will enable to localize the targets.

**Simulations made/needed to verify science case or feasibility:** No simulations are needed in addition to the ones obtained from the ETC tool.

**Origin of the targets:** Lists will be available in the GCN Circulars or ATels.

**NGS:** GRBs Latitude distribution (to be checked)

**Acquisition:** our targets are point sources so 20" FoV will be selected. Finding charts available.

**Calibrations:** Standard. No special astrometric requirement. The presence of photometric calibrators in the field has to be verified. The relative photometry precision derived with the ETC is sufficient to reach the main science goal.

**Data Processing Requirements:** The PSF modelling should be feasible with stars in the FoV.

**Any other comment:** none.

# 7.6 Unveil the formation of Globular Clusters and their contribution to cosmic Reionization

**Authors:** Vanzella E., Calura F., Meneghetti M., Gilli R., Mignoli M., Comastri A. (INAF - OABO), Mercurio A. (INAF - OACN), Rosati P. (University of Ferrara), Castellano M., Fontana A., Giallongo E., Grazian A., Merlin E., Pentericci L. (INAF - OAR), Cristiani S., Cupani G., Nonino M. (INAF - OAT), Tozzi P. (INAF - OAA), Vignali C., Brusa. M. (University of Bologna)

**Brief description of the science case:**

There are two intriguing questions that need to be answered with future cosmological observations: (1) what sources reionized the universe and (2) what formation mechanisms are behind the densest and most ancient stellar systems known today, the Globular Clusters (GC). These two issues may coalesce into a common answer. The general consensus on the first point is that the plethora of faint star-forming systems are the main producers of the ultraviolet non-ionizing background at z>3 (1500 Å), however it is not clear if they dominate also the ionizing field (Lambda<912 Å) and eventually reionized the Universe (e.g., Bouwens et al. 2015, ApJ, 811, 140; Robertson et al. 2015, ApJ, 802,19); the second point is a long-standing problem (e.g., Renzini et al., 2015, MNRAS, 454, 4197). The most popular formation scenarios suggest that GC progenitors (during the formation of the first-generation stars) were more massive than today, up to a factor of 10 or more, i.e., several $10^6$ M$\odot$ (based on the so-called mass budget argument, e.g., D'Ercole et al., 2010, MNRAS, 407, 854; Renzini et al. 2015). In the local universe about 0.1% of the stellar mass in galaxies is contained inside GCs (Harris, Harris & Alessi, 2013 ApJ, 772, 82), on the other hand, if the most popular formation scenarios are correct, they would contribute to the ~50% of the total mass in stars at z>4 (Katz & Ricotti, 2013, MNRAS, 432, 3250; Renzini 2017, MNRAS, 469L, 63). It has already been proposed that forming GCs are promising candidate reionizing sources (Ricotti et al. 2002, MNRAS, 336, 33; Katz & Ricotti 2013; Boylan-Kolchin M., 2017, MNRAS in press, arXiv/1705.01548). With a typical mass of ~(1-2) X $10^6$ M$\odot$ a proto-GC is expected to have a peak UV (1500A) magnitude of M(AB) ~ -17 (-19 for the most massive ones) with an age of ~ 3-5Myr (Schaerer & Charbonnel 2011, MNRAS, 413, 2297). This absolute magnitude suggests that the recent search for extremely faint (M<-13) high redshift star-forming ``galaxies'' at z>3 (e.g. Castellano et al., 2016, ApJ, 823, 40; Bouwens et al. 2017b., ApJ in press, arXiv/160800966; Alavi et al. 2016, ApJ, 832, 56; Vanzella et al. 2017a, MNRAS, 456, 3803), plausibly meets the proto-GCs observational conditions. While this was an anecdotal thinking until recently, new results in strongly lensed fields confirm the detections of forming super-star clusters at z=3-6.5 with resolved effective radii of R$_e$~30-60 parsec (e.g., Vanzella et al. 2016a, ApJ, 821, 27; 2017c, ApJ in press, arXiv/17030.2044; Bouwens et al. 2017a, ApJ in press, arXiv/161000283), some of them representing the best candidate proto-GCs ever discovered (Re ~ 17 pc at z=6.15, Vanzella et al. 2017b, MNRAS, 467, 4304). MAORY+MICADO (**MAORY+** hereafter) will embrace the different modes of star-formation in high-z galaxies on different scales, including sub-clumps, giant HII regions, super-star clusters and the best candidates proto-GCs. The next frontier is therefore the exploration of new low-luminosity and mass regimes, and this requires telescopes 30-40m($\varnothing$) class. The scientific case is therefore twofold:

(1) Access new **luminosity** (-13<M$_{uv}$<-17), **stellar mass** (few $10^5$ M$_\odot$) and **size** (of a few parsec scale) domains at redshift 3-12, and address the nature of super-star clusters, their stellar populations, the feedback processes, and directly/indirectly the LyC leakage (e.g., Vanzella et al. 2016a, 2017b, Vanzella et al., 2016b, ApJ, 825, 41).

(2) Address the still open question of the globular cluster formation by observing directly the **proto-GC phase** (e.g., Vanzella et al. 2017b).

MAORY+ will allow us to perform in the **unlensed fields** what we have started performing in the lensed fields, though in relatively small cosmic volumes (Vanzella et al. 2016b, 2017b). Figure 7.6-1 shows a pair of lensed star-clusters at z=3.222 with $R_e \approx 30$ pc, from HST observations that resemble MAORY+ capabilities. MAORY+ will be crucial to eventually explore the extremely small sizes and detect emission lines in the ultraviolet (z>6) or optical (z<4) rest-frame of star-forming star-clusters of 30-60 pc effective radius at z>3. Spectroscopy will explore the nature of stellar population and the status of the interstellar medium (e.g., opacity to ionizing radiation). The addition of gravitational telescopes (MAORY+LENS) will allow us to spatially resolve details of 1-2 pc scale at z>3 for the first time, and reach deeper flux limits than the unlensed fields. Furthermore, the absorption-line-science for sources fainter than magnitude 30 will be feasible only with

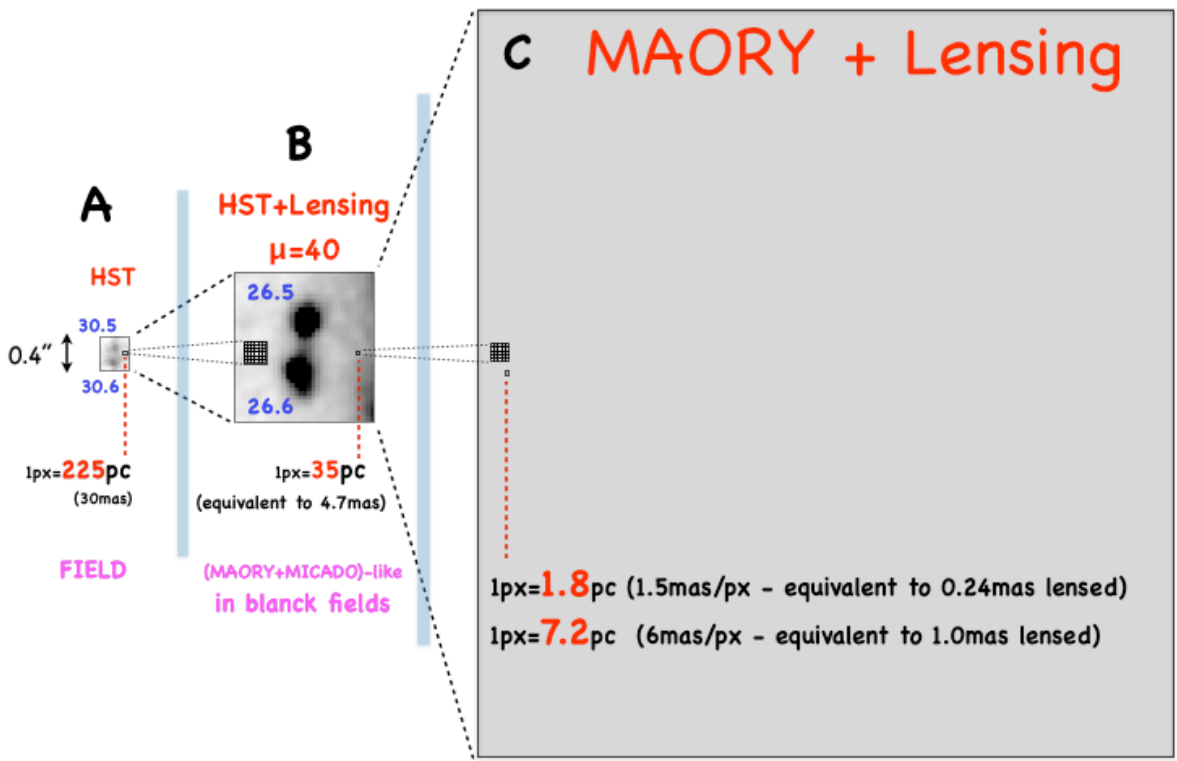

Figure 7.6-1 An example extracted from an observed HST+lensing system (named ID14, Vanzella et al. 2017c) showing the MAORY/MICADO capabilities. Panel A: the HST cutout (0.4'') of a spectroscopically confirmed (z=3.2222) pair of super-star clusters of 30 parsec effective radius each and intrinsic magnitude ~30.5 (26.5 after magnification). The system is barely detected. The single (non-lensed) pixel (30 mas) probes 225 pc. The same pair of clusters strongly lensed (amplified by a factor 40) is shown in panel B. Thanks to the gravitational lensing effect the effective pixel scale in pane B is 4.7 mas, offering a preview of the MAORY+MICADO like capabilities in the unlensed fields. Panel C: the same area (of panel A) properly rescaled adopting the typical MAORY+ pixel scale: MAORY+ and strong lensing will allow us to investigate details at the parsec-scale (2-7pc/px). The size of the single pixel (in pc and mas) is also reported.

ELT+LENS.

**MICADO Pixel Scale / Fov**: *1.5 mas/px and 20 arcsec FoV.* 1.5mas/px is preferred and will allow us to spatially resolve star-clusters of 1-2 pc radius at z~3-10 in lensed fields, and 30-60 pc in the unlensed fields.

**MICADO Observation mode**: Standard Imaging

**MICADO Spectral set-up**: TBD.

**Filters required**: Y, J, H and K-bands are useful. The K-band probes rest-frame 5500/3100/1800A at z=3/6/10; the Y-band will follow the rest-frame UV (lambda>1216A) up to z=6.

**Estimate Survey Area/Sample Size/ Number of Images/Epochs**: Specific targets (dwarf, super star-cluster, proto-GC) will be selected from HST and JWST observations. Typical target size would be ≪5". We already started to collect promising candidate proto-GC at redshift 3-6.6 with magnitudes 28.0-31.5 (Vanzella et al. 2017b) and others will be possibly identified at z>6.5 with JWST. The probability to observe at least one proto-GC associated to a typical z>3 LBG is higher than 20% (Vanzella et al. 2017b), also in line with the calculations reported in Renzini (2017). We therefore expect a large number of targets, the exact number will be better defined in the next future.

**Average Integration time per image (magnitude of targets; S/N required)**:

• IMAGING: There will be two classes of targets:

- **unlensed fields**: in this case we aim to explore sizes of star-clusters down to 30-60 pc

- **lensed fields**: thanks to lensing magnification in addition the spatial resolution offered, we aim to open a completely new regime covering effective radii >1 pc at z>3.

The typical target magnitude in the unlensed field is >29.

The typical target magnitude in the lensed field is >26.

The required S/N necessary to derive a solid morphological modelling is S/N>5. The estimated average integration time per target in the unlensed field is therefore ~5-10 h; the estimated average integration time per target in the lensed fields is 1-2h.

• SPECTROSCOPY: we aim at investigating the physical properties of high-z star-clusters (and/or proto-GC). At z<3.5 the optical rest-frame is accessible from the ground (like Oxygen and Balmer lines) and will provide information about the status of the ISM (ionization or density-bounded), outflows, and dynamical masses. The expected integration time in long-slit mode: at z<3.5 the [OIII]5007 is a prominent line with an expected flux of A×f. Where A is the amplification factor provided by the strong lensing (A=1 means no lensing) and f ~ a few $10^{-19}$ erg/s/cm$^2$ (e.g., see Vanzella et al. 2016a, 2017b,c, where some prototype sources are presented). The factor "A" typically spans the interval 2-30. At z>6 the ultraviolet spectrum (λ<2000 Å) rest-frame is accessible from the ground and therefore high-ionization lines and doublets like CIV1548-1550, HeII1640, OIII]1661-1666 and CIII]1906-1909 with fluxes of the order of A×f~$10^{-19}$ erg/s/cm$^2$). Requested amount of time: TBD.

**Observation requirements**: Small dithers + Large dithers. In the imaging mode, the positioning is not very relevant. In the spectroscopic mode, the centering of the target into the slit is crucial.

**Strehl or EE required**: Probe spatial elements as small as possible. The best spatial resolution performances the most effective is the probe of the stellar mass density in the core of proto-GC at z>3.

**Image Stability Required**: No special requirements.

**Astrometric Accuracy**: No special requirement.

**SCAO vs. MCAO:** While SCAO mode would provide the highest performances in terms of spatial resolution it is very unlikely to find good candidates within 5-10 arcsec of a bright star that can be used as SCAO NGS. Hence MCAO should be the mode of choice in the vast majority of the cases.

**Comparison with JWST or other facilities:** MAORY+MICADO will outperform any other current ground-based telescope and future space facilities, like JWST, in terms of **spatial** resolution, **spectral** resolution and **depth** in the near-infrared bands. JWST will provide optimal targets with full coverage at lambda>2.5um, probing the optical and near-infrared rest-frame out of z=6-12. Sizes as small as a few parsecs (< 10-20 pc) and dynamical masses as low as $10^7$ M determined from rest-frame optical lines (like [OIII]5007 at z<3.5) are attainable **only** with the E-ELT.

**Synergies with other facilities:** HARMONI@ELT will also add the multiplexing mode allowing us to observed multiple targets at the same time. This is relevant since we expect some level of clustering if proto-GC are considered (e.g., Katz & Ricotti 2013). ALMA will possibly offer another perspective of the same forming star clusters, possibly reaching competitive spatial resolution. ELT/HIRES would provide higher spectral resolution follow-up, if required, while  MOSAIC in the MOS mode may help addressing the issue of proto-GC luminosity function.

**Simulations made/needed to verify science case or feasibility:** Figure 7.6-1 shows an example mimicking the MAORY+MICADO performances. A strongly lensed pair of super star-clusters at z=3.222 (of ~30.5 magnitude each) are magnified by a gravitational telescope (MACS J0416) by a factor of ~40 (Vanzella et al. 2017c) The resulting equivalent pixel scale is 4.7 mas. Besides the similarity with lens systems, dedicated simulations in unlensed and lensed fields will be very useful to estimate the level of accuracy in the light profile studies and/or extraction of structural parameters form the real PSF by performing simulations with the, e.g., SimCADO software or by our own tool that includes the gravitational lensing effect, already used in Vanzella et al. (2017b).

**Origin of the targets:** Targets can be already extracted from the lists provided, e.g., by Vanzella et al. (2017b,c); Bouwens (2017a,b). Several other targets will appear in the next 1-2 years based on Hubble Frontier Fields and other lensed fields (like CLASH, Postman et al., 2012, ApJ, 199, 25; and RELICS, Coe et al. in preparation). JWST will produce **candidate** proto-GCs in the unlensed fields (through statistical analysis) but cannot spatially resolve them (if significantly smaller than 30 mas), while MAORY+MICADO will access their real sizes and perform light profiles (at least for the most massive ones). The dense stellar systems not spatially resolved in the lens fields by JWST will be unique targets for ELT spatial resolution capabilities. We will apply on GO JWST time for the search of such proto-GC at z>3.

**NGS:** TBD

**Acquisition:** In the imaging mode, the pointing of a single target is not crucial, being the target much smaller (≤1") than the 20" × 20" FoV. The pointing can be verified with a finding chart.

**Calibrations:** Standard. No special astrometric requirements. Photometry will be supported by HST/JWST data, and the presence of photometric calibrators in the field has to be verified. A relative photometric precision of 0.01-0.05 in the near-infrared bands (Y, J, H) would be sufficient to reach the main science goal.

**Data Processing Requirements:** Pre-reduced stacked images are the natural starting point for the scientific analysis. The knowledge of the PSF is crucial when deriving basic light profiles and sizes. The effective radius of the target at the level of a few mas is one of the main key measures, achievable only with MAORY+. The modeled-PSF will be deconvolved from the image. This will be accessed through simulations (e.g., SimCADO) and/or using real stars or compact objects (AGNs or QSO) in the FoV.

**Any other comments:** none.

## 7.7 Probing the assembly of high redshift early-type galaxies

**Authors**: M. Gullieuszik; R. Falomo; L. Greggio; E. Portaluri; B. Poggianti; A. Moretti (INAF - OAPD)

**Brief description of science case:**

The assembly history of galaxies is still one of the most lively debated topics in the general context of the study of galaxy formation and evolution. HST observations gave evidence of a strong evolution in the mass-size relation of early-type galaxies. High-redshift (z~3) quiescent galaxies have sizes ~5 times smaller than their counterpart in the local Universe (see e.g. Buitrago et al. 2008, ApJ, 687, L61; Cassata et al, 2011, ApJ, 743, 96; van der Wel et al, 2014, ApJ, 788, 28). Minor dry mergers seem to be the most likely mechanisms driving the evolution (Oser et al 2012; ApJ, 744, 63) but also adiabatic expansion could play a role (Fan et al. 2008, ApJ, 689, 101). Poggianti et al (2013) however, pointed out that if high-z galaxies are compared with local galaxies with old stellar populations, the size evolution between redshift ~2 and 0 results to be mild (a factor of ~1.6). A key aspect to take into consideration to address the problem of galaxy mass assembly, is the evolution of the stellar populations in normal and compact galaxies from the early Universe to the present epoch. To this end, it turns out that the study of population gradients in high-redshift galaxies is of paramount importance to characterize the distribution of the stellar content and the star formation history in galaxies (see e.g. Gargiulo et al. 2011, 412, 1804; Guo et al. 2011, ApJ, 735, 18).

The study of size evolution and the scaling relations of high-redshift galaxies (e.g. van der Wel et al. 2014) are limited to the most massive and largest galaxies; the limit is set by the spatial resolution of HST. The mass-size relation at z=2 is to date defined for galaxies with masses $>\sim 10^{10.5} M_\odot$ and $R_e>0.5$ kpc (the spatial resolution of HST is ~100 mas, which corresponds to ~0.8 Kpc at z~2).

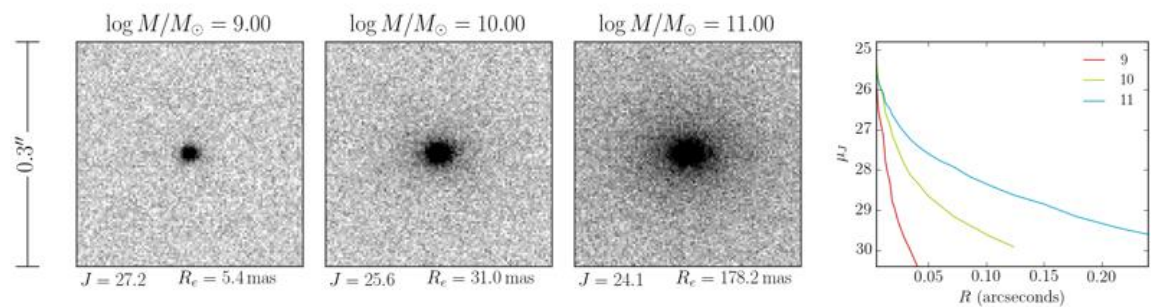

Figure 7.7-1 Simulated MICADO+MAORY J-band observations of early-type galaxies at z=2 of three different masses. The right panel shows the corresponding surface brightness profiles. The exposure time is 3h. Parameters are from Phase A study (see Gullieuszik et al. 2016).

A leap forward in this research field is expected thanks to the unprecedented spatial resolution of E-ELT+MAORY+MICADO. Gullieuszik et al. (2016, A&A, 593, A24) explored the expected performances of E-ELT imaging in the measurements of the structural parameters of high redshift galaxies using AETC (http://aetc.oapd.inaf.it). Simulations carried out adopting the specifications for the Phase A design show that with 3 h exposure time it will be possible to recover the radius of early-type galaxies with $R_e$ of a few milliarcseconds with ~10-20 % accuracy; extrapolating the scaling relations of van der Wel et al. (2014), this is the size expected for galaxies of $M=10^9$ M_sun at z~2-3 (see Gullieuszik et al. 2016). An example is shown in Figure 7.7-1.

These simulations also show that it will be possible to obtain reliable measurements of U-V restframe colour gradients as small as 0.1 mag/dex in the inner regions of early-type galaxies at z=2. Using stellar isochrones, we calculated that, for relatively young stellar populations (≲3 Gyr), a gradient of 0.1 mag/dex corresponds to a variation of ~25% in age or of 0.3 dex in metallicity.

JHK filters are suitable to observe the young stellar populations of high-redshift galaxies as they map the UV rest-frame bands; E-ELT imaging will therefore provide key data to study the mass assembly of early-type galaxies and the mechanisms driving galaxy formation and the evolution of high redshift compact galaxies in a wide range of masses.

**Keywords:** imaging – photometry – high-redshift galaxies

**MICADO Pixel Scale / Fov:** 4mas/px and 50 arcsec FoV

**MICADO Observation mode:** Standard Imaging

**Filters required:** J, H and K, to map UV rest-frame bands at z=2-3

**Estimate Survey Area/Sample Size/ Number of Images/Epochs:**

We estimated the galaxy number density using catalogs from the Hubble Ultra Deep field (Coe et al. 2006). We selected early type galaxies with a Sersic index>2.5 at 1.7<z<2.3, and found ~10 sources per arcmin$^2$ with a $R_e$<5pix (see Figure 7.7-2). We therefore expect a significant number of galaxies with size smaller than HST spatial resolution in the MICADO FoV. A survey area of ~10 arcmin$^2$ will therefore provide a statistically significant number of sources to build reliable mass-size relations in different redshift bins. We do not need multi-epoch observations.

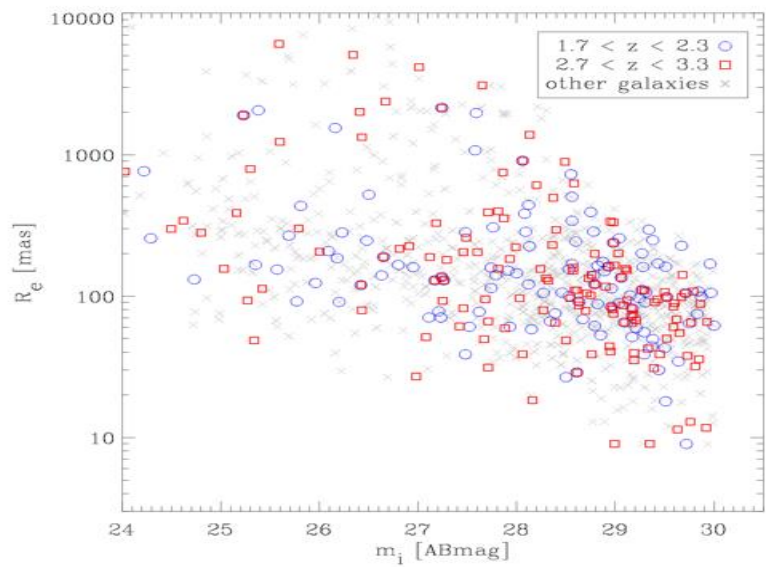

Figure 7.7-2 Size-magnitude relation of early-type galaxies (selected as those with n ≥ 2.5) observed in the Hubble Ultra Deep field (11.97 arcmin^2) as resulted from Coe et al. (2006), based on ACS i'-band photometry. We selected objects having i'<30 mag (Abmag), stellarity < 0.4, and error on Re < 10 px. Red squares are galaxies with 1.7 < z < 2.3, blue circles are galaxies with 2.7 < z < 3.3.

**Average Integration time per image (magnitude of targets; S/N required):** The simulations in Gullieuszik et al. (2016) assumed an integration time of 3 h. This value should be checked using new simulations with updated MICADO+MAORY prescriptions.

**Observation requirements**: Large dithers. No special observation requirements.

**Strehl or EE required:** Modeling galaxy surface density strongly depends on the accuracy of the assumed PSF model, particularly for small galaxies, with size of the order of the size of the PSF core. According to our experience (based on simulations), the accuracy of structural parameters recovery does not depends only on the SR value; the presence of substructures in the PSF faint and/or the shape of the PSF wings can have a significant impact on the profile fitting (Portaluri et al. 2017 MNRAS, 466, 3569).

**Image Stability Required:** Stability is required to retrieve a reliable PSF model, that is a crucial aspect for this science case.

**Astrometric Accuracy:** No special requirements

**SCAO vs. MCAO:** MCAO to take full advantage of the large corrected FoV and obtain stable correction across the whole FoV

**Comparison with JWST or other facilities:** Only the extraordinary spatial resolution of EELT+MICADO+MAORY would allow to probe even the smallest galaxies, that have sizes of the order of (or even smaller than) the spatial resolution of JWST.

**Synergies with other facilities (4MOST/MOONS, LSST/ALMA/HARMONI/METIS, HIRES/MOSAIC), but also VLT or other smaller telescope instruments:** Future deep spectroscopic surveys to be carried out e.g. with MOONS will provide important information on the redshift of possible target galaxies. HARMONI will complement spectral information on the same targets.

**Simulations made/needed to verify science case or feasibility:** The feasibility of this science case has been assessed in Gullieuszik et al. (2016) using simulations carried out adopting MICADO+MAORY Phase A specifications. Simulations with updated specifications and PSFs are required.

**NGS:** Sky coverage should not be a concern.

**Acquisition:** No precise pointing is required.

**Calibrations:** Standard. No special astrometric calibration is required. Absolute photometric calibration with 10% accuracy would match the scientific goals.

**Data Processing Requirements:** no particular requirements

**Any other comments:** In principle there is no preferential fields to be selected as targets for this science case; the choice will be done to optimize the availability of NGS and any available (future) spectroscopic observation.

# 7.8 Structure, stellar populations and properties of stellar halos and intracluster light.

**Authors:** *Spavone M., Iodice E., Napolitano N. R. (INAF - OACN), Cantiello M. (INAF - OATE)*

**Brief description of science case:**

The hierarchical structure formation at all scales is one of the strongest prediction of the ΛCDM model. In this framework, cluster of galaxies are expected to grow over time by accreting smaller groups. The galaxies at the centre of the clusters continue to undergo active mass assembly and, in this process, the gravitational interactions and merging among systems of comparable mass and/or smaller objects play a fundamental role in defining the observed structures. Typical signs of a major merger or strong encounter are star-forming tidal tails (Toomre & Toomre, 1972, ApJ, 178, 623); the accretion of a smaller companion by a pre-existing galaxy forms stellar streams and/or shells (Quinn, 1984, ApJ, 279,596). Due to the relatively long dynamical time with increasing galactocentric distance, the imprints of the mass assembly reside in the galaxy outskirts: these are the regions of the stellar halo. This is an extended and diffuse component made of stars stripped from satellite galaxies, with multiple stellar components and complex kinematics, which is still growing at the present epoch. The stellar halos have an average surface brightness below 29 mag/arcsec$^2$, in the V band, and can host substructures, in the form of shells, streams and tidal tails. On the cluster scale, during the infall of groups of galaxies to form the cluster, the material stripped from the galaxy outskirts builds up the intracluster light (ICL), a diffuse faint stellar component, at which the relics of the interactions among galaxies (stellar streams and tidal tails) also contribute, that grows over time with the mass assembly of the cluster (Mihos et al., 2015, ApJ, 809L, 21).

From the theoretical side, semi-analytic models combined with cosmological N-body simulations have become very sophisticated, with detailed predictions about the structure of stellar halos, the ICL formation and the number of substructures in various kinds of environment (e.g. Cooper et al., 2015, MNRAS, 454, 3185). Recent studies have demonstrated that the overall structure of stellar halos, their stellar populations and the properties of their dynamical substructure directly probe two fundamental aspects of galaxy formation in the ΛCDM model: the hierarchical assembly of massive galaxies and their dark matter halos, and the balance between in situ star formation and accretion of stars through mergers.

The rich environments as groups and clusters of galaxies are therefore the appropriate sites to study the mass assembly processes that lead to the observed galaxy structures, stellar halos and ICL in order to test the hierarchical formation theories at all scales, since one has the chance to catch them "in act". For nearby galaxies (<10 Mpc), the properties of the outer stellar halos were traced by studying the resolved stars out to several kpc from the galaxy centre (Greggio et al., 2014, A&A, 562A, 73). Recently, the first results of the "Panoramic Imaging Survey of Centaurus and Sculpotor (PISCeS)" provided the most extended view of the stellar halo in Centaurus A, down to the surface brightness of $\mu_V \sim 32$ mag/arcsec$^2$ (Crnojevic et al., 2016, ApJ, 823, 19): the spatial density map of stars has revealed new streams, shells and disrupting satellites that trace well the accretion history of this galaxy. Deep images of the Virgo cluster have revealed several faint streams of ICL among the galaxy members and the stellar halos of the bright cluster galaxies (BCGs) are traced at very large distances from the centre (Mihos et al., 2017, ApJ, 834, 16). Groups and clusters of galaxies also usually host dwarf and ultra-compact galaxies, intra-cluster populations of GCs and planetary nebulae. The study and the characterization of such discrete tracers at high redshifts, historically hampered by their faintness, could be fostered by MAORY+MICADO. Stellar halos are still almost unexplored in NIR bands. However, such kind of study could help in better constraining the in-situ component of ETGs, since NIR bands trace the most evolved stellar population. With relatively little ambiguity, the empirically-defined stellar halo can be identified with the bulk of the accreted stellar mass. From an observational perspective, in ETGs the connections between different mechanisms of mass growth and the structural components inferred from images are much less straightforward. For this reason in-situ stars in ETGs will be extremely difficult to distinguish by using only optical photometry.

The large collecting area of ELT, together with the high spatial resolution (6 times better than JWST), will allow to obtain very deep photometry (at least 1 mag better than JWST) and to detect faint substructures in galaxy halos, as well as to resolve other intracluster populations, such as UCDs, GCs, dwarfs and so on.

**The main goal is to study the mass assembly in cluster of galaxies at z~0.1 by mapping the diffuse light component and detecting the discrete tracers in the stellar halos of bright cluster members and in the intracluster regions.**

We will carry out deep and detailed photometric mapping of our ETG sample taking advantage of the high resolution of MAORY+MICADO in the J and K bands. By fitting the light profiles, and comparing the results to simulations of elliptical galaxy assembly, we will be able to identify signatures of the transition between "relaxed" and "unrelaxed" accreted components and to constrain the balance between in situ and accreted stars. Such comparison with the predictions from theoretical simulations will allow us to use the distribution of diffuse light as a robust statistical probe of the hierarchical assembly of massive galaxies (see Spavone et al. 2017 for details).

As starting sample we propose to select 5 objects from the sample used by Gonzalez et al. (2005, ApJ, 618, 195) since, with data acquired with a 1 m telescope, they provided an estimate of the ICL amount for the BCG of each cluster out to about 100 kpc from the center. The much larger collecting area of E-ELT will allow deeper photometry with MICADO, and should give the possibility to detect faint substructures. This may be considered as a "control sample" to test methods and estimates, therefore this proposal can be considered as a pilot project for future investigation on this topic.

At z~0.1, 1 arcsec ~ 2kpc, thus, to reach at least 100 kpc from the center of the dominant BGC, we need 4 pointings with MAORY+MICADO.

**Keywords:** imaging – photometry – stellar halos – LSB structures

**MICADO Pixel Scale / Fov:** 4mas/px and 50 arcsec FoV, motivated by the need to have the larger FoV possible.

**MICADO Observation mode:** Standard Imaging

**MICADO Spectral set-up:** none

**Filters required:** J and K seem the best filters for this science case, since the J − K colour gives a good estimate of the metallicity and is quite insensitive to the presence of a young stellar population.

**Estimate Survey Area/Sample Size/ Number of Images/Epochs:** since to achieve our goal we need 5hrs per field (see below), to cover 4 fields around each BGC, we need an integration time of 20hrs for each cluster and for each filter. Therefore, for 5 clusters in the sample and two filters (J and K), we need in total about 120 hrs.

**Average Integration time per image (magnitude of targets; S/N required):** S/N~9 is reached at J=26 with 5 h per filter. These estimates have been obtained on Mar 28, 2017, with the ESO-ELT ETC, assuming a px scale of 5 mas (the smallest scale currently available in the ETC) and an A0V star as a template.

**Observation requirements:** Small dithers + Large dithers [refer to table 8 of MICADO OCD]

**Strehl or EE required:** SR stability is much important than maximum Strehl.

**Astrometric Accuracy:** no special requirement

**SCAO vs. MCAO:** MCAO is preferred, since we need an homogeneous Strehl correction over a wider FoV.

**Comparison with JWST or other facilities:** JWST lacks the necessary resolution. ELT+MAORY+MICADO will have a factor of 6 better resolution. Moreover, the much larger collecting area of E-ELT allows deeper photometry with MICADO compared to NIRcam+JWST, by more than 1 mag.

**Synergies with other facilities:** Most of targets in the sample have been mapped in the Xray domain.

**Simulations made/needed to verify science case or feasibility:** There is no special need for simulations for this specific science case.

**Origin of the targets:** The targets are part of the list given by Gonzalez et al. (2005).

**NGS:** Targets cover the entire range of galactic longitudes (7°< l < 359°) and an almost wide range of galactic latitudes (-88°< b < -35°), thus the presence of NGS is not a concern.

**Acquisition:** Pointing precision is not critical, since we deal with extended objects. Finding charts will be sufficient. PSF characterization across the field is also required.

**Data Processing Requirements:** The most critical operation in such deep photometric analysis is the estimation and the subtraction of the sky background.

**Any other comment:** none.

# 7.9 Confirming the high-z nature of z>10 galaxy candidates. Studying their structure and shape.

**Authors**: Saracco P. (INAF - OABR), Mannucci F. (INAF - OAA), Ciliegi P. (INAF - OABO)

**Brief description of science case:**

The first galaxies are thought to have formed in the first 300-400 Myr since the Big Bang. Their detection and the analysis of their structure will be fundamental in understanding the processes of the formation of galaxies. Over the last few years, remarkable progress has been made in extending samples back to this time, with 20-30 candidates selected at redshifts z~9-11 with HST 11 (Bouwens et al. 2016, ApJ 803, 34; Ellis et al. 2013, ApJ 763, L7; Oesch, et al., 2013, ApJ 773, 75; Oesch, et al., 2015, ApJ 808, 104; Zitrin, et al. 2014, ApJ 793, L12; Ishigaki, et al. 2015, ApJ 799, 12; McLeod et al. 2015, MNRAS 450, 3032).

The advent of JWST will significantly boost this search both in terms of number of candidates and in terms of number of confirmed high-redshift galaxies. For instance, according to the simulations of Cowley et al. (2017, arXiv:170202146) it is expected one galaxy at z>10 per NIRcam field (~10 arcmin$^2$) at a depth of about H~29. Other sources of galaxy candidates at z>10 will be the wide surveys planned with LSST and Euclid that, even if they will be much shallower than JWST images, will cover much larger areas probing the population of the brightest/most massive galaxies at these redshifts.

MAORY+MICADO would be very effective to probe the high redshift nature of pre-selected candidates. The confirmation of the z>10 nature of these galaxies could be fixed through the detection of the Lyman-break at ~1200Å. Deep J-, H- or K-band imaging (~29-30 mag) combined with intermediate (K1[2.06 μ] and K2[2.22 μ]) filters for source at z>13, will allow us to constrain the amplitude of the break and to determine the slope of the UV-continuum redward of the break, establishing the high redshift nature. The high angular resolution, 6 times better than JWST, and the depth reachable in few hours (H~K~29 in five hours) of exposure, make MAORY+MICADO an excellent instrument to follow-up high-z candidates pre-selected from other fields/surveys (e.g. Euclid, LSST, JWST deep fields).

Most importantly, it is worth to note that it is expected that galaxies at z>10-11 have sizes <0.07-0.06 arcsec (see e.g. the simulations by Cowley et al. 2017), i.e. comparable or smaller than the PSF of JWST in H (~0.064 arcsec, sampled by 2 pixels). Hence, primordial galaxies will not be resolved and their shape and structure cannot be studied with JWST. On the contrary, MCAO (and even more SCAO) observations at MAORY+MICADO, reaching an angular resolution ~0.012 arcsec, would allow us to determine the shape and the structural parameters of primordial galaxies. However, we have to keep in mind that, even in the best case of a SR SR=0.5, the MAORY PSF is such that ~50% of the flux of a point source is distributed over the seeing-limited PSF, that is 0.32 arcsec in K. Hence, a good knowledge of the MCAO PSF during the observations will be fundamental to allow us to derive the intrinsic profile of galaxies (if regular) through the convolution of the PSF with various analytic profiles and multi-component profile models (e.g. with GALFIT). This, in turn, will allow us to derive intrinsic structural parameters such as radius, shape and concentration of the light profile and to determine the regular or irregular/clumpy nature of primordial galaxies on scale of about 70-80 pc.

*This Science Case is part of the MAORY ST initiative on high-redshift galaxies.*

**Keywords:** imaging – high-redshift galaxies

**MICADO Pixel Scale / Fov** – 4mas/px and 50arcsec FoV (or 1.5mas/px and 20arcsec)

**MICADO Observations mode:** Standard imaging

**MICADO Spectral set-up:** None

**Filters required:** according to the aim of the observation: J, H, and K1+K2 are needed to confirm the high redshift nature of the target and can be used to study the structure; K filter is sufficient to study the structure if the high redshift nature is already known.

**Estimate Survey Area/Sample Size/ Number of Images/Epochs** - Follow-up of galaxy candidates at z>10-11 pre-selected from other fields/surveys, e.g. JWST, Euclid, LSST.

**Average Integration time per image (magnitude of targets; S/N required)**: 10 hours to reach a limiting magnitude (S/N>3) H~29 and K1&K2~29 (AB magnitudes) according to the ELT ETC.

**Observation requirements**: Standard dithering. Scale stability recommended.

**Strehl or EE required**: The best angular resolution in MCAO (SR=0.5) resulting in a FWHM~0.012 arcsec. This would make MICADO unique to describe the structure of primordial galaxy on scale of 70-80 pc. The best choice would be SCAO observations if a suited NGS was present.

**Astrometric Accuracy**: no special requirement.

**SCAO vs. MCAO**: MCAO would allow us to observe all the candidates independently of the presence of a bright NGS. SCAO would be preferable to derive accurate structure of primordial galaxies if a suited NGS was present.

**Comparison with JWST or other facilities**: The PSF of NIRcam at K is 0.064", comparable or larger than the expected size of galaxies at z>10-11 (<0.07 arcsec, Cowley et al. 2017). Hence, primordial galaxies will not be resolved by JWST. On the contrary, the expected PSF of MCAO (under best conditions, SR=0.5) will be ~0.012 arcsec. Even better is the resolution expected for the SCAO mode. Hence, contrary to JWST, MAORY+MICADO will allow us to determine the shape and the structural parameters of primordial galaxies.

**Synergies with other facilities**: This program requires the pre-selection of targets from other surveys (e.g. Euclid, LSST, JWST deep fields).

**Simulations needed to verify science case or feasibility**: Simulations would be useful to study the effect of the PSF on the reconstruction of the intrinsic profile of small galaxies (<0.06 arcsec). The MAORY-MICADO PSF is such that for a SR SR=0.5, ~50% of the flux of a point source falls within the diffraction limited PSF (0.012 arcsec) while the remaining ~50% is distributed over the seeing-limited PSF, that is within 0.32 arcsec in K. Hence, detailed and dedicated simulations based on real images of clumpy and irregular galaxies convoluted with the PSF of MAORY will allow us to verify the actual capabilities of MICADO in resolving the structure of these galaxies.

**Origin of the targets**: The targets should come from deep/wide surveys (e.g. Euclid, LSST, JWST deep fields).

**NGS**: Given the nature of the targets, sky coverage should not be a concern in MCAO mode.

**Acquisition:** No precise pointing is required.

**Calibrations:** Standard.

**Data Processing Requirements**: Detailed PSF knowledge would be extremely useful to perform morphological and structural analysis of the sources in the fields.

**Any other comments**: None.

# 7.10 Searching for the first galaxies: LBGs at z>13 (Assessment and feasibility study)

**Authors**: *Saracco P. (INAF - OABR), Mannucci F. (INAF - OAA), Ciliegi P. (INAF - OABO)*

**Brief description of science case:**

The first galaxies are thought to have formed in the first 300-400Myr of the universe. Over the last few years, remarkable progress has been made in extending samples back to this time, with more than ~700 galaxies identified at z>6 with HST (e.g. Bouwens et al. 2015, ApJ 830, 67; Finkelstein et al. 2015, ApJ 810, 71) and 20-30 candidates at redshifts z~9-11 (Bouwens et al. 2016, ApJ 803, 34; Ellis et al. 2013, ApJ 763, L7; Oesch, et al., 2013, ApJ 773, 75; Oesch, et al., 2015, ApJ 808, 104; Zitrin, et al. 2014, ApJ 793, L12; Ishigaki, et al. 2015, ApJ 799, 12; McLeod et al. 2015, MNRAS 450, 3032).

However, the spectroscopic confirmation of such high-z candidates is extremely difficult: the visibility of the Lyman-alpha (Lyα) line (1216 A) during the reionization epoch is expected to be quenched by the neutral gas surrounding star-forming galaxies; this deficit has already been observed at z>7 (Pentericci, et al. 2014, ApJ 793, 113, and references therein). Therefore, the confirmation of galaxies at z>10 would require the detection of the continuum break close to the Lyα line, where the intergalactic medium strongly attenuate the ultraviolet flux. While the spectroscopic detection of this break is challenging even for E-ELT, intermediate band photometry (K1[2.06 μ] and K2[2.22 μ]) combined with deep H-band filter would allow us to select galaxies at z~13-16 (assuming that galaxies exists at these redshifts) using the color-color diagram (H-K1) vs. (K1-K2). Good candidate galaxies at z~13-16 should show the sharp Lyman-break at <1200 A and a blue UV-continuum redward of the break. For a galaxy at redshifts z~13, the Lyman-break falls in the H filter yielding red (H-K1) color. The (K1-K2) color is fundamental to determine the slope of the UV-continuum redward of the break establishing the blue UV color and hence the high redshift nature. The use of the single K-band filter would not allow to determine the slope of the UV continuum producing a large contamination by lower redshift (e.g. dust reddened) interlopers.

For those cases in which the Lyα emission survives (if any) and falls in the K1 or K2 bands then the spatial distribution of the Lyα emission will be captured thanks to the high spatial resolution. A Lyα with a rest-frame equivalent width of 100A will have an observed equivalent width of 1400A and 1600A at redshift 13 and 15, respectively. Therefore, the line emission would dominate the entire K1 or K2 band fluxes, providing info on the spatial distribution of the Lyα emission and hence on the star forming regions within the first galaxies in the universe. The presence of such a line in the K1 band would also facilitate the selection of the object, because it would mimic an even bluer K1-K2 continuum.

The search for candidates at such high redshift is a tricky procedure depending on some factors. From the luminosity function (LF) of galaxies at z>9-10, it is clear that the number of bright UV galaxies seen at H≤27 drops more than a factor of five from z=7-8 to z~10 (e.g. Bouwens et al. 2016). It is expected that their number further decrease at higher redshift because galaxies are expected to be less massive. Hence, to test for the presence of primordial galaxies at z~13-16 it is fundamental to reach fainter H-band magnitudes, i.e. lower stellar masses, than those reached at z~9-10, at least two magnitudes fainter, i.e. H~29.

*This Science Case is part of the MAORY ST initiative on high-redshift galaxies.*

**Keywords:** imaging – high-redshift galaxies

**MICADO Pixel Scale / Fov** - 4mas/px and $50 \times 50$ arcsec$^2$ FoV

**MICADO Observations mode:** Standard imaging

**MICADO Spectral set-up:** None

**Filters required:** H, K1, K2 (13.5<z<15.5)

**Estimate Survey Area/Sample Size/ Number of Images/Epochs** - The number of galaxies expected at 13.5<z<15.5 in a MICADO-MAORY field is given by the relation $N_{exp} = \Phi_{gal} \, \Omega_{MICADO}/4\pi \, V[z_{13.5}, z_{15.5}]$. $\Phi_{gal}$ is the expected volume density of galaxies at that redshift, $\Omega_{MICADO} = 6.6 \times 10^{-8}$ sr is the solid angle subtended by the $50 \times 50$ arcsec$^2$ FoV; $V[z_{13.5}, z_{15.5}] \sim 335 \times 10^9$ Mpc$^3$ is the volume within 13.5<z<15.5, defined by the z at which the Ly break (<1200 A) exits the H filter (~1.75 mu) and the z at which the break reaches the K-band filter. At H~29 we should reach $M_{160} \sim -19.0$ at z~13. From the LF of Bowens et al. (2016; see also Cowley, et al. 2017, arXiv:170202146) the volume density of galaxies with magnitude $M_{160} \sim -19.0$ is <$3 \times 10^{-4}$ Mpc$^3$ at z~10. According to the simulations of Cowley et al. (2017), this number density should decrease by almost an order of magnitude (a factor 7-8) at z~13. However, the actual density is critically dependent on the evolution of the faint-end slope of the LF. The expected volume density of galaxies at M160~-19 (H~29) is in the range $\Phi_{gal}$<5-25x10$^{-5}$ Mpc$^{-3}$. Thus, the expected number density of z>13 galaxies in a MICADO field is $N_{exp}$~0.1-0.5 gal. *A blind search for z>13 galaxies would require at least 10 MICADO fields (~50 hr/filter) to provide a detection or to put a significant upper limit to the number density of z>13 galaxies.*

**Average Integration time per image (magnitude of targets; S/N required):** 5 hours to reach a limiting magnitude (S/N>3) H~29 and K1&K2~29 (AB magnitudes) according to the ELT ETC.

**Observation requirements:** Small dithering. Scale stability recommended.

**Strehl or EE required:** The highest possible to maximize the efficiency of MICADO in detecting primordial galaxies. For a SR SR=0.5, the best case, ~50% of the flux of a point source fall within the diffraction limited PSF, i.e. within 0.012", while the remaining 50% within the seeing limited PSF, 0.32". Given the unknown size and shape of primordial galaxies, it is difficult to quantify how this will affect their detectability (see also Comparison with JWST).

**Astrometric Accuracy:** no special requirement.

**SCAO vs. MCAO:** MCAO is the only viable mode to perform a search for z>13 galaxies, because of the larger FoV and for the Sky Coverage at high Galactic latitudes.

**Comparison with JWST or other facilities:** The FoV of NIRcam at JWST is 9.7 arcmin$^2$ i.e. 10 times the f.o.v. of MICADO (0.8 arcmin$^2$) while their sensitivities at 2.1 μ are similar for point sources. This would make MAORY+MICADO less efficient than JWST in searching for z>10 galaxies. However, the PSF of NIRcam at K is 0.064", ~6 times the FWHM of the diffraction limited component of the PSF expected for MCAO mode (best conditions) and comparable to or larger than the expected size of z>10 galaxies (<0.07-0.06 arcsec; Cowley et al. 2017). Hence, primordial galaxies will not be resolved by JWST. Their (low) surface brightness will be dimmed by JWST because of the larger PSF and this will penalize their detectability. It is difficult to quantify the gain of MAORY+MICADO since it is based on the (unknown) size and shape of primordial galaxies. Simulations would be needed.

**Synergies with other facilities**: A complementary approach would be that to proceed through follow-up of high-z candidates pre-selected from other fields/surveys, e.g. Euclid, LSST, JWST deep fields.

**Simulations needed to verify science case or feasibility**: Simulations could be useful to understand the effect of the PSF on the detectability of very small galaxies (<0.06 arcsec) and to define the best combination of filters/exposure times to optimize the selection.

**Origin of the targets**: The survey does not require preselected targets.

**NGS**: Given the nature of the blind search, sky coverage should not be a concern.

**Acquisition:** No precise pointing is required.

**Calibrations:** Standard.

**Data Processing Requirements**: Detailed PSF knowledge would be extremely useful to perform morphological and structural analysis of the sources in the fields.

**Any other comments**: This is an assessment and feasibility study of a very important scientific case that will dominate the scene of the galaxy formation in the next few years. MAORY+MICADO results to be less efficient in this venture with respect to JWST. On the other hand, it is important to evaluate its contribution in this frontier field of research.

# 7.11 The galaxy central regions and the stellar mass growth of z>1.5 galaxies.

**Authors**: Saracco P. (INAF - OABR), Gargiulo A. (INAF - IASFMI), Mannucci F. (INAF - OAA), Ciliegi P. (INAF - OABO), Zibetti S., Gallazzi A. (INAF - OAA)

**Brief description of science case:** Over the last few years, both models and observations seem to attribute to the galaxy central regions a role in regulating the stellar mass growth of galaxies. Observations have established the existence of relationships between the central stellar mass density $\Sigma_{1kpc}$ (the mass density within a radius of 1 kpc) and other main properties of galaxies and have shown the presence, in these regions, of stellar populations with properties different from those of the stars in the outer regions. The choice of 1 kpc as radius to define the central region is imposed by the limiting angular resolution of HST (FWHM>0.12 arcsec) that, at z>0.7-0.8, cannot resolve linear scales smaller than ~1 kpc.

The central mass density $\Sigma_{1kpc}$ correlates with the stellar mass M* of early-type galaxies (ETGs; Saracco, Gargiulo & Longhetti, 2012, MNRAS, 422, 3107; Bai, et al. 2014, ApJ, 789, 134) and of star forming galaxies (Fang, et al. 2013, ApJ, 776, 63; Tacchella, et al. 2015, Science 348, 314). On the contrary, $\Sigma_{1kpc}$ is anti-correlated to the star formation rate (SFR) of galaxies (Fang, et al. 2013; Mosleh M., et al. 2017, ApJ, 837, 2; Whitaker et al. 2017, arXiv16007310). These relations have a break at certain values of the central mass density. The $\Sigma_{1kpc}$-M* relation changes the slop at a critical density $\Sigma_{1kpc}$~2-3×$10^3$ M pc$^{-2}$ (Saracco et al. 2017, A&A 567, 94), while the SFR seems to drop for densities larger than 5×$10^2$ M$\odot$ pc$^{-2}$ (Whitaker et al. 2017). Accordingly, simulations suggest that quenching occurs in conjunction with a certain value of the central stellar mass density (Tacchella et al. 2016).

Even though there may not be a direct causal connection between the above quantities (e.g. quenching and central density may be the result of some underlying process), the above relations tends to disappear for larger radii, showing that the regions and the mechanisms possibly affecting the galaxy as a whole take place on scales not larger than 1 kpc. Moreover, studies of colour gradients show an excess of UV emission just from the central regions of ETGs at z~1.4 (Gargiulo, Saracco & Longhetti, 2011, MNRAS 412, 1804; Ciocca, et al. 2017, MNRAS 466, 4492), suggesting the presence either of a steady weak star formation or of a He-rich population of stars in their center. Therefore, observations suggest that the central regions of galaxies store information on the earliest phases of galaxy mass growth and host the stellar populations resulting from the early stages of their formation.

Resolving galaxies at scales <<1 Kpc would allow us to establish what is the region responsible of the observed relations and to put constraints on the possible mechanisms occurring in such regions. MAORY-MICADO, reaching an angular resolution ~0.012 arcsec, will probe scales of the order of <100 pc for z>1 galaxies. Complementary imaging and spectroscopic observations would provide the data set required to probe in depth the galaxy central regions.

**Imaging** - Broad-band H or K observations of z>1-1.5 galaxies will trace their inner structure at such high spatial resolution allowing to study the dependence of the above scaling relations on the central regions. For galaxies at z>2.3 (z>3) the two filters J (H) and K will bracket the Balmer break (D4000, $\lambda_{rest}$<4000 A) mainly sensitive to the age of the stellar population. Hence, observations in these two bands will also provide the spatial distribution of relative abundance of younger to older stars inside galaxies. In particular, we will probe whether the galaxy central regions host different stellar populations already at this early epoch. The fitting to the spatially resolved color(s) with stellar population synthesis models will constrain their main properties, age and star formation history. In this regard, simulations based on real images of galaxies as templates convoluted with the MAORY PSF will be useful to estimate the feasibility and the reliability of the measured color variation inside the galaxies. Indeed, the AO PSF is characterized also by an important seeing-limited component (FWHM~0.32 in K in the best conditions) that, for a point source, encloses ~50% of the flux.

**Spectroscopy**: Single slit spectroscopy (0.05" width), at the angular resolution of MICADO, will allow to resolve stellar metallicity and age in the central regions and in the outskirts of the galaxies. Starting from z>1.4, it will be possible to measure the metallicity through Mgb(5173). Going to higher redshift, other features will enter the spectral range of MICADO (Hb(4860A) FeI(4380A), G-band(4300A), D4000, (H+K)CaII(3950A)) allowing us to measure the age and the metallicity of the stellar populations in the center of the galaxies and in the outer regions. This, in turn, will allow us to put constraints on the star formation history that regulate the mass growth of galaxies and to understand whether a causal link exists between the central density of galaxies and their mass growth.

*This Science Case is part of the MAORY ST initiative on high-redshift galaxies.*

**Keywords:** imaging – photometry – spectroscopy - high-redshift galaxies

**MICADO Pixel Scale / Fov**: 4mas/px and 50arcsec FoV

**MICADO Spectral set-up**: 0.05 slit width, IzJ grism for targets at z~2; HK grism, 0.05 slit width, (R=2500) for targets at z~3.

**Filters required**: J and K for galaxies up to z~2.3; H and K for targets at z>2.3

**Estimate Survey Area/Sample Size/ Number of Images/Epochs:**
*Imaging -* Optimal targets would be galaxies belonging to clusters or to overdensities at z>2.3 in order to maximize the number of galaxies with multiband observations at the same rest-frame wavelength. The main targets would be passive galaxies. In the case of a cluster or of an overdensity at z>2, on the basis of the known z~2 clusters and overdensities (e.g. Newman, 2014, ApJ 788, 51; Mei, et al. 2015, ApJ 804, 117), we estimate to have about 10-15 passive member galaxies per arcmin$^2$ to H(AB)<22-23, a more uncertain but larger number of star forming galaxies. Five to eight clusters/overdensities should be observed to collect a sample of about 50-100 passive galaxies (and a larger sample of star forming ones). Observing in two filters, this part of program would require ~32 hours.

**Spectroscopy:** In fact, given the constraints of the MICADO spectroscopic mode, it will be possible to observe one target per exposure. Hence, a subsample of ETGs will be extracted from the above photometric sample. Spectroscopic targets will be selected to cover the whole range of stellar mass density at fixed mass (~5x10$^{10}$ M). We expect to observe a sample of about 10 ETGs at two different redshift intervals, z~2.3 and z~3.3, summing up to 20. This part of program would require ~40 hours.

**Average Integration time per image (magnitude of targets; S/N required):**

**Imaging:** $t_{exp}$=2 h, to reach a surface brightness SB H(AB)=26 mag/arcsec$^2$, S/N=10 (over 1 arcsec$^2$) (ESO-ETC). This SB would allow us to sample the galaxy profile up to at least 2-3 effective radii.

**Spectroscopy**: $t_{exp}$ = 2 h to reach a S/N>10 per resolution element in the core of ETGs.

**Observation requirements**: large dithers (≥ few arcsec) are required for an optimal sky removal (on-off) both for imaging and spectroscopy. Precise positioning is not important. No rotation. Stability of the PSF image is important but not critical.

**Strehl or EE required**: as high as possible (SR=0.5 in K). The drive is to achieve the smallest angular resolution (see also Simulations).

**Image Stability Required**: stability is required to achieve the actual SR SR=0.5 in the final stacked image.

**Astrometric Accuracy**: standard.

**SCAO vs. MCAO**: given the larger AO corrected FoV, MCAO is better suited to carry out this program.

**Comparison with JWST or other facilities**: it is expected that MAORY-MICADO will provide images with an angular resolution ~6 times better than JWST (NIRCam FWHM~0.064" in H) and hence, to sample the inner structure of galaxies at spatial scales 6 times smaller (see also Simulations).

**Synergies with other facilities**: the next generation of surveys (e.g. Euclid, LSST, JWST wide/deep surveys, etc.) will provide a large number of clusters and overdensities at z>2 that may be the targets of this program.

**Simulations made/needed to verify science case or feasibility**: simulations would extremely useful to assess the effect of the AO PSF on the galaxy profiles (see however Gullieuszik, et al. 2016, A&A 593, 24) and, most importantly, on the reliability of spatially resolved color measurements. Indeed, in the best case (SR=0.5), ~50% of the flux of a point source fall within the diffraction limited PSF of MAORY (FWHM~0.012 arcsec), while the remaining fraction is distributed within the seeing-limited PSF (FWHM~0.32 arcsec in K). Hence, simulations based on real galaxy images taken at different wavelengths as templates would be necessary to define how this affects spatially resolved color measurements.

**Origin of the targets**: targets will be clusters and overdensities already known or that will be discovered at z>2 especially from the next large surveys.

**NGS**: sky coverage should not be a concern as targets with suitable NGS asterisms can be selected from large available sets.

**Acquisition:** No precise pointing is required.

**Calibrations:** Standard, no special requirement.

**Data Processing Requirements:** detailed PSF knowledge.

**Any other comments**: Possible partial overlaps and complementarity with contributions of Zibetti et al. and of Gullieuszik et al. this white book.

# 7.12 From Dual to Binary SMBH

**Authors:** *Severgnini P., Della Ceca R., Saracco P., Ballo L., Cicone C. (INAF - OABR), Ciliegi P. (INAF - OABO), Landoni M. (INAF - OABR), Mapelli M. (INAF - OAPD)*

**Brief Description Of Science Case:**

The search for and the characterization of dual (kpc scale separation) and binary (pc separation) active supermassive black hole (SMBH) population is a hot topic of current astrophysics, given its relevance to understand galaxy formation and evolution. Since it is now clear that the most massive galaxies should harbour central SMBHs (Kormendy & Richstone 1995, A&A 33, 581; Ferrarese & Ford 2005, SSRv 116, 523) the formation of these dual/binary systems is the inevitable consequence of frequent galaxy mergers. The dynamical evolution of binary SMBH systems within the merged galaxy, and their interaction with the host (via both dynamical encounters and feedback during baryonic accretion onto one or both SMBHs) encode crucial information about the assembly of galaxy bulges and SMBHs. Moreover, if the binary SMBHs eventually coalesce, they will emit gravitational waves, hence these sources could be detected with incoming low frequency gravitational wave experiments. The different stages of gravitational interaction can be schematised in the following way: galaxy pairs (tens of kpc separation), dual SMBHs (kpc scale separation), binary SMBHs (pc separation) and the final collapse with gravitational wave emission (<1 pc separation). Unfortunately, observing directly the SMBH during different merger stages is still a challenging task due to not only the stringent resolution requirement but even to the same difficulty to identify SMBHs. In these merger states, SMBHs are actually expected to be embedded in a large amount of dust and gas and thus strongly obscured and elusive both in the UV and in the optical bands. While thousands of galaxy pairs are known and dozens of dual SMBHs at >1 kpc separation have been found, only a few definitive sub-kpc dual SMBHs have been discovered and studied so far (e.g. Rodriguez et al. 2006, ApJ 646, 49; Boroson & Lauer 2009, *Nat.* 458, 53; Valtonen et al. 2008, *Nat.* 452, 851). Thanks to the unprecedented spatial resolution and high sensitivity offered by MAORY+MICADO, it will be possible to detect and confirm systems of two SMBHs down to sub-kpc scale separations up to z~3, i.e. the redshift beyond which a very low SMBH coalescence rates is actually expected due to the minimum timescale predicted for the overall merger process (of the order of 2 Gyr, Khan et al. 2016, ApJ 828, 73). In particular, it will be possible to detect multiple SMBHs down to few dozen pc scale separation for z<1 and down to few hundred pc scale separation for z<3. By exploiting the MAORY+MICADO imaging capability it will be possible to provide for the first time an overall picture of the still lacking transition phase between dual and binary SMBHs. This will allow us not only to investigate some of the most relevant open astrophysical issues, such as the role of galaxy mergers as drivers of galaxy evolution and their relevance for SMBH growth and AGN triggering, but also to directly observe a class of sources that are expected to be extragalactic emitters of gravitational waves.

**Keywords:** imaging – photometry – extragalactic SMBH pairs

**Micado Pixel Scale / FoV:** 1.5 mas/px and 20 arcsec FoV motivated by the small angular separations that we want to probe.

**Micado Observation mode:** Standard Imaging.

**Micado Spectral set-up**: None. For the vast majority of the sub-kpc SMBH pair candidates, medium resolution optical/NIR spectra are already available or will be available in the next few years.

**Filters required**: We require the K band filter in order to minimize the effects of the intrinsic absorption on the SMBH detection.

**Estimate Survey Area/Sample Size/ Number of Images/Epochs:** The sample will be composed of sup-kpc dual SMBH *candidates* and, as quoted also in the "Origin of the targets", it will be defined over the next few years, in the light also of new results expected from JWST and LSST. The rate of real double sources among sup-kpc dual SMBH *candidates* is still an open issue. By considering only the double-peaked AGNs (see below "Oring of the targets), the real number of double SMBH could vary from 10% to 50% of the samples studied so far (e.g. Shen et al. 2011, ApJ, 735, 48, Fu et al. 2012, ApJ, 745, 67). Setting this rate on solid statistical basis is one of the aims of this project. To this end we need a large (hundreds of sources) and well defined sample of sup-kpc dual SMBH candidates. As a starting point of this project, we propose to observe the K-band bright end (K<22mag) of the most promising sub-kpc dual SMBH candidates that will be available at the time of observations (we note that at the time of writing hundreds of candidates are already available, see "Origin of the target", and they will be obviously increase with the advent of the next generation of instruments, e. g. JWST and LSST).

Taking into account the capabilities of ELT+MAORY+MICADO, assuming an exposure time of about 10 min for the faintest targets (S/N>50, K~22 mag, see "Origin of the target"), it will be possible to observe a sample of about 200 candidates in ~30h (including overheads). However, this sample size will be reconsidered after the advent of JWST, by focusing on those targets challenging even for the resolution of JWST.

**Average Integration time per image (magnitude of targets; S/N Required):** We require an average integration time per image ≤10 min to reach a S/N≥50.

**Observation requirements:** Dithering patterns are required.

**Strehl or EE required:** The project needs the best spatial resolution performances on each target in order to be able to separate point sources at few pc (z<1) or few hundreds of pc (z>1), i.e. the best angular resolution in MCAO (SR=0.5) resulting in a FWHM~0.012 arcsec (see also "Simulations").

**Astrometric Accuracy**: No special requirement.

**SCAO VS. MCAO:** We will use the MCAO as standard observing mode. SCAO mode will be used in the few cases in which there will be a bright star within 10 arcsec from the target.

**Comparison with JWST or other facilities:**

- *VLBI:* Nowadays, the technique of very long baseline interferometry (VLBI) provides the highest achievable angular resolution thus representing a powerful tool to confirm sub-kpc binary AGNs only if both SMBHs are radio-loud (about 10% of the AGN population). ELT+MAORY+MICADO will complement VLBI interferometry towards the radio quiet SMBHs, i.e. the remaining 90% of the overall AGN population. Moreover, we note that the presence of a double compact VLBI source could be also explained by the presence of a powerful jet or by star-formation knots, hence not necessarily implying a double AGN.

- *JWST:* In order to fill the gap between dual and binary SMBHs, achieving a very high

spatial resolution is a key issue: ELT+MAORY+MICADO will allow to reach a spatial resolution about a factor of 6 higher then JWST.

**Synergies with other facilities:**

- ***ALMA***: ALMA can be used to resolve and image cold dusty structures and molecular clouds in the immediate vicinity of the coalescing SMBHs and inform us about their extent and morphology. Furthermore, spectroscopic studies with ALMA can shed light onto the kinematics of such molecular gas and so help identify possible feeding (e.g. inflow of gas towards the SMBHs) and feedback (e.g. outflows of gas departing from the SMBHs) mechanisms associated with the final stages of the merger.

- ***LSST***: Thanks to its large sky coverage, high sensitivity, broad wavelength coverage (up to 1μm) and the valuable temporal information, LSST will be able to detect million of optically-selected AGN. Each region of the LSST sky will be pointed different times in each band (about 200 visits) by allowing to explore variability and outburst events on time-scales from minutes to a decade and thus providing new binary SMBH candidates to be followed up with MICADO+MAORY (see "Origin of the targets - *Light curve variability*").

- ***ELT+HIRES***: The sample of binary SMBH candidates that will be discovered by MICADO+MAORY, will be followed-up with the second generation high-resolution spectrograph HIRES (adaptive optics required) on the E-ELT. The high resolution and simultaneous wide wavelength coverage (4000Å to 2.4μm) of HIRES will allow us to constrain with high signal-to-noise (S/N>10) the SMBH kinematics, the host star formation rate, gas kinematics (velocity dispersion, inflows, outflows, line profiles and asymmetries) and thus the stellar and AGN feedback effects.

***Simulations made/needed to verify science case or feasibility***:

The MAORY-MICADO PSF is such that for a SR SR=0.5, ~50% of the flux of a point source falls within the diffraction limited PSF (0.012 arcsec) while the remaining ~50% is distributed over the seeing-limited PSF, that is within 0.32 arcsec in K. Hence, detailed and dedicated simulations of double point sources will be useful to asses the minimum SR needed to separate them, as a function of their angular separation.

**Origin of the targets:**

The starting point of this project will be to observe with MAORY+MICADO the K-band bright end (K<22mag) of the most promising sub-kpc dual SMBH candidates selected from literature (see below). This sample will be defined over the next few years, in the light also of new results expected from JWST and LSST.

The K-band limiting magnitude chosen will allow us to investigate QSOs with bolometric luminosities in the range of $10^{42}$-$10^{46}$ erg/sec, i.e. the bulk of the QSO population, up to high redshift.

MAORY+MICADO will represent the decisive step to fill the gap between dual and binary SMBH. The targets will be selected through the following techniques:

- ***Double-peaked emission line AGNs***: Searching for double-peaked narrow optical emission lines emerging from the two separate narrow-line regions (NLRs) was proposed as a method to select dual AGNs (e.g. Wang et al., 2009, ApJ 705, 76). Thanks to the already available large spectroscopic databases, double-peaked emission line AGN samples have been indeed recently built as possible candidates of sub-kpc scale SMBH systems. For example the Sloan Digital Sky Survey (SDSS) has already provided different samples of spectroscopically selected double-peaked AGN at different redshift (Zhou et al. 2004, ApJ 604, L33; Xu & Komossa 2009, ApJ 705,

L20; Liu 2016, A&A 592, L4; Smith et al. 2010, ApJ 716, 866; Comerford et al. 2012, ApJ 753, 42; Fu et al. 2012, ApJ 745, 67; Ge et al. 2012, ApJSS 201, 31; Barrows et al. 2013, ApJ 769, 95; Benítez et al. 2013, ApJ 763, 36). However, apart for the presence of dual/binary SMBHs, double-peaked spectral lines can also be due to other effects occurring in a single AGN, e.g. peculiar kinematics or jet–cloud interactions (Heckman et al., 1984, ApJ 281, 525, Gabányi et al. 2017 [arXiv:1701.04572]), a rotating, disk-like NLR (Xu & Komossa, 2009, ApJ 705, L20), and the combination of a blobby NLR and extinction effects (Crenshaw et al., 2010, ApJ 708, 419). Therefore, high-resolution images are essential to confirm true SMBH pairs.

- *Light curve variability*: Another proposed signature of sub-kpc SMBH is periodic modulation of the luminosity in AGNs or quasars (e.g. in optical, UV), induced by the orbital motion of the binary system and assuming that the accretion is modulated on time-scales of the order of the orbital period of the SMBHs. Thus far, these searches have identified candidates with periods of order of several years or less (e.g., Graham et al. 2015, MNRAS 453, 1562; Graham et al. 2015, *Nat.* 518, 74; Liu 2015, ApJ 803 L16; Charisi et al. 2016, MNRAS 463, 2145). A significant uncertainty in this approach arises from the fact that regular quasar light curves exhibit red noise (e.g., MacLeod et al. 2010, ApJ 721, 1014 and references therein), blurring the distinction between a periodic signal and normal quasar variability when only a few oscillations of the light curve are observed (see discussion by Vaughan et al. 2016, MNRAS 461, 3145). High-resolution images are essential in removing these uncertainties.

- *Galaxies with distorted morphology:* As already discussed above the formation of dual/binary SMBH systems is the inevitable consequence of frequent galaxy mergers. Thus a natural place where looking for them are galaxies with disturbed morphologies and tidal features.

**NGS**: Our sample will be built up in the next few years also taking into account the new data/results from JWST and/or LSST. The availability of Natural Guide Stars will be tested after the sample definition.

**Acquisition:** We will observe SMBH pairs separated by less than 10", thus obtaining an adequate pointing should not be a problem. Moreover, for each target, finding charts are available at different wavelengths.

**Calibrations:** Standard

**Data Processing Requirements:** Detailed PSF knowledge both for SCAO and MCAO modes.

**Any Other Comments:** *None*

# 7.13 Formation and evolution of galactic nuclei and black holes.

**Authors:** *Falomo R., Paiano S. (INAF - OAPD), Uslenghi M. (INAF - IASFMI), Fantinel D. (INAF - OAPD)*

**Brief description of science case:**
The relationship between the central black hole and the properties of their host galaxies is a fundamental ingredient for a comprehensive interpretation of the structure and evolution of the galaxies (e.g. Heckman and Best, 2014, ARAA). Of particular relevance is the understanding of how this relationship evolves over the cosmic time as it can offer important clues for the study of the formation processes of galaxies and their central massive BHs.

To investigate this issue one need to measure the black hole mass and the properties (luminosity, size, morphology, stellar population, ...) of a suitable sample of galaxies at different redshift. This requires exceptional observing capabilities in terms of sensitivity and spatial resolution. Unfortunately, at high redshift the sphere of influence of massive BH cannot be probed for inactive galaxies with any present and planned future instrumentation because of the extremely small angular size, However, this is possible for active galaxies using dynamical information from the regions emitting gas closest to the central source (broad line regions; e.g. Peterson 2014, JPh Conf.S. 372).

In order to explore this link it is needed to measure both the mass of the central BH and the properties of its host galaxy over a significant cosmic time. This can be done using spatially resolved spectroscopy of quasars and their host galaxies. The BH mass can be derived from the width of the broad emission lines and the continuum luminosity of the nucleus (using virial method from single epoch spectra or more accurately getting the size of BLR from reverberation mapping technique). At $z > 2$ quasar host galaxies are angularly very small (0.2"-0.3" half light radius) therefore it is extremely important to image the sources with a very narrow PSF (to reduce the emission from the nucleus) and high sensitivity to properly detect the faint extended emission from the host galaxies (see e.g. Falomo et al 2008, ApJ, 673, 694). These conditions are well matched by the future ELT camera (MAORY+MICADO) that thanks to the combination of high sensitivity and superb spatial resolution, when combined with MCAO corrections, is able to derive quasar host properties with high accuracy.

**Keywords:** imaging – photometry – active galactic nuclei — quasars – black holes
**MICADO Pixel Scale / FoV:** Depending on the effective shape of the PSF 1.5 mas px-scale could help to characterize the PSF and thus the image decomposition (nucleus + host).
**MICADO Observation mode:** Standard Imaging.
**MICADO Spectral set-up:** none (but NIR spectroscopy of targets can complement the study).
**Filters required:** J, H, Ks (optimal combination of bands and PSF performances).
**Estimate Survey Area/Sample Size/ Number of Images/Epochs:** There are no problems to choose suitable QSO in the sky. In order to map a significant range of BH masses and redshift about 50 objects are considered reasonable for this program.
**Average Integration time per image (magnitude of targets; S/N required):** The evaluation of the integration time required for this program need detailed simulations (see below).

**Observation requirements:** Given the small size of the objects small (few arcsec) dithers is adequate for sky subtraction during the on-target exposures.

**Strehl or EE required:** A high SR is required to reduce the flux of the bright nucleus in the innermost region and thus improve the contrast between the flux of the host galaxy and the wing of the PSF (nucleus).

**Astrometric Accuracy:** no special requirements for astrometry

**SCAO vs. MCAO:** for the brightest target SCAO could be a favorite option. However, the right choice between SCAO and MCAO remains on the accuracy to characterize the PSF in the whole spatial range of interest for the target. For the fainter targets, it could be possible to choose QSO that are relatively close to bright stars in order to determine the PSF from the same exposure on the QSO.

**Comparison with JWST or other facilities:** Spatial resolution combined with high sensitivity is key issue for the program. JWST will not be able to characterize the properties of the QSO host galaxies at very high redshift and in all cases where the nucleus-to-host ratio is so high

**Synergies with other facilities:** Observations at other frequencies of the chosen targets can complement the information on the properties of the active galaxies. In particular, those that can be obtained with comparable spatial resolution (e.g. ALMA, SKA).

**Simulations made/needed to verify science case or feasibility:** As an example, we show in Figure 7.13-1 simulations performed adopting the PSF provided by MAORY during the phase A study (assuming atmospheric seeing of 0.6" FWHM). The simulation includes sky and environment background, statistical and read out noise, and appropriate convolution of the galaxy models with the above PSF. We show here a representative example of a QSO at z = 2 with H = 20 and host galaxies (n = 2.5) 1.5 mag fainter.

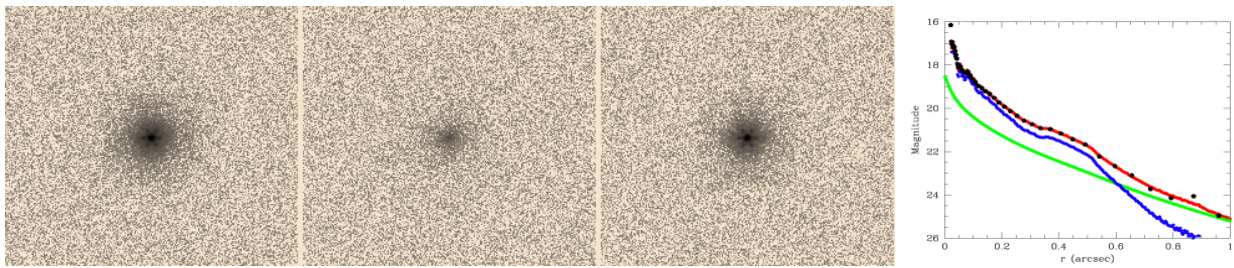

Figure 7.13-1 *Left*: Example of the simulated image (H band; $t_{exp}$=2) of a QSO at z =2. From left to right the images depicts: QSO=nucleus+host, only host, and only nucleus. The QSO has MR =-24 and its host galaxy has MR =-22.5, $R_e$ = 0.3" ~ 2 kpc, and Sersic index n=2.5.
*Right*: The azimuth-averaged radial brightness profile of the QSO (filled squares) compared with the fit (red line) of the two components model: point source (blue) and host galaxy (green) as derived from GALFIT.

**Origin of the targets:** QSO catalogs

**NGS:** No particular problems to find targets with close NGS / NGS asterisms

**Acquisition:** No relevant issues.

**Calibrations:** Standard calibration is sufficient.

**Data Processing Requirements:** The most critical issue for this study is the possibility to determine and characterize the PSF.

**Any other comment:** none.

# 7.14 The star formation history of galaxies at the peak of their compactness

**Authors**: Zibetti S., Gallazzi A., Mannucci F. (INAF - OAA), Saracco P., Gargiulo (INAF - OABR) A., Gullieuszik M. (INAF - OAPD)

**Brief description of science case**:

In the last decade, a number of studies on the size and compactness distributions of galaxies as a function of their mass and star formation activity at different cosmic epochs have converged to suggest a scenario for the evolution of massive galaxies. After an early period (z>2.5-3) of "regular" growth and star formation activity, a *fast compaction event before (or associated to) quenching* must have occurred in massive galaxies, which at z~2 appear to concentrate most of their stellar mass in cores of size <1kpc, reaching average mass densities about one order of magnitude larger than in galaxies of similar mass today. Two key questions arise: 1) which physical mechanism(s) is (are) responsible for the *compaction*? 2) what causes the *subsequent growth in size* (and decrease in stellar density) of the descendant population? In order to answer these questions, we need, first of all, to *characterize* the properties of the *galaxies at the peak of their compactness*. In particular, we require to characterize their *structural parameters* (type of surface brightness/mass density profile, concentration, effective radius etc.) and their *stellar content in terms of star formation history and chemical enrichment*. This will allow us to understand how quick and violent was the compaction mechanism, e.g. if it takes place all at once, possibly through a violent starburst and a "submm-galaxy" phase (e.g. by Toft et al. 2014) then imprinting the typical post-starburst spectral morphology, or through a more prolonged and continuous process, which would result in a more continuous star formation history and a different chemical enrichment pattern. The *comparison between compact star-forming and compact quiescent galaxies* will allow us to understand if the former are indeed consistent with being the progenitors of the latter through simple quenching. The comparison with the lower redshift counterparts will allow us to understand how much of the subsequent evolution occurs through gas-rich processes (associated to additional star formation and causing an alteration of the z~2 stellar population properties of the compact core) or through "dry" mergers/accretions (which substantially alter only the structure of the galaxy, with no significant change in the star formation history and chemical enrichment).

These tasks can only be fulfilled by *deep imaging and spectroscopy at high resolution in the rest-frame optical/NIR range*. Given the typical size of (the cores of) compact massive galaxies at z≥2, in order to resolve them a resolution of <0.01" (<~80pc) is required. Such an exquisite resolution is not attainable even with the JWST, while it is well within the reach of MAORY+MICADO. We propose to study in great detail a relatively small sample of a few tens of compact galaxies, both quiescent and star forming, at z=2 and at 2.5<z<3, in order to test the possible connection between the two classes and the temporal evolution of their population in that crucial time span. *Deep, high-resolution multi-band imaging* will provide the *structural* characterization and a rough indication of the possible *stellar population variations* inside these galaxies. The *long-slit spectroscopy* offered by MICADO will allow to *fully characterize the stellar population properties and the kinematics of the resolved compact core* and to identify the origin (age/metallicity/dust) of possible colour gradients in the outer parts, thus to understand the *build-up mechanisms* in these galaxies.

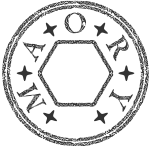

**MICADO Pixel Scale / Fov**: 4 mas/pix and 50" FoV. Although the high-resolution configuration (1.5mas/px and 20" FoV) would be even better for our study, we do not expect any bright NGS close enough to our targets to use SCAO and take advantage of the smallest pixel scale. We thus prefer to use the low-res setup and patrol a more extended environment neighbouring the targets (r~200 kpc) to identify possible signs of interactions. This offers also a synergy with the science cases proposed by Gullieuszik and Saracco.

**MICADO Spectral set-up**: JHK coverage is required for the z~2 subsample, where it allows to map from approximately the D4000n break (relevant for age determinations) to ~8000 Å, thus including the Balmer lines and several Fe and Mg features (relevant for metal enrichment estimates). In addition, we will map several emission lines (if present, e.g. in the blue compact galaxies), up to the H$\alpha$+[NII] complex and the [SII] doublet, which will allow us to perform a full characterization of the ionized interstellar medium and an accurate kinematic/dynamic study of these galaxies and of the associated gas flows. For the z>2.5 subsample, we shift the coverage to the H and K windows only, in order to provide similar rest-frame coverage. Ideal resolution is R~4000, which provides an optimal trade-off in terms of sky-line subtraction and SNR. This kind of resolution is more than sufficient for the spectral population modelling as it corresponds to velocity dispersions of ~30 km/s, while compact galaxies typically have velocity dispersions of several 100 km/s. We would need a slit width 30 mas to sample the light of the compact cores (r<~150pc) with the maximum efficiency.

**Filters required**: JHK to cover the UV-optical restframe dominated by stellar emission

**Estimate Survey Area/Sample Size/ Number of Images/Epochs**: Compact galaxies are quite rare and we estimate ~0.005 (~0.003) compact quiescent (star-forming) galaxies / arcmin$^2$ per 0.1 redshift interval at z~2. Thus, we will only be able to observe one target per field, plus all neighbouring galaxies in a radius of ~200 kpc. We propose to observe a minimum of 10 quiescent and 10 star-forming compact galaxies in the two redshift ranges, hence 40 galaxies in total.

***Average Integration time***: (cores of) typical compact galaxies of Mstar of a few $10^{10}$ M$_\odot$ at z~2 are H_AB<~22 mag (~23 mag) at z~2 (z~3).

*Imaging:* In order to map these galaxies beyond the brightest core, we aim at reaching SB in H(AB) ~25-26 mag arcsec$^{-2}$ with SNR>10 over 1 arcsec$^2$. This requires approximately a couple of hours per filter even with the E-ELT (see also cases by Saracco, Gullieuszik)

*Spectroscopy:* With two hours integration, we will get SNR per resolution element >40 (>20) in H band, out of the OH skylines, for z~3 (z~3) galaxies, sufficient for stellar population analysis. [Note that the z~3 sample might be actually brighter than assumed because of the younger ages, so we might be able to reach the same SNR as at z~2]

**Observation requirements:** *Imaging*: dither patterns of ~5 arcseconds required to avoid superpositions of the possible extended profiles (out to 20 kpc). *Spectroscopy*: dithering along the slit with ~5 arcsecond offsets. Important to precisely center the source in the slit, but should not be difficult given the compactness of the sources.

**Strehl or EE required:** As low as possible (S=0.6 in K). The drive is to achieve an angular resolution <~ 0.02 arcsec (70% of the flux within 0.02 arcsec) to resolve the compact cores.

**Image Stability Required**: stability is required to reach the requested SR in the combined images, given the long total exposure.

**Astrometric Accuracy**: No special requirements

**SCAO vs. MCAO**: MCAO, impossible to find bright NGS for SCAO in deep cosmological fields

**Comparison with JWST or other facilities**: Only the extraordinary spatial resolution of EELT+MICADO+MAORY would allow to probe the smallest galaxies, that have sizes of the order of (or even smaller than) the spatial resolution of JWST.

**Synergies with other facilities (4MOST/MOONS, LSST/ALMA/HARMONI/METIS, HIRES/MOSAIC), but also VLT or other smaller telescope instruments**: Spectroscopic confirmation of the targets and environmental characterization of the surroundings may be done in synergy with MOONS. The most interesting targets may be followed up with the IFU Harmoni, in order to obtain a complete spatial mapping of the optical/NIR SED of these galaxies looking for signs of interactions, accretions, disturbances etc. Gas and dust content characterization with ALMA desirable.

**Simulations made/needed to verify science case or feasibility**: Some simulations already available from Gullieszik et al. (2016). More could be useful for the spectroscopic part.

**Origin of the targets**: K-band selection from deep NIR imaging catalogs (e.g. UKIDSS-UDS) with accurate photometric or spectroscopic redshift, in the future also from EUCLID or LSST.

**NGS**: sky coverage should not be a concern.

**Acquisition**: Centering in slit is crucial for spectroscopy, targets should be bright enough to check the acquisition with a through-slit image

**Calibrations**: Standard. No special astrometric calibration is required. Absolute photometric calibration with 5% accuracy would match the scientific goals.

**Data Processing Requirements**: detailed PSF knowledge for 2D image decomposition.

**Any other comments**: the imaging part of this case overlaps with those of Gullieuszik and Saracco.

**\*\*\* End of document \*\*\***